\documentclass[12pt]{article}
\usepackage{amsmath,amsfonts,feynmp,epsf}
\usepackage{amssymb}
\usepackage{graphicx}
\usepackage{grffile}
\input epsf
\usepackage[nosort]{cite}
\usepackage[english]{babel}
\usepackage{fullpage}

\makeatletter\@addtoreset{equation}{section}\makeatother

\def\bra#1{\mathinner{\langle{#1}|}}
\def\ket#1{\mathinner{|{#1}\rangle}}
\def\braket#1{\mathinner{\langle{#1}\rangle}}

{\catcode`\|=\active
  \gdef\Braket#1{\left<\mathcode`\|"8000\let|\BraVert {#1}\right>}}
\def\BraVert{\egroup\,\mid@vertical\,\bgroup}
{\catcode`\|=\active
  \gdef\set#1{\mathinner{\lbrace\,{\mathcode`\|"8000\let|\midvert #1}\,\rbrace}}
  \gdef\Set#1{\left\{\:{\mathcode`\|"8000\let|\SetVert #1}\:\right\}}}
\def\midvert{\egroup\mid\bgroup}
\def\SetVert{\egroup\;\mid@vertical\;\bgroup}

\def\be{\begin{equation}}
\def\ee{\end{equation}}
\def\bea{\begin{eqnarray}}
\def\eea{\end{eqnarray}}
\def\ie{\begin{equation}\begin{aligned}}
\def\fe{\end{aligned}\end{equation}}

\makeatletter\@addtoreset{equation}{section}\makeatother

\renewcommand{\title}[1]{\vbox{\center\LARGE{#1}}\vspace{5mm}}
\renewcommand{\author}[1]{\vbox{\center#1}\vspace{5mm}}
\newcommand{\address}[1]{\vbox{\center\em#1}}
\newcommand{\email}[1]{\vbox{\center\tt#1}\vspace{5mm}}

\newcommand{\half}{\frac{1}{2}}
\newcommand{\Tr}{\text{Tr}}

\newcommand{\rD}{\overrightarrow{D}}
\newcommand{\lD}{\overleftarrow{D}}
\newcommand{\lrD}{\overleftrightarrow{D}}

\newcommand{\A}{{\alpha}}
\newcommand{\B}{{\beta}}
\newcommand{\C}{{\gamma}}
\newcommand{\D}{{\delta}}

\newcommand{\vo}{{\nu_1}}
\newcommand{\vt}{{\nu_2}}
\newcommand{\tr}{\text{tr}}

\begin{document}
\begin{titlepage}
\begin{center}
\vskip 1cm

\title{Chern-Simons Theory with Vector Fermion Matter}

\author{Simone Giombi$^{1,a}$, Shiraz Minwalla$^{2,b}$, Shiroman Prakash$^{2,c}$, \\
Sandip P. Trivedi$^{2,d}$, Spenta R. Wadia$^{2,3,e}$  and
Xi Yin$^{4,f}$}

\address{${}^1$Perimeter Institute for Theoretical Physics, Waterloo, Ontario, N2L 2Y5, Canada}
\address{${}^2$Dept. of Theoretical Physics, Tata Institute of Fundamental Research, \\Homi Bhabha Rd,
Mumbai 400005, India.}
\address{${}^3$International Centre for Theoretical Sciences, Tata Institute of Fundamental Research, Homi Bhabha Rd,
Mumbai 400005 and\\ TIFR Centre Building, Indian
Institute of Science, Bangalore 560012, India}
\address{
${}^4$Center for the Fundamental Laws of Nature,
Jefferson Physical Laboratory,\\
Harvard University,
Cambridge, MA 02138 USA}

\email{$^a$sgiombi@pitp.ca, $^b$minwalla@theory.tifr.res.in, $^c$shiroman@gmail.com, $^d$trivedi.sp@gmail.com,
$^e$wadia@theory.tifr.res.in,
$^f$xiyin@fas.harvard.edu}

\end{center}

\abstract{We study three dimensional conformal field theories described by 
$U(N)$ Chern-Simons theory at level $k$ coupled to massless fermions in the 
fundamental representation. By solving a Schwinger-Dyson equation in 
lightcone gauge, we compute the exact planar free energy of the theory at 
finite temperature on $\mathbb{R}^2$ as a function of the 't Hooft coupling $\lambda=N/k$. 
Employing a dimensional reduction regularization scheme, we find that 
the free energy vanishes at $|\lambda|=1$; the conformal theory does not 
exist for $|\lambda|>1$. We analyze the operator spectrum via the anomalous 
conservation relation for higher spin currents, and in particular show that 
the higher spin currents do not develop anomalous dimensions at leading 
order in $1/N$. We present an integral equation whose solution in principle 
determines all correlators of these currents at leading order in $1/N$ and 
present explicit perturbative results for all three point functions up to two 
loops. We also discuss a lightcone Hamiltonian formulation of this theory
 where a $W_\infty$ algebra arises. The maximally supersymmetric version of 
our theory is ABJ model with one gauge group taken to be $U(1)$, 
demonstrating that a pure higher spin gauge theory arises as a limit of 
string theory.}

\vfill

\end{titlepage}

\eject \tableofcontents

\section{Introduction}

Chern-Simons theories are fascinating from several points of view. The attractive features of these theories, unique to three-dimensions, had been recognized early on, notably in \cite{Deser:1982vy} and \cite{Deser:1981wh}. They arise, for example, in the  study  of knot invariants \cite{Witten:1988hf},
 the classification of
 two-dimensional rational conformal field theories \cite{Moore:1989vd},
 and in the study of the
 quantum Hall effect \cite{Frohlich}. More recently, superconformal Chern-Simons theories \cite{Ivanov:1991fn, Gaiotto:2007qi} have been shown to play a
crucial role in the AdS/CFT correspondence
\cite{Aharony:2008ug}. 

In this paper we will be interested in non-supersymmetric, but conformally invariant, $U(N)$ Chern-Simons theories
with matter. As we shall explain below, these theories have an important distinguishing feature; they effectively admit large $N$ non-supersymmetric 
lines of fixed points of the renormalization group.

Lines of fixed points parameterized by a coupling constant are especially
interesting from the viewpoint of the AdS/CFT correspondence. Such lines
of $d$ dimensional CFTs have the potential of interpolating between a
simple field
theoretic description at weak coupling and a relatively simple bulk
gravitational description at strong coupling, as demonstrated by the famous supersymmetric examples \cite{Maldacena:1997re, Aharony:2008ug}.

Now when $d \geq 4$, lines of fixed points appear to be rather exotic. 
The  examples we know of (like the large $N$ Banks-Zaks fixed line 
of QCD) involve theories with a parametrically large number of flavours
of  matter fields. It is interesting, on the other hand, that effective 
fixed lines of large $N$ Chern-Simons theories coupled to matter fields are 
plentiful and very easily constructed in 2+1 dimensions even with very 
simple matter content\footnote{We thank O. Aharony for discussions on this point.}.

Consider a level $k$, $U(N)$ Chern-Simons theory coupled to a single fermion in any representation of the gauge group. The only gauge invariant power counting relevant or marginal
operators in such a theory are the fermion kinetic term and mass term.
A continuum quantum theory built from such a Lagrangian depends on two
discrete parameters, $k$ and $N$, and a single continuous parameter, the
physical mass $m$ of the fermionic field.  At energies $E \gg m$, the dynamics of this
theory is scale invariant as well as nontrivial.
Nontriviality is ensured by the fact that the
discrete Chern-Simons coupling, which induces interactions among the
fermions,
cannot run and so is nonvanishing even at arbitrarily high energy.
The nontrivial CFT that controls the high energy
behaviour of this system is most directly constructed
by choosing the bare mass to
set the physical mass $m$ of this system to zero.

The parameters $k$ and $N$ labeling the CFT are discrete. However it is well known that the loop
counting parameter in a $U(N)$ Chern-Simons theory is the 't Hooft coupling
$\lambda =\frac{N}{k}$
whenever all matter fields transform in representations whose
dimension does not grow faster with $N$ than $N^2$.
In the large $N$ (and simultaneously large $k$)
limit $\lambda$ is effectively a continuous parameter (exactly as in ABJM theory \cite{Aharony:2008ug}). For this reason the discretum of
CFTs described by integer values of $k$ and $N$
coalesces into a fixed line in the large  $N$ limit.

We emphasize that a variety of such non-supersymmetric fixed lines exist in three dimensions. Choosing the fermions that transform in, say, the adjoint representation
of $U(N)$ gives one such fixed line of theories. Choosing fermions that transform
in the bifundamental of $U(N) \times U(M)$ yields another example --
one that preserves parity when $N=M$ and the two Chern-Simons levels are equal
and opposite  \footnote{The ease of construction of  conformal field
theories in $d=3$ intriguingly suggests that it is particularly easy to construct
non supersymmetric quantum theories of gravity in $d=4$.}.

The study of fixed lines of
large $N$ Chern-Simons theories with matter
and their bulk duals appears to be an interesting programme. For examples
with a large amount of supersymmetry this programme was spectacularly
initiated by ABJM \cite{Aharony:2008ug} and carried forward in several
follow up papers \cite{Benna:2008zy,Hosomichi:2008jb,Aharony:2008gk}. In this paper we initiate a detailed study of
perhaps the simplest of the non-supersymmetric fixed lines -- the theory of a single fundamental fermion coupled to
a $U(N)$ level $k$ Chern Simons theory
turns out to be surprisingly tractable at all values of $\lambda$, for
reasons we now explain.

The Chern-Simons coupled gauge field has $N^2$ components. Superficially,
the large $N$ limit of these theories is governed by the summation over a
complicated web of planar graphs. The complexity is illusory, as pure Chern-Simons theory has no propagating degrees of freedom -- the only propagating degrees of freedom in our system are the fundamental fermions. Consequently, the theory we investigate is a vector model with $N$ 
degrees of freedom.  Large $N$ limits of vector theories are much simpler
than their matrix counterparts, and sometimes prove to be exactly solvable. Indeed, both in terms of  diagrammatics and canonical structure, the theory we study bares a close resemblance to 't Hooft's solution of two-dimensional QCD in the large-$N$ limit using light-cone gauge \cite{'tHooft:1974hx}.\footnote{It has been conjectured recently \cite{Gaberdiel:2010pz} that the appropriate two dimensional analogs of large $N$ vector models are $W_N$ minimal models. Its duality with higher spin gauge theories and in particular Vasiliev's system in $AdS_3$ are further explored in \cite{Gaberdiel:2011wb, Ahn:2011pv, Gaberdiel:2011zw, Chang:2011mz, Papadodimas:2011pf}.}

The simplicity of the vector model in the large $N$ limit, combined with the crucial choice of light-cone gauge, allows us to derive
several exact results (valid to leading order in $N$ but to all orders in $\lambda$.)  
One of our key results is the  exact expression for the free energy of our
theory as a function of $\lambda$ and temperature. 
In obtaining this result we find that 
our line of fixed points  exists only if $0 \leq \lambda <1$.
We  also enumerate
the spectrum of ``single trace'' primary operators and demonstrate
that their scaling dimensions are not renormalized as a function of $\lambda$. We present  
an integral equation whose solution in
principle determines
all correlation functions of all these non renormalized operators - though
we leave a detailed study of this integral equation and its solutions for future work. Based on the analysis of the spectrum of primary operators and explicit
computation of some three point functions at low orders in perturbation 
theory, we also make several statements about the as-yet-unidentified holographic dual 
to our theory, which, as we discuss, must be a higher spin gauge theory.
In the rest of this introduction we will summarize the key results of our paper in more detail.  

\subsection{Outline of the paper and summary of results}
A summary of our main results is as follows.

In section \ref{free-energy-section}, we first calculate the exact fermion 
propagator for the theory on $\mathbb{R}^3$ in light-cone gauge. Our result, which 
is valid to leading order in the $\frac{1}{N}$ expansion but is exact 
in $\lambda$ is listed in \eqref{aef}.  We then turn to the finite 
temperature partition
function of our theory, i.e., the partition function of the theory on $\mathbb{R}^2 \times S^1$. We demonstrate that this partition function is completely determined by the self energy
in the exact fermion propagator on $\mathbb{R}^2 \times S^1$. This self energy
obeys a nonlinear Schwinger-Dyson integral equation \eqref{fgeqen}, which,
by an apparent accident of numbers, turns out to admit a remarkably simple exact solution \eqref{actsol2}. Using this solution, we find
that the free energy of our theory as a function of temperature and
$\lambda$, in a box of volume $V_2$ (which is taken to be very large) is
given by
\begin{equation}\label{feint}
F=-\frac{N V_2 T^3 }{6 \pi} \left[ {\tilde c}^3
\frac{1-\lambda}{\lambda}
+ 6 \int_{\tilde c}^\infty  dy ~ y
\ln \left ( 1+e^{-y} \right) \right]\,,
\end{equation}
where
${\tilde c}$ is the unique real solution to the equation\footnote{Though it is not manifest
the free energy in eq.(\ref{feint}) is  an even function of $\lambda$, as is required by the invariance of the free energy under parity.}
\begin{equation}\label{cansint}
\tilde c =  2 \lambda
\ln \left(2 \cosh \frac{{\tilde c}}{2}  \right)\,.
\end{equation}
\eqref{cansint} has no real solutions for $|\lambda|>1$; indeed, our fixed
line of theories exists only in the interval $|\lambda| \in [0, 1)$.

The fact that the free energy is proportional to $-N V_2T^3$. 
is an immediate consequence of the extensivity, conformality and 
large $N$ counting (disc diagrams dominate the free energy at large $N$), 
and could have been asserted on general grounds. The nontrivial part of 
\eqref{feint} is the function of $\lambda$, 
$$h(\lambda)=\frac{1}{6 \pi} \left[ {\tilde c}^3
\frac{1-\lambda}{\lambda}
+ 6 \int_{\tilde c}^\infty  dy ~ y
\ln \left ( 1+e^{-y} \right) \right], $$
that multiplies the factor $-V_2 T^3 N$. About $\lambda=0$ this 
function has an analytic expansion in even powers of $\lambda$. 
As $|\lambda|$ increases in $[0,1)$, this function  
 decreases monotonically from the free value
$\frac{3}{4 \pi}\zeta(3)$ to zero (see Fig \ref{plotF} in section \ref{free-energy-section}). $h(\lambda)$  is a measure of the number of available states in 
our system; the fact that it monotonically decreases implies a 
thinning of degrees of freedom at stronger coupling that is taken to an 
extreme  as $\lambda \to 1$, when $h(\lambda)$ vanishes. Recall that the ABJM
theory exhibits a similar phenomenon as $\lambda$ is scaled to $\infty$.

In the strict large $N$ limit we expect all transport properties of 
the finite temperature phase to be governed by a {\it collision-less} 
Boltzmann transport equation (the analogue of equation 9.66 of \cite{Sachdev} 
in the critical $O(N)$ sigma model in 3 dimensions, see Section 
\ref{discussion-section}) \footnote{We thank K. Damle  and S. Dutta for discussions on 
this point.}. In other words the finite temperature system appears to behave 
like a collection of free fermions 
with a $\lambda$ and temperature dependent dispersion relation. This
conclusion is also suggested by the fact that the exact finite temperature 
fermion two point function \eqref{actsol2} has an extremely simple analytic 
structure as a function the Lorentzian frequency $\omega$. The propagator 
has no cuts on the real $\omega$ axis The only singularity in this propagator
in the complex $\omega$ plane, are the poles located at  
$$\omega=\pm \sqrt{k^2+{\tilde c}^2 T^2}\, .$$

In section \ref{operator-section}, we study the spectrum of those operators whose
dimension stays fixed in the large $N$ limit. All such operators
are constructed as a product of ``single trace'' operators, where by
``single trace'' we mean operators such as $\bar{\psi}^i \psi_i$, which are formed out of the contraction of a single
fundamental fermionic index with a single antifundamental fermionic
index. We demonstrate that the spectrum of single
trace operators is not renormalized as a function of $\lambda$ at leading
order in $N$. The set of single trace
primaries is given, as in the free theory  \cite{Leigh:2003gk},
by a single ``current'' $J^{(s)}_{\mu_1 \ldots \mu_s}$ of dimension $s+1$ at every  spin $s=1,2, \ldots, \infty$ together with the single dimension two scalar
${\bar \psi \psi}$.

In the strict large $N$ limit, all the operators listed above except for
the scalar are primaries of short representations of the conformal algebra
as we now explain. Short representations always
have null states: the null states for a dimension $s+1$, spin $s$
current are simply the states formed out of the divergence of the
current. While the divergence of
$J^{(s)}_{\mu_1 \ldots \mu_s}$ does vanish in the free theory, at nonzero coupling the
currents obey an equation of the schematic form
\begin{equation}\label{divint}
\partial \cdot J^{(s)}\sim  {1\over k} J J + {1\over k^2} J J J
\end{equation}
Although the RHS of \eqref{divint} is nonvanishing, it is
a multitrace contribution, and so contributes only at subleading order in
$\frac{1}{N}$ when inserted into a two point function. In other words
the currents $J^{(s)}$ are effectively conserved, hence
protected, within two point functions, at leading order in large $N$. 
The operators $J^{(s)}$ develop anomalous dimensions at first subleading order
in $\frac{1}{N}$ (see subsection \ref{anom-curr} for details). 

A complete solution of our theory requires an algorithm to compute
all correlation functions of $J^{(s)}$ at every value of $\lambda$.
In section \ref{correlation-function-section} we demonstrate that
the three point functions of the spin operators, with all free indices
chosen to lie in the $x^-$ direction\footnote{$x^-$ refers to a null direction, and
we are working in the light-cone gauge $A_{-}=0$.}, are determined once we solve the integral
equation \eqref{iiee} that determines ``quantum corrected'' versions of the corresponding
operators. In section \ref{Formal-Solution} we also 
derive the integral equation \eqref{sathiksir} whose solution determines
the full quantum effective action of the theory in terms of the fermionic
fields. This effective action may be used to compute arbitrary correlators
of the spin operators with all polarizations in the $x^-$ direction.
We leave a study of the relevant integral equations, and of the correlation
functions of currents with arbitrary polarizations, to future work.
In this paper (see section \ref{correlation-function-section})
we have contented ourselves with a study of three point functions
of the current operators at tree, one-loop and two-loop level obtained
using perturbative techniques. In particular, we demonstrate that certain parity odd tensor structures of three point functions recently found in 
\cite{Maldacena:2011nz, Giombi:2011rz} arise at one-loop order.

In section \ref{hamiltonian-section}, motivated by the natural appearance 
of bilocal variables in the large-$N$ limit (see, e.g., \cite{Wadia:1980rb}),  we briefly discuss a light-cone Hamiltonian formulation that enables one to express the theory entirely in terms of bilocal variables, $M(x,y)$, that satisfy a $W_\infty$ algebra. Using the method of coadjoint orbits, we then derive an action for these bilocal variables and study its large-$N$ saddle point, which turns out to be particularly simple.

What is the bulk dual description of the line of CFTs studied in this paper? In order to address this question, we first
note that it has been conjectured in \cite{Sezgin:2003pt}, extending \cite{Klebanov:2002ja} (see also \cite{Sezgin:2002rt} for earlier related work), that our
field theory in the free $\lambda \to 0$ limit admits a dual description
as the parity-preserving ``type-B" Vasiliev theory \cite{Vasiliev:1999ba}. We will assume the
correctness of this conjecture in what follows. Clearly, then,
the bulk description of our finite $\lambda$ theory is some deformation
of the type B Vasiliev theory. It is also clear that deformation must
continue to preserve higher spin symmetry in the bulk at the classical
level; any bulk deformation that explicitly breaks Vasiliev's higher spin
symmetry classically would require the introduction of
new bulk fields (representing the longitudinal polarizations). The
fact that the spectrum
of single trace operators in our theory is not renormalized
as a function of $\lambda$ implies that the dual theory does not have
new fields corresponding to longitudinal polarizations, and so
presumably does not break the higher spin invariance classically. It follows that
the dual description is some higher spin theory in $AdS_4$ at every value
of $\lambda$, in which higher spin symmetry is broken at loop level in the bulk.

Vasiliev has written down a family of higher spin theories in $AdS_4$, parameterized by a single function of one variable \cite{Vasiliev:1999ba}. The cubic interactions in this family of theories, for instance, are governed by a phase (as well as the overall coupling constant). The
parity preserving type B theory (dual to the parity preserving
free fermion theory) is a special one in this family, where the phase is equal to $\pi/2$ in some conventions, see e.g. \cite{Sezgin:2003pt}. The only other parity preserving theory, known as type A model, is such that the phase vanishes, and is conjecturally dual to the free/critical 
bosonic vector model \cite{Klebanov:2002ja}. Theories with generic phase different from $0$ or $\pi/2$ do not preserve
parity. It is natural to suggest that
the dual description of the line of theories studied
in this paper is given by a one parameter (the parameter is $\lambda$)
set of such generalized Vasiliev theories with a $\lambda$
dependent phase. However, the evidence for this suggestion appears mixed
(see section \ref{holographic-dual-section}). A
preliminary
bulk study of 3 point functions of operators computed using the phase
generalized Vasiliev theory, suggests that the result should be given
by the schematic form
\begin{equation}\label{vp}
\sin^2  \theta \langle JJJ\rangle_F + \cos^2 \theta \langle JJJ\rangle_B
\end{equation}
where $\langle JJJ\rangle_F$ is the current-current-current 3 point function in the large N
free Fermi theory and $\langle JJJ\rangle_B$ is the corresponding 3 point function in the
large N free boson theory. As we describe in section \ref{correlation-function-section} below,
the  explicit perturbative 3 point functions
described in the paragraph above appear to agree impressively
with this prediction at two loops if we set $\theta = {\pi\over 2}(1-\lambda) +{\cal O}(\lambda^3)$.\footnote{One may speculate that this relation between $\theta$ and $\lambda$ is in fact exact, and that our theory in the $\lambda\to 1$ limit could be a theory of weakly coupled bosons! There is an analog in two dimensions: the $W_N$ minimal model \cite{Gaberdiel:2010pz} has a 't Hooft coupling $\lambda$, which also ranges from 0 to 1. In the $\lambda\to 0$ limit it becomes a theory of free fermions, while in the $\lambda\to 1$ limit it becomes a theory of free bosons (modulo subtleties with twisted sectors). } On the other
hand this identification, together with \eqref{vp}, implies that all
one loop contributions to current-current-current three point functions vanish.
As far as we can tell this is not the case. Conformal invariance allows
parity violating structures for the three point functions of currents \cite{Giombi:2011rz}, and
these structure appear to be generated, in field theory,
 with nonzero coefficient at
order $\lambda$ (see section \ref{correlation-function-section}). We are
not sure what to make of this
disagreement. If we accept it in the most straightforward way,
it would suggest that our theory is not dual to the general phase
Vasiliev model and that our line of Chern-Simons theory
is dual to a yet-to-be-constructed higher spin theory that reduces,
in the $\lambda \to 0$ limit, to the parity preserving Vasiliev theory.
It would be especially interesting to identify the bulk dual to our
field theory in the strong coupling limit $ \lambda \to 1$.
However we leave the resolution of these issues (and several others)
to future work. 

In subsection \ref{string} we also note that the maximally 
supersymmetric extension of the theory studied in this paper is 
precisely the ABJ model \cite{Aharony:2008gk} with one of the gauge 
groups taken to be $U(1)$. It follows that the higher spin bulk dual to the 
 maximally supersymmetric version of the model studied in this paper 
is a particular limit of string theory. 

In summary, in this paper we have initiated the task of determining
an exact solution to a simple but nontrivial fixed line of quantum
field theories, namely $U(N)$ Chern-Simons theories coupled to fermionic fundamental matter.
We have been able to exactly compute the finite temperature partition
function.  It is possible that
correlation functions in this theory are also exactly computable both
at zero and at finite temperature. 
Exact results for current--current and stress tensor--stress tensor
two point functions, as functions of temperature, would be particularly fascinating to obtain. Our work also
suggests several obvious generalizations, for instance to the theory
of fundamental bosons \cite{oferetal} or supersymmetric Chern-Simons theory. It also
throws up a sharp question for the dual theory: what precise deformation
of the parity-preserving Vasiliev theory is our system dual to?
We hope to return to all these fascinating questions in the near future.

\bigskip

{Note added: O. Aharony, G. Gur-Ari and R. Yacoby have recently demonstrated that massless scalar fundamental matter coupled to Chern-Simons theory
also leads to a large $N$ fixed line of CFTs \cite{oferetal}. Using arguments 
very similar to those employed in part of section 3 of our paper, they have 
also argued that the higher spin currents in their theory are not renormalized 
as a function of the coupling constant. We thank O. Aharony for 
sharing his results with us prior to publication and also for several 
discussions on these issues over the last year.}

\section{Free Energy on $\mathbb{R}^2$ at Finite Temperature}
\label{free-energy-section}
In this section we will evaluate the free energy of our theory (a single
species of massless fundamental fermions coupled to a Chern-Simons gauge
field) in the t' Hooft large $N$ limit. The Euclidean action for our theory is
\begin{equation}\label{euclidean-action}
\begin{split}
S &= \frac{ik}{4 \pi}\int
\text{Tr} \left(A d A + \frac{2}{3} A^3\right)+ \int
 {\bar \psi}\gamma^\mu D_\mu  \psi,
\end{split}
\end{equation}
Our theory is taken to be
at temperature
$T$, and lives on a spatial $\mathbb{R}^2$ whose regulated volume we denote
as $V_2$.

In order to evaluate the free energy of our theory, we first fix a gauge.
We work in the lightcone gauge $A_-=0$. This gauge is defined in terms of
an analytic continuation from Lorentzian space. For this reason $A_-=0$ does not imply $A_+=0$. Alternatively one can work in Lorentzian space and analytically continue to Euclidean space (after integrating out the gauge fields) in \eqref{eam}. The gauge boson
self interaction term vanishes in this gauge, a feature that enormously
simplifies analysis. \footnote{This feature is also true of the more
straightforward ``temporal'' or axial gauge $A_3=0$. Feynman diagrams in this
gauge are, however, plagued by logarithmic divergences that we have found
difficult to interpret and deal with. In contrast the divergences
in the lightcone gauge employed in this subsection are relatively tame, and 
are easy to interpret
and deal with. We thank S. Bhattacharyya and J. Bhattacharya for 
extensive discussions on perturbation theory and its 
divergences in Axial gauge.} Computations in pure Chern-Simons theory 
on $\mathbb{R}^3$ in this 
Euclidean continuation of the light-cone gauge were done previously in 
\cite{Frohlich:1989gr}, \cite{Labastida:1997uw}.

In our analysis below we will encounter divergent integrals that need
to be regulated. We choose to regulate all integrals in the scheme of
dimensional reduction. More specifically we evaluate all integrals as follows.
We evaluate $\gamma$ traces,  $\epsilon$ contraction etc in $d=3$.
This process leaves us with a set of scalar integrals. We then evaluate
the resultant integrals by analytic continuation from $d=3-\epsilon$
dimensions. This regularization scheme is widely employed in the previous
literature on Chern-Simons matter theories (see e.g. \cite{susydimred, Chen:1992ee, Ivanov:1991fn, Gaiotto:2007qi}). It is  manifestly Lorentz
invariant, and  also respects  gauge invariance at least up to two-loops (see \cite{Chen:1992ee}).
We will assume without proof in what follows  that our regularization
scheme is indeed gauge invariant. If this is indeed the case then
the theory defined by this regularization scheme must be Lorentz
invariant even though we work in a gauge that breaks Lorentz invariance.
We will find some evidence for the Lorentz invariance of our final results
giving some a posteriori evidence for our assumption that our regularization
scheme respects gauge invariance.

As we will explain below, the finite temperature free energy of  our theory
is completely determined by the fermion self energy on
$\mathbb{R}^2  \times S^1$ (see \eqref{sft}). In order to evaluate the
free energy we proceed as follows. As a preliminary step
to our analysis, we first
determine the exact fermion propagator of our theory on $\mathbb{R}^3$. We then
determine the exact fermion propagator on $\mathbb{R}^2 \times S^1$. Finally, we  proceed
to use this result to determine the free energy of our theory.

\subsection{Exact fermion propagator on $\mathbb{R}^3$}
\bigskip
In the planar limit, the exact fermion propagator in light-cone gauge receives contributions from a very simple class of diagrams. Indeed, the gluon vertex vanishes in our gauge, and moreover at large $N$ we can neglect fermion loops since the fermions are in the fundamental representation. Therefore, only rainbow diagrams contribute to the exact propagator. These can be resummed exactly by solving a Schwinger-Dyson equation which may be depicted diagrammatically as

\centerline{\begin{fmffile}{SD_self_energy_1}
        \begin{tabular}{c}
            \begin{fmfgraph*}(85,65)
                \fmfleft{i1}
                \fmfright{o1}
                \fmfv{l=1PI,l.a=0,l.d=-.13h,decoration.shape=circle, decoration.size=.5h, decoration.filled=empty}{v}
                \fmf{plain}{i1,v,o1}
             \end{fmfgraph*}
        \end{tabular}
        \end{fmffile}
        =
        \begin{fmffile}{SD_self_energy_2}
        \begin{tabular}{c}
            \begin{fmfgraph*}(85,65)
                \fmfleft{i1}
                \fmfright{o1}
                \fmfv{decoration.shape=circle, decoration.size=.3h, decoration.filled=shaded}{v}
                \fmf{wiggly,left=1,tension=0}{v1,v2}
                \fmf{plain}{i1,v1,v,v2,o1}
             \end{fmfgraph*}
        \end{tabular}
        \end{fmffile}
}
\centerline{The planar fermion self energy.}\bigskip
\noindent
This picture may be transcribed into equations as follows.
Let the exact fermion propagator be given by
\begin{equation}\label{ae}
\langle \psi(p)_m {\bar \psi(-q)}^n \rangle
=\delta_m^n
 \frac{1}{i p_\mu \gamma_\mu + M_{bare}+\Sigma} \times (2 \pi)^3 \delta(p-q).
\end{equation}
Then
\begin{equation}\label{seeq}
\Sigma(p)= \frac{N}{2}\int \frac{d^3q}{(2 \pi)^3} \left(  \gamma^\mu \frac{1}
{i\gamma^\alpha q_\alpha+ M_{bare}+\Sigma(q)} \gamma^\nu \right) G_{\mu \nu}(p-q)
\end{equation}
where
$$\langle A^a_\mu(p) A^b_\nu(-q) \rangle=
(2\pi)^3 \delta(p-q) G_{\mu\nu}(p) \delta^{ab}$$
is the bare gluon propagator.\footnote{The factor of $\frac{1}{2}$ in \eqref{seeq} has its origin
in the $\frac{1}{2}$ on the RHS of \eqref{comp}.}
Here $M_{bare}$ is the mass term that appears in the bare Lagrangian.
In what follows we will adjust $M_{bare}$ to ensure that the physical
fermion mass vanishes (this choice corresponds to tuning the
theory to the conformality). Note that with our definition of $\Sigma$ in \eqref{ae}, the sum of 1PI diagrams is given by $-\Sigma$.

The equation \eqref{seeq} applies in fact in any ghost free  gauge in which all gluon interactions
vanish. In this section we will solve \eqref{seeq} in the lightcone gauge.

\subsubsection{Exact solution of the gap equation}\label{ge}

In the this subsection we will solve the gap equation \eqref{seeq} in light-cone gauge (see Appendix \ref{appC} for a detailed listing of our
conventions, propagators etc
in lightcone gauge):
\begin{equation}
\label{gapsh}
\Sigma(p)=-i 2\pi \lambda \int \frac{d^3q}{(2 \pi)^3} \left(
\gamma^3 \frac{1}
{i \gamma^\mu q_\mu+ M_{bare}+  \Sigma(q)} \gamma^+ -\gamma^+ \frac{1}
{i \gamma^\mu q_\mu+ M_{bare}+ \Sigma(q)} \gamma^3 \right)
\frac{1}{(p-q)^+}.
\end{equation}

Let us first better understand the matrix structure of the self energy
$\Sigma$. Let $A$ represent
an arbitrary $2 \times 2$ matrix
$$A=A_II + A_+ \gamma^+ + A_- \gamma^- + A_3 \gamma^3$$
For use below and in later sections we define the matrix
valued functions of $A$, $H_+(A)$ and $H_-(A)$ as
\begin{equation}\label{mana}
H_+(A) \equiv \gamma^3 A \gamma^+-\gamma^+ A \gamma^3
= 2 \left(  A_I \gamma^+ -  A_- ~I \right)
\end{equation}
\begin{equation}\label{mano}
H_-(A)\equiv \gamma^3 A \gamma^- -\gamma^- A \gamma^3
= 2 \left(  -A_I \gamma^- + A_+ I \right)
\end{equation}
The gap equation \eqref{gapsh} may be rewritten as
\begin{equation}
\label{gapshn}
\Sigma(p)=-i 2 \pi \lambda \int \frac{d^3q}{(2 \pi)^3} \left(
H_+ \left[ \frac{1}
{i \gamma^\mu q_\mu+ M_{bare} + \Sigma(q)}  \right]
\frac{1}{(p-q)^+} \right).
\end{equation}
In the discussion which follows we will sometime  abbreviate the notation $\Sigma(p)$ to $\Sigma$ for
brevity.  Now  let
\begin{equation} \label{exps}
\Sigma= i \Sigma_\mu \gamma^\mu + \Sigma_I I -M_{bare}I
\end{equation}
Using
$$\frac{1}{i \gamma^\mu (q_\mu+\Sigma_\mu) + \Sigma_I}
= \frac{-i \gamma^\mu (p_\mu +\Sigma_\mu)+\Sigma_I}{(p+\Sigma)^2+\Sigma_I^2}$$
together with \eqref{mana}, we may rewrite \eqref{gapshn} as
\begin{equation}\label{gapshnn}
\Sigma(p) =-i 4 \pi \lambda
\int \frac{d^3q}{(2 \pi)^3}
\frac{\gamma^+\Sigma_I+i I (q +\Sigma(q))_-}{(q_\mu+\Sigma_\mu(q))
(q^\mu+\Sigma^\mu(q)) +\Sigma_I(q)^2}
\frac{1}{(p-q)^+}
\end{equation}
Plugging \eqref{exps} into the LHS of \eqref{gapshnn} and equating
the coefficients of linearly independent matrices,  it follows
immediately that
\begin{equation}\label{ps}
\Sigma_-=\Sigma_3=0.
\end{equation}
And that
\begin{equation}\label{speq}
\begin{split}
\Sigma_+(p)&=-4 \pi  \lambda
\int \frac{d^3q}{(2 \pi)^3}
\frac{\Sigma_I}{\left( (q+\Sigma(q))^2+\Sigma_I(q)^2 \right)}~\frac{1}{(p-q)^+}\\
\Sigma_I(p)-M_{bare}&=4 \pi \lambda \int  \frac{d^3q}{(2 \pi)^3} \frac{q_-}
{\left( (q+\Sigma(q))^2+\Sigma_I(q)^2\right)}
~\frac{1}{(p-q)^+}
\end{split}
\end{equation}

What can we say about the dependence of $\Sigma_+(p)$ and $\Sigma_I(p)$ on
$p$? First note that the that the RHS of the two equations in
 \eqref{speq} is independent of $p^3$. It follows that $\Sigma$ is a
function only of the in plane momenta $p^1$ and $p^2$ but is independent of
$p^3$. Rotational invariance in the 12 plane and requirement
of conformality (i.e. the requirement that no mass scale enter the
physical propagator) then together completely fix the momentum dependence of
$\Sigma_+$ and $\Sigma_I$:
\begin{equation}\label{ansa}
\begin{split}
\Sigma_I(p)&=f_0 p_s\\
\Sigma_+(p)&= p_+ g_0=p^- g_0\\
\end{split}
\end{equation}
where
\begin{equation} \begin{split} \label{mpd}
p_s&=\sqrt{p_1^2+p_2^2}=\sqrt{2}|p^-|=\sqrt{2}|p^+|
\end{split}
\end{equation}
and $f_0$ and $g_0$ are dimensionless numbers (that are functions of the
coupling constant $\lambda$).

Plugging \eqref{mpd} and \eqref{ps} into \eqref{speq} we find
\begin{equation}\label{speqnt}
\begin{split}
g_0&=-\frac{4 \pi \lambda}{p^-}
\int \frac{d^3q}{(2 \pi)^3}~
\frac{q_sf_0}{q_3^2+ q_s^2(1+g_0+f_0^2)} ~ \frac{1}{(p-q)^+} \\
f_0|p|-M_{bare}&=4 \pi \lambda \int \frac{d^3q}{(2 \pi)^3}~
\frac{q^+}{q_3^2 + q_s^2(1+g_0+f_0^2)}~
\frac{1}{(p-q)^+}
\end{split}
\end{equation}
We will now proceed to determine the numbers $g_0$ and $f_0$ as functions of
$\lambda$.

We first note that the integrals on the RHS of \eqref{speqnt} are
(power counting) linearly divergent. In order to proceed we need to
regulate these divergences. We adopt a regulator that is manifestly
Lorentz invariant as well as plausibly gauge invariant. Our regularization
procedure is simply the following: we analytically continue all diagrams
to $3-\epsilon$ dimensions. Two of these dimensions span the 1-2 plane.
We integrate over the remaining $1-\epsilon$ dimensions using
the formula
\begin{equation}\label{intf}
\int_{-\infty}^\infty \frac{d^{1-\epsilon}x}{a^2+x^2}=
\frac{A^\epsilon \pi}{|a|^{1+\epsilon}}
\end{equation}
Here $A$ is a number of order unity which is easily computed. However, as
we will see below, none of the integrals we compute in this paper have
a $\frac{1}{\epsilon}$ type divergence (this corresponds to an absence of
logarithmic divergences, were we to use a momentum cutoff). The only effect
of the dimensional regularization cut off procedure, employed in our paper,
is to discard linear (and below also cubic) divergences in a gauge and
Lorentz invariant manner. For that reason, when we take $\epsilon \to 0$
at the end of the calculation, we effectively set $A$ to unity. Hence
we immediately set $A$ to unity instead of carrying it around in
our computation. The regularization procedure we employ here is essentially the dimensional
reduction scheme used in \cite{Chen:1992ee}, adapted to our light-cone gauge.

Using \eqref{intf} in \eqref{speqnt} yields
\begin{equation}\label{speqn}
\begin{split}
g_0&=-\frac{2 \pi \lambda}{p^-}
\frac{f_0}{\sqrt{1+g_0+f_0^2}}
\int \frac{d^2 q}{(2 \pi)^2}
 \frac{1}{q_s^\epsilon(p-q)^+} \\
f_0 p_s-M_{bare}&={2 \pi \lambda}  \frac{1}{\sqrt{1+g_0+f_0^2} }
\int \frac{d^2 q}{(2 \pi)^2}
\frac{q^+}{q_s^{1+\epsilon}(p-q)^+}
\end{split}
\end{equation}
In order to do the integrals we move to polar coordinates in the 12 plane.
The integrals we need to evaluate are
\begin{equation}\label{intodo} \begin{split}
\int \frac{d^2 q}{(2 \pi)^2}
\frac{q^+}{q_s^{1+\epsilon}(p-q)^+}&= \frac{1}{(2 \pi)^2}
\int_0^\infty q_s^{1-\epsilon}dq \int_0^{2 \pi}
d \theta \frac{p_s \cos \theta -q_s}{q_s^2+p_s^2-2 p_s q_s \cos \theta} \\
\int \frac{d^2 q}{(2 \pi)^2 p^-}
 \frac{q_s^{1-\epsilon}}{q_s^\epsilon(p-q)^+}&=\frac{2}{(2 \pi)^2 p_s}
 \int_0^\infty q_s^{1-\epsilon} dq \int_0^{2 \pi}
d \theta \frac{p_s -q_s \cos \theta }{q_s^2+p_s^2-2 p_s q_s \cos \theta} \\
\end{split}
\end{equation}
Using contour techniques it is not difficult to verify that for $q>p$
\begin{equation}\label{te}\begin{split}
\int_0^{2 \pi}  \frac{d\theta}{q^2+p^2-2 pq \cos \theta}
& =\frac{2 \pi}{q^2-p^2}\\
\int_0^{2 \pi}  \frac{d\theta \cos \theta}{q^2+p^2-2 pq \cos \theta}
&= \frac{p}{q}\frac{2 \pi}{q^2-p^2}\\
\end{split}
\end{equation}
It follows that
\begin{equation}\label{actint}
\begin{split}
\int_0^{2 \pi}
d \theta \frac{p_sq_s \cos \theta -q_s^2}{q_s^2+p_s^2-2 p_s q_s \cos \theta}&=0 ~~~(q_s<p_s)\\
\int_0^{2 \pi}
d \theta \frac{p_sq_s \cos \theta -q_s^2}{q_s^2+p_s^2-2 p_s q_s \cos \theta}&=-2 \pi
 ~~~(q_s>p_s)\\
\int_0^{2 \pi}
d \theta \frac{p_sq_s -q_s^2 \cos \theta }{q_s^2+p_s^2-2 p_s q_s \cos \theta}&= 2 \pi ~~~(q_s<p_s)\\
\int_0^{2 \pi}
d \theta \frac{p_sq_s  -q_s^2 \cos \theta}{q_s^2+p_s^2-2 p_s q_s \cos \theta}&=0
 ~~~(q_s>p_s).\\
\end{split}
\end{equation}
It follows that \eqref{speqn} reduces to
\begin{equation}\label{speqn2}
\begin{split}
g_0&=
-\frac{\lambda f_0}{\sqrt{1+g_0+f_0^2}}\\
f_0p_s-M_{bare}&
=-\frac{\lambda}{\sqrt{1+g_0+f_0^2}}
\int_{p_s}^\infty q^{-\epsilon} dq =
\frac{\lambda}{\sqrt{1+g_0+f_0^2}}p_s
\end{split}
\end{equation}
It follows from \eqref{speqn2} that $M_{bare}=0$ (this is a consequence
of our use of dimensional regularization; $M_{bare}$ is linearly divergent
in a cut off regulator, as is clear from the second of \eqref{speqn2}).
The remaining equations reduce to
\begin{equation}\label{speqnn}
\begin{split}
g_0&=
-\frac{\lambda f_0}{\sqrt{1+g_0+f_0^2}}\\
f_0&=\frac{\lambda}{\sqrt{1+g_0+f_0^2}}\\
\end{split}
\end{equation}
The solution to \eqref{speqnn} is remarkably simple
\begin{equation}\begin{split}\label{mfs}
f_0&=\lambda\\
g_0&=-\lambda^2 \\
g_0+f_0^2&=0\\
\end{split}
\end{equation}
In other words, the self energy receives contributions from one and
two loop graphs but not at any higher order in perturbation theory!
This completes our solution of the gap equation.

We emphasize that our final result \eqref{mfs} depends in a crucial way
on our choice of regularization scheme. Were we, for instance to regulate
the integrals \eqref{speqnt} by modifying the gauge boson propagator with
a Gaussian damping factor $e^{-\frac{(p-q)^2}{2 \Lambda^2}}$, then we would have found $g_0=0$ (from the
integral over the angle in the vector $p-q$). It is of course quite
clear that a crude cut off on the gauge boson momentum does not
preserve either gauge or Lorentz invariance. We hope on the other hand that
the more sophisticated dimensional regularization scheme preserves both these
symmetries. This assumption, which is as yet unproved, is the main weakness
in the analysis presented in this section. We will return to this
point at the end of the section.

In summary, the exact fermion propagator, at leading order in large $N$,
is given by
\begin{equation}\label{aef}
\langle \psi(p)_m {\bar \psi(-q)}^n \rangle
=\delta_m^n
 \frac{1}{ip_3 \gamma^3 +i p_{-} \gamma^{-} + i (1-\lambda^2)p_{+}\gamma^{+} +\lambda p_s} \times (2 \pi)^3 \delta(p-q).
\end{equation}
(Here $m,n$ are colour indices and we have suppressed the spinor indices).

\subsubsection{Rederivation of the gap equation as a Schwinger-Dyson equation}

The starting point of our analysis in this subsection
is the path integral representation of the partition function
$$Z=\int D \psi D \bar{\psi} D A_\mu e^{-S}$$
where the action $S$ is the Euclidean space lightcone gauge action for our
field theory, listed explicitly in \eqref{ea2}. The gauge fields appear
quadratically in in \eqref{ea2} and may be integrated out.
Integrating out the gauge field
from \eqref{ea2} yields the path integral\footnote{In integrating out the gauge field, we absorb the factor coming from the determinant of the gauge field kinetic operator in the normalization of the path integral. In other words, we normalize the pure Chern-Simons partition function to 1.}
$$Z=\int D \psi D\bar{\psi} e^{-S}$$
where the action $S$ is now given by
\begin{equation}\label{eam}
\begin{split}
S &= i \int \frac{d^3 p}{(2 \pi)^3}
 {\bar \psi}(-p)  \gamma^\mu p_\mu  \psi(p)\\
&+\frac{2 \pi i}{k}
\int \frac{d^3p}{(2 \pi)^3} \frac{d^3 r}{(2 \pi)^3}
\frac{d^3 q}{(2\pi)^3}  \frac{1}{q^+}
{\bar \psi^m(-p)} \gamma^+ \psi_n(p-q)
{\bar \psi}^n(-r) \gamma^3 \psi_m(r+q)\\
\end{split}
\end{equation}

Let us pause to note that, starting from the bilocal action \eqref{eam}, one can conveniently derive the Schwinger-Dyson equation for the fermion self-energy (as in \cite{Wadia:1980rb}) via 
\begin{eqnarray}
0 & = &  \int D\psi D\bar{\psi} 
\frac{\delta}{\delta \bar{\psi}^m(-p)} 
\left( e^{-S}  \bar{\psi}^n (p') \right)  
\\
  & = & \int D\psi D\bar{\psi}  
\left( \delta^n_m \delta^3(p'+p)
 - \frac{\delta S}{\delta \bar{\psi}^m(-p)}\bar{\psi}^n(p') \right)  e^{-S} 
\end{eqnarray}
which gives the following relation involving the exact fermion propagator and four-point functions,
\begin{equation}
\begin{split}
  (2 \pi)^3\delta^3(p'+p)  = &    i p_\mu \gamma^\mu\braket{\psi_m(p)\bar{\psi}^n(p')}
\\
&  + \frac{2 \pi i}{k}
  \int \frac{d^3 r}{(2 \pi)^3}
\frac{d^3 q}{(2\pi)^3}  \frac{1}{q^+}
 \gamma^+ \braket{\psi_a(p-q)
{\bar \psi}^a(-r) \gamma^3 \psi_m(r+q){\bar \psi^n(p')}} 
\\
&
- \frac{2 \pi i}{k}
 \int\frac{d^3 r}{(2 \pi)^3}
\frac{d^3 q}{(2\pi)^3}  \frac{1}{q^+}
 \gamma^3 \braket{\psi_a(p-q) \bar \psi^a(-r) \gamma^+  {\psi_m(r+q) \bar \psi}^n(p')} \label{exact-fermion-propagator-sd}.
\end{split}
\end{equation}
In the large-$N$ limit, this factorizes to yield
\begin{equation}
\begin{split}
  \braket{\psi_m(p)\bar{\psi}^n(p')}  = &  \frac{1}{i p_\mu \gamma^\mu} (2 \pi)^3\delta^3(p'+p) 
\\
&  - \frac{1}{i p_\mu \gamma^\mu}\frac{2 \pi i}{k}
  \int \frac{d^3 r}{(2 \pi)^3}
\frac{d^3 q}{(2\pi)^3}  \frac{1}{q^+}
 \gamma^+ \braket{\psi_a(p-q)
{\bar \psi}^a(-r)} \braket{\gamma^3 \psi_m(r+q){\bar \psi^n(p')}} 
\\
&
+ \frac{1}{i p_\mu \gamma^\mu}\frac{2 \pi i}{k}
 \int\frac{d^3 r}{(2 \pi)^3}
\frac{d^3 q}{(2\pi)^3}  \frac{1}{q^+}
 \gamma^3 \braket{\psi_a(p-q) \bar \psi^a(-r)} \gamma^+ \braket{{\psi_m(r+q) \bar \psi}^n(p')} \label{exact-fermion-propagator-sd-2}
\end{split}
\end{equation}
which, upon substituting \eqref{ae}, gives the gap equation:
\begin{equation}
\Sigma(p) =
-  2 \pi i \lambda
 \int
\frac{d^3 q}{(2\pi)^3}  \frac{1}{p^+ - q^+}
H_+ \left(\frac{1}{i (q_\mu) \gamma^\mu + \Sigma(q)}\right).  
\end{equation}

\subsubsection{Rewriting the field theory as a path integral over
singlet fields}\label{gesp}

In this subsection we will reformulate the path integral that evaluates
the partition function of our field theory as a path integral over
singlest fields. The new path integral is weakly coupled in the large
$N$ limit (the action in terms of the new variables is proportional to
$N$). The gap equation \eqref{gapsh} follows as the classical equation of
motion of this large $N$ action.

While the work out
of this subsection is considerably more complicated than that of
subsection \eqref{ge} it has one significant advantage; it reveals how
the solution of the gap equation is related to the value the partition
function of the theory. While the value of the partition function is of no
physical significance for the theory on $\mathbb{R}^3$, it is of great significance
on $\mathbb{R}^2 \times S^1$ (as it determines the thermal partition function of the
theory on $\mathbb{R}^2$). For this reason the  results of this subsection
will prove very useful in our discussion of the finite temperature partition
function in the next section.

We now introduce some convenient shorthand notation. Let
\begin{equation} \label{Mdef}
M(P, q)=  \frac{1}{N} \int \frac{dq^3}{2\pi} \psi_m(\frac{P}{2}+q){\bar \psi}^m(
\frac{P}{2}-q)\,.
\end{equation}
$M$ is a $2 \times 2$ matrix in spinor space but a singlet in colour space.
While one of its arguments, $P$,
is a 3 momentum, its second argument $q$ is a 2 momentum (in integral on the
RHS of \eqref{Mdef} is over the 3 component of $q$). \eqref{eam}
may be rewritten as
\begin{equation}\label{eamm}
\begin{split}
S &= i \int \frac{d^3 p}{(2 \pi)^3}
 {\bar \psi}(-p)  \gamma^\mu p_\mu  \psi(p)\\
&-\frac{2 \pi i N^2}{k}
\int \frac{d^3 P}{(2 \pi)^3} \frac{d^2 q}{(2 \pi)^2}
\frac{d^2 q'}{(2 \pi)^2}
\frac{1}{(q-q')^+} Tr \left(M (P, q) \gamma^+
M(-P, q') \gamma^3  \right)\\
 &= i \int \frac{d^3 p}{(2 \pi)^3}
 {\bar \psi}(-p)  \gamma^\mu p_\mu  \psi(p)\\
&-\frac{\pi i N^2}{k}
\int \frac{d^3 P}{(2 \pi)^3} \frac{d^2 q}{(2 \pi)^2}
\frac{d^2 q'}{(2 \pi)^2}
\frac{1}{(q-q')^+}
Tr \left[ \left(M (P, q) \gamma^+
M(-P, q') \gamma^3  \right)-\left(M (P, q) \gamma^3
M(-P, q') \gamma^+  \right) \right]\\
\end{split}
\end{equation}
(the flip in sign is due to the fact that we had to take one fermionic
field through three others).
Expanding the matrix $M$ in a complete basis of $2 \times 2$ matrices
\begin{equation}\label{sexp}
M=M_+ \gamma^+ + M_- \gamma^-+ M_3 \gamma^3+
M_I I
\end{equation}
we find that \eqref{eamm} reduces to
\begin{equation}\label{eammm}
\begin{split}
S &= i \int \frac{d^3 p}{(2 \pi)^3}
 {\bar \psi}(-p)  \gamma^\mu p_\mu  \psi(p)\\
&+\frac{8\pi i N^2}{k}
\int \frac{d^3P}{(2 \pi)^3} \frac{d^2 q}{(2 \pi)^2}
\frac{d^2 q'}{(2 \pi)^2}
~\frac{1}{(q-q')^+}~ M_-(P, q)
{M_I}(-P, q')~ \\
\end{split}
\end{equation}
where we have used
\begin{equation}\label{trid}
Tr \left( [\gamma^{-}, \gamma^{+}]\gamma^3 \right)=-4.
\end{equation}
Note in particular that $M_+$ and $M_3$ drop out of this expression.

We will now rewrite the interaction term (the term quadratic in $M$)
in \eqref{eammm} in terms of a Lagrange multiplier field
\begin{equation}\label{defs}
\Sigma=\Sigma_+ \gamma^+ + \Sigma_I I.
\end{equation}
where $\Sigma$ will turn out to be the self energy
of the fermion field. To this end we define the ``inverse'' Greens function
$G^{-1}(p)$ by the requirement that
\begin{equation}\label{ginvdef}
\int \frac{d^2 q}{(2 \pi)^2 }
G^{-1}(p-q) \frac{1}{(q-r)^+}= (2 \pi)^2 \delta^2(p-r)
\end{equation}
Note that $G^{-1}$ is an odd function of its argument. Note also that
\begin{equation}\label{ginvdef2}
\int \frac{d^2 r}{(2 \pi)^2 }
\frac{1}{(q-r)^+} G^{-1}(r-p) = (2 \pi)^2 \delta^2(q-p)
\end{equation}
These are the only properties of $G^{-1}$ that we will need in this paper;
in particular we will never need the explicit form of the function $G^{-1}$.

Now it is obvious that
\begin{equation}\label{pfff}
Z= \frac{\int D\psi D \Sigma_- D\Sigma_I e^{-(S+E)}}
{\int D \Sigma_- D \Sigma_I e^{-E}}
\end{equation}
where we have chosen
\begin{equation}\label{exp} \begin{split}
E& = 2 \times \frac{N}{4 \pi i \lambda}
\int \frac{d^3P}{(2\pi)^3} \frac{d^2q}{(2 \pi)^2}
 \frac{d^2q'}{(2 \pi)^2} \\
&  \bigg[
\left( \Sigma_+(P, q)- 4 \pi i \lambda
\int \frac{d^2r}{(2 \pi)^2} M_I(P,r)\frac{1}{(r-q)^+}
\right) \times  G^{-1}(q-q') \\
 & \times \left( \Sigma_I(-P, q')- 4 \pi i \lambda
\int \frac{d^2r'}{(2 \pi)^2}\frac{1}{(q'-r')^+} M_{-}(-P,r')
\right) \bigg]
\end{split}
\end{equation}
Note that $E$ is a function of the two new Lagrange multiplier fields
$\Sigma_-$ and $\Sigma_I$. The path  integral in the denominator in
\eqref{pfff} is simply a number of order unity and we will
omit to write it in the equations that follow.
The effective action in the numerator, $S+E$, evaluates to
\begin{equation}\label{eammmm}
\begin{split}
S &= i \int \frac{d^3 p}{(2 \pi)^3}
 {\bar \psi}(-p)  \gamma^\mu p_\mu  \psi(p)\\
&+ \int \frac{d^3P}{(2\pi)^3} \frac{d^3q}{(2 \pi)^3}
{\bar \psi}(\frac{P}{2}-q)\Sigma(-P, q) \psi(\frac{P}{2}+q)   \\
&+\frac{N}{2 \pi i \lambda}
\int \frac{d^3P}{(2\pi)^3 }\frac{d^2q}{(2 \pi)^2}\frac{d^2q'}{(2 \pi)^2}
\Sigma_+(P, q) G^{-1}(q-q') \Sigma_I(-P,q')
\end{split}
\end{equation}
In the second line above we have used the fact that
$$-2 (\Sigma_+ M_- + \Sigma_I M_I)= -Tr \Sigma M=
-\frac{1}{N}\int \frac{d q_3}{2 \pi} Tr \Sigma \psi {\bar \psi}=
\frac{1}{N}\int \frac{d q_3}{2 \pi} {\bar \psi} \Sigma \psi$$
Using \eqref{trid} the last line in \eqref{eammm} may be rewritten as
a trace, yielding
\begin{equation}\label{eammmm2}
\begin{split}
S &= i \int \frac{d^3 p}{(2 \pi)^3}
 {\bar \psi}(-p)  \gamma^\mu p_\mu  \psi(p)\\
&+ \int \frac{d^3P}{(2\pi)^3} \frac{d^3q}{(2 \pi)^3}
{\bar \psi}(\frac{P}{2}-q)\Sigma(-P, q) \psi(\frac{P}{2}+q)   \\
&-\frac{N}{8 \pi i \lambda}
\int \frac{d^3P}{(2\pi)^3 }\frac{d^2q}{(2 \pi)^2}\frac{d^2q'}{(2 \pi)^2}
G^{-1}(q-q')Tr \left( \gamma^- \Sigma(P, q) \gamma^3 \Sigma(-P,q')
\right)
\end{split}
\end{equation}

The action \eqref{eammmm2} may be rewritten as
\begin{equation}\label{em3}
\begin{split}
S &=  \int \frac{d^3P}{(2\pi)^3} \frac{d^3q}{(2 \pi)^3}
{\bar \psi}(\frac{P}{2}-q) \left( (2 \pi)^3
\delta^3(P) i \gamma^\mu q_\mu  + \Sigma(-P, q) \right)\psi(\frac{P}{2}+q)   \\
&-\frac{N}{8 \pi i \lambda}
\int \frac{d^3P}{(2\pi)^3 }\frac{d^2q}{(2 \pi)^2}\frac{d^2q'}{(2 \pi)^2}
G^{-1}(q-q')Tr \left( \gamma^- \Sigma(P, q) \gamma^3 \Sigma(-P,q')
\right)
\end{split}
\end{equation}
The dependence of (\ref{em3}) on fermionic fields is quadratic, so the
later may be integrated out. Performing this operation yields
\begin{equation}\label{pfn}
Z= \int D \Sigma_- D\Sigma_I e^{-S}
\end{equation}
where
\begin{equation}\label{em4}
\begin{split}
S &=  - N {\rm Tr} \ln \left(  (2 \pi)^3
\delta^3(P) i \gamma^\mu q_\mu + \Sigma(-P, q) \right)   \\
&-\frac{N}{8 \pi i \lambda}
\int \frac{d^3P}{(2\pi)^3 }\frac{d^2q}{(2 \pi)^2}\frac{d^2q'}{(2 \pi)^2}
 G^{-1}(q-q')Tr \left( \gamma^- \Sigma(P, q) \gamma^3 \Sigma(-P,q')
\right)
\end{split}
\end{equation}
Notice that \eqref{em4} is written purely in terms of singlet fields, and
is multiplied by an overall factor of $N$. \eqref{em4} represents an exact
rewriting of the partition function of the original theory as a partition
function over the singlet fields $\Sigma$; this path integral is weakly
coupled in the large $N$ limit.

The action, \eqref{em4}, is somewhat formal, as it is written in terms
of a determinant over an infinite dimensional matrix. However the
equivalent of \eqref{em4} is much simpler for translationally
invariant $\Sigma$ configurations of the form
$$\Sigma(P, q) = (2 \pi)^3 \delta(P) \Sigma(q).$$
The form of this action is perhaps most clearly obtained by retreating
to \eqref{em3}, which reduces, for translationally $\Sigma$ configurations to
\begin{equation}\label{eammmm3}
S = \int \frac{d^3 p}{(2 \pi)^3}
 {\bar \psi}(-p)  \left[i \gamma^\mu p_\mu + \Sigma(p) \right]  \psi(p)
-\frac{N V}{8 \pi i \lambda}
\int \frac{d^2q}{(2 \pi)^2}\frac{d^2q'}{(2 \pi)^2}
G^{-1}(q-q')Tr \left( \gamma^- \Sigma(q) \gamma^3 \Sigma(q')
\right)
\end{equation}
$V$ here is a factor of the volume of spacetime, and we have used
$$ \left[ (2 \pi)^3 \delta(P) \right]^2 = V (2 \pi )^3 \delta(P)$$
in the last term of the last line.

Integrating out the fermions in \eqref{eammmm3} yields a very explicit
special case of \eqref{em4}
\begin{equation}\label{eammmmm}
S = -NV \int \frac{d^3 q}{(2 \pi)^3} {\rm Tr} \ln\left( i \gamma^\mu q_\mu +
\Sigma(p) \right) -\frac{NV}{8 \pi i \lambda}
\int \frac{d^2q}{(2 \pi)^2}\frac{d^2q'}{(2 \pi)^2}
G^{-1}(q-q')Tr \left( \gamma^- \Sigma(q) \gamma^3 \Sigma(q')
\right)
\end{equation}
While all terms in the action in \eqref{eammmmm} are proportional to $N$,
the fields in that action are gauge singlets. At leading order
in the large $N$ expansion it follows that the free
energy for our theory may evaluated simply by minimizing \eqref{eammmmm}
w.r.t. $\Sigma$. The variational equation we encounter in this minimization
process is
\begin{equation}\label{vareq}
\int
  {\rm Tr} \left[ \frac{d^2q}{(2 \pi)^2}
\delta \Sigma[q] \int \left(\frac{d^2q'}{(2 \pi)^2}
\frac{-1}{8 \pi i \lambda} G^{-1}(q-q')
\left( \gamma^3 \Sigma(q') \gamma^- -\gamma^- \Sigma(q') \gamma^3 \right)
-\int \frac{dq^3}{2 \pi} \frac{1}{i \gamma^\mu q_\mu + \Sigma}  \right)
\right]=0
\end{equation}
In terms of the function $H_{-}$ defined \eqref{mano}
\begin{equation}\label{vareqn}
\int
  {\rm Tr} \left[ \frac{d^2q}{(2 \pi)^2}
\delta \Sigma[q] \int \left(\frac{d^2q'}{(2 \pi)^2}
\frac{-1}{8 \pi i \lambda} G^{-1}(q-q')
\left( H_-\left(\Sigma(q')\right) \right)
-\int \frac{dq^3}{2 \pi} \frac{1}{i \gamma^\mu q_\mu + \Sigma}  \right)
\right]=0
\end{equation}

The equation \eqref{vareqn} is of the form
\begin{equation}\label{vareqn2}
\int
  {\rm Tr} \left[ \frac{d^2q}{(2 \pi)^2}
\delta \Sigma[q] B(q)
\right]=0
\end{equation}
where
$$B(q)=  \int \left(\frac{d^2q'}{(2 \pi)^2}
\frac{-1}{8 \pi i \lambda} G^{-1}(q-q')
\left( H_-\left(\Sigma(q')\right) \right)
-\int \frac{dq^3}{2 \pi} \frac{1}{i \gamma^\mu q_\mu + \Sigma}  \right) $$
As $\delta \Sigma$ is an arbitrary matrix of the form \eqref{defs}
it follows that
$$B_{-}(q)=B_I(q)=0$$
i.e. that
$$H_+(B)=0$$ (see \eqref{mana})
Using the fact that
$$H_+(H_-(\Sigma))=4 \Sigma$$
and integrating both sides of \eqref{vareqn} against the
kernel $\frac{1}{(p-q)^+}$ and  using the defining property of the function
$G^{-1}$, it follows from $H_+(B))=0$ that
\begin{equation}\label{vareqmm}
\Sigma(p)=-2 \pi i \lambda \int \frac{d^3q}{(2 \pi)^3}
\left( \gamma^3 \frac{1}{i \gamma^\mu q_\mu + \Sigma} \gamma^+
- \gamma^+ \frac{1}{i \gamma^\mu q_\mu + \Sigma} \gamma^3 \right)
\frac{1}{(p-q)^+}
\end{equation}
in precise agreement with  \eqref{gapsh}.

The value of the Euclidean action on the saddle point,
\eqref{eammmmm},  may be rewritten as
\begin{equation}\label{sf}
\begin{split}
S &= -NV \int \frac{d^3 q}{(2 \pi)^3} {\rm Tr} \ln\left( i \gamma^\mu q_\mu +
\Sigma(q) \right) +\frac{N V}{16 \pi i \lambda}
\int \frac{d^2q}{(2 \pi)^2}\frac{d^2q'}{(2 \pi)^2}
G^{-1}(q-q')Tr \left(H_{-}[\Sigma(q)] \Sigma(q') \right)\\
&=-NV \int \frac{d^3 q}{(2 \pi)^3} {\rm Tr}
\left[
\ln\left( i \gamma^\mu q_\mu +
\Sigma(q) \right) +\frac{1}{8}
H_{-}(\Sigma(q))
H_{+}\left( \left( \frac{1}{i \gamma^\mu q_\mu + \Sigma(q)} \right)
\right) \right]\\
&=-NV \int \frac{d^3 q}{(2 \pi)^3} {\rm Tr}
\left[
\ln\left[ i \gamma^\mu q_\mu +
\Sigma(q) \right] -\frac{1}{2} \Sigma(q)
 \left( \frac{1}{i \gamma^\mu q_\mu + \Sigma(q)} \right)  \right]
\\
\end{split}
\end{equation}
where we have used the equation of motion in going from first to
the second line. In going from the second to the third line we have
used the fact that for an arbitrary matrix $A$
$${\rm Tr} \left( H_{-}(\Sigma) H_{+}(A) \right)=-4 {\rm Tr}
\left( \Sigma A \right).$$

\subsubsection{Diagrammatic expansion for the vacuum energy}

The result (\ref{sf}), expressing the vacuum energy in terms of the exact fermion self-energy $\Sigma(p)$, has a clear diagrammatic interpretation. The perturbative planar diagrammatic expansion takes the form
\bigskip\bigskip

\centerline{
vacuum energy~~ =\begin{fmffile}{vac_1}
        \begin{tabular}{c}
            \begin{fmfgraph*}(40,40)
                \fmfsurroundn{w}{2}
                \fmf{plain,tension=0,right=1}{w1,w2,w1}
                \end{fmfgraph*}
        \end{tabular}
        \end{fmffile}
        + $1\over 2$
        \begin{fmffile}{vac_2}
        \begin{tabular}{c}
            \begin{fmfgraph*}(40,40)
                \fmfsurroundn{w}{2}
                \fmf{plain,tension=0,right=1}{w1,w2,w1}
                \fmf{wiggly,tension=0}{w1,w2}
             \end{fmfgraph*}
        \end{tabular}
        \end{fmffile}
        + ${1\over 2}$
        \begin{fmffile}{vac_3}
        \begin{tabular}{c}
            \begin{fmfgraph*}(40,40)
                \fmfsurroundn{w}{4}
                \fmf{plain,tension=0,right=1}{w1,w3,w1}
                \fmf{wiggly,tension=0,left=.3}{w1,w2}
                \fmf{wiggly,tension=0,left=.3}{w3,w4}
                \end{fmfgraph*}
        \end{tabular}
        \end{fmffile}
        + ${1\over 3}$
        \begin{fmffile}{vac_4}
        \begin{tabular}{c}
            \begin{fmfgraph*}(40,40)
                \fmfsurroundn{w}{12}
                \fmf{plain,tension=0,right=1}{w1,w7,w1}
                \fmf{wiggly,tension=0,left=.3}{w1,w4}
                \fmf{wiggly,tension=0,left=.3}{w5,w8}
                \fmf{wiggly,tension=0,left=.3}{w9,w12}
                \end{fmfgraph*}
        \end{tabular}
        \end{fmffile}
}
\centerline{
+ ${1\over 2}$\begin{fmffile}{vac_5}
        \begin{tabular}{c}
            \begin{fmfgraph*}(40,40)
                \fmfsurroundn{w}{12}
                \fmf{plain,tension=0,right=1}{w1,w7,w1}
                \fmf{wiggly,tension=0}{w1,w5}
                \fmf{wiggly,tension=0}{w12,w6}
                \fmf{wiggly,tension=0}{w11,w7}
                \end{fmfgraph*}
        \end{tabular}
        \end{fmffile}
        + $1\over 4$
        \begin{fmffile}{vac_6}
        \begin{tabular}{c}
            \begin{fmfgraph*}(40,40)
                \fmfsurroundn{w}{16}
                \fmf{plain,tension=0,right=1}{w1,w9,w1}
                \fmf{wiggly,tension=0,left=.3}{w1,w4}
                \fmf{wiggly,tension=0,left=.3}{w5,w8}
                \fmf{wiggly,tension=0,left=.3}{w9,w12}
                \fmf{wiggly,tension=0,left=.3}{w13,w16}
             \end{fmfgraph*}
        \end{tabular}
        \end{fmffile}
        +
        \begin{fmffile}{vac_7}
        \begin{tabular}{c}
            \begin{fmfgraph*}(40,40)
                \fmfsurroundn{w}{14}
                \fmf{plain,tension=0,right=1}{w1,w8,w1}
                \fmf{wiggly,tension=0,left=.3}{w1,w4}
                \fmf{wiggly,tension=0,left=.3}{w14,w5}
                \fmf{wiggly,tension=0,left=.3}{w6,w9}
                \fmf{wiggly,tension=0,left=.3}{w10,w13}
                \end{fmfgraph*}
        \end{tabular}
        \end{fmffile}
        + ${1\over 2}$
        \begin{fmffile}{vac_8}
        \begin{tabular}{c}
            \begin{fmfgraph*}(40,40)
                \fmfsurroundn{w}{16}
                \fmf{plain,tension=0,right=1}{w1,w9,w1}
                \fmf{wiggly,tension=0}{w1,w8}
                \fmf{wiggly,tension=0}{w2,w7}
                \fmf{wiggly,tension=0}{w16,w9}
                \fmf{wiggly,tension=0}{w15,w10}
                \end{fmfgraph*}
        \end{tabular}
        \end{fmffile}
        + $\cdots$
}
\bigskip
\noindent
One may verify that, in terms of exact planar fermion self energy, the above expansion can be written as

\bigskip
\centerline{
vacuum energy~~ =\begin{fmffile}{vac_1}
        \begin{tabular}{c}
            \begin{fmfgraph*}(40,40)
                \fmfsurroundn{w}{2}
                \fmf{plain,tension=0,right=1}{w1,w2,w1}
                \end{fmfgraph*}
        \end{tabular}
        \end{fmffile}
        $-$
        \begin{fmffile}{vac_9}
        \begin{tabular}{c}
            \begin{fmfgraph*}(40,40)
                \fmfsurroundn{w}{2}
                \fmf{plain,tension=0,right=1}{w1,w2,w1}
                \fmffixed{(.25w,0)}{v,w1}
                \fmfv{d.sh=circle,d.f=empty,d.si=.45w}{w1}
                \fmflabel{$\Sigma$}{v}
             \end{fmfgraph*}
        \end{tabular}
        \end{fmffile}
        + ${1\over 2}\!\!\!$
        \begin{fmffile}{vac_10}
        \begin{tabular}{c}
            \begin{fmfgraph*}(50,50)
                \fmfsurroundn{t}{2}
                \fmffixed{(0,0)}{w1,t1}
                \fmffixed{(-.225w,0)}{w2,t2}
                \fmf{plain,tension=0,right=1}{w1,w2,w1}
                \fmffixed{(.2w,0)}{v1,w1}
                \fmffixed{(-.2w,0)}{v2,w2}
                \fmfv{d.sh=circle,d.f=empty,d.si=.45w}{w1}
                \fmflabel{$\Sigma$}{v1}
                \fmfv{d.sh=circle,d.f=empty,d.si=.45w}{w2}
                \fmflabel{$\Sigma$}{v2}
                \end{fmfgraph*}
        \end{tabular}
        \end{fmffile}
        }
\centerline{
        $-$ ${1\over 3}\!\!\!$
        \begin{fmffile}{vac_11}
        \begin{tabular}{c}
            \begin{fmfgraph*}(50,50)
                \fmfsurroundn{t}{3}
                \fmfsurroundn{u}{2}
                \fmffixed{(0,0)}{w1,t1}
                \fmffixed{(-.225w,0)}{w2,t2}
                \fmffixed{(0,-.16w)}{w3,t3}
                \fmf{plain,tension=0,right=1}{u1,u2,u1}
                \fmffixed{(.2w,0)}{v1,w1}
                \fmffixed{(-0.01w,0.18w)}{v2,w2}
                \fmffixed{(-0.16w,-0.17w)}{v3,w3}
                \fmfv{d.sh=circle,d.f=empty,d.si=.45w}{w1}
                \fmflabel{$\Sigma$}{v1}
                \fmfv{d.sh=circle,d.f=empty,d.si=.45w}{w2}
                \fmflabel{$\Sigma$}{v2}
                \fmfv{d.sh=circle,d.f=empty,d.si=.45w}{w3}
                \fmflabel{$\Sigma$}{v3}
                \end{fmfgraph*}
        \end{tabular}
        \end{fmffile}
       ~~ $+$~ $\cdots\cdots$
~~ $+$ ${1\over 2}\!\!\!$
        \begin{fmffile}{vac_12}
        \begin{tabular}{c}
            \begin{fmfgraph*}(50,50)
                \fmfsurroundn{t}{2}
                \fmffixed{(0,0)}{w1,t1}
                \fmffixed{(-.225w,0)}{w2,t2}
                \fmf{plain,tension=0,right=1}{w1,w2,w1}
                \fmffixed{(.2w,0)}{v1,w1}
                \fmffixed{(-.2w,0)}{v2,w2}
                \fmfv{d.sh=circle,d.f=shaded,d.si=.45w}{w1}
                \fmfv{d.sh=circle,d.f=empty,d.si=.45w}{w2}
                \fmflabel{$\Sigma$}{v2}
                \end{fmfgraph*}
        \end{tabular}
        \end{fmffile}
 }
\bigskip
\noindent
where the shaded blob represents the exact planar fermion propagator $( i{\slash\!\!\! q}+\Sigma)^{-1}$.
In particular, the second term in the last line of (\ref{sf}), $-{1\over 2} {\rm Tr}[\Sigma( i{\slash\!\!\! q}+\Sigma)^{-1}]$ is precisely such that the correct symmetry factors are restored.

\subsection{The finite temperature theory}

In this section we study the logarithm of the path integral of our system on
$R^{2} \times S^1$ where the circumference of the $S^1$ is taken to
be $\beta$.  This path integral determines the free energy of
the field theory at temperature $T= \beta^{-1}$.

The formulas that determine the path integral on $\mathbb{R}^2 \times S^1$
are straightforward generalizations of the formulas on $\mathbb{R}^3$.
Every equation in subsection \ref{gesp} carries through
with the replacement
$$\int \frac{dp_3}{(2 \pi)} f(p_3) \rightarrow \frac{1} {\beta}
\sum_{n \in Z+\frac{1}{2}} f(\frac{2 \pi n}{\beta})$$
$$ V \rightarrow V_2 \beta $$
so that
$$ V \int \frac{d^3p}{(2 \pi)^3} \rightarrow V_2  \int \frac{d^2p}{(2 \pi)^2}
\sum_{n}.$$
In particular the Euclidean action is given by
\begin{equation}\label{sft}
S=NV_2 \sum_{n} \int \frac{d^2 q}{(2 \pi)^2} {\rm Tr}
\left[
\ln\left[ i \gamma^\mu q_\mu + \Sigma_T(q) \right] -\frac{1}{2} \Sigma_T(q)
 \left( \frac{1}{i \gamma^\mu q_\mu + \Sigma_T(q)} \right)  \right]
\end{equation}
where $T$, the temperature is $\beta^{-1}$ and
the function $\Sigma_T(q)$ obeys the gap equation
\begin{equation}\label{get}
\Sigma_T(p)=-2 \pi i \lambda ~ \frac{1}{\beta}\sum_{n}
\int \frac{d^2q}{(2 \pi)^2}
\left( \gamma^3 \frac{1}{i \gamma^\mu q_\mu + \Sigma_T(q)}
\gamma^+
- \gamma^+ \frac{1}{i \gamma^\mu q_\mu + \Sigma_T(q)} \gamma^3 \right)
\frac{1}{(p-q)^+}
\end{equation}
where $n$ is an integer and
$$q^3= \frac{2 \pi (n+ \frac{1}{2})}{\beta}.$$

In order to determine the free energy at finite temperature $T$, we need to
solve the gap equation \eqref{get} and plug the solution into
\eqref{sft}. We take up these exercises in turn.

\subsubsection{The finite temperature gap equation}

As in subsection \ref{ge} it follows immediately that $\Sigma_T$ is
a linear combination of $\gamma^+$ and $I$, and that it is
independent of $p_3$. Rotational symmetry and the constraints of
conformality then imply
$$\Sigma_T(p)+M_{bare}I=f( \beta p_s ) p_s I
+ i g(\beta p_s ) p^- \gamma^+ $$
for some as yet unknown functions $f(\beta p_s)$ and $g(\beta p_s)$.
Note that the new dimensionful scale $\beta$, now allows $f$ and $g$ to
 be functions of $p_s$, generalizing the pure numbers $f_0$ and $g_0$
of the previous section.
The zero temperature results of the previous subsection imply that
\begin{equation}\label{ly1}
\begin{split}
\lim_{y \to \infty} f(y)&= f_0=\lambda\\
\lim_{y \to \infty} g(y)&=g_0=-\lambda^2
\end{split}
\end{equation}

The analogue of \eqref{speq} is
\begin{equation} \label{feq}
p_s f(p_s \beta) -M_{bare}= 4 \pi \frac{\lambda}{\beta}
\int \sum_n \frac{d^{2+ \epsilon}q }{(2 \pi)^2}
\frac{q^+}{ \left( \frac{2 \pi (n+\frac{1}{2})}{\beta} \right)^2
+ q_s^2(1+g(q_s \beta) +|f(q_s \beta)|^2)} \frac{1}{(p-q)^+}
\end{equation}
\begin{equation} \label{geq}
g(p_s \beta) p^- = -4 \pi \frac{\lambda}{\beta}
\int \sum_n \frac{d^{2+\epsilon}q}{(2 \pi)^2} q_s
\frac{f(q_S \beta)}{ \left( \frac{2 \pi \left(n + \frac{1}{2} \right)}
{\beta} \right)^2
+ q_s^2(1+g +f^2(q_s\beta))}  \frac{1}{(p-q)^+}
\end{equation}
The summations in these equations are easily carried out using the formula
\begin{equation}\label{form}
\int dq^{\epsilon}
\sum_{n= -\infty}^\infty \frac{1}{\left(n+\frac{1}{2} \right)^2+a^2 + q^2}=
\frac{\pi}{|a|^{1-\epsilon}} \tanh(\pi |a|)
\end{equation}
(this is the analogue of \eqref{intf} in the previous section - as in the
previous section we have set $\epsilon$ to zero in every place where it
will be inessential for regularization) yielding
\begin{equation} \label{feq2}
f p_s -M_{bare}={2 \pi \lambda}
\int \frac{d^2q q_s^{-\epsilon}}{(2 \pi)^2}
  \tanh \left( \frac{\beta q_s }{2}\sqrt{1+g+f^2} \right)
\frac{q^{+}}{q_s(p-q)^{+}\sqrt{1+g+f^2}}
\end{equation}
\begin{equation} \label{geq2}
g p^- = -2 \pi \lambda
\int \frac{d^2q q^{-\epsilon}}{(2 \pi)^2}
\tanh \left( \frac{\beta q_s}{2}\sqrt{1+g+f^2} \right)
\frac{f}{\sqrt{1+g +f^2}}
\frac{1}{(p-q)^+}
\end{equation}
where we have left implicit the fact that the $f$ and $g$ are functions of
$p_s$ on the LHS of \eqref{feq2}\eqref{geq2}, but are functions of $q_s$ on the
RHS of the same equations.

In each of \eqref{feq2} and \eqref{geq2}  we move to
polar coordinates and use use \eqref{te} to perform the angular integrals
to obtain
\begin{equation} \label{feqb}
f p_s =-\lambda
\int_p^\infty q^{-\epsilon}
d q_s   \tanh \left( \frac{\beta q}{2}\sqrt{1+g+f^2} \right)
\frac{1}{\sqrt{1+g+f^2}}
\end{equation}
\begin{equation} \label{geqb}
g =  -2\lambda
\int_0^p \frac{q_s^{1-\epsilon} d q_s}{p^2_s}
\tanh \left( \frac{\beta q_s}{2}\sqrt{1+g+f^2} \right)
\frac{f}{\sqrt{1+g +f^2}}
\end{equation}
Adding and subtracting $q^{-\epsilon}$ from the integrand of \eqref{feqb}
and doing the integral on the trivial piece we find
 \begin{equation} \label{feqd}
f =\lambda
- {\lambda} \int_p^\infty \frac{d q_s}{p_s}   \left(
\frac{\tanh \left( \frac{\beta q_s}{2}\sqrt{1+g+f^2} \right) }
{\sqrt{1+g+f^2}} - 1
\right)
\end{equation}
\begin{equation} \label{geqd}
g =- 2 \lambda
\int_0^p \frac{q_s d q_s}{p_s^2}
\frac{\tanh \left( \frac{\beta q_s}{2}\sqrt{1+g+f^2} \right)f }
{\sqrt{1+g +f^2}}
\end{equation}
In terms of the variable
$$x=\frac{q}{p}.$$
\begin{equation} \label{feqe}
f(y)=\lambda
- {\lambda} \int_1^\infty d x
\left( \frac{\tanh \left( \frac{y x}{2}\sqrt{1+g(y x) +f(y x)^2}
\right) }{\sqrt{1+g(y x)+f(y x)^2}} - 1
\right)
\end{equation}
\begin{equation} \label{geqe}
g(y) = -2  {\lambda}
\int_0^1 x dx
\left(  \frac{ \tanh \left( \frac{yx}{2}\sqrt{1+g(yx)+f(yx)^2} \right) f(yx) }
{\sqrt{1+g(yx) +f(yx)^2}} \right)
\end{equation}
where the variable
$$y =p \beta.$$
Equivalently
\begin{equation} \label{fgeqen} \begin{split}
f(y)&=\lambda
- \frac{\lambda}{y} \int_y^\infty d x
\left(  \frac{ \tanh \left( \frac{x}{2}\sqrt{1+g(x) +f(x)^2} \right)}
{\sqrt{1+g(x)+f(x)^2}} - 1 \right)\\
g(y) &= -2  \frac{\lambda}{y^2}
\int_0^y x dx  \frac{\tanh \left( \frac{x}{2}\sqrt{1+g(x)+f(x)^2}  \right)
f(x)}{\sqrt{1+g(x) +f(x)^2}}
\end{split}
\end{equation}

\subsubsection{The exact solution}
Quite remarkably it is possible to find the exact solution to
\eqref{fgeqen}. We start
with \eqref{fgeqen} written in the form
\begin{equation} \label{fgeqenn} \begin{split}
y \left( f(y)-f_0 \right)
&=-{\lambda} \int_y^\infty d x
\left( \frac{\tanh \left( \frac{x}{2}\sqrt{1+g(x) +f(x)^2}
\right)}{\sqrt{1+g(x)+f(x)^2}} - 1
\right)\\
y^2 g(y) &= -2  {\lambda}
\int_0^y x dx
\frac{\tanh \left( \frac{x}{2}\sqrt{1+g(x)+f(x)^2} \right)f(x)}
{\sqrt{1+g(x) +f(x)^2}}.
\end{split}
\end{equation}
Differentiating both equations w.r.t. $y$ we obtain
\begin{equation} \begin{split}
\label{fg}
& y f'(y) + f(y) = \lambda {\tanh ({y\over 2} \sqrt{1+g(y)+f(y)^2})\over \sqrt{1+g+f^2}},
\\
& y g'(y) + 2g(y)=
-2\lambda f(y) {\tanh ({y\over 2} \sqrt{1+g(y)+f(y)^2})\over \sqrt{1+g+f^2}}.
\end{split}
\end{equation}

Multiplying the first equation by $2f$ and add it to the second equation, we cancel the RHS and obtain
\begin{equation}\label{fgn}
y {d\over dy}(g+f^2) + 2(g+f^2) = 0.
\end{equation}
From this we solve
\begin{equation}\label{sss}
g(y) + f(y)^2 = {c\over y^2},
\end{equation}
where $c$ is a constant. Now the first equation in (\ref{fg}) becomes simply
\begin{equation}
y f'(y) + f(y) = \lambda {\tanh ({y\over 2}
\sqrt{1+{c\over y^2}})\over \sqrt{1+{c\over y^2}}}
\end{equation}

Integrating we have
\begin{equation}\begin{split}\label{actsol}
& f(y) = {\lambda\over y} \int_0^y dz {\tanh ({z\over 2}
\sqrt{1+{c\over z^2}})\over \sqrt{1+{c\over z^2}}} + {\tilde c\over y}
= \frac{2 \lambda}{y}
\ln \left( \frac{\cosh\left[\frac{1}{2} \sqrt{c + y^2} \right]}
{\cosh\left[\frac{\sqrt{c}}{2}  \right]} \right) +{ \tilde c \over y}\\
& g(y) = {c\over y^2} - f(y)^2.
\end{split}
\end{equation}
Here $\tilde c$ is another integration constant.

While the solution to a differential equation depends on integration
constants, the solution to an integral equation is unique (it does not
have undetermined integration constants). The appearance of $c$ and
${\tilde c}$ in our solutions above is an artifact of our having solved
by converting the integral equation into a differential equation.
The integral equations \eqref{fgeqen} are actually solved by \eqref{actsol}
only for a particular choice of $c$ and ${\tilde c}$.

We first note that the function $f$ must tend to $f_0$ at large $y$.
This is automatic in all our solutions (it does not impose
any constraints on $c$ or ${\tilde c}$). However a further requirement
is that the expansion of $f$ about this constant value (at large $y$)
should start at $\frac{1}{y^2}$ rather than $\frac{1}{y}$
(this follows immediately 
upon plugging \eqref{sss} into the RHS of the first of \eqref{fgeqenn}; 
the RHS of that equation is manifestly $\propto \frac{1}{y}$ ).
The requirement that
\begin{equation}\label{req}
f(y)=f_0+{\cal O}(1/y^2)
\end{equation}
determines
\begin{equation}\label{cans}
\tilde c =  2 \lambda
\ln \left(2 \cosh \frac{\sqrt{c}}{2}  \right).
\end{equation}
Let us now turn to the small $y$ behaviour of $f$ and $g$. At small $y$
\begin{equation}  \label{sybehav}
\begin{split}
f(y)&=\frac{{\tilde c}}{y} +{\cal O}(y), \\
g(y)&=\frac{c-{\tilde c}^2}{y^2} +{\cal O}(y^0). \\
\end{split}
\end{equation}
Plugging \eqref{sss} into the RHS of the second of \eqref{fgeqenn},
however, we find that the RHS evaluates to ${\cal O}(y^2)$ at small $y$ 
implying that $g(y)={\cal O}(y^0)$ at small $y$. It follows that  
\begin{equation}\label{scc}
c= {\tilde c}^2.
\end{equation}
Plugging this relation into \eqref{cans} yields the following
equation for ${\tilde c}$
\begin{equation}\label{cans2}\begin{split}
\frac{\tilde c}{2 \lambda} &=
\ln \left(2 \cosh \frac{\tilde c}{2}  \right). \\
\end{split}
\end{equation}
Note that ${\tilde c}$ is an odd function of $\lambda$.   
$|{\tilde c}|$ is a monotonically increasing function of $\lambda$, which
diverges at $|\lambda|=1$. \eqref{cans2} has no solution for $|\lambda|>1$,
indicating that the theory does not exist for $|k| >N$.
At leading order in at small $\lambda$ we have
\ie
{\tilde c}=2 \lambda \ln 2 +{\cal O}(\lambda^3).
\fe
As $\lambda$ approaches unity we have
\ie\label{lambdaunityexp}
{\tilde c}= \ln \frac{2}{(1-\lambda)}  -\ln \left(
\ln \frac{2}{1-\lambda} \right)
+ {\cal O}\left(\ln \ln \ln \frac{2}{1-\lambda} \right).
\fe
In order to physically interpret the divergence of ${\tilde c}$
as $\lambda \to 1$ note that the exact thermal propagator has a pole
whenever
$$p^2+ {\tilde c}^2 T^2=0.$$
In other words ${\tilde c}$ has a simple physical interpretation; it is
the thermal mass of the field $\psi$ in units of the temperature. It
follows that the fermion thermal mass diverges in this limit $\lambda \to 1$.

Using \eqref{cans2} and \eqref{scc} we may rewrite our solutions for
$f$ and $g$ as
\begin{equation}\begin{split}\label{actsol2}
& f(y) = \frac{2 \lambda}{y}
\ln \left( 2 \cosh{\sqrt{{\tilde c}^2 + y^2}\over 2} \right), \\
& g(y) = {{\tilde c}^2\over y^2} - f(y)^2.
\end{split}
\end{equation}
with ${\tilde c}$ given by \eqref{cans2}.
In the large $y$ limit
\begin{equation}\label{ly}\begin{split}
f(y)&
=\lambda \sqrt{1+\frac{{\tilde c}^2}{y^2}} +{\cal O}(e^{-y}), \\
g(y)&=-\lambda^2 + \frac{{\tilde c}^2 \left(1-\lambda^2 \right)}{y^2}
+{\cal O}(e^{-y}), \\
\end{split}
\end{equation}
while at small $y$
\begin{equation}\label{sy}\begin{split}
f(y)&= \frac{\tilde c}{y} +  \frac{\lambda y}{2 {\tilde c} }
\tanh \left( \frac{{\tilde c}}{2} \right)
+{\cal O}(y^3), \\
g(y)&= -\lambda \tanh \left( \frac{{\tilde c}}{2} \right)
+{\cal O}(y^2).
\end{split}
\end{equation}

\subsubsection{Free energy as a function of temperature}

As we have explained above, the path integral of our theory on the manifold
$\mathbb{R}^2 \times S^1$, with the circumference of the $S^1$ equal to $\beta$,
is given by $e^{-S_T}$ where
\begin{equation}\label{sfta}
S_T 
=NV_2 \sum_{n} \int \frac{d^2 q}{(2 \pi)^2} {\rm Tr}
\left[
\ln\left[ i \gamma^\mu q_\mu + \Sigma_T(q) \right] -\frac{1}{2} \Sigma_T(q)
 \left( \frac{1}{i \gamma^\mu q_\mu + \Sigma_T(q)} \right)  \right].
\end{equation}
Let us define
\begin{equation}\label{sfn}
S_0 
=-NV_2 \beta \int \frac{d^3 q}{(2 \pi)^3} {\rm Tr}
\left[
\ln\left[ i \gamma^\mu q_\mu +
\Sigma(q) \right] -\frac{1}{2} \Sigma(q)
 \left( \frac{1}{i \gamma^\mu q_\mu + \Sigma(q)} \right)  \right].
\end{equation}
Then the partition function
$$Z={\rm Tr} e^{-\beta H}$$
of our system in a flat spatial box of volume
$V_2$ is given by
$$\ln Z=S_0 -S_T,$$
so that the finite temperature free energy, $F(T)$, of the theory is given
by
$$F(T)=\frac{S_T-S_0}{\beta}.$$
We will now proceed to use our exact solution to the finite temperature
gap equation to compute $S_T-S_0$.

\subsubsection{Explicit evaluation of the free energy}

\begin{figure}
\begin{center}
\includegraphics[width=100mm]{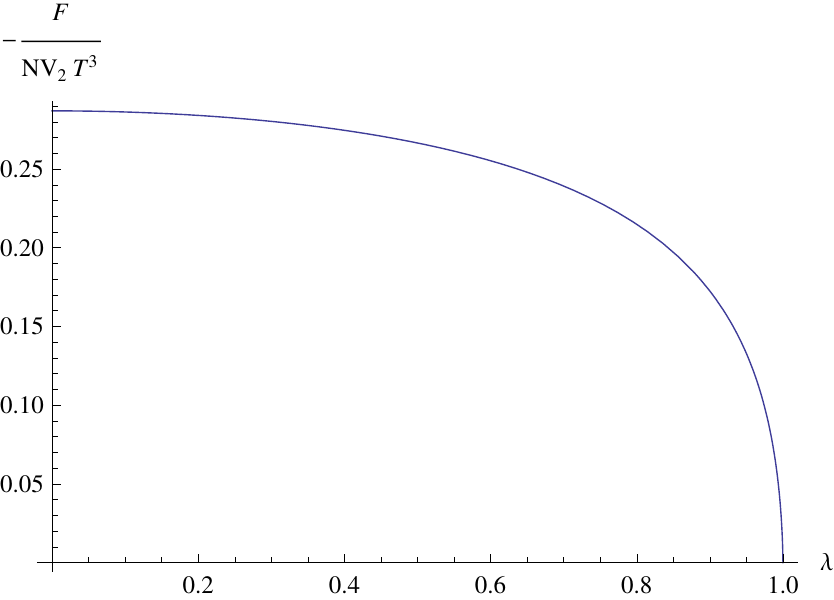}
\parbox{13cm}{\caption{The free energy on $S^1\times \mathbb{R}^2$ as a function of $\lambda$.}
\label{plotF}}
\end{center}
\end{figure}

In order to find an explicit expression for the free energy,
we find it convenient to use the expressions in the second line
of \eqref{sfta} and \eqref{sfn}. $S_T-S_0$ may be written as
\begin{equation}\label{sfst} \begin{split}
& S_T-S_0= -NV_2 \sum_{n} \int \frac{d^2 q}{(2 \pi)^2} {\rm Tr}
\ln\left( i \gamma^\mu q_\mu+\Sigma_T(q) \right)
+ NV_2 \int \frac{d^3 q}{(2 \pi)^3} {\rm Tr}
\ln\left( i \gamma^\mu q_\mu + \Sigma_T(q)  \right) \\
&- NV_2 \beta \int \frac{d^3 q}{(2 \pi)^3}
{\rm Tr}
\ln \frac{\left( i \gamma^\mu q_\mu +\Sigma_T(q)  \right)}
{\left( i \gamma^\mu q_\mu +\Sigma(q)  \right)} \\
&+\frac{N V_2}{2} \sum_n \int \frac{d^2 q}{(2 \pi)^2} {\rm Tr}
 \left( \frac{ q_s f(\beta q_s) + i g(\beta q_s) q_+ \gamma^+}{i \gamma^\mu q_\mu + \Sigma_T(q)} \right)
- \frac{NV_2 \beta}{2} \int \frac{d^3 q}{(2 \pi)^3} {\rm Tr}
 \left( \frac{f_0 q_s+ i g_0 p_{+} \gamma^{+}}{i \gamma^\mu q_\mu + \Sigma(q)} \right).
\end{split}
\end{equation}
The integral on the first line of \eqref{sfst} is convergent and
evaluates to
\begin{equation}\label{firstline}\begin{split}
&-NV_2 \sum_{n} \int \frac{d^2 q}{(2 \pi)^2}
\ln\left(q_3^2 +q_s^2 + {\tilde c}^2 T^2  \right)
+ NV_2 \beta  \int \frac{d^3 q}{(2 \pi)^3}
\ln\left(q_3^2+q_s^2 + {\tilde c}^2 T^2 \right)\\
&=-2 N V _2\int \frac{d^2 q}{(2 \pi)^2}
\ln \left ( 1+e^{-\beta\sqrt{{\tilde c}^2 T^2 + q_s^2}} \right)\\
&=-2 N T^2 V _2\int \frac{d^2 x}{(2 \pi)^2}
\ln \left ( 1+e^{-\sqrt{{\tilde c}^2  +x_s^2}} \right)\\
&=-\frac{2 N T^2 V _2}{2 \pi} \int_{|\tilde c|}^\infty  dy ~ y
\ln \left ( 1+e^{-y} \right).
\end{split}
\end{equation}
The second line of \eqref{sfst} has a linear divergence, which
however disappears in our dimensional reduction scheme
\begin{equation}\label{secondline}\begin{split}
& -NV_2\beta  \int \frac{d^{3-\epsilon} q}{(2 \pi)^3}
\ln\frac{\left(q^2+ {\tilde c}^2 T^2 \right)}
{\left(q^2 \right)} \\
&= -NV_2T^2 |{\tilde c}|^3
\int \frac{d^{3-\epsilon} y}{(2 \pi)^3}
\ln\frac{y^2+ 1}
{y^2}  \\
&= -\frac{NV_2T^2 |{\tilde c}|^3}{2 \pi^2}
\int_0^\infty dy
y^{2-\epsilon} \ln\left( 1+\frac{1}{y^2} \right)=
\frac{NV_2T^2 |{\tilde c}|^3}{6 \pi}, \\
\end{split}
\end{equation}
where in the last step we integrated by parts and used
$$\int_0^\infty \frac{y^{2-\epsilon}}{1+y^2}=-\frac{\pi}{2}$$
in the limit of small $\epsilon$.
Let us now turn to the third line of \eqref{sfst}. The second
term in the third line simply vanishes under dimensional regularization.
The first term in the third line may be evaluated as follows
\begin{equation}\label{fttl}\begin{split}
&\frac{N V_2}{2} \sum_n \int \frac{d^2 q}{(2 \pi)^2}
\, \frac{ 2 q^2_s f^2(\beta q_s) + q_s^2 g(\beta q_s)}
{q_3^2+q_s^2 +  T^2 {\tilde c}^2} 
\\
&
=\frac{N V_2}{4} \int \frac{d^2 q}{(2 \pi)^2}\,
\frac{ \left(2 q^2_s f^2(\beta q_s) + q_s^2 g(\beta q_s) \right) \tanh
\frac{\sqrt{q^2+{\tilde c}^2 }}{2}}
{\sqrt{{q_s^2 +  T^2 {\tilde c}^2}}} \\
&=\frac{N V_2 T^2}{4} \left[\int \frac{d^2 q}{(2 \pi)^2}
\, \frac{ \left(2 q^2_s f^2(q_s) + q_s^2 g(q_s) \right) \tanh
\frac{\sqrt{q^2+{\tilde c}^2 }}{2}
-\lambda^2 \left({\tilde c}^2 + q^2 \right) -{\tilde c}^2}
{\sqrt{{q_s^2 +  {\tilde c}^2}}}
\right.
\\
&\left. ~~~~~+\int \frac{d^2 q}{(2 \pi)^2}
 \frac{\lambda^2 \left({\tilde c}^2 + q^2 \right)+ {\tilde c}^2}
{\sqrt{{q_s^2 +  {\tilde c}^2}}} \right].
\end{split}
\end{equation}
The first term in \eqref{fttl} is finite and evaluates to
\begin{equation}\label{fty}
\frac{N V_2 T^2}{4\pi} \,|{\tilde c}|^3
\left( \frac{\lambda^2}{6}-\frac{1}{6 |\lambda|} -\frac{1}{2|\lambda|}
+ \frac{1}{2} \right). 
\end{equation}
The second term in \eqref{fttl} is divergent. In dimensional regularization
it evaluates to
\begin{equation}\label{sty}
-\frac{N V_2 T^2|\tilde c|^3}{4\pi} \left(\frac{\lambda^2}{6} + \frac{1}{2} \right).
\end{equation}
The third line of \eqref{sfst} is given by the sum of \eqref{fty} 
and \eqref{sty} and evaluates to 
\begin{equation}\label{tlr}
-\frac{N V_2 T^2 |\tilde c|^3 }{6\pi |\lambda|}.
\end{equation}
Putting it all together we find that the free energy is given by
\begin{equation}\label{fe}
F=-\frac{N V_2 T^3 }{6 \pi} \left[ {|\tilde c|}^3
\frac{1-|\lambda|}{|\lambda|}
+ 6 \int_{|\tilde c|}^\infty  dy ~ y
\ln \left ( 1+e^{-y} \right) \right]\,.
\end{equation}
Although it is not manifest, the free energy can also be written in the form \eqref{feint}, and is an analytic function of $\lambda$ in the interval 
$(-1,1)$.\footnote{Theories with massless bosons usually do not have an analytic  expansion of their
 free energy in terms of the coupling constant. Such non-analytic behavior in Chern-Simons-matter theories with massless bosons was observed explicitly e.g. in \cite{Gaiotto:2007qi}, \cite{Smedback:2010ji}. This non-analyticity has its origin in infrared divergences which are absent in our theory because the only propagating degrees of freedom are the fermions which lack a zero-mode along the thermal circle.} To see this more explicitly, note that we can write the term in square brackets in \eqref{fe} as 
\begin{equation}
\begin{aligned}
&\frac{{|\tilde c|}^3}{|\lambda|}-|\tilde c|^3+6\int_0^{\infty}dy ~ y \ln \left ( 1+e^{-y} \right)
+6\int_0^{|\tilde c|}dy \left(\frac{y^2}{2}-y \ln \left(2\cosh\frac{y}{2}\right)\right)\\
&=\frac{9}{2}\zeta(3)+\frac{{\tilde c}^3}{\lambda}-6\int_0^{|\tilde c|}dy ~ y \ln \left(2\cosh\frac{y}{2}\right),
\end{aligned}
\end{equation}
where in the last line we used $|\tilde c|^3/|\lambda|={\tilde c}^3/\lambda$, as follows from \eqref{cans2}. Note that the integral in the last term is clearly an even function of $|\tilde c|$, and so the absolute value can be omitted. This shows that one may effectively rewrite \eqref{fe} as in \eqref{feint}, which is manifestly analytic. Indeed its small $\lambda$ expansion is given by 
\begin{equation}
F=-N V_2 T^3 \left[\frac{3 \zeta(3)}{4 \pi} - \frac{2 (\log2)^3}{3\pi} \lambda^2
-\frac{(\log2)^4}{2\pi} \lambda^4+{\cal O}(\lambda^6) \right].
\label{F-pert}
\end{equation}
Note that it contains only even powers of $\lambda$, consistently with parity.
 A plot of the free energy as a function of $\lambda$ is given in Fig. \ref{plotF}.
We see that $-F$ decreases monotonically from the free field value $-F=\frac{3N V_2 T^3 }{4 \pi}\zeta(3)$ to zero at $\lambda=1$.

In the limit that $|\lambda| \to 1$, $|{\tilde c}|$ is large and the integral
in \eqref{fe} may be approximated by
$$\int_{|{\tilde c}|}^\infty  dy ~ y  \ln \left ( 1+e^{-y} \right)
={|{\tilde c}|} e^{- |{\tilde c}|} + e^{-|{\tilde c}|} +{\cal O}(e^{-2 {\tilde c}|}).$$
At leading nontrivial order this evaluates to
$${\tilde c}^2 \frac{1-|\lambda|}{2},$$
and the free energy is given by
\begin{equation}\label{felargec}
F=-\frac{N V_2 T^3 (1-|\lambda|)}{6 \pi} \left[ {|{\tilde c}|}^3
 + 3 {\tilde c}^2 + {\cal O}(\tilde c) \right]\,,\qquad |\lambda| \rightarrow 1\,,
\end{equation}
where ${\tilde c}$ is given by \eqref{lambdaunityexp}.

\subsection{Consistency of our gauge and regularization scheme}

As we have emphasized on multiple occasions, we have obtained the beautiful
result \eqref{fe} by employing a rather unusual gauge (a lightcone like
gauge in Euclidean space) and the regularization scheme of dimensional
reduction, which we have assumed preserves the gauge invariance of our theory.
In this section we list the independent evidence that the procedure
employed in this section defines a sensible and Lorentz invariant theory.

\begin{enumerate}
\item The exact fermion propagator \eqref{aef} develops poles only when
$p^2=0$, a condition that is Lorentz invariant. This is a necessary condition
for the Lorentz invariance of fermion scattering processes in our theory.
\footnote{Particle scattering is, strictly speaking, not well defined in 
our theory (or any conformal theory) due to infrared divergences. We 
could cure these divergences in our theory softly breaking conformal invariance
with a fermion mass. Scattering processes in this deformed theory are 
presumably well defined. The poles of the mass deformed theory are clearly 
physical and must be Lorentz invariant. As our computation of the 
propagator of massless theory, presented in this paper, exhibits no 
IR divergences, it must equal the zero mass limit of the mass deformed 
propagator, and hence must be Lorentz invariant.}
The Lorentz invariance
of the poles of the propagator is far from automatic, and is in fact
violated in several regularization schemes. In, for instance, the regularization
scheme described just below \eqref{mfs} we find $g_0=0$ but $f_0 \neq 0$
(in fact $f_0$ obeys the equation $f_0=-\frac{\lambda}{\sqrt{1+f_0^2}}$
with this regulator). The poles of the fermion propagator with this regulator
occur  at $p_3^2+ p_s^2(1+f_0)$ and are not Lorentz invariant.
\item In Appendix \ref{rotational-invariance-appendix} we have demonstrated that the expectation value,
at one loop,  of the gauge invariant Wilson line operator is Lorentz
invariant and moreover agrees with the result obtained in the manifestly 
Lorentz invariant Feynman gauge. 
\item Our results above indicate that our theory is infinitely strongly
coupled at $\lambda=1$, and that the theory does not exist (at least as
a conformally invariant theory) for $\lambda>1$. 
There is in fact a very simple interpretation of this result. It is well-known that the bare Chern-Simons level 
can acquire a finite shift at 1-loop order in perturbation theory. This effect is regularization dependent. 
It was demonstrated by \cite{Chen:1992ee} that the Chern-Simons matter theory defined using the dimensional reduction
scheme, employed in this paper, does not acquire a 1-loop shift of $k$. On the other hand, 
if the theory is regulated by the addition of a small Yang Mills term, the bare level, which we denote $k_{YM}$, gets 
shifted by $sign(k_{YM})\,N$. Therefore the results of \cite{Chen:1992ee} imply that the two regularizations yield the same physical theory provided $$|k|=|k_{YM}|+N.$$
Let us define $\lambda= \frac{N}{k}$ as we have done in this paper,
and $|\lambda_{YM}|=\frac{N}{|k_{YM}|}$. It follows that
$$|\lambda_{YM}|=\frac{|\lambda|}{1-|\lambda|}$$
and
$$|\lambda|=\frac{|\lambda_{YM}|}{1+|\lambda_{YM}|}$$
In particular as $|\lambda| \to 1$, we have $|\lambda_{YM}| \to \infty$.
Now a Chern-Simons theory regulated by the addition of a small Yang Mills
term clearly exists at all values of $|\lambda_{YM}|$. However it follows
from the discussion above that this only requires the dimensionally regulated
theory to exist for $|\lambda| \leq 1$. Moreover the limit $|\lambda| \to 1$
should be interpreted as the approach to strong coupling. Our exact
result for the finite temperature free energy is perfectly consistent
with this interpretation. The theory ceases to exist for $|\lambda| >1$.
Moreover the limit $|\lambda| \to 1$ displays the extreme thinning of
degrees of freedom that is plausible in an extreme strong coupling limit.\footnote{Note that (\ref{felargec}) expressed in terms of $k_{YM}$ takes the form
$$
F = - {|k_{YM}| V_2T^3\over 6\pi} \left(\ln {N\over |k_{YM}|}\right)^3 + \cdots, ~~~~\lambda \to 1.
$$
This behavior in the $\lambda\to 1$ limit could be consistent with a weakly coupled boson description (see \cite{Gaiotto:2007qi}).}

\item In Appendix \ref{two-loop-anomalous-dimension-appendix} we have explicitly computed the two loop
anomalous dimension, at first subleading order in $\frac{1}{N}$,
of the scalar operator ${\bar \psi}\psi$. The result so obtained
agrees perfectly with the result of the same computation performed in
Feynman gauge.
\end{enumerate}

The points listed above make it at least plausible that our gauge choice is
free of problems, and that our regularization scheme preserves gauge invariance
and so defines a Lorentz invariant theory. While the direct evidence for these
claims is substantial, it is not overwhelming. Significant
additional evidence for the Lorentz invariance of our final theory
could be obtained by demonstrating the Lorentz invariance of four Fermi
scattering at one loop. We will not, however, attempt that consistency
check in this paper.

\section{Operator Spectrum}
\label{operator-section}
In this section, we study the spectrum of gauge invariant
operators in our theory. We focus mainly on operators whose dimension
is held fixed as $N$ is taken to infinity. By large $N$ factorization,
all such operators are products of ``single trace'' operators. By a single
trace we mean an operator that is constructed from the contraction of a
fundamental (derivative of $\psi$) and an antifundamental (derivative
of ${\bar \psi}$). We will argue below that the spectrum of single
trace operators in our theory includes one scalar and one current of
spin $s$ for $s =1 \ldots \infty$. The dimension of the scalar operator
is given, at every value of $\lambda$, by $2 + {\cal O}(1/N)$, while the
dimension of the spin $s$ current $J^{(s)}$ is given by $\Delta =
s+1+{\cal O}(1/N)$. We will also argue that the currents $J^{(s)}$ are ``almost''
conserved; more precisely that these currents obey the anomalous conservation
law listed schematically in \eqref{div}. These nonlinear anomalous
conservation equations carry a great deal of information, and may eventually
prove very useful in ``solving'' the theory.

\subsection{The free limit}\label{free}

In this subsection we review several well known properties of Chern-Simons
theory coupled to fundamental fermions in the free ($ \lambda \to 0$) limit.
In this limit our theory reduces to a theory of free fundamental fermions
subject to a $U(N)$ singlet Gauss law constraint; this limit has, of course,
been studied in the previous literature (see e.g. \cite{Leigh:2003gk}). In
this section we gather those results that will help us study the
theory at finite $\lambda$.

\subsubsection{``Single trace'' conformal representation content of the free theory } \label{st}

\footnote{This subsection was worked out in collaboration with 
J. Bhattacharya.}
In this subsection we compute the ``single trace'' operator content of the
theory of free fermions. By ``single traces'' we mean those operators
in our theory
that are given by contracting a single (but arbitrary) fundamental field with
an anti-fundamental field. The results of this subsection were already
presented in \cite{Leigh:2003gk}; however we rederive them here for
completeness.

Let $\Delta$ represent the scaling dimension and $J_3$ the $z$ component
of the angular momentum of any operator. We will now compute the
partition function
\begin{equation}\label{pfnat}
{\rm Tr} x^{\Delta}\mu^{J_3}
\end{equation}
over all single trace operators in our theory.
As ``single traces'' are obtained by multiplying an arbitrary fundamental field
with an arbitrary anti-fundamental, the trace over ``single traces'' is simply
the product of the partition function for elementary fundamental fields with
the partition function for elementary antifundamental fields.

The partition function over elementary fundamental
fermionic fields (including arbitrary numbers of derivatives but modulo
equations of motion) is simply the character of the $(1,\half)$ representation
of the conformal group
(see Appendix \ref{unit}) and is  given by
\begin{equation}
\label{fca}
F_F(x,\mu)= \frac{x(\mu^{\half} +\mu^{-\half})}{(1-\mu x)(1-\mu^{-1}x)}
\end{equation}
The partition function over antifundamentals is given by the same formula, and
so the ``single trace'' partition function is given by $F_F^2$.

It is not difficult to decompose this partition function into the contribution
of primary operators and their descendents. In order to accomplish
this we note that
\begin{equation}\begin{split}
&\left[ \frac{x(\mu^{\half} +\mu^{-\half})}{(1-\mu x)(1-\mu^{-1}x)}\right]^2\\
&= \frac{1}{(1-\mu x)(1-\mu^{-1}x) (1-x)}
\left[ \frac{x^2(1-x)(2+\mu +\mu^{-1})}{(1-\mu x)(1-\mu^{-1}x)} \right]\\
&=\frac{1}{(1-\mu x)(1-\mu^{-1}x) (1-x)}
\left[ x^2 + \sum_{s=1}^\infty
\left(x^{s+1} \chi_{s}(\mu)-x^{s+2}\chi_{s-1}(\mu) \right) \right]\\
&=\chi_{2,0}(x,\mu)+ \sum_{s=1}^\infty \chi_{s+1,s}(x, \mu)\\
\end{split}
\end{equation}
where $\chi_{\Delta, s}(x, \mu)$ is the character of a representation of the
conformal algebra with dimension $\Delta$ and spin $s$ and we have used the
character formulae of Appendix \ref{unit}. It follows
that single trace operators are given by primaries that transform in
the representations
$$(2, 0)+ \sum_{j=1}^\infty (j+1, j)$$
together with their descendents. It will be important below that the only
long operator that appears in this decomposition is $(2,0)$ (see Appendix
\ref{unit}); all other operators in the list above appear in short
representations of the conformal algebra.

\subsubsection{Explicit form of the primary operators}\label{expfree}

As we have explained above, the primary field content of the free fermion
theory is given by the $(2,0)$ operator ${\bar \psi} \psi$ plus
symmetric traceless currents $J^{(s)}_{\mu_1 \ldots \mu_s}$ for all
$s$. As we have seen above, the currents are primaries that head short
representations of the conformal algebra; the shortening condition is simply
the statement that the currents obey the  conservation equation
\begin{equation} \label{conserv}
\partial^\mu J^{(s)}_{\mu \mu_1 \ldots \mu_{s-1}}=0
\end{equation}
Single trace currents $J^{(s)}$ that have dimension $s+1$, spin $s$ and are
conserved are unique up to a choice of scale in the free fermion theory.
We have found explicit expressions for each of the currents $J^{(s)}$. Following
\cite{Giombi:2009wh} we
find it convenient to package these expressions in the form of a generating
function $O(x, \epsilon)$ defined by
\begin{equation} \label{opo}
 \mathcal O(x;\epsilon) = \sum J^{(s)}_{{\mu_1} {\mu_2} \ldots {\mu_s}}\epsilon^{\mu_1}\ldots \epsilon^{\mu_s}
\end{equation}
where $\epsilon^\mu$ is an arbitrary vector. As all currents are bilinear in
the fermions, the generating function is
given by an expression of the form
\begin{equation}
 \mathcal O(x;\epsilon) =\bar{\psi} F(\vec{\gamma},
\overrightarrow{\partial_\mu},\overleftarrow{\partial_\mu},\vec{\epsilon})
\psi
\end{equation}
Making a convenient choice for the overall scale of each $J^{(s)}$ we find
\begin{equation} \label{opt}
 F = \vec{\gamma}.\vec{\epsilon} f(\overrightarrow{\partial_\mu},\overleftarrow{\partial_\mu}, \vec{\epsilon})
\end{equation}
where
\begin{equation} \label{fincur}
 f(\vec{u}, \vec{v}, \vec{\epsilon}) = \frac{\exp{\left( \vec{u}\cdot \vec \epsilon-\vec{v}\cdot \vec \epsilon \right)} \sinh \sqrt{2 \vec{u}\cdot \vec{v} \vec \epsilon \cdot \vec \epsilon -   4\vec{u}\cdot\vec{\epsilon} \vec{v}\cdot\vec{\epsilon}}}{\sqrt{2 \vec{u}\cdot \vec{v} \vec \epsilon \cdot \vec \epsilon -   4\vec{u}\cdot\vec{\epsilon} \vec{v}\cdot\vec{\epsilon}}}
\end{equation}
The Taylor expansion
\begin{eqnarray*}
f &  = & 1 + \epsilon
    (u-v) + \frac{1}{6} \epsilon ^2
   \left(3 u^2-10 u v+3 v^2+2
   w\right)  \\ & \phantom{=} & +  \epsilon ^3 \left(\frac{u^3}{6}-\frac{7 u^2
   v}{6}+\frac{7 u v^2}{6}+\frac{u
   w}{3}-\frac{v^3}{6}-\frac{v
   w}{3}\right) \\ & \phantom{=} & +\frac{1}{120} \epsilon ^4
   \left(10 (u-v)^2 (2 w-4 u v)+(4 u
   v-2 w)^2+5 (u-v)^4\right) +{\cal O}(\epsilon^5)
\end{eqnarray*}
(above, $w=\vec{u}\cdot\vec{v}$, $u=\vec{u}\cdot \vec \epsilon$, $v=\vec v \cdot \vec \epsilon$.)
implies the following explicit expressions for the first four currents
\begin{eqnarray*} \label{expexp}
 J_{\mu}  & = & \bar{\psi} \gamma_\mu \psi \\
 J_{\mu_1 \mu_2} & = & \bar{\psi}\gamma_{\mu_1} \left( \overrightarrow{\partial_{\mu_2}} - \overleftarrow{\partial_{\mu_2}} \right) \psi \\
J_{\mu_1 \mu_2 \mu_3} & = & \frac{1}{6} \bar{\psi}\gamma_{\mu_1} \left( 3 \overleftarrow{\partial_{\mu_2}} \overleftarrow{\partial_{\mu_3}}  - 10 \overleftarrow{\partial_{\mu_2}} \overrightarrow{\partial_{\mu_3}} + 3 \overrightarrow{\partial_{\mu_2}} \overrightarrow{\partial_{\mu_3}} + 2  ( \overleftarrow{\partial_{\sigma}}\overrightarrow{\partial^{\sigma}}) \eta_{{\mu_2}{\mu_3}} \right) \psi \\
J_{\mu_1 \mu_2 \mu_3 \mu_4} & = & \frac{1}{6} \bar{\psi}\gamma_{\mu_1} \Big( \overleftarrow{\partial_{\mu_2}} \overleftarrow{\partial_{\mu_3}} \overleftarrow{\partial_{\mu_4}}
-7 \overleftarrow{\partial_{\mu_2}} \overleftarrow{\partial_{\mu_3}} \overrightarrow{\partial_{\mu_4}}
+ 7 \overleftarrow{\partial_{\mu_2}} \overrightarrow{\partial_{\mu_3}} \overrightarrow{\partial_{\mu_4}}
- \overrightarrow{\partial_{\mu_2}} \overrightarrow{\partial_{\mu_3}} \overrightarrow{\partial_{\mu_4}} \\
& \phantom{=} &
+2 (\overleftarrow{\partial_{\sigma}}\overrightarrow{\partial^{\sigma}}) \overleftarrow{\partial_{\mu_2}} \eta_{{\mu_3}{\mu_4}}
- 2
(\overleftarrow{\partial_{\sigma}}\overrightarrow{\partial^{\sigma}}) \overrightarrow{\partial_{\mu_2}} \eta_{{\mu_3}{\mu_4}} \Big ) \psi
\end{eqnarray*}
where all indices above are understood to be symmetrized.

The existence of spin $s$ conserved currents gives rise to a large symmetry
algebra, called the higher spin algebra, of the free fermion theory.

\subsubsection{The full multitrace partition function}\label{sphere}

In this section we present a computation of the partition function of the free
theory on $S^2 \times S^1$, i.e. the free theory on $S^2$ at finite
temperature, along the lines of \cite{Shenker:2011zf}. By the state operator map this is the same as the partition
function over all operators (single and multi trace) of the free theory.

The exact partition function on $S^2 \times S^1$ is given by
\cite{Sundborg:1999ue, Aharony:2003sx}
\begin{equation} \label{epf}
Z=\int DU\exp \left( -\sum_n \frac{ N^2 |\rho_n|^2 - N(-1)^{n+1} F_F(x^n)(\rho_n+\rho_n^*)}{n}
\right)
\end{equation}
where $x=e^{-\beta}= e^{-\frac{1}{T}}$, $F(x)$ is the letter partition
function of the
fundamental fermions, $U$ is a unitary matrix (the holonomy of $A_\mu$
around the time circle) and
$$\rho_n= \frac{1}{N} Tr U^n$$

Let us first study \eqref{epf} in the limit that $x$ is held fixed as we take
the large $N$ limit. In this case the integral \eqref{epf} is very simply
evaluated by completing the square in the $\rho_n$ variables and yields
\begin{equation}\label{sbf}
Z=e^{\sum_n \frac{F_F(x^n)^2}{n}}
\end{equation}
In the high temperature limit $F_F(x)= 2 T^2$ and \eqref{sbf}
reduces approximately to
\begin{equation}\label{htoo}
\ln Z= 4 T^4 \zeta(5)
\end{equation}

\eqref{sbf} is simply the Bose exponentiation the single particle
partition function $F_F(x)^2$. However we have already argued above that
$F(x)^2$ is precisely  the single trace operator partition function.
It follows that \eqref{sbf} represents the partition function of a
noninteracting gas of single trace operators (or particles), with
single trace partition function $F_F(x)^2$.

In order to obtain the result \eqref{sbf} we shifted the variable
$\rho_n$ by $\frac{F(x^n)}{N}$. When $x$ is of order unity this shift is of
order $\frac{1}{N}$ and is negligible. In the high temperature limit, however,
$$F_F(x) \approx 2 T^2$$
and the large $N$ saddle point value of the integral \eqref{sbf}
is given by
$$ \rho_n=(-1)^{n+1}\frac{ 2 T^2}{N n^2}.$$
If we set
$$T= \sqrt{N} t$$
and hold $t$ fixed as $N$ is sent to $\infty$, the saddle point
value of $\rho_n$ is independent of $N$
\begin{equation}\label{spr}
 \rho_n=(-1)^n\frac{ 2 t^2}{n^2}
\end{equation}

Let $\rho(\theta)$ denote the density of eigenvalues of the unitary matrix
$U$. It follows that
$$ \rho_n= \frac{1}{N} Tr U^n= \int d\theta \rho(\theta) e^{i n \theta}$$
The saddle point eigenvalue distribution may be reconstructed from its
Fourier modes via
$$ \rho(\theta)= \frac{1}{2 \pi} \left[1+  \sum_{n=1}^\infty
\left( \rho_{-n} e^{i n \theta} +   \rho_{n} e^{-i n \theta} \right) \right]$$
The eigenvalue distribution that corresponds, in particular, to \eqref{spr}
is given by
$$  2 \pi \rho(\theta) =
1+ 4t^2  \sum_{n=1}^\infty  \frac{ (-1)^{n+1}}{n^2} \cos n \theta.$$
This eigenvalue distribution takes its minimum value at $\theta=\pi$.
$\rho(\pi)$ is negative when
\begin{equation} \label{ned}
t \geq  \frac{\sqrt{3}}{\sqrt{2}\pi} .
\end{equation}
Now a negative value for the eigenvalue density function is meaningless;
when the inequality \eqref{ned} is obeyed what really happens is that
our system undergoes a Gross-Witten-Wadia phase transition \cite{Gross:1980he, Wadia:1980cp} to a gapped phase.

The saddle point solution for $\rho(\theta)$ is quantitatively complicated
above the phase transition temperature \eqref{ned}. In broad qualitative
terms, however, the eigenvalue distribution gets increasingly peaked as
$t$ increases. In the limit $t \gg 1$ the eigenvalue distribution tends
to a $\delta$ function. In this limit $\rho_n=1$ for all $n$, and the
logarithm of the partition function is given by
\begin{equation}\label{ht}
\ln Z=4 N T^2 \sum_{n=1}^\infty \frac{(-1)^{n+1}}{n^3}= 3 N \zeta(3) T^2
\end{equation}
In this limit we recover the partition function of the free theory
on $\mathbb{R}^2$ (the leading term in \eqref{F-pert}), with the volume factor $V_2$ replaced by $4 \pi$, the
volume of the unit two sphere.

In summary, the partition function of our theory on an $S^2$ of unit
radius interpolates between the partition function of
a gas over single trace operators \eqref{sbf} and \eqref{htoo}
(when $T \ll \sqrt{N}$)  to the gas of
deconfined fermions (when $t \gg \sqrt{N}$). Note that when
$\ln Z$ is of order $N^2$ when $T$ approaches $\sqrt{N}$ both from
below (see \eqref{htoo}) and from above (see \eqref{ht}). The interpolation
between these behaviours is not smooth; it goes through a third
order Gross-Witten-Wadia transition.


\subsection{Non-renormalization of the scaling dimension of
the current operators} \label{nonrenom}

As we have explained above, the single trace
operator content, in the free limit, is given in terms of representations
of the conformal algebra by
$$(0,2) + \sum_{j=1}^\infty (j+1, j).$$
As we have explained in Appendix \ref{unit},
representations of the sort $(s+1, s)$ are short. Operators that transform
in these representations can develop anomalous dimensions only after
combining with a long representation with quantum numbers $(s+2, s-1)$
(see \eqref{charactersso}).

Now the operators in the $(2,1)$ and $(3,2)$ conformal representations
are the conserved currents for the $U(1)$ fermion flavour symmetry and
the fermionic stress tensor respectively. These operators are exactly
conserved currents so cannot develop anomalous dimensions at any value
of $\lambda$. Let us now consider the operator that transforms in the $(4, 3)$
representation of the conformal algebra in the free theory. This operator can
develop an anomalous dimension only upon combining with an operator in
the representation $(5,2)$. However the only spin two single trace operator,
the stress tensor, transforms as $(3,2)$ for all values of $\lambda$.

We now make a key assumption that we justify in much more detail below:
we assume that, as far as the analysis of leading large $N$ scaling
dimensions is concerned, we can simply ignore all mixing of single-trace
and multi-trace operators. It follows that, at leading order in
$\frac{1}{N}$, the operator in the $(4,3)$ representation cannot develop
an anomalous dimension at any value of $\lambda$, simply because at
no value of $\lambda$ can there exist a single-trace operator in the $(5,2)$
representation with
which the $(4,3)$ operator can combine to form a long representation.

This argument can now be repeated recursively, to demonstrate that the scaling
dimensions of all ``single trace'' spin $s$ operators are exactly protected
(at leading order in $N$) at all values of $\lambda$. The only way this
argument could fail is if the theory underwent a severe phase transition
at a finite value of $\lambda$, across which the spectrum of the theory was
not continuous.

Note that the non-renormalization theorem relies on the assumption of
non mixing of single and multi trace operators (which we will justify
in detail below) and the the sparsity of single trace operators in our
system.  As we will see explicitly below, our non-renormalization theorem
will be violated by $\frac{1}{N}$ corrections.

Finally note that nothing in the argument we have presented so far
prevents the operator ${\bar \psi} \psi$ from developing an anomalous
dimension. However we will proceed to argue below that this is also
impossible; the scaling dimension of this operator is also protected (as
a function of $\lambda$) in the interacting theory.

\subsection{Explicit form of the current operators}\label{expint}

In subsection \ref{expfree} above we have determined the explicit form of
the primary operators that transform in the $(s+1, s)$ representation
of the conformal algebra. In the previous subsection we have argued
that the interacting theory has primaries with the same quantum numbers
at all values of $\lambda$. In this subsection we will determine the
explicit form of these primary operators, in the interacting theory, in
terms of the bare fermionic fields $\psi$.

In the interacting theory at finite $\lambda$ let
${\hat J}^{(s)}$ denote the currents obtained from the following procedure:
replace all derivatives in the expression for $J^{(s)}$ in the free theory
(see subsection \ref{expfree}) with covariant derivatives. Explicitly,
for  $s=1 \ldots 4$ we have
\begin{eqnarray*}
 {\hat J}^{(1)}_{\mu}  & = & \bar{\psi} \gamma_\mu \psi \\
 {\hat J^{(2)}}_{\mu_1 \mu_2} & = & \bar{\psi}\gamma_{\mu_1} \left( \overrightarrow{D_{\mu_2}} - \overleftarrow{D_{\mu_2}} \right) \psi \\
{\hat J^{(3)}}_{\mu_1 \mu_2 \mu_3} & = & \frac{1}{6} \bar{\psi}\gamma_{\mu_1} \left( 3 \overleftarrow{D_{\mu_2}} \overleftarrow{D_{\mu_3}}  - 10 \overleftarrow{D_{\mu_2}} \overrightarrow{D_{\mu_3}} + 3 \overrightarrow{D_{\mu_2}} \overrightarrow{D_{\mu_3}} + 2  ( \overleftarrow{D_{\sigma}}\overrightarrow{D^{\sigma}}) \eta_{{\mu_2}{\mu_3}} \right) \psi \\
{\hat J^{(4)}}_{\mu_1 \mu_2 \mu_3 \mu_4} & = & \frac{1}{6} \bar{\psi}\gamma_{\mu_1} \Big( \overleftarrow{D_{\mu_2}} \overleftarrow{D_{\mu_3}} \overleftarrow{D_{\mu_4}}
-7 \overleftarrow{D_{\mu_2}} \overleftarrow{D_{\mu_3}} \overrightarrow{D_{\mu_4}}
+ 7 \overleftarrow{D_{\mu_2}} \overrightarrow{D_{\mu_3}} \overrightarrow{D_{\mu_4}}
- \overrightarrow{D_{\mu_2}} \overrightarrow{D_{\mu_3}} \overrightarrow{D_{\mu_4}} \\
& \phantom{=} &
+2\overleftarrow{D_{\mu_2}} (\overleftarrow{D_{\sigma}}\overrightarrow{D^{\sigma}})  \eta_{{\mu_3}{\mu_4}}
- 2
(\overleftarrow{D_{\sigma}}\overrightarrow{D^{\sigma}}) \overrightarrow{D_{\mu_2}} \eta_{{\mu_3}{\mu_4}} \Big ) \psi
\end{eqnarray*}
where all indices on the RHS are understood to be symmetrized.

The currents ${\hat J}^{(s)}$ are not
automatically tracelessness in their vector indices. The reason for this
lack of tracelessness is that covariant derivatives, unlike their ordinary
counterparts, do not commute. However the commutator of two covariant
derivatives is a field strength. Now according to the classical Chern-Simons
equation of motion
\begin{equation}\label{cseom}
(F_{\mu \nu})^i_j=\frac{ \pi}{k} \epsilon_{\mu \nu \rho}
{\bar \psi}^i \gamma^\rho \psi_j
\end{equation}
where the $i$ and $j$ indices are colour indices. Now consider a factor
of the field strength $F$ inserted inside a ``single trace'' fermion bilinear.
By the equation of motion cited above\footnote{While the equation \eqref{cseom}
was derived classically, we believe it also applies quantum mechanically,
in an appropriate regulator scheme,
as the current $J^{(1)}$ is the unique dimension two, spin one operator in the
theory.}, this insertion splits the ``single trace''
into a ``double trace'' operator divided by $k$. Further factors of $F$ inside
any of the new resultant ``traces'' repeats this operation.
It follows that the spin $s-2$, $s-4$ etc components of the current
${\hat J^{(s)}}$ is given (via the equations of motion) by
`multi trace' operators that are schematically take the form
$\frac{s^m}{k^{m-1}}$ where $s$ represents any ``single trace'' operator
so $s^m$ stands for an $m$ trace operator.

The fact that the operators ${\hat J}^{(s)}$ are not traceless means that these
currents are not of definite spin; they include components of spin $s$,
spin $s-2$ etc. Let us define the interacting currents $J^{(s)}$ as the projection
of ${\hat J}^{(s)}$ to its spin $s$ component, i.e. the projection that removes
all traces from ${\hat J}^{(s)}$. $J^{(s)}$ is, by definition, a spin $s$ current,
and is of power counting dimension $s+1$. As we have explained above, $J^{(s)}$ and
${\hat J}^{(s)}$ differ only by ``multi trace'' expressions.

Now the primary operator that transforms in the $(s+1, s)$ representation
of the conformal algebra is necessarily an expression of spin $s$ and
of power counting dimension $s+1$ (we use a renormalization scheme in which
operators can mix only if they have equal classical dimension). The full
set of such operators is easily enumerated; in the free theory it consists
of those descendents of the primaries $(j+1, j)$ that are of dimension
$s+1$ and spin $s$ (clearly this requires  $j \leq s$). Note that (in
the free theory) these are all single trace operators. All
spin $s$ multitrace operators in the free theory have dimension
greater than or equal to $s+2$.

In the interacting theory, the full set of operators of spin $s$ and classical dimension $s+1$,
is given by replacing all derivatives by covariant derivatives in the free
answer and then projecting onto the spin $s$ component (i.e. removing all
traces). As above, the projection onto spin $s$ leaves the ``single trace''
part of the operator untouched, but adds ``multi trace'' operators to the mix.

Now let us compute the divergence of the most general possible spin $s$,
classical dimension $s+1$ current listed above in the interacting theory.
The computation of this divergence differs from the same computation
in the free theory in three ways. First covariant derivatives do not commute,
and that results in extra factors of the field strength; as we have
explained above such factors modify the multi trace parts of the answer but
leave the ``single trace'' part of the answer untouched. Second, as explained
above, the current $J^{(s)}$ has extra multitrace operators as compared to the
free current. This additional complication also affects only the multi trace
parts of the answer. Finally the fermion equation of motion could be quantum
mechanically modified, but any such modification is necessarily in terms of
`multi trace' operators.

In other words the single trace part of the divergence of the general spin $s$
current in the interacting theory is identical to the result of the
same computation in the free theory (after replacing derivatives with
covariant derivatives). However the interacting divergence includes,
in addition,  several multi trace contributions that are absent in the
free theory.

Now in subsection \ref{expfree} we
computed the unique operator of dimension $s+1$ and spin $s$, in the free
theory, that is also conserved. The interacting theory possesses no
exactly conserved operator of this dimension.  The operator $J^{(s)}$
comes closest to a conserved current, in that it is the unique current
that obeys the schematic equation
\begin{equation}\label{div}
\partial \cdot J^{(s)}\sim  {1\over k} J J + {1\over k^2} J J J
\end{equation}
(on the RHS of the equation above the symbol $J$ refers either to a
current or to a descendent of a current; the important point in this equation
is that the RHS contains no single trace pieces).
\footnote{The absence of higher trace operators on the RHS of this equation
follows from a consideration of quantum numbers. The is of spin $s-1$
and classical dimension $s+2$. Recall that $\Delta -s \geq 1$ for all
single trace operators. It follows that $\Delta -s \geq m$ for $m$ trace
operators, and so $m \leq 3$.}
In other words $J^{(s)}$ is the unique spin $s$ and classical dimension $s+1$
field in the interacting theory whose divergence has no single trace
component.

As we will see below, the operator $J^{(s)}$ constructed above,
may be identified with unique dimension $s+1$ spin $s$ primary of
the interacting theory. Before proceeding we will first, however,
give an example of how our abstract and exact construction of $J^{(s)}$,
(as the projector onto the spin $s$ sector of the current ${\tilde J}^s$)
may actually be practically implemented at leading order in $\lambda$,
for the special case of the current $J^{(3)}$.  For this purpose
in the next subsection we evaluate the current $J^{(3)}$ and
its divergence using the classical (but interacting) field equations.
In the subsequent subsection we return to the demonstration that
the dimension of $J^{(s)}$ is $s+1$.

\subsubsection{$J^{(3)}$ as an example}

The trace of $\hat J^{(3)}$ is given by
\begin{equation}
{\hat J^{(3)}}_{\mu \nu}{}^{\nu} = {1\over 6}\left[  \bar{\psi} \gamma_\mu (\rD^2+\lD^2) \psi +  \bar{\psi} \gamma^\nu ([\rD_\nu,\rD_\mu]+[\lD_\mu, \lD_\nu]) \psi \right].
\end{equation}
Using the identities listed in Appendix \ref{appB}
\begin{eqnarray*}
 {\hat J^{(3)}}_{\mu \nu}{}^{\nu} & = &  -\frac{1}{12} \bar{\psi} F^{\rho \sigma} \epsilon_{\rho \sigma \nu} (\gamma^\nu  \gamma_\mu + \gamma_\mu \gamma^\nu) \psi \\
& = & - {1\over 6} \bar{\psi} F^{\rho \sigma} \epsilon_{\rho \sigma \mu} \psi \\
& = & \frac{\pi}{3k} (\bar{\psi}\psi)(\bar{\psi} \gamma^\mu \psi).
\end{eqnarray*}
In the last line we used the equation of motion for the field strength $F$
(see Appendix \ref{appB}).
Upon projecting out the trace we find
\begin{equation}
{J^{(3)}}_{\mu_1 \mu_2 \mu_3} - {\hat J^{(3)}}_{\mu_1 \mu_2 \mu_3}= - \frac{\pi}{5k} \eta_{(\mu_1 \mu_2} (\bar{\psi}\psi)( \bar{\psi} \gamma_{\mu_3)} \psi).
\end{equation}

As we have explained, the currents $J^{(s)}$ are not expected to be exactly
divergence free for $s\geq 3$.
On the other hand the divergence of the currents $J^{(1)}$ and $J^{(2)}$
vanishes even in the interacting theory (this is obvious for $J^{(1)}$ and
is also true for the stress tensor $T^{\mu\nu}$ as it can be explicitly checked).
In Appendix \ref{appB} we have also explicitly
computed the divergence of $J^{(3)}$ in the classical but interacting fermion
theory. Our result is (see Appendix \ref{divjt})
\begin{equation}\label{jtanom}
\partial^\mu {J}^{(3)}_{\mu \nu_1 \nu_2} =
-\frac{16\pi}{5k} \left[  \eta_{\nu_1 \nu_2} \left( \partial^\mu J^{(0)} \right) J^{(1)}_\mu
-3 \left( \partial_{(\nu_1} J^{(0)} \right) J^{(1)}_{\nu_2)}
+ 2  J^{(0)} \partial_{(\nu_1} J^{(1)}_{\nu_2)}  \right]
\end{equation}
We have presented our result in terms of the scalar ``current''
$J^{(0)}= {\bar \psi } \psi$ and the vector current $J^{(1)}_\mu ={\bar \psi} \gamma_\mu \psi$.

The equation \eqref{jtanom} has been derived classically; it could certainly
receive quantum corrections. However all such corrections are necessarily
of higher order in $\lambda$; at small $\lambda$ \eqref{jtanom} gives the
leading order contribution to the divergence of $J^{(3)}$.

\subsection{Anomalous dimensions of the current operators} \label{anom-curr}

We will now use the fact that the operators $J^{(s)}$ obey equations of the
form \eqref{div} to argue that the scaling dimension of the operator $J^{(s)}$
is $s+1$ up to corrections of order $\frac{1}{N}$. We will also explain
how the knowledge of the precise form of the RHS of \eqref{div}
can immediately be converted into a computation of the anomalous dimensions
of $J^{(s)}$ at leading order in $\frac{1}{N}$, and systematically in the
$\frac{1}{N}$ expansion.

Let us first review how it follows that a conserved spin $s$ current
has $\Delta=s+1$. For this purpose we use the state operator map, and
denote the state dual to the operator $O$ by $| O \rangle$. The
standard argument takes the schematic form
$$ \langle \partial J| \partial J \rangle
=\langle J | [K, P] | J \rangle =(\Delta -s-1)\langle J | | J \rangle
$$
(see below for the argument with all indices in place). If
$\langle \partial J| \partial J \rangle $ vanishes then it follows immediately
that $\Delta=s+1$. In the our theory $\langle \partial J| \partial J \rangle $
does not quite vanish (see \eqref{div}). However as we will argue below,
\eqref{div} determines it to be a $\frac{1}{N}$ times a product of known
two point functions.

In order to all see this in detail we start with some preliminaries.
In  radial quantization, $J_{\underline{\mu}}^{(s)}$ inserted at the origin corresponds to a state $|J_{\underline{\mu}}^{(s)} \rangle$ on the sphere.
For convenience we introduce a set of differently normalized currents, $j^{(s)}_{\underline{\mu}}$, whose corresponding states $|j^{(s)}_{\underline{\mu}}\rangle$ obey
\ie
\langle j_{\underline{\mu}}^{(s)} | j_{\underline{\nu}}^{(s')}\rangle = \delta_{ss'} \delta_{\underline{\mu},\underline{\nu}}
\fe
where $\delta_{\underline{\mu},\underline{\nu}}$ is 1 if $\underline{\mu},\underline{\nu}$ are the same set of indices up to permutation and 0 otherwise. $J^{(s)}$ and $j^{(s)}$ are related by $J^{(s)}_{\underline{\mu}} = a_s j^{(s)}_{\underline{\mu}}$, where the tree level result for the normalization constant $a_s$ is computed in Appendix B.2.2. Using the inversion $x'_\mu = x_\mu/x^2$, the two point function in position space is related by
\ie
\langle j_{\underline{\mu}}^{(s)}(x) j_{\underline{\nu}}^{(s)}(0) \rangle
& = (-)^s x^{-2-2\delta_s} {\partial x'^{\sigma_1}\over\partial x^{\mu_1}} \cdots {\partial x'^{\sigma_s}\over\partial x^{\mu_s}} \langle j_{\underline{\sigma}}^{(s)}(x') | j_{\underline{\nu}}^{(s)}(0) \rangle
\\
&= (-)^s x^{-2s-2-2\delta_s}\prod_{i=1}^s (\delta^{\sigma_i}_{\mu_i}-{2 x^{\sigma_i} x^{\mu_i}\over x^2}) \cdot\langle j_{\underline{\sigma}}^{(s)}| j_{\underline{\nu}}^{(s)} \rangle.
\fe
where $\delta_s$ is a possible anomalous dimension for $J^{(s)}$. In the second step we moved $j_{\underline{\sigma}}^{(s)}(x')$ to $j_{\underline{\sigma}}^{(s)}(0)$ at no cost, since the difference is a conformal descendant, which is orthogonal to $|j(0)\rangle$. One can analogously work out the momentum space two point function.

We can also translate \eqref{div} into the language of states in radial quantization, schematically of the form
\ie
P^\mu | j_{\mu\mu_1\cdots\mu_{s-1}}^{(s)}\rangle = {1\over \sqrt{N}} A |jj\rangle + {1\over N} B |jjj\rangle.
\fe
Taking its norm (using that $P^{\dagger}=K$), we have
\ie\label{innsch}
\langle j^{(s)}_{\mu\mu_1\cdots\mu_{s-1}}| K^\mu P^\nu | j_{\nu\nu_1\cdots\nu_{s-1}}^{(s)}\rangle
= {1\over N}A^2 \langle jj|jj\rangle + {\cal O}({1\over N^2})
\fe
Since $| j_{\nu\nu_1\cdots\nu_{s-1}}^{(s)}\rangle$ is a conformal primary, it is annihilated by $K_\mu$. Using the conformal algebra, the LHS of (\ref{innsch}) is equal to
\ie
&\langle j^{(s)}_{\mu\mu_1\cdots\mu_{s-1}}| [K^\mu, P^\nu] | j_{\nu\nu_1\cdots\nu_{s-1}}^{(s)}\rangle
= 2\langle j^{(s)}_{\mu\mu_1\cdots\mu_{s-1}}| (\delta^{\mu\nu} D + M^{\mu\nu}) | j_{\nu\nu_1\cdots\nu_{s-1}}^{(s)}\rangle
\\
&= 2\bigg[ (s+1+\delta_s)\langle j^{(s)}_{\mu\mu_1\cdots\mu_{s-1}} | j_{\mu\nu_1\cdots\nu_{s-1}}^{(s)}\rangle +(\delta_{\mu\nu} \delta_{\nu\rho}-\delta_{\nu\nu}\delta_{\mu\rho}) \langle j^{(s)}_{\mu\mu_1\cdots\mu_{s-1}} | j_{\rho\nu_1\cdots\nu_{s-1}}^{(s)}\rangle
\\
&~~~\left.
+ \sum_{i=1}^{s-1} (\delta_{\mu\nu_i} \delta_{\nu\rho}-\delta_{\nu\nu_i}\delta_{\mu\rho}) \langle j^{(s)}_{\mu\mu_1\cdots\mu_{s-1}} | j_{\nu\nu_1\cdots\nu_{i-1}\rho\nu_{i+1}\cdots\nu_{s-1}}^{(s)}\rangle \right]
\\
&= 2\delta_s \langle j^{(s)}_{\mu\mu_1\cdots\mu_{s-1}} | j_{\mu\nu_1\cdots\nu_{s-1}}^{(s)}\rangle.
\fe
In the last step we used the fact that $j_{\mu_1\cdots\mu_s}^{(s)}$ is traceless. The order $1/N$ contribution to the RHS of (\ref{innsch}) comes from the disconnected four-point function, i.e. schematically
\ie
\langle jj|jj\rangle = \langle j|j\rangle \langle j|j\rangle + {\cal O}({1\over N}).
\fe
This relates the anomalous dimension $\delta_s$ at order $1/N$ to the product of two two-point functions of $J$ of lower spins in the {\it free} theory. Note that this method cannot be used in the $s=0$ case. In Appendix E, an explicit two-loop calculation for the ${\cal O}(1/N)$ anomalous dimension of the scalar operator $\bar\psi\psi$ is performed.

As an explicit example, let us consider the spin 3 current $J^{(3)}_{\mu\nu\rho}$, which obeys the anomalous current conservation relation (\ref{jtanom}). In terms of states in radial quantization, we have the (tree level) relation
\ie
|\psi_{\nu_1\nu_2}\rangle \equiv P^\mu | j^{(3)}_{\mu\nu_1\nu_2}\rangle = c \left[ \delta_{\nu_1\nu_2} P^\mu|j^{(0)}\rangle
\otimes |j^{(1)}_\mu\rangle - 3 P_{(\nu_1}|j^{(0)}\rangle
\otimes |j^{(1)}_{\nu_2)}\rangle + 2|j^{(0)}\rangle\otimes P_{(\nu_1}|j^{(1)}_{\nu_2)}\rangle \right]
\fe
where $c=-{16\pi\over 5k} {a_0 a_1\over a_3}$. Taking its norm and using the conformal algebra commutation relations, we find, in particular,
\ie
&\langle \psi_{\mu\nu}|\psi_{\mu\nu}\rangle = 252 |c|^2
\\
&=2\delta_3 \langle j^{(3)}_{\mu\nu\rho}| j^{(3)}_{\mu\nu\rho}\rangle= 20 \delta_3,
\fe
and so
\ie
\delta_3 = {63\over 5}|c|^2 = {252\over 625}{\lambda^2\over N},
\fe
to the leading nontrivial order, i.e. two-loop order. There are both planar and non-planar corrections at higher orders in $\lambda$ and in $1/N$.

A relation of the form (\ref{innsch}) holds to higher order in $1/N$ as well; we would however need the connected four-point functions etc., as well as potential $1/N$ corrections to the relation (\ref{div}), in order to compute the next order contribution to $\delta_s$ in $1/N$.

\subsection{Anomalous current conservation relations within correlation
functions}\label{anom}

As we have explained above, the primary operators that transform in the
$(s+1, s)$ representation (in the large $N$ limit) obey anomalous
conservation equations of the form \eqref{div}. The equation
obeyed, in particular, by $J^{(3)}$ (in the classical interacting theory) is
listed in \eqref{jtanom}. The nonlinear equations of form \eqref{jtanom}
carry a lot of information, as we will explore in this subsection.

Let us first investigate the implication of anomalous conservation equations
on three point functions of spin $J^{(s)}$ operators at leading order in
$N$. \footnote{We have seen
in the previous section that the operators $J^{(s)}$ are effectively conserved
within two point functions at leading order in $N$}. We find the schematic
equation
\begin{equation}\label{anonconsthp}
\partial^\mu \langle J_{\mu \ldots} J J \rangle
= \frac{1}{k} \langle J J J J\rangle + \frac{1}{k^2} \langle JJJJ J \rangle
\end{equation}
While the leading behaviour of the second term in \eqref{anonconsthp} (which
comes from factorizing the 5 pt function into the product of a 2 and 3 pt
function) is ${\cal O}(1)$ (and so subleading in the large $N$ limit)
the leading behaviour of the first term on the RHS (this comes from factoring
the 4 pt function into a product of two point functions) is ${\cal O}(N)$
and so of leading order. It follows that our current operators are not,
in general, conserved within three point functions even at leading order
in the ${\cal O}(1/N)$ expansion.
By equating scaling dimensions on the LHS and RHS of \eqref{anonconsthp} 
it follows immediately that the
RHS of \eqref{anonconsthp} can be nonzero only when $s \geq s_1+s_2$.
In other words the three point function
\begin{equation}\label{tpm}
\langle J^{(s)} J^{(s_1)} J^{(s_2)} \rangle
\end{equation}
can obey anomalous conservation equations (rather than genuine conservation
equations) only if $s_1$, $s_2$ and $s_3$ violate or saturate
the triangle inequality. In all the explicit computations we perform 
below, it will turn out that the RHS of \eqref{anonconsthp} is nonzero 
only when the triangle inequality is explicitly violated (i.e. it vanishes
when $s_1=s_2+s_3$). It is possible that this is a general exact 
result that could perhaps be proved by an analysis of allowed structures
for 3 point functions, but we will not attempt such an analysis in this paper. 

\subsection{Non-renormalization of the scalar operator $J^{(0)}$ }\label{snr}

Let us now apply the arguments of the previous subsection to the three
point correlator $\langle J^{(3)}(x) J^{(1)}(y) J^{(0)}(z) \rangle$. According to the arguments
of the previous subsection, and \eqref{jtanom}, the divergence of
this three point function w.r.t the variable $x$ is schematically proportional
(in the large N limit) to a term proportional $\langle \partial J^{(0)} J^{(0)}\rangle \langle J^{(1)} J^{(1)}\rangle$
plus another term proportional to  $\langle J^{(0)} J^{(0)}\rangle \langle \partial J^{(1)} J^{(1)}\rangle$. The
weight under overall scaling of each of these expressions is
$5 +2 \Delta_0$ where $\Delta_0$ is the as yet unknown scaling dimension of the
current operator $J^{(0)}$. On the other hand the weight under scaling of the
divergence of the three pt function is $6+\Delta_0$. Equating these weights
we find that $\Delta_0=2$. We conclude that, just like their spin $s$
counterparts, the scaling dimension of the scalar current $J^{(0)}$ is
not renormalized as a function of $\lambda$ \footnote{We thank O. Aharony
for discussions on this topic.}

This non-renormalization result may be argued for more intuitively
as follows: the LHS of \eqref{jtanom}
is of dimension 5. The RHS of the same expression is also of dimension
5 if and only $J^{(0)}$ has dimension 2. The slightly technical argument of the
previous paragraph simply makes the argument precise.

\section{Three Point Functions}
\label{correlation-function-section}

\subsection{Quantum corrected operators and their 3-point functions}

In the previous section we demonstrated that the dimension
$s+1$ spin $s$ operators primary operators $J^{(s)}_{\mu_1 \ldots \mu_s}$ are
given by the following construction. First replace all derivatives in
\eqref{opo}, \eqref{opt} and \eqref{fincur} by covariant derivatives.
Then project the resultant currents, ${\tilde J}^{s}$, onto the spin
$s$ subsector (i.e. we project all lower spins out of the currents).
While this procedure is formally exact, a practical implementation of the
projection procedure requires use of equations of motion to express
$F_{\mu\nu}$ in terms of fermion bilinears. This replacement could receive
quantum corrections; moreover it gives rise to quite complicated expressions
involving a summation over multi trace contributions.

All complications having to do with the projection operation
disappear when we focus on the particular component of $J^{(s)}$ that
is obtained by dotting all of its free indices with any constant null
vector $\epsilon$. This is because all multitrace
contributions to $J^{(s)}$ come multiplied with factors of the metric with
free indices and vanish when all these indices are contracted
with $\epsilon$ (we use $(\epsilon)^2=0$ ). Consequently, in this subsection,
we will work exclusively with the operators obtained by contracting all free
indices of $J^{(s)}$ with a constant null vector. 

In the gauge employed in this paper there is an additional simplification
if the null vector in question is chosen to lie in the $x^-$ direction.
In that case the expression for $J^{(s)}$ involves only $D_-$ derivatives of
$\psi$ and ${\bar \psi}$. In the lightcone gauge employed throughout this
paper and in much of this subsection, $A_-$ vanishes, so that
$D_-= \partial_-$. It follows that this particular
component of the spin $s$ current is given directly by the expressions
\eqref{opo}, \eqref{opt} and \eqref{fincur} upon setting
$\epsilon =\epsilon^-$. Explicitly we have
\ie
j^{(s)}(p|\varepsilon) = \int {d^3q\over (2\pi)^3}  \bar\psi^{i\A}(p-q)j^{(s)}_{\A\B}(q;p|\varepsilon) \psi_i{}^\B(q)
\fe
where
\ie
& j^{(0)}_{\A\B}(q;p|\varepsilon) = \epsilon_{\A\B},\\
& j^{(s)}_{\A\B}(q;p|\varepsilon) = (\gamma^+)_{\A\B} \left. e^{\varepsilon^-(
2q^+ - p^+)} {\sin \left( 2\varepsilon^-\sqrt{q^+(p^+-q^+)} \right)\over 2\varepsilon^-\sqrt{q^+(p^+-q^+	)}} \right|_{(\varepsilon^-)^{s-1}} ,~~~~s\geq 1.
\fe
(see \eqref{fincur}).
In this particular case, it is easy to define and compute
${\cal J}^{(s)}(p|\varepsilon)$, the exact quantum corrected
version of the operator
$j^{(s)}(p|\varepsilon)$. We define
\ie
{\cal J}^{(s)}(p|\varepsilon) = \int {d^3q\over (2\pi)^3}  \bar\psi^{i\A}(p-q){\cal J}^{(s)}_{\A\B}(q;p|\varepsilon) \psi_i{}^\B(q)
\fe
where ${\cal J}^{(s)}_{\A\B}(q;p|\varepsilon)$ is defined to be the sum over all
the graphs depicted in Fig. 2.
Clearly, ${\cal J}^{(s)}_{\A\B}(q;p|\varepsilon)$ obeys the Schwinger-Dyson
equation
\ie \label{iiee}
{\cal J}^{(s)}_{\A\B}(q;p|\varepsilon) =  j^{(s)}_{\A\B}(q;p) - \int {d^3\ell\over (2\pi)^3} \left[\gamma^\mu{1\over i{\slash\!\!\!\ell}+\Sigma(\ell)}\right]_{\A\C}  {\cal J}^{(s)}_{\C\D}(\ell;p|\varepsilon) \left[{1\over i({\slash\!\!\! \ell}-{\slash\!\!\! p})+\Sigma(\ell-p)} \gamma^\nu\right]_{\D\B} G_{\mu\nu}(\ell-q).
\fe
where $j^{(s)}_{\A\B}(q;p|\varepsilon)$ is the classical spin-$s$ vertex.

In the special case of $s=1$, it is not difficult to define, and set up
an integral equation (analogous to \eqref{iiee})
for the exact quantum corrected operator for arbitrary polarization vector
(spin 1 is special because the expression for the current contains no
derivatives).

Once we have solved the integral equation \eqref{iiee}, three point functions
of the corresponding operators are determined simply by Wick contraction
using the exact fermion propagator. The exact expression for two point
function $\langle O_1 O_2 \rangle$  may also be obtained by Wick contracting
the exact operator $O_1$ with the bare operator $O_2$ (or vica versa).

We postpone all discussion of the solutions to the
integral equations \eqref{iiee}, to future work


\begin{equation}\nonumber
\begin{array}{l}
\begin{fmffile}{SD_vert_1}
        \begin{tabular}{c}
            \begin{fmfgraph*}(70,65)
                \fmfleft{i1,i2}
                \fmfright{o1,o2}
                \fmfv{decoration.shape=circle, decoration.size=.4h, decoration.filled=shaded}{v}
                \fmf{phantom,tension=4}{i1,v,i2}
                \fmf{plain_arrow}{o2,v,o1}
             \end{fmfgraph*}
        \end{tabular}
        \end{fmffile}
=
        \begin{fmffile}{SD_vert_2}
        \begin{tabular}{c}
            \begin{fmfgraph*}(70,65)
                \fmfleft{i1,i2}
                \fmfright{o1,o2}
                \fmffixed{(0,0)}{v,w}
                \fmfv{decoration.shape=circle, decoration.size=.1h, decoration.filled=empty}{v}
                \fmfv{decoration.shape=cross, decoration.size=.1h, decoration.filled=empty}{w}
                \fmf{phantom,tension=4}{i1,v,i2}
                \fmf{plain_arrow}{o2,v,o1}
             \end{fmfgraph*}
        \end{tabular}
        \end{fmffile}
        +
        \begin{fmffile}{SD_vert_3}
        \begin{tabular}{c}
            \begin{fmfgraph*}(70,65)
                \fmfleft{i1,i2}
                \fmfright{o1,o2}
                \fmfv{decoration.shape=circle, decoration.size=.3h, decoration.filled=shaded}{v}
                \fmfv{decoration.shape=circle, decoration.size=.15h, decoration.filled=shaded}{w1}
                \fmfv{decoration.shape=circle, decoration.size=.15h, decoration.filled=shaded}{w2}
                \fmf{phantom,tension=4}{i1,v,i2}
                \fmf{wiggly,tension=0}{v1,v2}
                \fmf{plain_arrow}{o2,v2}
                \fmf{plain}{v2,w2}
                \fmf{plain,tension=.7}{w2,v,w1}
                \fmf{plain}{w1,v1}
                \fmf{plain_arrow}{v1,o1}
             \end{fmfgraph*}
        \end{tabular}
        \end{fmffile}
\\        
~~~~=
        \begin{fmffile}{SD_vert_2}
        \begin{tabular}{c}
            \begin{fmfgraph*}(70,65)
                \fmfleft{i1,i2}
                \fmfright{o1,o2}
                \fmffixed{(0,0)}{v,w}
                \fmfv{decoration.shape=circle, decoration.size=.1h, decoration.filled=empty}{v}
                \fmfv{decoration.shape=cross, decoration.size=.1h, decoration.filled=empty}{w}
                \fmf{phantom,tension=4}{i1,v,i2}
                \fmf{plain_arrow}{o2,v,o1}
             \end{fmfgraph*}
        \end{tabular}
        \end{fmffile}
+
    \begin{fmffile}{SD_vert_ex1}
        \begin{tabular}{c}
            \begin{fmfgraph*}(60,65)
                \fmfleft{i1}
                \fmfright{o1,o2}
                \fmffixed{(-.1h,0)}{t1,i1}
                \fmffixed{(0,-.05h)}{t2,o1}
                \fmffixed{(0,.05h)}{t3,o2}
                \fmffixed{(0,0)}{t1,w1}
                \fmffixed{(0,0)}{t2,w2}
                \fmffixed{(0,0)}{t3,w3}
                \fmfv{decoration.shape=circle, decoration.size=.1h, decoration.filled=empty}{t1}
                \fmfv{decoration.shape=cross, decoration.size=.1h, decoration.filled=empty}{w1}
                \fmffixed{(0,.5h)}{o1,v}
                \fmffixed{(0,.5h)}{v,o2}
                \fmffixed{(-.5h,0)}{z,t1}
                \fmf{wiggly,left=.2,tension=.0}{x,xp}
                \fmf{plain}{t3,x,t1,xp,t2}
             \end{fmfgraph*}
        \end{tabular}
        \end{fmffile}
+        
\begin{fmffile}{SD_vert_ex2}
        \begin{tabular}{c}
            \begin{fmfgraph*}(60,65)
                \fmfleft{i1}
                \fmfright{o1,o2}
                \fmffixed{(-.1h,0)}{t1,i1}
                \fmffixed{(0,-.05h)}{t2,o1}
                \fmffixed{(0,.05h)}{t3,o2}
                \fmffixed{(0,0)}{t1,w1}
                \fmffixed{(0,0)}{t2,w2}
                \fmffixed{(0,0)}{t3,w3}
                \fmfv{decoration.shape=circle, decoration.size=.1h, decoration.filled=empty}{t1}
                \fmfv{decoration.shape=cross, decoration.size=.1h, decoration.filled=empty}{w1}
                \fmffixed{(0,.5h)}{o1,v}
                \fmffixed{(0,.5h)}{v,o2}
                \fmffixed{(-.5h,0)}{z,t1}
                \fmf{wiggly,left=1.1,tension=.0}{y,w}
                \fmf{wiggly,left=.2,tension=.0}{x,xp}
                \fmf{plain}{t3,x,y,u,w,t1,wp,up,yp,xp,t2}
             \end{fmfgraph*}
        \end{tabular}
        \end{fmffile}
+ \begin{fmffile}{SD_vert_ex3}
        \begin{tabular}{c}
            \begin{fmfgraph*}(60,65)
                \fmfleft{i1}
                \fmfright{o1,o2}
                \fmffixed{(-.1h,0)}{t1,i1}
                \fmffixed{(0,-.05h)}{t2,o1}
                \fmffixed{(0,.05h)}{t3,o2}
                \fmffixed{(0,0)}{t1,w1}
                \fmffixed{(0,0)}{t2,w2}
                \fmffixed{(0,0)}{t3,w3}
                \fmfv{decoration.shape=circle, decoration.size=.1h, decoration.filled=empty}{t1}
                \fmfv{decoration.shape=cross, decoration.size=.1h, decoration.filled=empty}{w1}
                \fmffixed{(0,.5h)}{o1,v}
                \fmffixed{(0,.5h)}{v,o2}
                \fmffixed{(-.5h,0)}{z,t1}
                \fmf{wiggly,left=.2,tension=.0}{x,xp}
                \fmf{wiggly,left=.2,tension=.0}{y,yp}
                \fmf{plain}{t3,x,y,t1,yp,xp,t2}
             \end{fmfgraph*}
        \end{tabular}
        \end{fmffile}
+\,\cdots        
\end{array}
\end{equation}
\bigskip\bigskip
\centerline{Figure 2: The quantum corrected vertex.}
\noindent

\subsection{Three point functions in perturbation theory}

In the rest of this section, we postpone the attempt to compute correlators
exactly in $\lambda$ (at leading order in large $N$) but instead turn
to study of three point functions of fermion
bilinear operators/currents $J^{(s)}$ perturbatively in $\lambda$.

We will begin by writing down the three point functions of currents in the free fermion theory, and describe some general structures of three point functions of higher spin currents as constrained by conformal symmetry. While the full tensor structure of the three point functions $\langle JJJ\rangle$ of currents of nonzero spin is somewhat complicated, it is constrained by conformal symmetry to be the linear combination of finitely many independent structures. In fact, as shown in \cite{Giombi:2011rz}, for conserved higher spin currents all parity preserving structures of $\langle J^{(s_1)}J^{(s_2)}J^{(s_3)}\rangle$ can be realized by free scalars and by free fermions, while a parity odd structure of $\langle J^{(s_1)}J^{(s_2)}J^{(s_3)}\rangle$ exists when the three spins $s_1,s_2,s_3$ obey triangular inequality.\footnote{Recall that when the spins $s_1, s_2, s_3$ obey the triangular inequality, the non-conservation of the currents in our theory does not show up in 3-point functions at leading order at large $N$.} The latter cannot be realized with free fields. One of the main results of this section is that the parity odd structure is indeed present in the interacting Chern-Simons-fermion theory.

The parity symmetry of Chern-Simons-fermion theory may be restored by simultaneously taking $k\to -k$. Therefore, in perturbation theory, the parity invariant tensor structures of $\langle JJJ\rangle$ occur at even loop order, while the parity odd tensor structures occur at odd loop order. We will analyze the one-loop contribution to the parity odd structure of $\langle J^{(1)} J^{(1)} J^{(2)}\rangle$ and $\langle J^{(2)} J^{(2)} J^{(2)}\rangle$,  and the two-loop contribution to the parity even structure of $\langle J^{(s_1)}J^{(s_2)}J^{(s_3)}\rangle$ for general nonzero $s_1,s_2,s_3$. The results will be summarized in section 4.3, while the details of the computation are left to Appendix G.
These results will be very useful in comparing with potentially holographically dual higher spin gauge theories in $AdS_4$ (as will be discussed in section 5).

\subsection{Three point functions in the free fermion theory}

As seen before, a generating function for the conserved higher spin currents in the free fermion theory is
\ie
J(x,\varepsilon) = \bar\psi(x) \gamma\cdot\varepsilon f(\varepsilon\cdot \overleftarrow\partial,\varepsilon\cdot\overrightarrow\partial) \psi(x)
\fe
where
\ie
f(u,v) = e^{u-v} {\sin (2\sqrt{uv})\over 2\sqrt{uv}}
\fe
and $\varepsilon$ is a null vector. By expanding to $\varepsilon^s$ terms, this gives the spin-$s$ currents for all nonzero $s$. The scalar operator $\bar\psi\psi$ need to be considered separately. It is useful to write the null polarization vectors in a bispinor form, $\varepsilon_{\A\B}=\lambda_\A \lambda_\B$. Since $\partial^2$ and ${\slash\!\!\!\partial}$ annihilate $\psi$ by the free equation of motion, we can treat $\partial_\mu$ as a null vector as well when acting on $\psi$, and make the substitutions
\ie
\overleftarrow \partial_{\A\B} \to \xi_\A\xi_\B,~~~~\overrightarrow\partial_{\A\B} \to \eta_\A \eta_\B,
\fe
where $\xi_\A, \eta_\A$ are formal spinorial symbols.
We can write the generating current as
\ie
J(x,\lambda) = (\bar\psi \lambda) {e^{(\lambda (\xi+i\eta))^2}-e^{(\lambda (\xi-i\eta))^2} \over 4 (\lambda\xi) (\lambda\eta)} (\lambda\psi).
\fe
The equation of motion can be implemented by $\eta\psi = 0$ and $\bar\psi\xi=0$. We may then make a further replacement
\ie
\psi_\A \to \eta_\A \phi,~~~~\bar\psi_\A \to \bar\phi\xi_\A
\fe
where $\phi$ is a formal object which behaves as a free massless scalar field, while keeping in mind that an extra minus sign comes with a fermion loop. We can then write
\ie
J(x,\lambda) = {1\over 4} \bar\phi \left[ e^{(\lambda (\xi+i\eta))^2}-e^{(\lambda (\xi-i\eta))^2} \right] \phi.
\fe
As a simple check, the divergence of the current is obtained from
\ie
(\partial_\lambda{\slash\!\!\!\partial}\partial_\lambda) J(x,\lambda) \to \left[(\xi\partial_\lambda)^2
+(\eta\partial_\lambda)^2\right] J(x,\lambda) = ((\xi+i\eta)\partial_\lambda)((\xi-i\eta)\partial_\lambda) J(x,\lambda)
\fe
which obviously vanishes.

Now let us compute the three point function $\langle JJJ\rangle$ for free fermions. The fermion propagator is
\ie
\langle \psi_\A(x) \bar\psi_\B(x') \rangle = {1\over 4\pi}{({\slash\!\!\! x}-{\slash\!\!\! x}')_{\A\B}\over |x-x'|^3} = -{\slash\!\!\!\partial}_{\A\B} {1\over 4\pi|x-x'|} \to i\eta_\A \xi'_\B \langle \phi(x) \bar\phi(x')\rangle
\fe
Here we are making the identification $\xi_\B'\sim i\eta_\B$ when acting on $\langle \phi(x)\bar\phi(x')\rangle$, so as to be consistent with $\overleftarrow\partial_{\A\B}'=\xi_\A'\xi_\B' \sim - \eta_\A\eta_\B = -\overrightarrow\partial_{\A\B}$. The three point functions for the generating current, $\langle J(x_1,\lambda_1)J(x_2,\lambda_2)J(x_3,\lambda_3)\rangle$, can be then written as
\ie
&{1\over (4\pi)^3}\overrightarrow{\prod_{i=1}^3} \left[ e^{(\lambda_i (\xi_i + i \eta_i))^2} - e^{(\lambda_i (\xi_i - i \eta_i))^2} \right]  {1\over |x_{i,i+1}|} + (1\leftrightarrow 2)
\\
&= {1\over (4\pi)^3}\sum_{r_i=\pm 1} \overrightarrow{\prod_{i=1}^3} r_i\exp\left[\left(\sqrt{-\varepsilon_i \cdot \overleftarrow\partial_i} +  r_i \sqrt{\varepsilon_i\cdot\overrightarrow\partial_i}\right)^2 \right] {1\over |x_{i,i+1}|}+ (1\leftrightarrow 2),
\fe
where $\overleftarrow\partial_i$ and $\overrightarrow\partial_i$ act cyclically along the product.
We can further rescale $\varepsilon_i\to t_i \varepsilon_i$, expand this generating function in $t_i$, and obtain up to normalization the three point functions of three currents of nonzero spins
\ie
&\sum_{s_1,s_2,s_3} t_1^{s_1}t_2^{s_2}t_3^{s_3} \langle J_{s_1}(x_1,\varepsilon_1)
J_{s_2}(x_2,\varepsilon_2)J_{s_3}(x_3,\varepsilon_3) \rangle.
\fe
It follows from essentially the same proof as in the appendix of \cite{Giombi:2010vg} that an alternative generating function can be written in a simple closed form
\ie
&\sum_{s_1,s_2,s_3}  {2^{s_1+s_2+s_3}s_1!s_2!s_3!\over (2s_1)!(2s_2)!(2s_3)!} t_1^{s_1}t_2^{s_2}t_3^{s_3} \langle J_{s_1}(x_1,\varepsilon_1)
J_{s_2}(x_2,\varepsilon_2)J_{s_3}(x_3,\varepsilon_3) \rangle
\\
&= {2\over (4\pi)^3} {1\over |x_{12}||x_{23}||x_{31}|} \sinh{Q_1+Q_2+Q_3\over 2} \sinh P_1 \sinh P_2 \sinh P_3.
\fe
The latter is naturally reproduced from the tree level three point function in the bulk dual type B Vasiliev theory, as shown in \cite{Giombi:2010vg}.

\subsection{A few comments on the general structure of three point functions}

Let us begin with an example: the three point function of the stress energy tensor $T_{\mu\nu}$, a quantity of interest in all CFTs. There are three linearly independent tensor structures for $\langle T_{\mu_1\nu_1}(x_1)T_{\mu_2\nu_2}(x_2)T_{\mu_3\nu_3}(x_3)\rangle$ that are allowed by conformal symmetry (together with current conservation) \cite{Osborn:1993cr},\cite{Maldacena:2011nz},\cite{Giombi:2011rz}. In other words,
\ie
\langle TTT\rangle = c_B \langle TTT\rangle_B + c_F \langle TTT\rangle_F + c_{odd} \langle TTT\rangle_{odd}.
\fe
Here $\langle TTT\rangle_B$ stands for the three point function of the stress energy tensor of a free massless scalar, $\langle TTT\rangle_F$ that of a free massless fermion, both of which are parity preserving; $\langle TTT\rangle_{odd}$ is a parity odd structure that is present only in parity violating interacting CFTs.

When the fermions are coupled to Chern-Simons gauge fields, both $\langle TTT\rangle_B$ and $\langle TTT\rangle_{odd}$ show up in the three point function of $T_{\mu\nu}$.
As we will see later, $\langle TTT\rangle_{odd}$ occurs already at one-loop order in perturbation theory, while $\langle TTT\rangle_B$ occurs at two-loop order (together with the free fermion structure $\langle TTT\rangle_F$). The relevant Feynman integrals are difficult to perform directly for generic configuration of operator insertions and polarizations. Instead, we make use of the conformal symmetry, and examine special limits in the configuration space where only a single tensor structure of interest survives; then all we need to do is to compute the overall coefficient of the surviving tensor structure, which is manageable.

We will make use of both conformal symmetry and current conservation to constrain the three point function $\langle J_{s_1}(x_1,\varepsilon_1)J_{s_2}(x_2,\varepsilon_2)J_{s_3}(x_3,\varepsilon_3)\rangle$. At leading order in $1/N$, this assumption is valid, up to the anomalous current conservation relation, of the form $\partial\cdot J \sim {1\over k} JJ + \cdots$, discussed in section 2. The relevant part of the divergence of the higher spin currents is expressed in terms of double trace operators, namely product of two currents, of lower spins. This leads to violating of current conservation in the three point function, but only is one of the three spins, $(s_1,s_2,s_3)$, is greater than the sum of the other two, in other words $(s_1,s_2,s_3)$ violating the triangular inequality. Also note that this divergence of the three point function is necessarily parity odd (due to the $1/k$ factor in the anomalous current conservation relation). So for the parity even structures, and for parity odd structures with $(s_1,s_2,s_3)$ obeying triangular inequality, we may proceed by assuming current conservation.
It is shown in \cite{Giombi:2011rz} under this assumption that $\langle J_{s_1}(x_1,\varepsilon_1)J_{s_2}(x_2,\varepsilon_2)J_{s_3}(x_3,\varepsilon_3)\rangle$ must be of the form $|x_{12}|^{-1}|x_{23}|^{-1}|x_{31}|^{-1}$ times a polynomial of appropriate degrees in the conformally invariant objects $P_i, Q_i, S_i$. The latter are defined as
\ie
&P_1 = {\lambda_2{\slash\!\!\! x}_{23}\lambda_3\over x_{23}^2},~~~
P_2 = {\lambda_3{\slash\!\!\! x}_{31}\lambda_1\over x_{31}^2},~~~
P_3 = {\lambda_1{\slash\!\!\! x}_{12}\lambda_2\over x_{12}^2},
\\
&Q_1 = -\lambda_1 ({{\slash\!\!\! x}_{12}\over x_{12}^2}+{{\slash\!\!\! x}_{31}\over x_{31}^2})\lambda_1,~~~
Q_2 = -\lambda_2 ({{\slash\!\!\! x}_{23}\over x_{23}^2}+{{\slash\!\!\! x}_{12}\over x_{12}^2})\lambda_2,~~~
Q_3 = -\lambda_3 ({{\slash\!\!\! x}_{31}\over x_{31}^2}+{{\slash\!\!\! x}_{23}\over x_{23}^2})\lambda_3,
\\
& S_{1} = {i\over 4} {|x_{12}|\over |x_{23}||x_{31}|} \left[ {(\lambda_2{\slash\!\!\! x}_{12}{\slash\!\!\! x}_{23}\lambda_2) (\lambda_3{\slash\!\!\! x}_{23}\lambda_3)\over x_{12}^2 x_{23}^2} + (\lambda_2\lambda_3)\lambda_2({ {\slash\!\!\! x}_{12}\over x_{12}^2}+{{\slash\!\!\! x}_{23}\over x_{23}^2})\lambda_3 \right],
\\
& S_{2} = {i\over 4} {|x_{23}|\over |x_{31}||x_{12}|} \left[ {(\lambda_3{\slash\!\!\! x}_{23}{\slash\!\!\! x}_{31}\lambda_3) (\lambda_1{\slash\!\!\! x}_{31}\lambda_1)\over x_{23}^2 x_{31}^2} + (\lambda_3\lambda_1)\lambda_3({{\slash\!\!\! x}_{23}\over x_{23}^2}+{{\slash\!\!\! x}_{31}\over x_{31}^2})\lambda_1 \right],
\\
& S_{3} = {i\over 4} {|x_{31}|\over |x_{12}||x_{23}|} \left[ {(\lambda_1{\slash\!\!\! x}_{31}{\slash\!\!\! x}_{12}\lambda_1) (\lambda_2{\slash\!\!\! x}_{12}\lambda_2)\over x_{31}^2 x_{12}^2} + (\lambda_1\lambda_2)\lambda_1({{\slash\!\!\! x}_{31}\over x_{31}^2}+{{\slash\!\!\! x}_{12}\over x_{12}^2})\lambda_2 \right].
\fe
where again we wrote the null polarization vector $\varepsilon_i$ in the bispinor form, in terms of a spinor $\lambda_i$. The polynomial in $P,Q,S$ is such that it is homogeneous of degree $2s_i$ in $\lambda_i$. Because all quadratic monomials in $S_i$'s can be expressed in terms of $P_i, Q_i$, we can assume the polynomial to be linear in the $S_i$'s.

Note that under the exchange $(x_1,\lambda_1)\leftrightarrow (x_2,\lambda_2)$,
\ie
& P_1\to - P_2, ~~~P_2\to -P_1,~~~P_3\to -P_3,
\\
&Q_1 \to -Q_2,~~~Q_2\to-Q_1,~~~Q_3\to -Q_3.
\fe
In case of correlators involving currents of identical spins, the symmetry under exchange of those current should be imposed correspondingly.

The $L$-loop contribution should transform under parity with the sign $(-)^L$, namely
\ie
\langle J_{s_1}(-x_1,-\varepsilon_1)J_{s_2}(-x_2,-\varepsilon_2)J_{s_3}(-x_3,-\varepsilon_3) \rangle_{L{\rm -loop}}
=(-)^L \langle J_{s_1}(x_1,\varepsilon_1)J_{s_2}(x_2,\varepsilon_2)J_{s_3}(x_3,\varepsilon_3) \rangle_{L{\rm-loop}}
\fe
for $s_1,s_2,s_3>0$. The spin 0 case is special because $J_0$ is parity odd unlike $J_s$ for $s>0$. We have therefore
\ie
&\langle J_{s_1}(-x_1,-\varepsilon_1)J_{s_2}(-x_2,-\varepsilon_2)J_{0}(-x_3) \rangle_{L}
=(-)^{L+1} \langle J_{s_1}(x_1,\varepsilon_1)J_{s_2}(x_2,\varepsilon_2)J_{0}(x_3) \rangle_L,
\\
&\langle J_{s_1}(-x_1,-\varepsilon_1)J_{0}(-x_2)J_{0}(-x_3) \rangle_{L}
=(-)^{L} \langle J_{s_1}(x_1,\varepsilon_1)J_{0}(x_2)J_{0}(x_3) \rangle_L,
\\&\langle J_0(-x_1)J_{0}(-x_2)J_{0}(-x_3) \rangle_{L}
=(-)^{L+1} \langle J_0(x_1)J_{0}(x_2)J_{0}(x_3) \rangle_L.
\fe
For example, $\langle J_0 J_0 J_0\rangle$ can only receive contribution from odd loops. Under $x\to -x, \varepsilon\to -\varepsilon$, $P_i, Q_i$ are invariant while $S_{ij}$ changes sign. So we expect the $L$-loop contribution to the three point function of currents of nonzero spins to be of the form
\ie
&\langle J_{s_1}(x_1,\varepsilon_1)J_{s_2}(x_2,\varepsilon_2)J_{s_3}(x_3,\varepsilon_3) \rangle_L
= {1\over |x_{12}||x_{23}||x_{31}|} F({s_1,s_2,s_3}|P, Q),~~~L~{\rm even},\\
&\langle J_{s_1}(x_1,\varepsilon_1)J_{s_2}(x_2,\varepsilon_2)J_{s_3}(x_3,\varepsilon_3) \rangle_L
= {1\over |x_{12}||x_{23}||x_{31}|} \sum_{i}S_{i,i+1} G_{i+2}(s_1,s_2,s_3|P, Q),~~~L~{\rm odd}.
\fe

Let us now examine the structure of the three point function of the spin 2 current, i.e. the stress energy tensor $T_{\mu\nu}$, in some detail. $F(2,2,2|P,Q)$ contain two possible structures which are that of the free scalar and free fermion theories. At odd loop level, the contribution to $\langle TTT\rangle$, if nonzero, must be proportional to
\ie
{  S_1(P_1^2 Q_1^2 +5P_2^2 P_3^2) + S_2(P_2^2 Q_2^2 +5P_3^2 P_1^2) + S_3(P_3^2 Q_3^2 +5P_1^2 P_2^2)   \over |x_{12}||x_{23}||x_{31}|}
\fe
We would like to see whether this structure appears at one loop (at planar level). If such a term is present at one-loop, it would be the entire one-loop contribution to $\langle TTT\rangle$ since it is the only tensor structure allowed by conformal symmetry. Therefore, it suffices to determine the overall coefficient by taking some limit in which the tensor structure simplifies but stays nonzero.

Similarly, there is a parity violating structure for $\langle jj T\rangle$, where $j$ is the spin 1 flavor current. It is proportional to
\ie\label{tjj}
{   2S_1 P_2^2 + 2S_2 P_1^2 + S_3Q_3^2   \over |x_{12}||x_{23}||x_{31}|}.
\fe

Let us consider the special configuration in which all three polarization vectors (or spinors) are equal, namely $\lambda_i=\lambda$ ($i=1,2,3$), so that
\ie
Q_1 = -P_2-P_3,
\fe
and
\ie
S_1 = {i\over 4}{P_1\Delta \over |x_{12}||x_{23}||x_{31}|}
\fe
etc., where we defined
\ie
\Delta \equiv \lambda {\slash\!\!\! x}_{31} {\slash\!\!\! x}_{12} \lambda = \lambda {\slash\!\!\! x}_{12} {\slash\!\!\! x}_{23} \lambda = \lambda {\slash\!\!\! x}_{23} {\slash\!\!\! x}_{31} \lambda.
\fe
We can write (\ref{tjj}) as
\ie
{i\over 4} \Delta {P_3 (P_1+P_2)^2 + 2 P_1 P_2 (P_1+P_2) \over x_{12}^2 x_{23}^2 x_{31}^2}
\fe
Now consider the limit in which $x_1$ and $x_2$ are very close, while $x_3$ is kept a finite distance away. For instance, $x_{12}=\delta$, $x_{31}=x$, $x_{32}=x+\delta$, and $x$ is kept finite. We have now
\ie\label{plim}
P_1= -{\varepsilon\cdot (x+\delta)\over (x+\delta)^2},~~~P_2= {\varepsilon\cdot x\over x^2} , ~~~P_3 = {\varepsilon\cdot \delta\over \delta^2},~~~\Delta =  \lambda {\slash\!\!\! x} {\slash\!\!\! \delta} \lambda,
\fe
To simplify things further, we may choose $x\cdot\varepsilon=0$, say by taking $x=t\hat e_0$ along Euclidean time direction ($\hat e_0$ is the unit vector in Euclidean time direction) and $\delta,\varepsilon$ purely spatial. We can choose $\varepsilon$ to lie along light cone direction $x^-$. In this case, (\ref{plim}) becomes
\ie\label{sslim}
P_1= -{\delta^+\over t^2+\delta^2},~~~P_2= 0 , ~~~P_3 = {\delta^+\over \delta^2},~~~\Delta =  \epsilon_{ijk} x^i \delta^j \varepsilon^k = t \delta^+,
\fe
The parity odd tensor structure for $\langle j_{\varepsilon}(0)j_{\varepsilon}(-\delta) T_{\varepsilon}(x) \rangle$ then reduces to
\ie
{i\over 4}  {t(\delta^+)^6\over \delta^4 t^2 (t^2+\delta^2)^3} \to {i\over 4}  {(\delta^+)^4\over \delta^4 t^7}
\fe
where we took the $\delta\to 0$ limit on the RHS.

Similarly, we can write the parity odd structure of $\langle TTT\rangle$ in this special case $\lambda_i=\lambda$ ($i=1,2,3$) as
\ie
&{i\over 4}\Delta { P_3(P_3^2 (P_1+P_2)^2 +5P_1^2 P_2^2) + P_1(P_1^2 (P_2+P_3)^2 +5P_2^2 P_3^2) + P_2 (P_2^2 (P_3+P_1)^2 +5P_3^2 P_1^2)  \over x_{12}^2 x_{23}^2 x_{31}^2}
\\
&= {i\over 4}{\Delta\over x_{12}^2 x_{23}^2 x_{31}^2} (P_1+P_2+P_3) (P_1 P_2+P_2P_3+P_3P_1)^2
\fe
In the limit of (\ref{sslim}), this becomes
\ie
{i\over 4}{t (\delta^+)^6\over \delta^8  (t^2+\delta^2)^4}  \to {i\over 4}{(\delta^+)^6\over \delta^8 t^7}
\fe
where the RHS is the $\delta\to 0$ limit.
Note that if we had taken a generic polarization configuration, in the $\delta\to 0$ limit we would expect the three point function of $T$ to scale like $\delta^{-3}x^{-6}$. To determine the parity odd structure in our limit, we need to go one order higher in $\delta$ than the naive leading order. In the next subsection, we will compute the coefficient of this term.

\subsection{One-loop and two-loop results}

In this subsection, we summarize the planar one-loop and two-loop perturbative results for the three point functions, leaving details of the computation to Appendix G.

First we consider the one-loop contribution to the scalar three point function $\langle J^{(0)}(x_1)J^{(0)}(x_2)J^{(0)}(x_3)\rangle$. Note that $J^{(0)}=\bar\psi\psi$ is parity odd, and its three point function can only receive contributions at odd loop order, a priori. We find that it in fact vanishes at one-loop, as shown in Appendix G.1. The computation is performed in both Feynman gauge and lightcone gauge; in Feynman gauge there are nontrivial cancellations between different diagrams, whereas in lightcone gauge each diagram vanishes individually.

In fact, we suspect that $\langle J^{(0)}(x_1)J^{(0)}(x_2)J^{(0)}(x_3)\rangle$ vanishes identically at planar level. This is suggested by the holographic dual, since $J^{(0)}$ is dual to a bulk scalar of mass $-2$ in $AdS$ units, and such a scalar should have vanishing cubic coupling, otherwise if one imposes the alternative boundary condition in which the dual scalar operator has dimension 1, the tree level three point function would diverge. This is in particular the case for the general parity violating version of Vasiliev's theory. In field theoretic terms, one may be able to see this by considering the UV fixed point obtained from a double trace deformation by $J^{(0)}J^{(0)}$, and relate $\langle J^{(0)}J^{(0)}J^{(0)}\rangle$ to the corresponding scalar three point function in the UV fixed point theory.

Next, we analyze the one-loop parity odd structure in two examples, $\langle jjT \rangle$ and $\langle TTT\rangle$, where $j=J^{(1)}$ is the $U(1)$ flavor current. Since there is only a single parity odd tensor structure allowed by conformal symmetry and current conservation, we only need to extract the overall coefficient. The computation is mostly easily done in temporal gauge, and by taking the special limit described in the previous subsection. In both cases, we find the answer to be nonzero. The results, expressed in terms of the conformally invariant objects $P_i, Q_i, S_i$ defined in \cite{Giombi:2011rz} and in the previous subsection, are
\ie
& \langle j(x_1,\varepsilon_1)j(x_2, \varepsilon_2)T(x_3, \varepsilon_3)\rangle_{1-loop} = -{2i\lambda\over\pi^3}\cdot {2S_1 P_2^2 + 2S_2 P_1^2+S_3 Q_3^2\over |x_{12}||x_{23}||x_{31}|},
\\
& \langle T(x_1, \varepsilon_1)T(x_2, \varepsilon_2)T(x_3, \varepsilon_3)\rangle_{1-loop} \\
&~~~= {18i\lambda\over \pi^3} \cdot {  S_1(P_1^2 Q_1^2 +5P_2^2 P_3^2) + S_2(P_2^2 Q_2^2 +5P_3^2 P_1^2) + S_3(P_3^2 Q_3^2 +5P_1^2 P_2^2)   \over |x_{12}||x_{23}||x_{31}|}.
\fe

Finally, we compute the two-loop parity even structure for $\langle J^{(s_1)} J^{(s_2)} J^{(s_3)}\rangle$ with general spins. As remarked earlier, this three point function is conserved to leading order in $1/N$ if the three spins obey triangular inequality, and when the triangular inequality is not obeyed, the non-conserving structure is necessarily parity odd and does not show up at two-loop. So at two-loop order the result is a linear combination of the free scalar and free fermion answer. The computation in Appendix G.3 focuses on extracting the coefficients of the free scalar structure $\langle JJJ\rangle_B$, by working in the temporal gauge and taking a special limit in which the free fermion structure $\langle JJJ\rangle_F$ while the free scalar structure survives. Our results may be summarized in the following form:
\ie\label{peven}
&\langle j^{(s_1)}(x_1, \varepsilon_1)j^{(s_2)}(x_2, \varepsilon_2)j^{(s_3)}(x_3, \varepsilon_3)\rangle_{2-loop} \\
&= c_B \langle j^{(s_1)}(x_1, \varepsilon_1)j^{(s_2)}(x_2, \varepsilon_2)j^{(s_3)}(x_3, \varepsilon_3) \rangle_B +c_F^{s_1s_2s_3} \langle j^{(s_1)}(x_1, \varepsilon_1)j^{(s_2)}(x_2, \varepsilon_2)j^{(s_3)}(x_3, \varepsilon_3) \rangle_F,
\fe
where the currents are normalized by their two point functions, with
\ie
c_B = {\pi^2\over 4}{\lambda^2\over \sqrt{N}}.
\fe
A nontrivial feature of this result is that $c_B$ is independent of the spins $s_1,s_2,s_3$. As will be discussed in section 5, this property agrees with what one expects from the tree level three point function in the parity violating version of Vasiliev's higher spin gauge theory in $AdS_4$. The agreement also suggests that $c_F^{s_1s_2s_3}$ should be independent of the spins as well, though we have not computed this explicitly at two-loop order.

\section{The Formal Solution at Large $N$}\label{Formal-Solution}


\subsection{Exact effective action for $\psi$ and $\Sigma$}

In the large $N$ limit, the exact quantum effective action for the $\Sigma$ field, obtained after integrating out the fermions, is given by \eqref{em4}. 
The field $\Sigma$ referred to above is an auxiliary field introduced
as a computational device. The effective action that we are really interested
in, in our problem, is the effective action for $\psi$. This quantum effective
action contains all information needed to compute scattering of
$\psi$ quanta, and also allows the exact computation of correlators
of some operators (see the previous subsection). In this subsection
we  outline how the exact quantum effective action for the fields
$\psi$ and $\Sigma$ may be computed in the large $N$ limit.

All interactions in the action \eqref{eammmm2} are of the form
${\bar \psi} \Sigma \psi$; consequently interactions always involve a
$\Sigma$ field. However the propagator of the $\Sigma$ field
is ${\cal O}(\frac{1}{N})$. It follows that the only interactions
that survive at leading order in large $N$ are those that compensate
for every $\Sigma$ propagator with a  fermion loop. With this in mind
it is easy to convince oneself that the exact quantum effective action
of our theory has the following structure at leading order in large $N$.
\begin{itemize}
\item{1.} The part of the effective action that involves only $\psi$
and ${\bar \psi}$ (i.e. is independent of $\Sigma$) is quadratic in
fermion fields. The quadratic form is the inverse of the fermion propagator.
\item{2.} The bare ${\bar \psi} \Sigma \psi$ interaction in \eqref{eammmm2}
is not renormalized. There are no additional interactions between
$\Sigma$ and the fermion fields.
\item{3.} All $\Sigma$ self interactions are obtained from the determinant
in \eqref{em4} (see the previous subsection). The expansion
of this determinant in powers of $\Sigma$ yields $\Sigma$ self interactions
of the form $\Sigma^k$ for all $k \geq 2$. The coefficient of the
$\Sigma^k$ term is diagrammatically described by a one loop $\psi$ graph
with $k$ sigma lines sticking out of it.
\end{itemize}

The net upshot is that, in order to compute the full quantum effective
jointly for the $\psi$ and $\Sigma$ fields,
we need only compute the exact fermion propagator (we have already
done this) together with the effective action \eqref{em4} o the
$\Sigma$ fields.

While the quantum effective action of our system is evaluated rather easily,
it is not straightforward to determine the $\Sigma$ propagator by inverting
its quadratic action. This process is complicated by the fact that
$\Sigma$ is a bilocal field, and relative momentum (the difference between
the two momenta dual to the two insertion points of $\Sigma$) is  not
conserved along $\Sigma$ propagators. The quadratic term in the action
for $\Sigma$ should be thought of as a ``matrix'' in relative momentum
space, and the inversion of this matrix requires some work. We turn to this
problem in the next subsection.

\subsection{Integral equation for the four Fermi vertex} \footnote{
The axial gauge version of the integral equation described in this 
section was worked out in collaboration with J. Bhattacharya and 
S. Bhattacharyya.}

Rather than evaluate the $\Sigma$
propagator through its effective action it is actually more straightforward
to set up the Schwinger Dyson equations that govern the off shell four
fermion scattering graph. This scattering process proceeds
via the exchange of a $\Sigma$ propagator, and so is precisely the
$\Sigma$ propagator with two fermions sewn onto each of the legs of this
propagator.
The four fermion scattering process is,
in fact a more useful object to compute than the $\Sigma$ propagator itself.
The Schwinger Dyson equation that determines this object is easily
derived; we find
\ie
\int {d^3p d^3 q d^3k\over (2\pi)^9} V_{\A\B\C\D}(p,q;k) \bar\psi^{i\A}(-k-p) \psi_j{}^\B(p) \bar\psi^{j\C}(k-q) \psi_i{}^\D(q).
\fe

\bigskip

\begin{equation}\nonumber
\begin{array}{l}
\begin{fmffile}{SD_fourpoint_1}
        \begin{tabular}{c}
            \begin{fmfgraph*}(85,85)
                \fmfleft{i1,i2}
                \fmfright{o1,o2}
                \fmfv{decoration.shape=circle, decoration.size=.5h, decoration.filled=shaded}{v}
                \fmf{plain_arrow}{i1,v,i2}
                \fmf{plain_arrow}{o2,v,o1}
                \fmfv{l=j}{i1}
                \fmfv{l=j}{o1}
                \fmfv{l=i}{i2}
                \fmfv{l=i}{o2}
             \end{fmfgraph*}
        \end{tabular}
        \end{fmffile}
        =
        \begin{fmffile}{SD_fourpoint_2}
        \begin{tabular}{c}
            \begin{fmfgraph*}(85,85)
                \fmfleft{i1,i2}
                \fmfright{o1,o2}
                \fmf{plain_arrow}{i1,v1,i2}
                \fmf{plain_arrow}{o2,v2,o1}
                \fmf{wiggly}{v1,v2}
                \fmfv{l=j}{i1}
                \fmfv{l=j}{o1}
                \fmfv{l=i}{i2}
                \fmfv{l=i}{o2}
             \end{fmfgraph*}
        \end{tabular}
        \end{fmffile}
        +
        \begin{fmffile}{SD_fourpoint_3}
        \begin{tabular}{c}
            \begin{fmfgraph*}(85,85)
                \fmfleft{i1,i2}
                \fmfright{o1,o2}
                \fmfv{decoration.shape=circle, decoration.size=.3h, decoration.filled=shaded}{v}
                \fmfv{decoration.shape=circle, decoration.size=.1h, decoration.filled=shaded}{w1}
                \fmfv{decoration.shape=circle, decoration.size=.1h, decoration.filled=shaded}{w2}
                \fmf{plain_arrow,tension=.7}{i1,v}\fmf{plain,tension=.7}{v,w1}\fmf{plain}{w1,v1}\fmf{plain_arrow}{v1,i2}
                \fmf{plain_arrow}{o2,v2}\fmf{plain}{v2,w2}\fmf{plain,tension=.7}{w2,v}\fmf{plain_arrow,tension=.7}{v,o1}
                \fmf{wiggly,tension=0}{v1,v2}
                \fmfv{l=j}{i1}
                \fmfv{l=j}{o1}
                \fmfv{l=i}{i2}
                \fmfv{l=i}{o2}
             \end{fmfgraph*}
        \end{tabular}
        \end{fmffile}
\\
\\
\\
~~~=  \begin{fmffile}{SD_fourpoint_ex1}
        \begin{tabular}{c}
            \begin{fmfgraph*}(70,70)
                \fmfleft{i1,i2}
                \fmfright{o1,o2}
                \fmf{plain}{i1,v1,i2}
                \fmf{plain}{o2,v2,o1}
                \fmf{wiggly}{v1,v2}
             \end{fmfgraph*}
        \end{tabular}
        \end{fmffile}  
+
\begin{fmffile}{SD_fourpoint_ex2}
        \begin{tabular}{c}
            \begin{fmfgraph*}(70,70)
                \fmfleft{i1,i2}
                \fmfright{o1,o2}
                \fmf{plain}{i1,v1,x,i2}
                \fmf{plain}{o2,xp,v2,o1}
                \fmf{wiggly,tension=.6}{v1,v2}
                \fmf{wiggly,tension=0}{x,xp}
             \end{fmfgraph*}
        \end{tabular}
        \end{fmffile}         
+
\begin{fmffile}{SD_fourpoint_ex3}
        \begin{tabular}{c}
            \begin{fmfgraph*}(70,70)
                \fmfleft{i1,i2}
                \fmfright{o1,o2}
                \fmf{plain}{i1,e,v1,a,b,x,i2}
                \fmf{plain}{o2,xp,c,d,v2,f,o1}
                \fmf{wiggly,tension=.4}{v1,v2}
                \fmf{wiggly,tension=0}{x,xp}
                \fmf{wiggly,tension=0,right=1.5}{a,b}
             \end{fmfgraph*}
        \end{tabular}
        \end{fmffile} 
+
\begin{fmffile}{SD_fourpoint_ex4}
        \begin{tabular}{c}
            \begin{fmfgraph*}(70,70)
                \fmfleft{i1,i2}
                \fmfright{o1,o2}
                \fmf{plain}{i1,e,v1,y,x,i2}
                \fmf{plain}{o2,xp,yp,v2,f,o1}
                \fmf{wiggly,tension=.5}{v1,v2}
                \fmf{wiggly,tension=0}{x,xp}
                \fmf{wiggly,tension=0}{y,yp}
             \end{fmfgraph*}
        \end{tabular}
        \end{fmffile} 
+\,\cdots        
\end{array}
\end{equation}


\centerline{The four-fermion vertex.}
\bigskip

\noindent
$V_{\A\B\C\D}(p,q;k)$ obeys the Schwinger-Dyson equation
\ie\label{sathiksir}
&V_{\A\B\C\D}(p,q;k) = -\lambda \gamma^\mu_{\A\B} \gamma^\nu_{\C\D} G_{\nu\mu}(k)
-\lambda \int {d^3\ell\over (2\pi)^3} \left[\gamma^\mu{1\over i({\slash\!\!\! p}+{\slash\!\!\! k}-{\slash\!\!\!\ell})+\Sigma(p+k-\ell)}\right]_{\A\sigma}
\\
&~~~\times  V_{\sigma \B\C\tau}(p,q;\ell)  \left[\gamma^\nu{1\over i({\slash\!\!\! q}-{\slash\!\!\!\ell})+\Sigma(q-\ell)}\right]_{\tau\D} G_{\nu\mu}(\ell).
\fe
Substituting in the lightcone gauge gluon propagator, and after a Fierz rearrangement, we obtain the integral equation
\ie\label{sathiksira}
&V_{\A\B\C\D}(p,q;k) = -8\pi i \left[\delta_{\C\B} (\gamma_-)_{\A\D}-\delta_{\A\D}(\gamma_-)_{\C\B}\right] {k^-\over k_s^2+i\epsilon}
-\lambda \int {d^3\ell\over (2\pi)^3} V_{\sigma \B\C\tau}(p,q;\ell) {\ell^-\over \ell_s^2+i\epsilon }
\\
&~~~\times  \left\{ \left[{1\over i({\slash\!\!\! p}+{\slash\!\!\! k}-{\slash\!\!\!\ell})+\Sigma(p+k-\ell)}\right]_{\tau\sigma}   \left[\gamma_-{1\over i({\slash\!\!\! q}-{\slash\!\!\!\ell})+\Sigma(q-\ell)}\right]_{\A\D}
\right.
\\
&~~~~~~~\left.- \left[\gamma_-{1\over i({\slash\!\!\! p}+{\slash\!\!\! k}-{\slash\!\!\!\ell})+\Sigma(p+k-\ell)}\right]_{\tau\sigma}   \left[ {1\over i({\slash\!\!\! q}-{\slash\!\!\!\ell})+\Sigma(q-\ell)}\right]_{\A\D}
\right\}.
\fe

\subsection{Effective action for $\psi$ fields alone}

In this brief subsection we describe an exact way of computing
the scattering of an arbitrary number of $\psi$ quanta (or correlators
built out of $\psi$ fields) working only with $\psi$ fields alone. The
building blocks of our construction are the exact $\psi$ propagator
and the exact four Fermi interaction vertex.

The basic rule is simple. One first draws all the Feynman diagrams that
contribute to the process in question using the exact quantum
effective action involving both $\Sigma$ field and $\psi$, described
above. In computing a $\psi$ scattering process, for instance, we must
restrict ourselves to tree graphs with the exact effective action.
Graphs for correlators have some loops, involving the fermions that
explicitly appear in the operators under study.

The diagrams described in the previous paragraph will in general
 have some $\Sigma$ internal lines, as well as some $\Sigma$ 3 or higher
point vertices. We then proceed to blow up every $\Sigma$ interaction
vertex into a fermion loop. Once we have done this, the only appearance
of $\Sigma$ in graphs is via propagators. Each propagator is sewn,
at each end, to two fermion lines. As a consequence each $\Sigma$
internal line carries contributes a factor of the fermion four point
function, described in the previous subsection. The exact answer
is obtained by enumerating all diagrams that occur in this process,
writing down the expressions for these diagrams, and then doing
the loop momenta (these loops arise from the blow up of $\Sigma$ interaction
vertices, as well as possible loops involving fields inserted in
external operators). At every stage, the fermion propagator that participates
in this process is the exact propagator.

\subsection{Exact evaluation of correlators}

The effective action of the previous section allows us, quite easily,
to compute the correlators singlet operators, each of which is bilinear in the fermion field (see the previous section for an extensive discussion on
operators whose correlators we evaluate). In the previous subsection
we have already outlined a method to evaluate 3 point functions of
the appropriate operators. The methods of this subsection generalize
to the evaluation of arbitrary $n$ point functions. We will make
no use of the exact operators defined in the previous section, but
use the effective action outlined in the previous subsection.

In the rest of this subsection we describe in some detail how we can
evaluate the three point functions of fermion bilinear operators using
the effective action method (our description generalizes relatively
straightforwardly to four and higher point correlators).

In order to perform this computation we only need the fermion effective
action expanded to sextic order in the fermions. The exact fermion
propagator is determined by \eqref{ae} and \eqref{seeq}. The 4 Fermi
interaction is determined by \eqref{sathiksir} or \eqref{sathiksira}. And the six fermion vertex
is given by a single fermion loop diagram with three attached fermion
bilinears. The fermion propagators to be used in this loop are the exact
propagators determined by \eqref{ae}-\eqref{seeq}, while the 4 Fermi vertex factors
associated with the attached fermion bilinears is the exact Fermi vertex
\eqref{sathiksir}.

Given these ingredients, the exact three point function of the three
singlet fermionic bilinears is given by the sum of four types of diagrams.
The first is a one loop triangle diagram, with the exact fermion propagators,
where each vertex in the triangle is one of the external operators.

\centerline{\begin{fmffile}{triangle_1}
        \begin{tabular}{c}
            \begin{fmfgraph*}(75,65)
                \fmfleft{i1,i2}
                \fmfright{o1,o2}
                \fmf{phantom,tension=10}{i2,v,o2}
                \fmffixed{(0,0)}{v,w}
                \fmffixed{(0,0)}{v1,w1}
                \fmffixed{(0,0)}{v2,w2}
                \fmffixed{(0,-.15h)}{v1,i1}
                \fmffixed{(0,-.15h)}{v2,o1}
                \fmfv{decoration.shape=circle, decoration.size=.08h, decoration.filled=empty}{v}
                \fmfv{decoration.shape=circle, decoration.size=.08h, decoration.filled=empty}{v1}
                \fmfv{decoration.shape=circle, decoration.size=.08h, decoration.filled=empty}{v2}
                \fmfv{decoration.shape=cross, decoration.size=.08h, decoration.filled=empty}{w}
                \fmfv{decoration.shape=cross, decoration.size=.08h, decoration.filled=empty}{w1}
                \fmfv{decoration.shape=cross, decoration.size=.08h, decoration.filled=empty}{w2}
                \fmfv{decoration.shape=circle, decoration.size=.8h, decoration.filled=shaded}{vc}
                \fmf{phantom,tension=.2}{v,vc}
                \fmf{phantom,tension=.2}{v1,vc}
                \fmf{phantom,tension=.2}{v2,vc}
                \fmf{plain,tension=.2}{v,v1,v2,v}
             \end{fmfgraph*}
        \end{tabular}
        \end{fmffile}
        =
        \begin{fmffile}{triangle_2}
        \begin{tabular}{c}
            \begin{fmfgraph*}(75,65)
                \fmfleft{i1,i2}
                \fmfright{o1,o2}
                \fmf{phantom,tension=10}{i2,v,o2}
                \fmffixed{(0,0)}{v,w}
                \fmffixed{(0,0)}{v1,w1}
                \fmffixed{(0,0)}{v2,w2}
                \fmffixed{(0,-.15h)}{v1,i1}
                \fmffixed{(0,-.15h)}{v2,o1}
                \fmfv{decoration.shape=circle, decoration.size=.08h, decoration.filled=empty}{v}
                \fmfv{decoration.shape=circle, decoration.size=.08h, decoration.filled=empty}{v1}
                \fmfv{decoration.shape=circle, decoration.size=.08h, decoration.filled=empty}{v2}
                \fmfv{decoration.shape=cross, decoration.size=.08h, decoration.filled=empty}{w}
                \fmfv{decoration.shape=cross, decoration.size=.08h, decoration.filled=empty}{w1}
                \fmfv{decoration.shape=cross, decoration.size=.08h, decoration.filled=empty}{w2}
                \fmfv{decoration.shape=circle, decoration.size=.2h, decoration.filled=shaded}{e1}
                \fmfv{decoration.shape=circle, decoration.size=.2h, decoration.filled=shaded}{e2}
                \fmfv{decoration.shape=circle, decoration.size=.2h, decoration.filled=shaded}{e3}
                \fmf{phantom,tension=.2}{v,vc}
                \fmf{phantom,tension=.2}{v1,vc}
                \fmf{phantom,tension=.2}{v2,vc}
                \fmf{plain,tension=.2}{v,e3,v1,e1,v2,e2,v}
             \end{fmfgraph*}
        \end{tabular}
        \end{fmffile}
   +
        \begin{fmffile}{triangle_3}
        \begin{tabular}{c}
            \begin{fmfgraph*}(85,75)
                \fmfleft{i1,i2}
                \fmfright{o1,o2}
                \fmf{phantom,tension=10}{i2,v,o2}
                \fmffixed{(0,0)}{v,w}
                \fmffixed{(0,0)}{v1,w1}
                \fmffixed{(0,0)}{v2,w2}
                \fmffixed{(0,-.15h)}{v1,i1}
                \fmffixed{(0,-.15h)}{v2,o1}
                \fmfv{decoration.shape=circle, decoration.size=.08h, decoration.filled=empty}{v}
                \fmfv{decoration.shape=circle, decoration.size=.08h, decoration.filled=empty}{v1}
                \fmfv{decoration.shape=circle, decoration.size=.08h, decoration.filled=empty}{v2}
                \fmfv{decoration.shape=cross, decoration.size=.08h, decoration.filled=empty}{w}
                \fmfv{decoration.shape=cross, decoration.size=.08h, decoration.filled=empty}{w1}
                \fmfv{decoration.shape=cross, decoration.size=.08h, decoration.filled=empty}{w2}
                \fmfv{decoration.shape=circle, decoration.size=.12h, decoration.filled=shaded}{e1}
                \fmfv{decoration.shape=circle, decoration.size=.12h, decoration.filled=shaded}{e2}
                \fmfv{decoration.shape=circle, decoration.size=.12h, decoration.filled=shaded}{e3}
                \fmfv{decoration.shape=circle, decoration.size=.12h, decoration.filled=shaded}{er1}
                \fmfv{decoration.shape=circle, decoration.size=.12h, decoration.filled=shaded}{er2}
                \fmfv{decoration.shape=circle, decoration.size=.22h, decoration.filled=shaded}{t}
                \fmf{phantom,tension=.2}{v,vc}
                \fmf{phantom,tension=.2}{v1,vc}
                \fmf{phantom,tension=.2}{v2,vc}
                \fmf{plain,tension=.1,left=.4}{v,er1,t}
                \fmf{plain,tension=.1,right=.4}{v,er2,t}
                \fmffixed{(-.3h,0)}{er1,er2}
                \fmf{plain,tension=.1}{t,e3,v1,e1,v2,e2,t}
             \end{fmfgraph*}
        \end{tabular}
        \end{fmffile}
   }
\centerline{            +
        \begin{fmffile}{triangle_4}
        \begin{tabular}{c}
            \begin{fmfgraph*}(90,80)
                \fmfleft{i1,i2}
                \fmfright{o1,o2}
                \fmf{phantom,tension=10}{i2,v,o2}
                \fmffixed{(0,0)}{v,w}
                \fmffixed{(0,0)}{v1,w1}
                \fmffixed{(0,0)}{v2,w2}
                \fmffixed{(0,-.15h)}{v1,i1}
                \fmffixed{(0,-.15h)}{v2,o1}
                \fmfv{decoration.shape=circle, decoration.size=.08h, decoration.filled=empty}{v}
                \fmfv{decoration.shape=circle, decoration.size=.08h, decoration.filled=empty}{v1}
                \fmfv{decoration.shape=circle, decoration.size=.08h, decoration.filled=empty}{v2}
                \fmfv{decoration.shape=cross, decoration.size=.08h, decoration.filled=empty}{w}
                \fmfv{decoration.shape=cross, decoration.size=.08h, decoration.filled=empty}{w1}
                \fmfv{decoration.shape=cross, decoration.size=.08h, decoration.filled=empty}{w2}
                \fmfv{decoration.shape=circle, decoration.size=.1h, decoration.filled=shaded}{e1}
                \fmfv{decoration.shape=circle, decoration.size=.1h, decoration.filled=shaded}{e2}
                \fmfv{decoration.shape=circle, decoration.size=.1h, decoration.filled=shaded}{e3}
                \fmfv{decoration.shape=circle, decoration.size=.1h, decoration.filled=shaded}{er1}
                \fmfv{decoration.shape=circle, decoration.size=.1h, decoration.filled=shaded}{er2}
                \fmfv{decoration.shape=circle, decoration.size=.1h, decoration.filled=shaded}{ner1}
                \fmfv{decoration.shape=circle, decoration.size=.1h, decoration.filled=shaded}{ner2}
                \fmfv{decoration.shape=circle, decoration.size=.2h, decoration.filled=shaded}{t2}
                \fmfv{decoration.shape=circle, decoration.size=.2h, decoration.filled=shaded}{t}
                \fmf{phantom,tension=.2}{v,vc}
                \fmf{phantom,tension=.2}{v1,vc}
                \fmf{phantom,tension=.2}{v2,vc}
                \fmf{plain,tension=.1,left=.4}{v,er1,t}
                \fmf{plain,tension=.1,right=.4}{v,er2,t}
                \fmf{plain,tension=.1,left=.4}{v2,ner1,t2}
                \fmf{plain,tension=.1,right=.4}{v2,ner2,t2}
                \fmffixed{(-.22h,0.05h)}{er1,er2}
                \fmffixed{(0.025h,.22h)}{ner1,ner2}
                \fmf{plain,tension=.1}{t,e3,v1,e1,t2}
                \fmf{plain,tension=.05}{t2,e2,t}
             \end{fmfgraph*}
        \end{tabular}
        \end{fmffile}
        +
        \begin{fmffile}{triangle_5}
        \begin{tabular}{c}
            \begin{fmfgraph*}(100,90)
                \fmfleft{i1,i2}
                \fmfright{o1,o2}
                \fmf{phantom,tension=10}{i2,v,o2}
                \fmffixed{(0,0)}{v,w}
                \fmffixed{(0,0)}{v1,w1}
                \fmffixed{(0,0)}{v2,w2}
                \fmffixed{(0,-.15h)}{v1,i1}
                \fmffixed{(0,-.15h)}{v2,o1}
                \fmfv{decoration.shape=circle, decoration.size=.08h, decoration.filled=empty}{v}
                \fmfv{decoration.shape=circle, decoration.size=.08h, decoration.filled=empty}{v1}
                \fmfv{decoration.shape=circle, decoration.size=.08h, decoration.filled=empty}{v2}
                \fmfv{decoration.shape=cross, decoration.size=.08h, decoration.filled=empty}{w}
                \fmfv{decoration.shape=cross, decoration.size=.08h, decoration.filled=empty}{w1}
                \fmfv{decoration.shape=cross, decoration.size=.08h, decoration.filled=empty}{w2}
                \fmfv{decoration.shape=circle, decoration.size=.1h, decoration.filled=shaded}{e1}
                \fmfv{decoration.shape=circle, decoration.size=.1h, decoration.filled=shaded}{e2}
                \fmfv{decoration.shape=circle, decoration.size=.1h, decoration.filled=shaded}{e3}
                \fmfv{decoration.shape=circle, decoration.size=.1h, decoration.filled=shaded}{er1}
                \fmfv{decoration.shape=circle, decoration.size=.1h, decoration.filled=shaded}{er2}
                \fmfv{decoration.shape=circle, decoration.size=.1h, decoration.filled=shaded}{ner1}
                \fmfv{decoration.shape=circle, decoration.size=.1h, decoration.filled=shaded}{ner2}
                \fmfv{decoration.shape=circle, decoration.size=.1h, decoration.filled=shaded}{mer1}
                \fmfv{decoration.shape=circle, decoration.size=.1h, decoration.filled=shaded}{mer2}
                \fmfv{decoration.shape=circle, decoration.size=.18h, decoration.filled=shaded}{t1}
                \fmfv{decoration.shape=circle, decoration.size=.18h, decoration.filled=shaded}{t2}
                \fmfv{decoration.shape=circle, decoration.size=.18h, decoration.filled=shaded}{t}
                \fmf{phantom,tension=.2}{v,vc}
                \fmf{phantom,tension=.2}{v1,vc}
                \fmf{phantom,tension=.2}{v2,vc}
                \fmf{plain,tension=.1,left=.4}{v,er1,t}
                \fmf{plain,tension=.1,right=.4}{v,er2,t}
                \fmf{plain,tension=.1,left=.4}{v1,mer1,t1}
                \fmf{plain,tension=.1,right=.4}{v1,mer2,t1}
                \fmf{plain,tension=.1,left=.4}{v2,ner1,t2}
                \fmf{plain,tension=.1,right=.4}{v2,ner2,t2}
                \fmffixed{(-.2h,0)}{er1,er2}
                \fmffixed{(0.08h,.2h)}{ner1,ner2}
                \fmffixed{(0.08h,-.2h)}{mer1,mer2}
                \fmf{plain,tension=.08}{t,e3,t1,e1,t2,e2,t}
             \end{fmfgraph*}
        \end{tabular}
        \end{fmffile}
}
\centerline{The exact planar three-point function.}\bigskip
\noindent

The second and third diagrams involve one or two four-fermion vertices.
The last diagram involves a single sextic fermion vertex; equivalently, it consists of three four-fermion vertices contracted along a fermion loop using the exact propagator.
It follows that, if we could solve both \eqref{seeq} and \eqref{sathiksir}
we would practically be able to compute all ``single trace'' three point
functions of operators that are quadratic in the fermion fields.

\section{Light-Cone Hamiltonian and $W_{\infty}$ Algebra}
\label{hamiltonian-section}
One of the most important features of the Euclidean field theory analysis in section \ref{gesp} of this paper is the exact representation of the  path integral entirely in terms of the bilocal field $\Sigma$. The use of bilocal fields in field theory and also in the formulation of the large $N$ limit has been discussed by various authors, see e.g. \cite{Yukawa:1950eq},\cite{Itzykson:1979fi},\cite{Wadia:1980rb},\cite{Dhar:1994ib}. Also, it
naturally appears in discussions of non-commutative field theory \cite{Das:1991qb},\cite{Dhar:1992hr},\cite{Gopakumar:2000zd}. See also \cite{Das:2003vw},\cite{Koch:2010cy},\cite{Jevicki:2011ss} for applications to holography and higher spin theories. These variables, besides highlighting the inherent non-locality of the theory, are natural to discuss the large $N$ limit. The large $N$ saddle point equation is the Schwinger-Dyson equation \eqref{seeq} for the propagator. One can then perform a systematic $1/N$ expansion around the saddle point solution. 

In the following, we will briefly explore the Hamiltonian formulation of the theory in light-cone time and in  $A_+=0$ gauge. Such a treatment of the path-integral allows one to express the theory in terms of `equal time' bilocal variables $M(x,y)$ satisfying a $W_\infty$ algebra. These operators are also constrained by a quadratic relation $M^2=M$, that has appeared before in discussions of non-commutative field theory. The Hamiltonian formulation throws a different light on the structure of this theory, and the $\lambda$ dependence in the exact fermion propagator \eqref{aef} arises in a simple way when evaluated at equal light-cone time.

We will first construct a light-cone Hamiltonian in terms a bilocal field $M(x,y)$ which is defined on a fixed $x^{+}$ slice, and then use this Hamiltonian to derive an action for $M(x,y)$ using the method of coadjoint orbits. Here $x^{+}$ will serve as a time coordinate and the Hamiltonian that we will construct is conjugate to this variable. 
Our analysis closely follows \cite{Dhar:1994ib}, applied to the large $N$ limit of $QCD_2$. 

We pause to remark that light-cone quantization of any theory has a number of important issues, and (a priori) is not necessarily equivalent to usual time-like quantization. See, e.g., \cite{Lenz:1991sa}.  While we leave careful consideration of these issues for future investigation, we will observe that, applied with appropriate qualification, the light-cone quantization appears to give some simplifying insights into the dynamics of the theory. 

We will be working entirely in Minkowski space-time for this section. We use the following explicit representation for gamma-matrices:
\begin{eqnarray}
&& \gamma^0 =  -i\sigma^2 = \begin{pmatrix} 0 & -1 \\ 1 & 0 \end{pmatrix}, ~~ \gamma^1 =  \sigma^1 = \begin{pmatrix} 0  & 1 \\ 1 & 0 \end{pmatrix}, ~~ \gamma^2  = \sigma^3 = \begin{pmatrix} 1 & 0 \\ 0 & -1 \end{pmatrix} \\ &&
\gamma^- = \frac{1}{\sqrt{2}} \left(-\gamma^0 + \gamma^1 \right) = \begin{pmatrix} 0 & \sqrt{2} \\ 0 & 0 \end{pmatrix}, 
~~ \gamma^- = \frac{1}{\sqrt{2}} \left( \gamma^0 + \gamma^1 \right) = \begin{pmatrix} 0 & 0 \\ \sqrt{2} & 0 \end{pmatrix}
\end{eqnarray}

The fermion is a two-component Dirac spinor which we write as
\begin{equation*}
\psi = \begin{pmatrix} \psi_- \\ \psi_+ \end{pmatrix}
\end{equation*}

These conventions are similar to the Euclidean conventions with a few minor differences. The lightcone coordinates are a sum of $x^0$ and $x^1$ and the non-light cone direction is $x^2$ (instead of $x^3_\text{Eucl.}$). In Minkowski space, it is natural to define $x^\pm = (\pm x^0 + x^1)/\sqrt{2}$, which happens to equal $x^\mp_\text{Eucl.}$, as used elsewhere in the paper.

The action in written in Minkowski space is
\begin{equation}
 S = - \frac{k}{4\pi} \int \tr \left(A \wedge d A + \frac{2i}{3} A \wedge A \wedge A \right) - \int d^3x \left( i\bar{\psi} \gamma^\mu D_\mu \psi + im \bar{\psi}\psi \right)
\end{equation}

We obtain the Hamiltonian for the theory in the gauge $A_+ = 0$. As already mentioned $x^+$ is the `time' coordinate, and $x^-$ and $x^2$ are the space coordinates. The Gauss constraint arising from the equation of motion of $A_+$ is:
\begin{eqnarray}
-\frac{k}{4\pi} \left(\partial_- A_2^c - \partial_2 A_-^c  - f^{abc} A_-^a A_2^b\right) 
+ \sqrt{2} \psi_-^\dagger  T^c \psi_- = 0 \label{gaugeConstraint}
\end{eqnarray}

The momenta conjugate to $\psi_+$ and $\psi_+^\dagger$ vanish -- this reduction in the number of degrees of freedom is a consequence of the choice of light-cone direction $x^+$ as the time variable -- the corresponding constraints arising from their equations of motion are:
\begin{eqnarray}
i \sqrt{2} \partial_- \psi_+ + i \partial_2 \psi_- - im \psi_- + \sqrt{2} A_-^a T^a \psi_+ + A_2^a T^a \psi_- & = & 0 \label{fermionConstraint1}\\
-i \sqrt{2} \partial_- \psi_+^\dagger - i \partial_2 \psi_-^\dagger + im \psi_-^\dagger + \psi_+^\dagger \sqrt{2} A_-^a T^a  + \psi_-^\dagger A_2^a T^a  & = & 0 \label{fermionConstraint2}
\end{eqnarray}

The momentum conjugate to $A^a_-$ is 
\begin{equation}
 \pi_{A_-} = \frac{\partial \mathcal L}{\partial (\partial_+ A^a_-)} = -\frac{k}{4\pi}  A_2^a 
\end{equation}
The momentum conjugate to $\psi_-$ is 
\begin{equation}
\pi_{\psi_-^i} =  \frac{\partial \mathcal L}{\partial (\partial_+ \psi_-)} =i \sqrt{2} \psi_{-}^\dagger
\end{equation}

The Hamiltonian then takes the following form
\begin{eqnarray}
 \mathcal H &=& 
 i \psi_-^\dagger \partial_2 \psi_+
+ im \psi_-^\dagger \psi_- \nonumber 
+ \psi_-^\dagger A^a_2 T^a \psi_+.
\end{eqnarray}

Using the residual gauge symmetry to set $A_-=0$, we can explicitly solve the constraints to express $A_2$, $\psi_+$ and $\psi^\dagger_+$ in terms of $\psi_-$ and its conjugate momentum $\psi_-^\dagger$. 

The resulting (spatially non-local) Hamiltonian, expressed in momentum space is:
\begin{equation}
\begin{split}
 H  = & \int \frac{d^2p}{4\pi^2}  \left( \frac{p_2^2 + m^2}{\sqrt{2} p_+} \right) \psi_-^\dagger(-p)\psi_-(p) \\
 &  + \frac{2\pi}{k}\int \frac{d^2p}{4\pi^2}\frac{d^2q}{4\pi^2}  \frac{d^2q'}{4\pi^2}  
 \left( \frac{(-ip_2-m) }{ p_-q_-} \right) \psi_-^\dagger(-p)\left(\psi_-^\dagger(q-q')\psi_-(p-q)\right) \psi_-(q') \\
&  - \frac{2\pi}{k}
\int \frac{d^2p}{4\pi^2}  \frac{d^2q}{4\pi^2}\frac{d^2p'}{4\pi^2} 
\left( \frac{ip_2-iq_2 -m}{ q_-(p_--q_-)} \right)
\psi_-^\dagger(-p) \left(\psi_-^\dagger(q-p')\psi_-(p-q)\right) \psi_-(p') \\
& +\sqrt{2}\frac{4\pi^2 }{k^2}
\int 
\frac{d^2p}{4\pi^2}\frac{d^2q}{4\pi^2}\frac{d^2p'}{4\pi^2}\frac{d^2q'}{4\pi^2}\frac{d^2q''}{4\pi^2}  \\ 
&
\left( \frac{1}{ p_- - q_- } \frac{1}{q'_-}\frac{1}{q_-} \right)
 \psi_-^\dagger(q-p')\left(\psi_-^\dagger(-p)\psi_-(p') \right)     \left(\psi_-^\dagger(q'-q'')\psi_-(p-q-q')\right)\psi_-(q'') 
\end{split}
\end{equation}

Note that, because we are using light-cone time, and we must choose some prescription to define the inverse of the light-cone momenta when one solves the constraints \eqref{fermionConstraint1} and \eqref{fermionConstraint2} in momentum space. Choosing the principal value prescription of regulating the infra-red divergence effectively eliminates the zero mode of $\psi_+$. The only dynamical degree of freedom in the theory is then $\psi_-$. As pointed out in \cite{Lenz:1991sa} in the limit of large $N$, using the principal value prescription in the definition of infra-red divergent quantities like $1/p_{-}$, in the related theory of two-dimensional $QCD_2$, one can obtain the correct 't Hooft equation that determines the meson spectrum. 

\subsection{Bilocal variables satisfying a $W_{\infty}$ algebra}
Let us define the following bilocal operator
\begin{equation}
 M(x,y) = \frac{\sqrt{2}}{N} \psi^\dagger_{j-}(x) \psi^j_-(y)
\end{equation}
These operators, which would be the gauge invariant Wilson lines along the $x^{-}$ direction in the $A_{-}$ gauge if the points $x$ and $y$ were not separated in the 2-direction, constitute the quantum phase space of the light-cone quantized formulation of the theory.

As we will be working in momentum space, it will be more convenient to use
\begin{equation}
 M(q,q') = \frac{\sqrt{2}}{(2\pi)^2N} \psi^\dagger_{j-}(-q) \psi^j_-(q')
\end{equation}

 It is easy to calculate the commutation relations using the basic fermion anti-commutation relations:
 \begin{equation}
\{ \psi_{-j}^\dagger(k), \psi_-^i(k) \} = \frac{1}{\sqrt{2}} \delta^i_j (2\pi)^2 \delta^2(k+k').
\end{equation}

These imply that the $M$'s satisfy the `quantum' $W_\infty$ algebra
\begin{equation}
 [ M(p,p') , M(q,q') ] = \frac{1}{N} \left( M(p,q')\delta^2(p'-q) - M(q,p')\delta^2(q'-p) \right)
\end{equation}

Notationally, we may often represent $M$ as an operator acting on a single particle Hilbert space, $\hat{M}$, with
\begin{equation}
 \bra{x}\hat{M} \ket{y} = M(x,y)
\end{equation}
though $M(x,y)$ is itself an operator acting on the Hilbert space of the quantum field theory, i.e., multi-particle states. Such constructions are familiar from non-commutative field theory.

In terms of $W_\infty$ variables, the Hamiltonian, using the principal value prescription, can be easily expressed as a `cubic polynomial' in $M$: 
\begin{equation}
\begin{split}
 H  = & N \int {d^2p}  \frac{p_2^2 + m^2}{2 p_-}  M(p,p) \\
&  - N \lambda \frac{1}{4\pi} 
\int {d^2p}{d^2q}{d^2q'}
 \frac{ip_2 + m}{ p_-q_-} 
 M(p,q')M(q'-q,p-q) 
\\ &
  - N \lambda \frac{1}{4\pi} 
\int {d^2p}{d^2q}{d^2q'}
 \frac{ip_2-iq_2  -m}{ q_-(p_--q_-)}
M(p,q')M(q'-q,p-q)  
\\
& - \frac{N \lambda^2}{8\pi^2k^2}
\int 
{d^2p}{d^2q}{d^2p'}d^2q'{d^2q''}
\frac{1}{q_-q'_-q''_-} 
 M(p'+q'',p)M(q+q'',p')M(p-q',q-q')
\end{split}
\end{equation}
The inverse of $p_-$ in the above manipulations is assumed to be defined using the principal value prescription.

\subsubsection{Constraint on the bilocal fields}
We now restrict the space of allowed states to be singlets under global gauge transformations i.e. the generators of global $SU(N)$ gauge transformations constrain the wave functions as follows:
\begin{equation}
 \int dx^- dx^2 \left(\psi_-^{\dagger b} \psi_-^a - \frac{1}{N} \delta^{ab} \psi_-^{\dagger c}\psi_- ^c\right) \ket{\Psi} = 0
\end{equation}

We also note that the baryon number operator 
\begin{equation}
 B = \frac{\sqrt{2}}{N(2\pi)^2} \int d^2p \psi_-^{\dagger a}(-p) \psi^a_-(p)
\end{equation}
commutes with the Hamiltonian and hence it is conserved and takes a definite value on physical states. This fact leads to a very important operator constraint in the phase space. From the definition of $M$, it follows that 
\begin{eqnarray}
\hat{M}^2(p,p') &=& \int dq M(p,q) M(q,p') = (1+B) M(p,p')\\
B &=&  \int d^2 p M(p,p)
\end{eqnarray} 

This completes the light-cone Hamiltonian formulation. The (light-cone quantized) phase space operators are the $M$'s. Their commutation relations satisfy the $W_{\infty}$ Lie algebra. Since the generators of this Lie algebra are represented in terms of the underlying fermions, they are not all independent and satisfy a constraint which is a consequence of gauge invariance and a conserved baryon number. It is clear that we are dealing with a $W_{\infty}$ Hamiltonian system, whose representation is fixed by the constraint $\hat{M}^2 = (1+B) \hat{M}$. We will restrict ourselves to the $B=0$ sector of this theory.

\subsection{The path integral and the large-$N$ classical solution}
From the light-cone Hamiltonian one can construct an action (and the path integral) for the theory using standard techniques of geometric quantization viz. the method of coadjoint orbits of the ($W_{\infty}$) Lie group. Equivalently the path integral can be derived using a basis of coherent states of the $M$ operators between adjacent time steps. The action one obtains in terms of the `c-number' bilocal fields is (see \cite{Dhar:1994ib, Dhar:1992hr} and references therein), 
\begin{equation}
 S_\text{classical} = N\frac{i}{4}\int ds dx^+ \tr \hat{M} [\partial_s \hat{M}, \partial_{+} \hat{M}] - \int dx^+ H \label{coadjoint-action}
\end{equation}

The first term of the action is the Kirilov-Berry phase. It is written using an auxiliary parameter $s$ besides the time $t$. If $t$ is periodic then $s$ parameterizes a unit disc. We need to impose the boundary condition $M(s=0,t)=0$ and $M(s=1,t)=M(t)$. The classical equations of motion are obtained by varying the action consistent with the $W_{\infty}$ group law and using the above mentioned boundary conditions. These equations are by construction the Hamilton equations of motion: $-i\partial_+ M = [H,M]$. 

The path integral measure over $M$ is constrained by the covariant condition $\hat{M}^2 = \hat{M}$, which ultimately arises from evaluating the operator constraint in the space of coherent states (see \cite{Dhar:1992hr}). In particular, to obtain the solution $\braket{M}$ that corresponds to the ground state expectation value of $M$, we solve the time-independent equation $[H,M]=0$. 

From the commutation relations of the $W_\infty$ variables, one can see that \textit{any} translationally invariant $\braket{M}$
\begin{equation}
\braket{M(p,q)} =  \delta^2(p-p') f(p)
\end{equation}
is a solution provided the constraint $\braket{\hat{M}}^2 = \braket{\hat{M}}$ is satisfied. This implies that $f(p)^2=f(p)$, which being independent of $\lambda$, is most naturally solved by $f(k) = \theta(-\frac{p_2^2+m^2}{p_-}) \approx= \theta(-p_-)$.\footnote{We thank J. David for discussion on this point.} 

It is important to note that this solution is the \textit{exact} solution in the large-$N$ limit. It allows one to read off the equal-time correlation function for $\psi_-$'s, valid to all orders in $\lambda$, 
\begin{equation}
 \langle \psi_{j-}^\dagger(q') \psi^j_-(q) \rangle = N \frac{1}{\sqrt{2}}(2\pi)^2 \braket{M(-q',q)}
\end{equation}
which implies:
\begin{equation}
 \langle \psi^j_-(q)\psi_{j-}^\dagger(q') \rangle =N \frac{1}{\sqrt{2}} \theta(q_-)(2\pi)^2\delta^2(q+q') \label{equaltimeW--}
\end{equation}
This is, in fact, one of the components of the equal-``time'' fermionic propagator. Using the constraints, we can use this solution to obtain other components of the exact propagator. The fermionic constraints in this large-$N$ solution read:
\begin{eqnarray}
\psi_+(q) 
& = & \frac{i}{{\sqrt{2} q_-}}
\left({iq_2}
-  \left(m + \lambda I_*(-q) \right) \right)\psi_-(q) \label{fCon1}
\\
\psi_+^\dagger(q') 
& = & \frac{i}{\sqrt{2} q'_-} \left( {iq'_2} - \left(m + \lambda I_*(q') \right) \right) \psi_-^\dagger(q') \label{fCon2}
\end{eqnarray}
where $I_*(-q) = \frac{1}{2\pi} \int {d^2p}  \frac{1}{p_+}  M(q-p)$.

Let us discuss the regularization of $(m+I_*(-q))$. First, let us note that if $q_- A(q) = J(q)$, then $A(q)= (\frac{1}{q_-} + h(q_2)\delta(q_-)) J(q)$ is a solution for any $h(q_2)$, because $q_- \delta(q_-)$ is effectively equal to zero, (assuming $J(q) \neq 1/q_-$  at $q_-=0$). This is ultimately an infrared ambiguity in the inverse of $\partial_-$.

Second, note that we will also regulate the vacuum solution $M(q-p) =\lim_{x^2 \rightarrow 0} \theta(q-p)e^{ik_2x^2}$ so that $\int d^2p M(p,p) = B = 0$. 

Taking both the infrared ambiguity and the regularization of $M$ into account, we have 
\begin{eqnarray*}
 I_*(q_-) &  = & \frac{1}{2\pi} \int dp_2 dp_- \left(\frac{1}{p_-}+h(p_2)\delta(p_-) \right)e^{i(q_2-p_2)x^2} \theta(q_- - p_-)\\
&  = & \frac{1}{2\pi} \int dp_2 dp_-  h(p_2)\delta(p_-) e^{i(q_2-p_2)x^2}\theta(q_- - p_-)\\
& = &  \frac{1}{2\pi} \int dp_2 h(p_2) e^{i(q_2-p_2)x^2}\theta(q_-)\\
& = & \frac{1}{2\pi} e^{iq_2x^2}\int dp_2 h(p_2) e^{-ip_2 x^2}\theta(q_-)
\end{eqnarray*}
Note that although $h(p_2)$ is not specified, it must be dimensionless. Let us take $h(p_2)= h_0 e^{-(p_2x^2)^2}\cos (p_2 x^2)$. Then
\begin{eqnarray*}
 I_*(q_-) &  = & \frac{h_0}{2{\pi}} e^{iq_2x^2}\int dp_2 e^{-(p_2x^2)^2}\cos (p_2 x^2) e^{-ip_2 x^2}\theta(q_-) \\
& = & \frac{h_0}{2{\pi}} e^{iq_2x^2}\int dp_2 e^{-(p_2x^2)^2}\cos^2 (p_2 x^2)\theta(q_-) \\
& = & e^{iq_2x^2}/x^2 \theta(q_-) \\
& = & (1/x^2 +iq_2) \theta(q_-)
\end{eqnarray*}
where, in the last line, we took the limit $x^2 \rightarrow 0$, and adjusted the numerical constant $h_0$.

In the limit $x^2 \rightarrow 0$ we get a divergent term independent of momentum, which we cancel with the mass counterterm, and a finite term remains, $iq_2$; hence $(m + I_*(-q)) = iq_2$. Regardless of the choice of $h$, by dimensional analysis the finite part of the integral $ \lim_{x^2 \rightarrow 0} e^{iq_2x^2}\int dp_2 h(p_2) e^{-ip_2 x^2}$ would have to be proportional to $q_2$, so our answer would be the unchanged up to a redefinition of $\lambda$ by a numerical factor, assuming $h(p_2)$ is real and reasonably well-behaved. 

Using \eqref{fCon1} and \eqref{fCon2}, the other components of the exact propagator are:
\begin{eqnarray}
\langle \psi_+(q) \psi_-^\dagger(q')\rangle
& = &
\frac{{-q_2}(1-\lambda)}{2 q_-}N \theta(q_-) (2\pi)^2\delta^2(q+q') \label{equaltimeW-+}
 \\
\langle \psi_-(q) \psi_+^\dagger(q')\rangle
 & = &
\frac{{-q_2}(1 + \lambda)}{2 q_-} 
N  \theta(q_-)(2\pi)^2\delta^2(q+q') \label{equaltimeW+-}
 \\
\langle \psi_+(q) \psi_+^\dagger(q')\rangle
& = & 
\frac{{q_2}^2(1- \lambda^2)}{2\sqrt{2} q_-^2}
N \theta(q_-) (2\pi)^2\delta^2(q+q') \label{equaltimeW++}
\end{eqnarray}

Let us compare with the equal time propagator derived earlier in section \ref{free-energy-section}, obtained from \eqref{aef}:
\begin{equation} \begin{split}
& \lim_{x^+ \rightarrow 0^+} \int \frac{dk_+}{2\pi} \left(\braket{\psi(k) \bar{\psi}(k)}\gamma^0 e^{ik_+x^+} \right) \\
& = \lim_{x^+ \rightarrow 0^+} \int \frac{dk_+}{2\pi}\frac{-i k_\mu\gamma^\mu -i\Sigma(k)_+\gamma^+ + \Sigma_I(k)}{k^2  + i \epsilon}\gamma^0 (2\pi)^2 \delta^2(k+k')
\end{split}
\end{equation}
Comparison with (\eqref{equaltimeW--}-\eqref{equaltimeW++}), immediately implies that $f_0^2+g_0=0$, and $f_0 \sim \lambda$. We see that, because of the $\lambda$ independence of the vacuum, all $\lambda$ dependence arises from the fermionic constraints \eqref{fermionConstraint1}, \eqref{fermionConstraint2}. 

Note that in this discussion we did not use dimensional regularization which was used in the Euclidean calculation and hence this agreement is significant.

\subsection{Large $N$ perturbation theory and a formal large $N$ solution}
 
Now that we have the classical solution at large $N$, we can construct a $W_{\infty}$ covariant perturbation theory by parameterizing
\begin{equation}
 M=e^{-iW/{\sqrt N}}M_{{\it class.}}e^{iW/{\sqrt N}} \label{fluct}
\end{equation}
 in terms of the fluctuations $W$. 

The $W$'s satisfy an algebra inherited from the commutation relations of the $M$'s, and break up into subalgebras when expanded around $M_{{\it class.}}$ exactly as in \cite{Dhar:1994ib}, where a similar treatment was carried out for $QCD_2$ that gave rise to 't Hooft's well known equation for the spectrum of mesons. 

The fluctuation equation obtained from substituting \eqref{fluct} into $-i\partial_+ M = [H,M]$ is a second order integral equation for $W$. It takes the following schematic form:
\begin{equation} 
\begin{split}
 \partial_+ \overline{W}(p,p';x^+) = & \frac{i}{4} \left( \frac{{p'}_2^2 + m^2}{p'_-} - \frac{p_2^2 + m^2}{p_-} \right) \overline{W}(p,p';x^+) \\
& + \lambda \int d^2q \left( \ldots W\right) + \lambda^2 \int d^2q d^2q' \left( \ldots W\right)  \label{fluct2}
\end{split}
\end{equation}
where $\overline{W}(p,p';x^+) = W(p,p')\theta(p_-)$

Any single-trace operator ($J_0$, $J_1$, $J_2$ etc.) can be expressed in terms of $M$'s using solution to the constraints, \eqref{gaugeConstraint}, \eqref{fCon1} and \eqref{fCon2}, and Hamilton's equation for $\partial_+ \psi_-$; hence, any two-point function of $J$'s can be expressed in terms of two-point functions of $M$'s. Because $\braket{M(p, p+k)} = 0$ for nonzero $k$, the Green's function for the fluctuation equation \eqref{fluct} is needed to calculate the leading nonzero contribution to a generic two-point function of $M$'s in the $1/N$ expansion. Thus, solving equation \eqref{fluct2} for $W_\infty$-covariant fluctuations around the classical solution would, in principle, enable one to calculate exact two-point functions of all single trace operators in the large $N$ limit (subject to, of course, subtleties associated with the light-cone Hamiltonian formalism.)

\section{Comments on the Holographic Dual}
\label{holographic-dual-section}

\subsection{Vector models are dual to higher spin gauge theories}

The non-renormalization of single trace currents in $U(N)$ Chern-Simons-fermion theory at leading order in $1/N$ indicates that its holographic dual has a free spectrum of purely higher spin gauge fields, of spin $s=0,1,2,3,\cdots$, and their bound states (which are dual to multi-trace operators). The divergences of the single trace currents are mixed with double and possibly triple trace operators, and the mixing is suppressed by $N^{-{1\over 2}}$. By comparing two and three point functions of the currents, we see that $N^{-{1\over 2}}$ should be identified with the bulk coupling constant.

Classically, the bulk nonlinear equations of motion must therefore preserve the higher spin gauge symmetries as well, for otherwise they would introduce extra longitudinal degrees of freedom for the higher spin fields even in the zero coupling limit. This means that the dual bulk theory is a {\it pure} higher spin {\it gauge} theory. On the other hand, the higher spin symmetry in the Chern-Simons-fermion theory is broken by $1/N$ corrections, and we expect the HS gauge symmetry to be broken in the bulk by boundary conditions through loop effects,\footnote{This has been shown to occur, for instance, in Vasiliev's A-type theory in $AdS_4$ with the $\Delta=2$ boundary condition, dual to the critical $O(N)$ model \cite{Giombi:2011ya}.} and the HS gauge fields become massive through the mixing of its longitudinal mode with bound states.

Our argument of non-renormalization of singlet bilinear currents extends to general vector models coupled to Chern-Simons gauge fields, such as the various supersymmetric extensions of the Chern-Simons-fermion theory discussed so far. We therefore anticipate all such theories to be dual to some higher spin gauge theory in the bulk. The bulk theory is generally not parity invariant.

\subsection{Parity violating Vasiliev theory}

More precisely, we conjecture that the three-dimensional critical $U(N)$ Chern-Simons theory at level $k$ coupled to a single fundamental fermion in the 't Hooft limit $N\gg 1$, $\lambda=N/k$ finite, is dual to a parity-violating higher spin gauge theory in $AdS_4$ that contains one gauge field for each non-negative integer spin $s=0,1,2,\cdots$.
A class of such theories has indeed been constructed by Vasiliev, generalizing the parity preserving A-type and B-type theories (which have been conjectured to be dual to free/critical $U(N)$ scalar and fermion vector models in the singlet sector). We will refer to them as parity-violating Vasiliev theories in $AdS_4$. The bulk interactions of the parity violating Vasiliev theory are governed by a single function
\ie
f(X) = 1+ X \exp(i\theta(X))
\fe
where $\theta(X)$ is a real, even function in $X$, but otherwise unconstrained (in the classical theory). Roughly, when expanded in powers of $X$, the order $X^n$ term in $f(X)$ governs the $(n+2)$-th order coupling of the bulk theory.

Vasiliev's theory has the feature that the bulk scalar field has mass $m^2=-2/\ell_{AdS}^2$, and the dual scalar operator (with one choice of boundary condition) has classical dimension 2. This does not follow from the linearized higher spin gauge symmetry, and is required in order to construct higher spin gauge invariant interactions. This feature is in precise agreement with the non-renormalization of the scalar operator $\bar\psi\psi$ at the planar level in the Chern-Simons-fermion vector model, which we have argued using the anomalous current conservation relation and verified explicitly at two-loop.

Now let us discuss the parity transformation in Vasiliev theory. The function $f(X)$ transforms to its complex conjugate $\overline{f(X)}$ under parity. One may either assign even or odd parity to the bulk scalar field, which corresponds to assigning even or odd parity to the variable $X$. In the former case, a parity invariant theory is given by $f(X)=\overline{f(X)}$, or $f(X)=1+X$; this is the A-type theory. In the latter case, where $X$ is parity odd, a parity invariant theory is given by $f(X)=1+iX$; this is the B-type theory. The B-type theory has been conjectured to the dual to the {\sl free} limit of the $U(N)$ Chern-Simons-fermion theory, where the Chern-Simons gauge fields decouple and their only effect is to restrict the operator spectrum to gauge singlets; parity is restored in this free limit.

It is thus natural to suspect that the Chern-Simons-fermion theory at large $N$ but finite 't Hooft coupling $\lambda=N/k$, is dual to the parity violating Vasiliev theory governed by a function $f(X)=1+X \exp(i\theta(\lambda, X))$, where
\ie
\theta(\lambda, X) = {\pi\over 2} + \beta(\lambda, X),
\fe
is real, even in $X$, and depends on $\lambda$ in some way. For instance, one should have $\beta(0,X)=0$, so that in the $\lambda=0$ case the conjecture reduces to the duality between B type theory and free fermions. While the CS-fermion theory violates parity symmetry, parity is restored if we flip the Chern-Simons level $k$ (i.e.  $\lambda\to -\lambda$) together with the standard parity action on the fields, under which the scalar operator $\bar\psi \psi$ is parity odd. So if the Chern-Simons-fermion theory is dual to the parity violating Vasiliev theory, we must also have $\beta(-\lambda, X)=-\beta(\lambda, X)$.


While there is a priori an infinite family of parity-violating Vasiliev theories in $AdS_4$ governed by the function $f(X)$, we may speculate that there is a particular choice of $f(X)$  determined by the 't Hooft coupling $\lambda$ in the above form, that describes the bulk theory dual to critical $U(N)$ CS-fermion theory at large $N$. It is then an interesting question to determine $\theta(\lambda, X)$ as a function of $\lambda$. In principle, this can be done perturbatively by comparing the $n$-point functions of the HS currents in CS-fermion theory and the bulk Vasiliev theory. We may expand the function $\beta(\lambda,X)$ as
$\sum_{k=0}^\infty \beta_k(\lambda) X^{2k}$. Then $\beta_k$ may in principle be determined by comparing with the $(2k+3)$-point function of the higher spin currents in the CFT at leading order in $1/N$. In particular, the leading phase $\beta_0(\lambda)$ can be determined by comparing with the three-point functions of currents.

The simplest check of such a conjecture would be the comparison of bulk tree level three point functions in parity-violating Vasiliev theory, and the planar three point function $\langle JJJ\rangle$ in the CS-fermion theory. The former computation, performed along the lines of \cite{Giombi:2010vg}, will be presented elsewhere. The result obtained using the gauge function method of \cite{Giombi:2010vg} (``$W=0$ gauge") indeed agrees with the parity-even part of $\langle JJJ\rangle$ in CS-fermion theory up to two-loop: in particular, the property that $c_B$ in (\ref{peven}) is independent of the three spins is a {\it prediction} from the bulk theory! However, we have failed to produce the parity-odd structure of $\langle JJJ\rangle$ from the bulk Vasiliev theory. This could be either due to the singular nature of the $W=0$ gauge in \cite{Giombi:2010vg} and that the parity-odd contribution is simply missed by the regularization of \cite{Giombi:2010vg}, or it could be that the correct dual bulk theory to the CS-fermion CFT is {\it not} described by Vasiliev's system, but rather some other higher spin gauge theory with the same field content in $AdS_4$. We shall leave the resolution of this puzzle to future work.

Let us comment on higher spin symmetry breaking as seen from the correlation functions.
While at tree level in the bulk, we expect the higher spin symmetry to be preserved, there can be subtle boundary contributions in computing correlators. Examples of similar nature in the bulk dual of critical $O(N)$ model have been discussed in \cite{Giombi:2011ya}. Recall the anomalous current conservation relation in the boundary theory, of the form
\ie
\partial^{\mu}j^{(s)}_{\mu\cdots} \sim {1\over \sqrt{N}} \sum_{s_1+s_2<s} [j^{(s_1)}] \otimes [j^{(s_2)}]
+{1\over N} \sum_{s_1+s_2+s_3<s}  [j^{(s_1)}] \otimes [j^{(s_2)}] \otimes [j^{(s_3)}]
\fe
On the right hand side, we have listed the conformal families that can appear in the divergence of the current $j^{(s)}$. The dependence on 't Hooft coupling $\lambda$ is omitted in this expression. As discussed in section 
\ref{operator-section}, though these multi-trace operators appear to be 
suppressed by powers of $1/N$, they can contribute to the corresponding
divergence of correlation functions at the {\it leading} nontrivial order 
in $1/N$.

For the three-point functions $\langle j^{(s_1)} j^{(s_2)} j^{(s_3)}\rangle$, the naive current conservation equation at leading order in $1/N$, namely ${\cal O}(N^{-{1\over 2}})$ when the two point functions are normalized to ${\cal O}(N^0)$,
\ie
\partial^{\mu_1} \langle j^{(s_1)}_{\mu_1\cdots} j^{(s_2)} j^{(s_3)}\rangle = 0
\fe
is necessarily satisfied if $s_1< s_2+s_3$.\footnote{In the 
explicit computations we have performed for the classical divergence of 
currents for spins up to 5 we in fact find that this the current is 
effectively conserved in correlators provided $s_1 <s_2+s_3+2$. It is possible 
that this relation is more generally true and can be proved from an analysis 
of structures allowed by conformal invariance in 3 point functions. If this 
is the case the three point function will be divergence free with respect to 
all three currents if the three spins $s_1,s_2,s_3$ obey this less strict triangular 
inequality.}

On the other hand, if $s_1\geq s_2+s_3$, then there is the 
possibility of a  nontrivial ${\cal O}(N^{-{1\over 2}})$ contribution,
\ie
\partial^{\mu_1} \langle j^{(s_1)}_{\mu_1\cdots} j^{(s_2)} j^{(s_3)}\rangle \sim {1\over\sqrt{N}}\langle j^{(s_2)} j^{(s_2)}\rangle \langle j^{(s_3)} j^{(s_3)}\rangle.
\fe
So even at leading order in $1/N$, the naive current conservation relation is violated by the three point function. Naively, the boundary-to-bulk propagator of the higher spin fields in $AdS_4$ is divergence free with respect to the position and polarization of the source current, and it may seem that such a non-conserved three point function cannot result from a tree level bulk computation. As pointed out in \cite{Giombi:2011ya}, however, this is not necessarily the case, due to subtle boundary contributions to the bulk integral.


\subsection{A higher spin gauge theory as a string theory}\label{string}

A long standing question is whether the pure higher spin gauge theory in $AdS$ can be embedded in string theory. We have seen that quite generally vector models coupled to Chern-Simons fields are dual to higher spin gauge theories. On the other hand, a large class of supersymmetric quiver Chern-Simons-matter theories are dual to string theories in $AdS_4$. A particularly interesting class of examples is that of the ABJ theory \cite{Aharony:2008gk}: the ${\cal N}=6$ $U(N)_k\times U(M)_{-k}$ Chern-Simons theory coupled to a pair of bifundamental hypermultiplets. It has been conjectured to be dual to type IIA string theory on $AdS_4\times \mathbb{CP}^3$, with a flat $B$-field obeying
\ie\label{flatb}
{1\over 2\pi\A'}\int_{\mathbb{CP}^1} B = {N-M\over k}+\frac{1}{2},
\fe
when $|N-M|<k$. The need for the shift by $\frac{1}{2}$ in the equation above was explained in \cite{Aharony:2009fc}.\footnote{We thank O. Aharony for bringing this to our attention.}

While the dual string theory is weakly coupled in the 't Hooft limit, the duality is believed to hold exactly for all $N,M$ and $k$. In the special case $M=1$ (or generally small $M$), ABJ theory reduces to a supersymmetric Chern-Simons {\it vector model}, in the following sense. The most straightforward ${\cal N}=3$ supersymmetric generalization of the vector model we considered so far is the ${\cal N}=3$ $U(N)$ or $SU(N)$ Chern-Simons theory coupled to a fundamental hypermultiplet. Most features of large $N$ vector models discussed in this paper still hold when a number of fundamental flavors are introduced (the number of flavors does not scale with $N$ in the large $N$ limit here). Now consider ${\cal N}=3$ $U(N)_k$ Chern-Simons theory with $2M$ hypermultiplets. By essentially the same argument as in section 2, at large $N$ this vector model is dual to a higher spin gauge theory in $AdS_4$ with $U(2M)$ Chan-Paton factors.\footnote{Note that Vasiliev's system can be generalized to a theory of nonabelian higher spin fields by introducing $U(N_f)$ Chan-Paton factors, i.e. replacing the $*$ algebra by its tensor product with the algebra of $N_f\times N_f$ matrices.} To obtain the ${\cal N}=6$ theory, one needs to further gauge the $U(M)$ flavor symmetry with ${\cal N}=3$ Chern-Simons coupling at level $-k$ (as opposed to level infinity). This however corresponds to simply modifying the $AdS$ boundary condition on the bulk $U(M)$ spin-1 gauge fields, from the purely magnetic boundary condition to an electric-magnetic mixed boundary condition \cite{Witten:2003ya}.

Let us comment briefly on the distinction between $U(N)$ and $SU(N)$ gauge groups here. Take the ABJ theory with $M=1$ as an example. The gauge group is $U(N)_k\times U(1)_{-k}$. Denote by $a$ and $\tilde a$ the diagonal $U(1)$ of the $U(N)_k$ and the $U(1)_{-k}$ gauge fields, respectively. Note that $a$ has Chern-Simons level $Nk$ as a $U(1)$ gauge field. The matter fields couple to $b=a-\tilde a$ only. The Chern-Simons action for $a,\tilde a$ takes the form
\ie
{k\over 4\pi} \int (N a \wedge da - \tilde a\wedge d\tilde a) = {k\over 4\pi} \int (N a \wedge da - (a-b)\wedge d (a-b)).
\fe
Since only $b$ couple to the matter fields, we can integrate out $a$ exactly, which sets $a=-{1\over N-1}b$. The resulting Chern-Simons action is that of the $U(1)$ gauge field $b$ at level $-{N\over N-1}k$. So the same theory can be expressed as $SU(N)_k\times U(1)_{-k'}$ Chern-Simons coupled to 2 bifundamental hypermultiplets, where $k'={N\over N-1}k$.

We are now naturally led to the conjecture that the $U(N)\times U(M)$ ABJ theory in the vector model limit, i.e. large $N$ and fixed finite $M$, is dual to an ${\cal N}=6$ version of the parity violating Vasiliev theory with $U(M)$ Chan-Paton factors. Furthermore, we conjecture that the type IIA string field theory on $AdS_4\times \mathbb{CP}^3$, with
a flat $B$-field as in (\ref{flatb}),
in limit $N\gg M$, is a higher spin gauge theory. Note that in the vector model limit $N\gg M$, the 't Hooft coupling $N/k$ is restricted to be less than 1, and the bulk IIA string theory is always in the highly stringy regime. 

In the ABJ theory, the bulk single string states are dual to single trace operators, of the form
\ie
{\rm Tr} (ABAB\cdots)
\fe
where $A$ and $B$ denote fields in the $(\Box,\overline\Box)$ and $(\overline\Box,\Box)$ representations respectively. In the $M=1$ case, the single trace trace breaks into the product of a number of bilinears which are singlets under the $SU(N)$ gauge group. So we learn that the general single string state should 
correspond a multi-particle state of higher spin fields. It suggests that the higher spin gauge fields are in fact more fundamental than strings in this case, as the latter  are composite objects made out of the former! Note that the coupling constant $g$ of the higher spin gauge theory is mapped to $\sim 1/\sqrt{N}$ of the vector model. Now $\lambda_{bulk} = g^2M \sim M/N$ is the {\it bulk} 't Hooft coupling. We see that the vector model limit $N\gg M$ is precisely where the nonabelian higher spin gauge theory is weakly coupled. As $M/N$ becomes of order 1, the higher spin fields interact strongly, and we expect their bound states to form the string excitations. If the dual higher spin gauge theory is indeed described by Vasiliev's system, it would be very interesting to find the explicit map from Vasiliev's master fields to the (closed) string fields, which could allow for a precise formulation of the closed string field theory in this limit.

\section{Discussion}
\label{discussion-section}

This paper has been devoted to the study of $U(N)$ Chern-Simons theory
coupled to a single massless fundamental fermion. We set out to
to solve this theory in the 't Hooft large $N$ limit and have been partly
successful in this task. In particular, we were able to compute the
partition function on $\mathbb{R}^2$ of the theory at finite temperature at all 
values of the 't Hooft coupling. In order to perform this computation we 
employed
a lightcone gauge, and used a dimensional reduction regularization scheme.
It would be useful to perform further checks of the consistency of
our gauge and regularization scheme.

In the course of our analysis of this theory we have encountered the 
fact that the higher spin currents obey anomalous conservation 
equations. In the classical theory we have computed the
explicit form of these conservation equation at low values of the spin 
(see e.g. \eqref{jtanom}). These nonlinear anomalous conservation equations 
contain a large amount of information; for example they encode the anomalous 
dimensions of the spin $s$ currents in a $\frac{1}{N}$ expansion 
(see subsection \ref{anom-curr}). It would be interesting to determine 
the explicit form of these anomalous conservation equations, if possible,
as a function of $\lambda$. 

We have also argued that the spectrum of local gauge invariant 
operators in our theory, at
dimensions of order unity, is not renormalized as a function of the
't Hooft coupling at leading order in large $N$. A class of
such operators (those that may be written entirely in terms of $x^{-}$
derivatives) are dressed due to  interactions only by rainbow graphs in 
the large $N$ limit. The  ``quantum dressed'' operator obtained in this 
way is given by the solution to an integral equation both at zero 
and at finite temperature. Exact expressions for all two and and three point 
functions are obtained by sewing these
quantum dressed operators together using the exact fermion propagator 
computed in this paper. Such a solution would allow us  to find exact 
expressions, for example, of fermion number current and stress tensor 
two point functions at finite temperature as a function of ${\vec k}$ and 
$\omega$.  As these quantities are of obvious physical interest, the 
corresponding integral equations deserve serious attention.

In this paper we have analyzed in detail the spectrum of local operators. But clearly the theory also admits interesting
non-local gauge invariant observables, namely the Wilson loop operators specified by a choice of a closed contour in space-time and a representation of the gauge group. In pure Chern-Simons theory, these are topological observables which compute certain knot invariants \cite{Witten:1988hf}. When Chern-Simons theory is coupled to matter, the Wilson loops are no longer topological. In a theory with fundamental matter, however, at leading order at large $N$ their expectation values are the same as in pure Chern-Simons theory, since diagrams with fundamental matter loops are suppressed. It may be interesting to explicitly compute the leading $1/N$ corrections for some Wilson loop in our theory. In standard examples of AdS/CFT, Wilson loop operators are dual to macroscopic open strings ending on the closed contour at the boundary. In this paper we have proposed that our theory should be dual to some higher spin gauge theory in $AdS_4$. It would be interesting to understand if Vasiliev's theory, or a deformation thereof, may possibly contain extended objects which could be dual to the Wilson loops of the gauge theory. 

It has recently been realized that the partition functions of supersymmetric
Chern-Simons theories on $S^3$ are often exactly computable via the techniques
of supersymmetric localization \cite{Pestun:2007rz, Kapustin:2009kz, Drukker:2010nc, Jafferis:2010un}. This quantity appears to play the role of a $c$ 
function under the renormalization group flow and so is of clear physical 
interest. It seems possible that the techniques employed in this paper will 
allow the exact computation of the $S^3$ partition function of our 
non-supersymmetric theory in the large $N$ limit, as a function of $\lambda$.

As we have explained in subsection \ref{sphere} above, in the limit 
$\lambda \to 0$,
the partition function of our theory on $S^2 \times S^1$ undergoes an
interesting phase transition at a temperature of order ${\cal O}(\sqrt{N})$ \cite{Shenker:2011zf}.
The low temperature phase may be thought of as a gas of single traces. 
At temperatures that are held fixed when $N$ is taken to infinity, the 
single trace anomalous dimension non renormalization theorems of this paper 
demonstrate that the free energy of this phase is not renormalized as a 
function of $\lambda$. At temperatures at of order $\sqrt{N}$ the free
energy of of the low temperature phase presumably does get renormalized 
as a function of $\lambda$. And the free energy of the 
high temperature phase of the partition function (which be thought of as a 
gas of the fundamental fermions rather than their mesonic bound states) 
certainly does get renormalized a function of $\lambda$. In this paper 
in particular we have computed the coefficient of the $T^3$ growth 
of the free energy in the very high temperature limit, and have shown that 
this coefficient, which is nonzero at generic $\lambda$, vanishes 
in the limit $\lambda \to 1$. It would be very interesting to
understand how the competition between the two phases changes as a function
of $\lambda$, especially in the limit $\lambda \to 1$. Does the vanishing 
of the free energy at extremely high temperatures, for instance, suggest 
that the phase transition temperature between the low and high temperature 
phases diverges as $ \lambda \to 1$?

It would  also be interesting to investigate transport properties in the high
temperature phase. The following argument suggests that the 
mean free path of any particular fermionic quantum, is of order $N$. 
Consider, for instance, a fermion of colour $i$ scattering off an anti 
fermion of colour $i$ to yield a fermion - anti-fermion pair of colour $j$. 
This scattering amplitude is of order $\frac{1}{N}$, as a consequence of
which the scattering probability is of order  $\frac{1}{N^2}$.
The number of potential scattering partners is of order unity in the thermal
bath, while there are $N$ possibilities for the colour $j$ of the final 
fermion pair. It follows that
the probability that any given fermionic quantum undergoes a scattering
process is of order $\frac{1}{N}$, giving rise to a mean free path
of order $N$. This result should apply at every value of $\lambda$.
All this very similar to the known behavior of transport properties 
in the critical O(N) model (see e.g. \cite{Sachdev}) and very different 
 from theories with matrix degrees of freedom
in the t' Hooft limit where the mean free time between collisions of a single 
parton is independent of $N$ \cite{Aharony:2005bm}. The fact that the fermion gas  effectively behaves as a free gas as far
as transport properties are concerned, suggests that fluctuations about
the high temperature phase do not decay over a time of order unity.
This discussion suggests that large $N$ vector models in the 't Hooft 
limit thermalize much less efficiently than their matrix model counterparts.
For this reason the dual description of such a high temperature phase
has properties that are very different from a conventional classical black 
hole. \footnote{When the ranks, $N_1$ and $N_2$ of the ABJ theory are both 
comparable, the theory undergoes a deconfinement transition, dual to 
a Hawking Page transition, at a temperature of order unity. The high 
temperature phase in this transition is a black hole in Einstein gravity. 
It follows from the discussion above, on the other hand, that when 
$N_1$ is lowered to a number of order unity with $N_2$ still large then 
the deconfinement phase transition temperature is much larger (it scales 
like $\sqrt{N_2}$). Moreover the dual black hole appears to change 
its nature qualitatively; it appears to stop absorbing.}

It should also be possible to generalize our discussion of the finite
temperature behaviour of our theory to include a chemical potential for
fermion number. The finite $\lambda$ behaviour of such a system
describes an interacting Fermi sea in three dimensions, and so may be of 
interest for various condensed matter problems.

It would be interesting to generalize
our computation of the finite temperature free energy to Chern-Simons theory
interacting with fundamental bosons \cite{oferetal}, and also to the supersymmmetric theory
with different numbers of supersymmetries.

An outstanding question about the theory studied in this paper
as well as the generalizations described above is ``what is its
bulk dual description?'' We have argued above that this dual description
is a higher spin theory. However we do not yet have a precise conjecture
for the nature of this dual; this is clearly an issue of immediate
interest, especially since the maximally 
supersymmetric version of our theory is embedded into string theory 
as a limit of the ABJ construction (see subsection \ref{string}). 

In this paper we have studied only the 't Hooft large $N$ limit of our 
theory. It is however clear from the example of ABJM theory, that 
large $N$ Chern Simons theories admit a wider class of large $N$ limits 
(e.g. $N \to \infty$, $\lambda \to \infty$, $\frac{N}{\lambda}$ fixed 
in the ABJM context). It would be interesting to investigate whether 
the theory studied in this paper admits an analogous double scaling limit, 
in which $\lambda$ is scaled to unity in a manner coordinated with 
the large $N$ limit\footnote{We thank O. Aharony for discussions on this 
point.}. The dimensions of the higher spin 
currents could be renormalized as a function of coupling in such a limit. 
The fact that we have been able to obtain exact 
results as a function of $\lambda$ perhaps makes such an investigation 
feasible.

To end this discussion we recall we could very easily argue that the 
spectrum of `single trace' operators in our theory is not renormalized as a 
function of $\lambda$, in the 't Hooft large $N$ limit, 
 by combining conformal representation theory with the 
sparseness of the single trace spectrum in vector models. The same non-renormalization
 argument does not apply in theories with adjoint 
or bifundamental matter fields as the single trace spectrum of these 
theories is not sparse.\footnote{In supersymmetric theories with 
such matter content, supersymmetric indices and representation theory 
give rise to a new source of non-renormalization theorems for 
single trace operators even in theories with bifundamental and adjoint 
matter, see e.g. the paper \cite{Minwalla:2011ma}.} On the 
other hand, as we have explained in the introduction, non supersymmetric 
effective large $N$ Chern-Simons fixed lines are easily constructed with 
matter fields in adjoint or bifundamental representations. The scaling 
dimension of single trace operators in such theories is protected neither by 
conformal nor supersymmetric representation theory. 
In these theories all single trace operators baring the stress tensor and 
currents for global symmetries can, and presumably 
do get renormalized. It is at least conceivable that some theory 
with adjoint or bifundamental fermions 
admits a strong coupling limit in which all but a finite number of 
single trace operators are infinitely renormalized, and the dual description 
of the theory is Einstein gravity in 
$AdS_4$, coupled to a minimal number of additional fields. It would 
clearly be very interesting to identify any theory with this property.

As is clear from the discussion above, the computations presented in this
paper have merely scratched the surface of a large and potentially
very interesting area of investigation. We hope to report on some of the
topics discussed above in the future.

\section*{Acknowledgements}

We are grateful to Ofer Aharony and Rajesh Gopakumar for crucial discussions
that led to the formulation of some of the questions addressed in this 
paper. We would like to acknowledge Sayantani Bhattacharyya, 
Jyotirmoy Bhattacharya and V. Umesh for collaboration at various stages 
of this project. We thank O. Aharony and J. Bhattacharya  for comments on the manuscript. We would also like to acknowledge useful discussions with 
C.-M. Chang, K. Damle, J. David, A. Dhar, S. Dutta,  D. Hofman,
N. Iizuka, D. Jafferis, S. Kim, S. S. Lee, G. Mandal, K. Papadodimas, 
M. Rangamani, S. Sachdev, A. Sen, C. Sonnenshein and E. Witten. 
S.G. is supported by Perimeter Institute
for Theoretical Physics. Research at Perimeter Institute is supported by the
Government of Canada through Industry Canada and by the Province of Ontario through the Ministry of Research $\&$ Innovation. The work of S.M. 
was supported by a Swarnajayanti fellowship. The work of S.M., S.P., 
S.T. and S.W. was supported by the DAE, government of India. S.G. would like to thank Princeton U., the Simons Center for Geometry and Physics, Mc Gill U., the Weizmann Institute, the organizers of the PhyMSI conference in Cargese and of the Simons Summer Workshop in Mathematics and Physics 2011 for hospitality during completion of this work. S.M. would 
like to thank the Weizmann Institute, Cambridge University, KIAS, ICTP,  
the organizers of GR19, PASCOS 2010, and the Indian Strings Meeting (ISM) 2011 for 
for their hospitality while this work was in progress. S.P. would like to thank the organizers of ISM 2011, and the Centre for High Energy Physics, Indian Institute of Science, Bangalore for their hospitality. X.Y. would like to 
thank the organizers of GR19 and of 
ISM 2011, Tata Institute of Fundamental Research, 
Berkeley Center for Theoretical Physics, Simons Center for Geometry and Physics, Fields Institute, and Aspen Center for Physics for their hospitality. The work of X.Y. was supported by the Fundamental Laws Initiative Fund at 
Harvard University and in part by NSF Award PHY-0847457.
S.M., S.P., S.T. and S.W.
would also like to acknowledge our debt to the people of India for
their generous and steady support to research in the basic sciences.

\section*{Appendices}
\appendix

\section{Conventions for Propagators and Gauge Conditions}
\label{appC}
The Euclidean action for our theory is
\begin{equation}\label{ea}
\begin{split}
S &= \frac{ik}{4 \pi}\int
\text{Tr} \left(A d A + \frac{2}{3} A^3\right)+ \int
 {\bar \psi}\gamma^\mu D_\mu  \psi,
\end{split}
\end{equation}
where
$$D_\mu \psi = \partial_\mu \psi -i A^a T^a \psi.$$ 
We have
\begin{equation}\label{otherconv}
\begin{split}
\epsilon_{123}&=\epsilon^{123}=1, \\
\gamma^i&=\sigma^i, ~~~(i=1 \ldots 3)\\
{\bar \psi}^\alpha&= (\psi_\alpha)^*,
\end{split}
\end{equation}
where $\sigma^i$ are the ordinary Pauli matrices. Note then that all $\gamma^\mu$ are Hermitian. This implies that
${\bar \psi} \psi$ and ${\bar \psi} \gamma^\mu \psi$ are
real, while $\int dx {\bar \psi} \gamma^\mu \partial_\mu \psi$ is imaginary. The gauge field will be taken to be $A^a T_a$ where $T_a$ is the
fundamental generator normalized so that $Tr T_a^2=\frac{1}{2}$.
Note that
\begin{equation} \label{comp}
\sum_{a} (T^a)_m^n (T^a)_p^q=\frac{1}{2} \delta_m^q \delta_p^n.
\end{equation}

\subsection{Lightcone gauge}
Let us define
\begin{equation}
\begin{split}
x^{\pm}&=\frac{x^1 \pm i x^2}{\sqrt{2}}, \\
A^{\pm}&=A_{\mp}=\frac{A^1 \pm i A^2}{\sqrt{2}}, \\
p^{\pm}&=p_{\mp}=\frac{p^1 \pm i p^2}{\sqrt{2}}, \\
p_s^2&\equiv p_1^2+p_2^2=2p^+p^-.
\end{split}
\end{equation}
We define the lightcone gauge in Euclidean signature by the condition $A_{-}=0$. This can be obtained from Wick rotation of the standard lightcone gauge in Lorentzian signature. However, in this paper we often think of $x^1,x^2$ as purely spatial coordinates. In particular, in the finite temperature calculation, the thermal time direction is orthogonal to the complex lightcone direction.

Note that under a rotation in the 12 plane
$A_- \rightarrow e^{i \alpha} A_-$. Consequently rotations in the 12 plane
commute with the condition $A_-=0$. 

Defining the momentum space fields
\begin{equation}
\begin{split}
A_\mu(x)&=\int \frac{d^3p}{(2\pi)^3} e^{i p\cdot x} A_\mu(p),\\
\psi_\alpha(x)&=\int \frac{d^3p}{(2\pi)^3} e^{i p\cdot x} \psi_\mu(p),\\
{\bar \psi}^\alpha(p)&=(\psi_\alpha(-p))^*,
\end{split}
\end{equation}
the momentum space action in lightcone gauge is
\begin{equation}\label{ea2}
\begin{split}
S &= \frac{-ik}{2\pi}\int \frac{d^3 p}{(2 \pi)^3}
\text{Tr} A_3(-p) p_- A_+(p) + \int \frac{d^3 p}{(2 \pi)^3}
i {\bar \psi}(-p)  \left( \gamma^\mu p_\mu  + M_{bare} \right) \psi(p)\\
&- i \int \frac{d^3 p}{(2 \pi)^3} \int \frac{d^3 q}{(2 \pi)^3}
 {\bar \psi}(-p) (\gamma^+ A_+(-q) + \gamma^3 A_3(-q))\psi(p+q),
\end{split}
\end{equation}
where $A= A^a T_a$.

It follows from this action that\footnote{The propagator for a theory
whose Euclidean action is
$$S=\frac{1}{2}\int \frac{d^3 p}{(2 \pi)^3} \phi_a(-p) Q^{ab}(p) \phi_b(p)$$
is given by
$$ \langle \phi_a(p) \phi_b(-q) \rangle= (2 \pi)^3 \delta(p-q)Q^{-1}_{ab}(p).$$
This rule is correct for both bosons as well as fermions.
In the case of the gauge field we have
$Q^{3 +}=-Q^{+3}=\frac{-i k p_-}{4 \pi}$.
In the case of the fermionic field we have
$Q^{{\bar \psi} \psi}= i p_\mu \gamma^\mu + M_{bare} $ and
$Q^{\psi {\bar \psi}}=i p_\mu (\gamma^\mu)^T + M_{bare} $.}
\begin{equation}\label{propagators}\begin{split}
\langle  \psi(p)_n {\bar \psi}^m(-q) \rangle&=
\delta_n^m \frac{-i \gamma^\mu p_\mu}{p^2} \times (2 \pi)^3 \delta(p-q), \\
\langle A^a_3(p) A^b_+(-q) \rangle &
=-\langle A_+^b(p) A_3^a(-q) \rangle= -\frac{4 \pi i}{k}
\frac{1}{p^+} \times (2 \pi)^3 \delta(p-q) \delta^{ab}.
\end{split}
\end{equation}
Adopting the notation
$$\langle A^a_\mu(p) A^b_\nu(-q) \rangle=
(2\pi)^3 \delta(p-q) G_{\mu\nu}(p) \delta^{ab}, $$
we have
\begin{equation}
\label{propnot}
G_{+3}(p)=-G_{3+}(p)=\frac{4 \pi i}{k p^+}.
\end{equation}
In coordinate space, the gauge propagator reads
\ie
&\langle A_3(x) A_+(0)\rangle = - \langle A_+ (x) A_3(0)\rangle = -{8\pi i\over k} \int {d^3p\over (2\pi)^3} {p^-\over p_s^2+i\epsilon}e^{ip\cdot x}
= -{8\pi \over k}\delta(x^3)\partial_+ \int {d^2p_s\over (2\pi)^2} {e^{ip_s\cdot x}\over p_s^2+i\epsilon}
\\
&= {2\over k}{\delta(x^0)\over x^+}.
\fe
\subsection{Temporal gauge}
The temporal gauge is defined by the condition $A_3=0$ (Wick rotating the Lorentzian temporal gauge $A_0=0$).
In this gauge, the gauge field propagator is written in position space as
\ie
\langle A_i(x) A_j(0)\rangle = {2\pi i\over k}\epsilon_{ij} {\rm sign}(x^3) \delta^2(\vec x),
\fe
and in momentum space,
\ie
\langle A_i(p) A_j(-q)\rangle &= {2\pi\over k}\epsilon_{ij}\left[{1\over p^3+i\epsilon}+{1\over p^3-i\epsilon} \right] (2\pi)^3\delta^3(p-q)
\\
&= {4\pi\over k}\epsilon_{ij} {p^3\over (p^3)^2+\epsilon^2}\, (2\pi)^3\delta^3(p-q).
\fe
\subsection{Feynman gauge}
We may add to the action a covariant gauge fixing term of the form
\ie
S_{F} = {k\over 4\pi} \int d^3x\, \xi {\rm Tr} (\partial_\mu A^\mu)^2.
\fe
The Feynman gauge is obtained in the limit $\xi\to \infty$, in which case the propagator for $A_\mu$ becomes simply
\ie
\langle A^a_\mu(p) A^b_\nu(-q)\rangle = -{4\pi\over k}\delta^{ab} \epsilon_{\mu\nu\rho} {p^\rho\over p^2}\,(2\pi)^3\delta^3(p-q).
\fe

\section{Perturbative Analysis of Fermion Self Energy}
In this Appendix we give some details on the perturbative computation of the fermion self-energy in the three different gauge choices defined in \ref{appC}: the Feynman gauge, temporal gauge, and lightcone gauge. The Feynman gauge has the advantage of being Lorentz covariant, but the cubic Chern-Simons interaction makes computations beyond one-loop complicated. The cubic coupling of the gauge field disappears in temporal and light cone gauges. The temporal gauge has the advantage of having a very simple form of the gauge field propagator, which makes explicit perturbative contributions easy. On the other hand, we will see below that it introduces unphysical logarithmic divergences in the fermion self energy at two loops (which should disappear in gauge invariant correlators). The light cone gauge does not suffer from this log divergence problem. While explicit perturbative calculations of correlation functions of gauge invariant operators are sometimes more conveniently done in temporal gauge, the lightcone gauge allowed us to partially solve the planar limit of the Chern-Simons-fermion theory (to all loop order).

\subsection{Feynman gauge}
In Feynman gauge, the 1-loop free energy is
\ie
&\Sigma(p) =2\pi \lambda\int {d^3 q\over (2\pi)^3} \epsilon_{\mu\nu\rho} {q^\rho\over q^2} \gamma^\mu{-i({\slash\!\!\! p}+{\slash\!\!\! q})\over (p+q)^2} \gamma^\nu
\\
&=-i 2\pi \lambda\int {d^3 q\over (2\pi)^3} {q^\rho(p+q)^\sigma \over q^2(p+q)^2}  \epsilon_{\mu\nu\rho}\gamma^\mu\gamma_\sigma \gamma^\nu
\\
&= -4\pi \lambda\int {d^3 q\over (2\pi)^3} {q\cdot(p+q) \over q^2(p+q)^2}  = -2\pi \lambda \int{d^3 q\over (2\pi)^3} {(p+q)^2+q^2-p^2\over q^2(p+q)^2}
\\
&= {\pi\over 4} \lambda |p|
\fe
The 1-loop corrected fermion propagator $\langle \psi_{i\A}(p) \bar\psi^j_\B(p) \rangle$ is then proportional to $\lambda \delta^j_i\epsilon_{\A\B}/|p|$.

\bigskip

\centerline{
\begin{fmffile}{self1}
        \begin{tabular}{c}
            \begin{fmfgraph*}(85,70)
                \fmfleft{i1}
                \fmfright{o1}
                \fmf{wiggly,right=1}{v1,v2}
                \fmffixed{(.6h,0)}{v2,v1}
                \fmf{plain}{i1,v1,v2,o1}
             \end{fmfgraph*}
        \end{tabular}
        \end{fmffile}
}

\subsection{Temporal gauge}
In the temporal gauge $A_3=0$, the one-loop fermion self energy is
\ie
&\Sigma(p) = 2\pi\lambda \int {d^3\vec q\over (2\pi)^3} \epsilon_{ij}\gamma^i {p^3-q^3\over (p^3-q^3)^2+\epsilon^2}{-i{\slash\!\!\! q} \over q^2}\gamma^j
\\
&= -i2\pi\lambda\int {d^3\vec q\over (2\pi)^3} \epsilon_{ij}\gamma^i {p^3-q^3\over (p^3-q^3)^2+\epsilon^2}{q^3 \gamma^3 \over q^2}\gamma^j
\\
&=-{8\pi^2  \lambda\over (2\pi)^3}\int_0^\infty dQ Q \int dq^3  {p^3-q^3\over (p^3-q^3)^2+\epsilon^2}{q^3 \over (q^3)^2+Q^2 }
\\
&={ \lambda} \int_0^\infty dQ {Q^2\over (p^3)^2 + Q^2}
\\
&={ \lambda}\left( \Lambda - {\pi\over 2}|p^3| \right)
\fe
After subtracting off the linear divergence by tuning the bare mass of the fermion, we end up with the renormalized 1-loop self energy
\ie
\Sigma_1(p) =  - {\pi \over 2} \lambda |p^3|.
\fe
We can carry on to the two-loop self energy in temporal gauge, which only receives the contribution from the rainbow diagram. It is given by
\ie
&\Sigma_2(p) = -2\pi\lambda \int {d^3q\over (2\pi)^3} {p^3+q^3\over (p^3 + q^3)^2+\epsilon^2} \epsilon_{ij} \gamma^i {1\over q^2} \Sigma_1(p) \gamma^j \\
&=i {(4\pi\lambda)^2\over 4}\gamma^3 \int {d^3q\over (2\pi)^3} {p^3+q^3\over (p^3 + q^3)^2+\epsilon^2}  {|q^3|\over q^2}
\\
&= {i\lambda^2}\gamma^3 \int_0^\infty dQ Q{p^3 \ln{(p^3)^2\over Q^2}\over (p^3)^2+Q^2}
\\
&= {i\lambda^2} p^3 \gamma^3 \left[ {\pi^2\over 12} - \left(\ln {\Lambda\over |p^3|}\right)^2 \right]
\fe
We believe that the $\log^2$ divergence is an artifact of the temporal gauge and should drop out in gauge invariant observables. A systematic treatment of regularization in the temporal gauge appears to be rather complicated. In the one and two loop corrections to three point functions of gauge invariant operators that we will perform later, such divergences are avoided by special choices of position and polarization configurations of the operator insertions.

\centerline{
\begin{fmffile}{self2}
        \begin{tabular}{c}
            \begin{fmfgraph*}(95,80)
                \fmfleft{i1}
                \fmfright{o1}
                \fmf{wiggly,right=1}{v1,v2}
                \fmf{wiggly,right=1}{v3,v4}
                \fmffixed{(.7h,0)}{v2,v1}
                \fmffixed{(.3h,0)}{v4,v3}
                \fmf{plain}{i1,v1,v3,v4,v2,o1}
             \end{fmfgraph*}
        \end{tabular}
        \end{fmffile}
}
\noindent

\subsection{Light cone gauge}
The 1-loop fermion self energy in the light cone gauge is (note the Gamma/Pauli matrix identities $\gamma^+\gamma^-=1+\gamma^3$, $\gamma^-\gamma^+=1-\gamma^3$)
\ie
& -4\pi i\lambda\int {d^3 q\over (2\pi)^3} \left(\gamma^3{-i\slash\!\!\! q\over q^2+i\epsilon} \gamma^+ - \gamma^+{-i\slash\!\!\! q\over q^2+i\epsilon} \gamma^3\right) {(p-q)^-\over (p-q)_s^2+i\epsilon}
\\
&= 8\pi\lambda\int {d^3 q\over (2\pi)^3} {q^+\over q^2+i\epsilon} {(p-q)^-\over (p-q)_s^2+i\epsilon}
\fe
We will now use dimensional reduction to regularize the integral, which will turn out to be convenient at two loops and higher. The integral can be done by separating out two lightcone directions $q^+, q^-$, and the remaining $1-\epsilon$ dimension. After performing the $1-\epsilon$ dimensional integral, we obtain (constant factors of the form $1+{\cal O}(\epsilon)$ are ignored, as there is no logarithmic divergence here)
\ie
& 4\pi\lambda\int {d^2 q_s\over (2\pi)^2} {q^+\over |q_s|^{1+\epsilon}} {(p-q)^-\over (p-q)_s^2+i\epsilon}
\\
&= - 2\pi\lambda (1+\epsilon)\int_0^1 dx\,(1-x)^{-{1\over 2}+{\epsilon\over 2}} \int {d^2q_s\over (2\pi)^2} {(q+xp)^+ (q-(1-x)p)^-\over [q_s^2+p_s^2x(1-x)+i\epsilon]^{{3\over 2}+\epsilon}}
\\
&= - \pi\lambda (1+\epsilon)\int_0^1 dx\,(1-x)^{-{1\over 2}+{\epsilon\over 2}} \int {d^2q_s\over (2\pi)^2} {q_s^2-x(1-x)p_s^2\over [q_s^2+p_s^2x(1-x)+i\epsilon]^{{3\over 2}+{\epsilon\over 2}}}
\\
&= - {\lambda\over 2} \int_0^1 dx\,(1-x)^{-{1\over 2}} \left[  - 3|p_s| x^{1\over 2}(1-x)^{1\over 2} \right]
\\
&= \lambda |p_s|.
\fe
So with dimensional reduction, the bare fermion mass is zero, and we find the renormalized 1-loop self energy
\ie
\Sigma_1(p) = {\lambda}|p_s|.
\fe
It is useful to write the 1-loop correction to the fermion propagator in position space,
\ie
-{\lambda} \int {d^3p\over (2\pi)^3} {|p_s|\over p^2+i\epsilon} e^{ip\cdot x}
= -{\lambda\over 4\pi} {|x^0|\over |x|^3}
\fe

The only potentially nontrivial planar contribution to the 2-loop fermion self energy comes from the rainbow diagram. It is given by
\ie
& -i{4\pi\lambda^2}\int {d^{3-\epsilon} q\over (2\pi)^{3-\epsilon}} [\gamma^3,\gamma^+] {|q_s|\over q^2+i\epsilon} {(p-q)^-\over (p-q)_s^2+i\epsilon}
\\
&\longrightarrow -i{4\pi\lambda^2\gamma^+} \int {d^2q_s\over (2\pi)^2}{1\over |q_s|^\epsilon}{(p-q)^-\over (p-q)_s^2+i\epsilon}
=- i \lambda^2 p^-\gamma^+,
\fe
where the integral in the last step is performed as in section 2.1.1. From this we find the two loop contribution to the fermion self energy,
\ie
\Sigma_2(p) =- i\lambda^2 p_+ \gamma^+.
\fe
As shown in section 2, the planar fermion self energy is in fact two-loop exact in the dimensional reduction scheme.

\section{Lorentz invariance of the Wilson line at one loop}
\label{rotational-invariance-appendix}\footnote{This appendix was 
worked out in collaboration with Sayantani Bhattacharyya.} 

The exact fermion propagator computed earlier in this paper in 
lightcone gauge is not Lorentz invariant. As the fermion propagator 
by is not gauge invariant, it is not physically observable. In this section 
we will compute a gauge invariant physical observable that is closely 
related to the fermion propagator, and demonstrate that this observable 
is Lorentz invariant (at one loop), and moreover takes the same value 
in lightcone gauge as in Feynman gauge. Our results may be taken as evidence
that the lightcone gauge employed in this paper defines a Lorentz invariant
theory, atleast at first order in $\lambda$. 

Consider the Wilson line
\be
\label{defwilson}
W= \frac{1}{N}
\langle 0|[Pe^{i \int_{x_i}^{x_f}\vec{A}\cdot d\vec{l}}]_m^n\psi(x_i)_n\bar{\psi}(x_f)^m|0\rangle
\ee
where $m, n$ are colour indices, we do not indicate the spinor indices of $\psi(x_i),\bar{\psi}(x_f)$ explicitly (the Wilson line is a matrix in spinor space,
in the same way that the propagator of the previous section is). The 
expectation value of the Wilson line operator described above is Lorentz 
invariant if and only if it takes the form 
$$(x_f-x_i)_\mu \gamma^\mu A(|x_f-x_i|) + I ~B(|x_f-x_i|).$$
In this subsection we will demonstrate, at the one loop level, that the
expectation value of the Wilson loop in lightcone gauge really does
take this form (at one loop we find that $A$ vanishes while $B$ is
nonzero).

At leading order, the Wilson line defined above is simply given by
$$\int \frac{d^3 p}{(2 \pi)^3}
e^{i(x_i-x_f).p} \frac{1}{i p_\mu \gamma^\mu}$$
and is manifestly rotationally invariant.

At next to leading order (or leading non-trivial order in $\lambda$)
the Wilson loop receives contributions from two graphs. The first arises
from the leading self energy correction to the $\psi {\bar \psi}$ 
propagator. In the lightcone gauge it is given by
\be
\label{selfen}
W_1=(4\pi \lambda) \int {d^3q \over (2\pi)^3} {d^3p \over (2\pi)^3} {e^{ip\cdot(x_i-x_f)} \over p^2 q^2}{q^+\over (p-q)^+}.
\ee
As we have seen above this diagram is easily evaluated,
however we will find it useful to leave the integral unevaluated at this
stage.

The second graph has one insertion of the ``Wilson line factor'' $\int_{x_i}^{x^f} \vec{A}\cdot d\vec{l}$ along the $\hat{y}$ direction (direction from
$x_i$ to $x_f$) ; this graph evaluates
to
\be
\label{wilsonline}
W_2={i N\over 2} \int {d^3p \over (2\pi)^3} {d^3p'\over (2\pi)^3}(e^{i p\cdot(x_i-x_f)}-e^{ip'\cdot(x_i-x_f)})
{p^\rho \gamma_\rho \gamma^\mu p'^{\sigma}\gamma_\sigma \over p^2 p'^2 ((p-p')\cdot \hat{y})} G_{\mu\hat{y}}(p-p')
\ee
(this result is correct both in lightcone gauge and in the Lorentz 
invariant gauge we will employ below).
In lightcone gauge, it is possible to show that this diagram evaluates to
\be
\label{expwg}
W_2= - 4\pi \lambda \int {d^3q \over (2\pi)^3} {d^3p \over (2\pi)^3}
{e^{ip \cdot (x_i-x_f)} \over p^2 q^2}
({q^+\over (p-q)^+}-{q . {\hat y}\over(p-q)\cdot {\hat y}})
\ee
In order to obtain \eqref{expwg} we have assumed that our regulator
respect the $3$ to $-3$ flip symmetry as well as rotational invariance
in the $12$ plane. These properties are both obviously true of the 
dimensional regulator employed in this paper.

It is obvious that neither $W_1$ nor $W_2$ is
rotationally invariant by itself. However
\be
\label{finalw}
W=W_1+W_2=4\pi \lambda \int  {d^3q \over (2\pi)^3} {d^3p \over (2\pi)^3}
{e^{ip \cdot (x_i-x_f)} \over p^2 q^2}{q\cdot {\hat y}\over(p-q)\cdot {\hat y}}
\ee
Note that a rotation of $x_f-x_i$ can be undone by a rotation
of the dummy variables $p$ and $q$ establishing that the Wilson line
depends only on $|x_f-x_i|$, establishing rotational invariance in any
regularization scheme - like the dimensional reduction scheme employed
in this paper - that respects rotational invariance.

It is interesting that the calculation of the Wilson loop proceeds along 
very similar lines and yields exactly the same answer in Feynman gauge. 
In this gauge, once again, we receive contributions from two graphs. 
The self energy graph gives
\be
\label{selfenFeyn}
W_1'=(4\pi \lambda) \int {d^3q \over (2\pi)^3} {d^3p \over (2\pi)^3} {e^{ip\cdot(x_i-x_f)} \over p^2 q^2}{q\cdot (p-q)\over (p-q)^2}
\ee
The second graph is given by \eqref{wilsonline} in which we must use the 
gauge propagator appropriate to the Feynman gauge, and evaluates to 
\be
\label{expwgp}
W_2'= - 4\pi \lambda \int {d^3q \over (2\pi)^3} {d^3p \over (2\pi)^3}
{e^{ip \cdot (x_i-x_f)} \over p^2 q^2}
({q\cdot(p-q)\over (p-q)^2}-{q \cdot {\hat y}\over(p-q). {\hat y}})
\ee
Clearly $W_1'+W_2'=W_1+W_2=W$ where $W$ is listed in \eqref{finalw}.

\section{Unitary Representations of the $d=3$ Conformal Group}\label{unit}

Unitary representations of the conformal group are labeled by the  spin $s$
and a scaling dimension $\Delta$ of their primary states. When $s\geq 1$ these
labels are subject to the inequality $\Delta \geq s+1$. In this case
representations that saturate the inequality are short; the null states
in this representation fall into a (long) representation with $\Delta=s+2$ and
spin =$s-1$.

In the special case $s=\half$ it turns out that $\Delta \geq 1$.
The representation with $\Delta=1$ is short and its null
states fall into a (long) representation with $\Delta=2$ and spin =$\half$.
The later is the representation of a free fermionic field, and the character
\eqref{pfnat} $F_F(x, \mu)$ of this field is given by
\begin{equation}\label{fcf}\begin{split}
F_F(x,\mu)=&\frac{x(\mu^{\half} +\mu^{-\half})}{(1-\mu x)(1-\mu^{-1}x)}
\end{split}
\end{equation}

Finally, when $s=0$ we have $\Delta=0$ or $\Delta\geq \half$. The
representation with $\Delta=0$ has a single state. The representation
with $\Delta=\half$ has null states in a representation with
$\Delta=\frac{5}{2}$ and $s=0$. The states of a free scalar field fall into
this representation; the character $F_S(x, \mu)$ of this representation is
given by
\begin{equation}
\label{fca-app}
F_S(x,\mu)=x^{\frac{1}{2}}\frac{(1+x)}{(1-\mu x)(1-\mu^{-1}x)}
\end{equation}

We now present character formulae for all unitary representations of the
3d conformal algebra. Let us define
\begin{equation}
\chi_s(\mu) = \sum_{j=-s}^s \mu^{j}
\end{equation}
$\chi_s(\mu)$ is, of course, the $SU(2)$ character at spin $s$ (here
$s$ is a positive integer or half integer). Let us also define
\begin{equation} \label{longmodule}
G_{\Delta, s}=\frac{x^{\delta}\chi_s(\mu)}{(1-x)(1-\mu x)(1-\mu^{-1}x)}
\end{equation}
$G_{\Delta, s}$ is the partition function over states that are obtained by
acting on an $SU(2)$ primary of spin $s$ with an arbitrary number of
derivatives, and so yields the character of any long representation of the
conformal algebra (i.e. any representation with $\Delta > s+1$ for $s\geq 1$
or $\Delta>1$ for $s=\half$ or $\Delta >\half$ for $s=0$).

Characters $\chi(x, \mu)$ of short representations of the conformal algebra
are obtained by evaluating the character of a hypothetical long representation
of that algebra and then subtracting out the character of its null states.
It follows that, for $s \geq 1$
\begin{equation}\label{charactersso}
\chi_{s+1, s}(x,\mu)=G_{s+1,s}(x, \mu)-G_{s+2, s-1}(x, \mu)
\end{equation}
For $s=\half$
\begin{equation}\label{charactersso2}
\chi_{1, \half}(x,\mu) =  G_{1,\half}(x, \mu)-G_{2, \half}(x, \mu)
= F_F(x, \mu)
\end{equation}
while for $s=0$
\begin{equation}\label{charactersso3}
\chi_{\half, 0}(x,\mu)=G_{\half, 0}(x, \mu)-G_{\frac{5}{2}, 0}(x, \mu)
= F_S(x, \mu)
\end{equation}

\section{Primary Operators in Free and Interacting Fermion Theories}
\label{appB}

In this appendix we present the details of some slightly tedious
computations involving free and interacting fermions.

\subsection{The generating function of conserved currents
for free fermions}

As we have explained in section \ref{operator-section} above, the single-trace primary operators of
the free fermion theory satisfy the
condition that their scaling dimension, $\epsilon$ and spin $s$ satisfy the
relation $\epsilon=s+1$, with the exception of $\bar{\psi}\psi$, which has
scaling dimension $2$ and spin $0$. Here, we will
determine explicit expressions for the corresponding primary operators.

How can we produce an operator of spin $s$ and dimension $s+1$ in a theory
of free fermions in $d=3$? We are interested in operators built out of
fermion bilinears (with colour indices contracted). As $\psi$ and ${\bar \psi}$
each have unit scaling dimension, the operator of interest must contain
exactly $s-1$ derivatives. As $s-1$ derivatives can give rise to at most
$s-1$ free traceless vector indices the remaining index (to make our operator
spin $s$) must come from a $\gamma$ matrix. 
\footnote{As $\gamma_{(\mu}\gamma_{\nu)} = \eta_{\mu\nu}$ we cannot have more than one of the current indices come from a $\gamma$ matrix. Also, the equations of motion tell us that a $\gamma$ matrix contracted with a derivative vanishes in its action on a fermion. The contraction of a derivative with a $\gamma$ matrix, sandwiched by other $\gamma$ matrices, can also be reduced to a form with
a single $\gamma$ matrix and derivatives using the equations of motion,
(e.g. $\gamma_{(\mu} \gamma_\rho \partial^\rho \gamma_{\nu)} = - \gamma_{(\mu}  \partial^\rho \gamma_{\nu)} \gamma_\rho + \gamma_{(\mu}  \partial^\rho \eta_{\nu) \rho} $ where
the first term on the RHS vanishes by the equations of motion and the last
term has a single $\gamma$ matrix). }
Consequently if we define the generating function $F$ such that
\begin{equation}
 \mathcal O(x;\epsilon) = \bar{\psi} F(\vec{\gamma},\overrightarrow{\partial_\mu},\overleftarrow{\partial_\mu},\vec{\epsilon}) \psi = \sum J^{(s)}_{{\mu_1} {\mu_2} \ldots {\mu_s}}\epsilon^{\mu_1}\ldots \epsilon^{\mu_s}
\end{equation}
then $F$ must take the form
\begin{equation}
 F = \vec{\gamma}.\vec{\epsilon} f(\overrightarrow{\partial_\mu},\overleftarrow{\partial_\mu}, \vec{\epsilon})
\end{equation}
Following \cite{Giombi:2009wh} we denote the arguments of $f$ as
vectors $u$ and $v$, so that $\vec{u}=\overleftarrow{\partial}$ and
$\vec{v}=\overrightarrow{\partial}$. The fermion equation of motion gives
\begin{equation}
 \vec{u}.\vec{\gamma}=\vec{v}.\gamma=u^2=v^2=0
\end{equation}

It is convenient to change variables to $\vec{y} \equiv \vec{u}-\vec{v}$ and $\vec{z} \equiv \vec{u}+\vec{v}$. Then we have:
\begin{equation}
 \vec{z} \cdot \vec{\gamma}=\vec{y} \cdot \vec{\gamma}=\vec{y} \cdot \vec{z} =0, ~ ~ -\vec{y}^2=\vec{z}^2=2 \vec{u}\cdot\vec{v} \neq 0
\end{equation}

Terms in $f$ will be of the form:
\begin{equation}
 f = A +  B \vec{\epsilon}\cdot \vec{y} + C \vec{\epsilon}\cdot \vec{z} + D \vec{u}\cdot \vec{v} \vec{\epsilon}\cdot \vec{\epsilon} + \ldots
\end{equation}
where each coefficient is a number.

If we define $w\equiv (\vec{u}\cdot \vec{v}) \vec{\epsilon}\cdot \vec{\epsilon}$, $z \equiv \vec{\epsilon}\cdot \vec{z}$, and $y=\vec{\epsilon}\cdot \vec{y}$, then $f$ can be thought of as a function of three variables $f(z,y,w)$.

The condition that each current is conserved can be expressed as:
\begin{equation}
 (\overleftarrow{\partial_\mu} + \overrightarrow{\partial_\mu}) {\partial \over \partial \epsilon_\mu} F =0
\end{equation}
which translates into the following condition on $f$:
\begin{equation}
 z {\partial_w f} + \partial_z f  =  0  \label{cons}
\end{equation}

The condition that each current is traceless can be expressed as:
\begin{equation}
{\partial \over \partial \epsilon^\mu } {\partial \over \partial \epsilon_\mu} F = 0
\end{equation}
which translates into:
\begin{equation}
\left(  5 \partial_w + 2w \partial_w^2-\partial_y^2+\partial_z^2+2z \partial_z \partial_w + 2 y \partial_y \partial_w \right) f = 0
\end{equation}

The solution to \eqref{cons} is  $f(w,y,z)=g(y,w- {z^2 \over 2})$. If we define $t=w- {z^2 \over 2}$, the equation for $g(y,t)$ is
\begin{equation}
 \left(4\partial_t+2t \partial_t^2-\partial_y^2+2y \partial_y \partial_t \right) g(y,t)=0
\end{equation}

The general solution satisfying $g = 1$ at $t=y=0$  is:
\begin{equation}
 g=e^{2ky} \frac{\sinh 2k\sqrt{2t+y^2}}{2k\sqrt{2t+y^2}}
\end{equation}
where $k$ is any constant, which we take to be $1/2$. 

The final form for $f$ is thus
\begin{equation}
 f(\vec{u}, \vec{v}, \vec{\epsilon}) = \frac{\exp{\left( \vec{u}\cdot \vec \epsilon-\vec{v}\cdot \vec \epsilon \right)} \sinh \sqrt{2 \vec{u}\cdot \vec{v} \vec \epsilon \cdot \vec \epsilon -   4\vec{u}\cdot\vec{\epsilon} \vec{v}\cdot\vec{\epsilon}}}{\sqrt{2 \vec{u}\cdot \vec{v} \vec \epsilon \cdot \vec \epsilon -   4\vec{u}\cdot\vec{\epsilon} \vec{v}\cdot\vec{\epsilon}}}
\end{equation}

Expanding the above expression in a power series around $\vec \epsilon$, we obtain:
\begin{eqnarray*}
f &  = & 1 + \epsilon
    (u-v) + \frac{1}{6} \epsilon ^2
   \left(3 u^2-10 u v+3 v^2+2
   w\right)  \\ & \phantom{=} & +  \epsilon ^3 \left(\frac{u^3}{6}-\frac{7 u^2
   v}{6}+\frac{7 u v^2}{6}+\frac{u
   w}{3}-\frac{v^3}{6}-\frac{v
   w}{3}\right) \\ & \phantom{=} & +\frac{1}{120} \epsilon ^4
   \left(10 (u-v)^2 (2 w-4 u v)+(4 u
   v-2 w)^2+5 (u-v)^4\right)
\end{eqnarray*}
(above, $w=\vec{u}\cdot\vec{v}$, $u=\vec{u}\cdot \vec \epsilon$,
$v=\vec v \cdot \vec \epsilon$.) which yields the currents reported in
\eqref{expexp}.

\subsection{Two-point functions of primary operators in the free theory}

In this subsection we explicitly compute the two-point function of conserved
currents
\begin{equation}
\langle \mathcal O(\vec{x};\vec{\epsilon}_1) \mathcal O(0;\vec{\epsilon}_2)\rangle
\end{equation}
determined in the previous subsection, and demonstrate that the two point
functions of currents of different spin are orthonormal. We take the two-point
function of the basic fermionic fields to be given by
\begin{equation}
\langle\psi(x) \bar{\psi}(0)\rangle = C_{\psi \psi} \frac{\gamma^\mu x_\mu}{x^3}.
\end{equation}

\subsubsection{Simple examples}
To set up notation and get intuition we first work out some simple examples.

As a first example, consider the two-point function of two scalar currents:
\begin{equation}
\langle\bar{\psi}(x)\psi(x) \bar{\psi}(0) \psi(0)\rangle
\end{equation}
Using Wick's theorem we rewrite this as:
\begin{equation}
\Tr \langle-\psi(0)\bar{\psi}(x)\rangle\langle\psi(x) \bar{\psi}(0)\rangle
\end{equation}
where the trace is over gamma matrix indices. We then compute it to be:
\begin{eqnarray*}
&&C_{\psi \psi}^2 \Tr \frac{\gamma^\mu x_\mu}{x^3} \frac{\gamma^\nu x_\nu}{x^3}  =  C_{\psi \psi}^2 \frac{x_\nu x_\mu}{x^6} \Tr \gamma^\mu \gamma^\nu \\
&& =   C_{\psi \psi}^2 \frac{x_\nu x_\mu}{x^6} 2 \eta^{\mu\nu} =   C_{\psi \psi}^2 \frac{2}{x^4}.
\end{eqnarray*}
To evaluate more complicated two-point functions, we make use of the identity
\begin{eqnarray}
 \Tr \gamma^\mu \gamma^\rho \gamma^\nu \gamma^\sigma 
& = & 2 \left(\eta^{\rho \nu} \eta^{\sigma \mu} + \eta^{\rho \mu} \eta^{\sigma \nu} - \eta^{\rho \sigma}\eta^{\mu\nu}  \right). \label{traceIdentity}
\end{eqnarray}
Note that \eqref{traceIdentity} is symmetric under interchange of $\rho$ and $\sigma$.

We next consider the two-point function of two spin-1 currents:
\begin{equation}
\langle \bar{\psi}(x) \gamma^\nu \psi(x) \bar{\psi}(y) \gamma^\mu \psi(y)\rangle
\end{equation}
We have:
\begin{eqnarray*}
\langle \bar{\psi}(x) \gamma^\mu \psi(x) \bar{\psi}(y) \gamma^\nu \psi(y)  \rangle & = &
\Tr \langle -\psi(y) \bar{\psi}(x)\rangle \gamma^\mu 
\langle\psi(x) \bar{\psi}(y)\rangle \gamma^\nu \\
& = & C_{\psi \psi}^2 \frac{x_\rho}{x^3} \frac{x_\sigma}{x^3} \Tr   \gamma^\rho \gamma^\nu \gamma^\sigma \gamma^\mu \\
& = & C_{\psi \psi}^2 \frac{x_\rho}{x^3} \frac{x_\sigma}{x^3} 2 \left(\eta^{\rho \nu} \eta^{\sigma \mu} + \eta^{\rho \mu} \eta^{\sigma \nu} - \eta^{\rho \sigma}\eta^{\mu\nu}  \right) \\
& = & 2 C_{\psi \psi}^2 \left( \frac{2 x^\mu x^\nu }{x^6} -\frac{\eta^{\mu\nu}}{x^4} \right).
\end{eqnarray*}
As above, we will often set $y=0$ in the last line.

\subsubsection{Results for all spins}

Let $\varepsilon$ be a {\it null} polarization vector. The two point function of the generating operator ${\cal O}(x;\varepsilon) = \sum J^{(s)}_{\mu_1\cdots\mu_s} \varepsilon^{\mu_1}\cdots \varepsilon^{\mu_s}$ is evaluated in the theory of $N$ free complex fermions to be
\ie
\langle {\cal O}(x;\varepsilon) {\cal O}(0;\varepsilon) \rangle = {N\over 32\pi^2 x^2}\left\{ \left[1-\left({4\varepsilon\cdot x\over x^2}\right)^2\right]^{-{1\over 2}} -1\right\} .
\fe
Expanding this in $\varepsilon$, we have
\ie
\langle J^{(s)}(x;\varepsilon) J^{(s)}(0;\varepsilon)\rangle = {N\over 32\pi^{5\over 2}} {2^{4s}\Gamma(s+{1\over 2})\over s!} {(\varepsilon\cdot x)^{2s}\over (x^2)^{2s+1}} ,
\fe
where $J^{(s)}(x;\varepsilon)$ is the spin-$s$ part of ${\cal O}(x;\varepsilon)$. The spin 0 case is special, where we have $\langle J^{(0)}(x) J^{(0)}(0)\rangle = {N\over 8\pi^2}|x|^{-4}$.

We have defined the set of currents $j^{(s)}$ with a different normalization convention, namely normalizing the norm of the corresponding state in radial quantization.
The relative normalization between $J^{(s)}$ and $j^{(s)}$ can be determined as follows.
If we define $j^{(s)}(x;\varepsilon) = j^{(s)}_{\mu_1\cdots\mu_s}(x) \varepsilon^{\mu_1}\cdots \varepsilon^{\mu_s}$, then
\ie
\langle j^{(s)}(x;\varepsilon) j^{(s)}(0;\varepsilon)\rangle = 2^s {(\varepsilon\cdot x)^{2s}\over (x^2)^{2s+1+\delta_s}},
\fe
for $s>0$. In the spin 0 case, we have $\langle j^{(0)}(x) j^{(0)}(0)\rangle = |x|^{-4}$.
From this we deduce
\ie
&J^{(s)}_{\underline{\mu}}(x) = a_s j^{(s)}_{\underline{\mu}}(x),
~~~~a_s = \left[ {N\over 32\pi^{5\over 2}} {2^{3s} \Gamma(s+{1\over 2})\over s!} \right]^{1\over 2},
\fe
for $s>0$. In the spin 0 case, $a_0 = {\sqrt{2N}\over 4\pi}$.
In the interacting Chern-Simons-fermion theory, $a_s$ receives quantum corrections.

\subsection{Explicit computation of the divergence of $J^{(3)}$}
\label{divjt}

In carrying out all our manipulations below, we use the fermion equation
of motion
\begin{equation}
 D_\mu \gamma^\mu \psi = D_\mu\bar{\psi}\gamma^\mu = 0.
\end{equation}
Some useful identities are:
\ie
& \gamma^\mu D_\mu D_\nu \psi = \gamma^\mu (D_\nu D_\mu -i F_{\mu \nu}) \psi =-i \gamma^\mu F_{\mu \nu} \psi,
\\
& D_\mu D_\nu \bar{\psi} \gamma^\mu=  (D_\nu D_\mu +i F_{\mu \nu}) \bar{\psi} \gamma^\mu = i\bar{\psi} \gamma^\mu F_{\mu \nu},
\\
& \gamma^\mu D_\mu \gamma^\nu D_\nu \psi  =  0, \\
& (\eta^{\mu \nu} + \gamma^{\mu \nu}) D_\mu D_\nu \psi = 0, \\
& D_\mu D^\mu \psi =  \frac{i}{2}\gamma^{\mu \nu} F_{\mu \nu} \psi,
\\
&  D_\mu  D_\nu \bar{\psi} \gamma^\nu \gamma^\mu = 0, \\
& D_\mu D^\mu \bar{\psi} =  \frac{i}{2} \bar{\psi} \gamma^{\mu \nu} F_{\mu \nu},
\fe
Note our convention is such that $[\rD_\mu, \rD_\nu] \psi = -i F_{\mu\nu} \psi$.
We now use the equation of motion for $F_{\mu \nu} =
F^a_{\mu \nu} T^a$, namely
\begin{equation}
 \epsilon^{\mu \nu \rho} F_{\nu \rho} = {4\pi\over k} J^{\mu},
~~~~{\rm or}~~~
 F^a_{\mu \nu} = {2\pi\over k} \epsilon_{\mu\nu\rho}J^{a \,\rho},
\end{equation}
where $J_\mu = J^a_\mu T^a$, $J_\mu^a = \bar{\psi}\gamma_\mu T^a \psi$.
It is also useful to have
\begin{eqnarray}
 J^a_\rho T^a \psi & = & (\bar{\psi}\gamma_\rho T^a \psi)\, T^a \psi \nonumber \\
 & = & -\frac{1}{4} \left( \gamma_\rho \psi (\bar{\psi}{\psi}) + \gamma^\mu \gamma^\rho \psi (\bar{\psi} \gamma_\mu \psi) \right) \nonumber \\
& = & -\frac{1}{4} \left( \gamma_\rho \psi J^{(0)} + \gamma^\mu \gamma_\rho \psi J^{(1)}_\mu \right),
\end{eqnarray}
and
\begin{equation}
 \bar{\psi}J^a_\rho T^a = -\frac{1}{4}\left( J^{(0)} \bar{\psi}\gamma_\rho + J^{(1)}_\mu \bar{\psi}\gamma_\rho \gamma^\mu \right).
\end{equation}
To derive these relations, we used $(T^a)^i_j (T^a)^l_m={1\over 2}\delta^i_m \delta^l_j$ and the 3d Fierz identity
$$\chi \bar{\lambda} = -\frac{1}{2}\bar{\lambda}\chi-\frac{1}{2}\bar{\lambda}\gamma_\mu\chi\gamma^\mu.$$

We can now proceed to explicitly compute $\partial^\mu J^{(3)}_{\mu \nu_1 \nu_2}$. First, consider the current $\hat J^{(3)}$ which is not traceless,
\begin{eqnarray*}
\hat{J}^{(3)}_{\mu_1 \mu_2 \mu_3} & = & \frac{1}{6} \bar{\psi}\gamma_{\mu_1} \left( 3 \overleftarrow{D_{\mu_2}} \overleftarrow{D_{\mu_3}}  - 10 \overleftarrow{D_{\mu_2}} \overrightarrow{D_{\mu_3}} + 3 \overrightarrow{D_{\mu_2}} \overrightarrow{D_{\mu_3}} + 2  ( \overleftarrow{D_{\sigma}}\overrightarrow{D^{\sigma}}) \eta_{{\mu_2}{\mu_3}} \right) \psi.
\end{eqnarray*}
Using the identities above we find that (before subtracting the trace) the divergence is given by:
\begin{eqnarray}
6 \partial^\mu \hat{J}^{(3)}_{\mu \nu_1 \nu_2} & = & -i\, \bar{\psi} \gamma^\mu \left( 16 \lD_\vo F_{\vt \mu} +16 F_{\mu \vo} \rD_\vt + 2 \eta_{\vo\vt} \left( \lD_\lambda F_{\mu \lambda} + F_{\lambda \mu} \rD_\lambda \right) \right) \psi \nonumber \\
& \phantom{=} &  -i \,\bar{\psi} \gamma^\vo \left( 16 (\lD^\mu F_{\vt \mu} + F_{\mu \vt} \rD^\mu) \right) \psi  \nonumber \\
& \phantom{=} & + ~ \bar{\psi} \gamma^\vo \left( 6 (\lD^2 \lD_\vt + \rD_\vt \rD^2) -10(\lD^2 \rD_\vt + \lD_\vt \rD^2) \right)\psi
\end{eqnarray}
We now substitute for $D^2$ and further simplify:
\ie
& 6 \partial^\mu \hat{J}^{(3)}_{\mu \nu_1 \nu_2}  = -i \bar{\psi} \gamma^\mu \left( 32 \lD_\vo F_{\vt \mu} +32 F_{\mu \vo} \rD_\vt + 2 \eta_{\vo\vt} \left( \lD_\lambda F_{\mu \lambda} + F_{\lambda \mu} \rD_\lambda \right) \right) \psi  \\
& ~~~-i \bar{\psi} \gamma^\vo \left( 16 (\lD^\mu F_{\vt \mu} + F_{\mu \vt} \rD^\mu) \right) \psi  -i  \bar{\psi} \left(2 \lD_\vo \tilde{F}_\vt + 2 \tilde{F}_\vo \rD_\vt - 6 (D_\vo \tilde{F}_\vt) \right) \psi,  \label{partIII}
\fe
where $\tilde{F}_\mu = \epsilon_{\mu \nu \rho} F^{\nu \rho}$.

Now substituting in $F$ and using Fierz identity, we have
\ie
& - 32 i \bar{\psi} \gamma^\mu \left( \lD_\vo F_{\vt \mu} +F_{\mu \vo} \rD_\vt \right) \psi
 = 
{32\pi\over k}\bigg[ \partial_\vo (\bar{\psi}\psi) \bar{\psi}\gamma_\vt \psi -  \partial_\vo (\bar{\psi}\gamma_\vt \psi) \bar{\psi}\psi \bigg], 
\\
& -2i \eta_{\vo\vt} \bar\psi \gamma^\mu \left( \lD_\lambda F_{\mu \lambda} + F_{\lambda \mu} \rD_\lambda \right) \psi
=
-\frac{2\pi}{k} \eta_{\vo\vt} \partial_\mu (\bar{\psi} \psi)\bar{\psi} \gamma^\mu \psi ,
\\
& -i\bar{\psi} \gamma^\vo \left( 16 (\lD^\mu F_{\vt \mu} + F_{\mu \vt} \rD^\mu) \right) \psi
=
 \frac{8\pi}{k} \Big[ -\eta_{\vo\vt} \partial_\mu (\bar{\psi} \psi)\bar{\psi} \gamma^\mu \psi + 2 \partial_\vo (\bar{\psi} \psi) \bar{\psi}\gamma_\vt \psi \\
& ~~~~~~~~~  -  \partial_\vo (\bar{\psi}\gamma_\vt \psi) \bar{\psi}\psi + \epsilon_{\vt \lambda \mu} (\bar{\psi}\lrD^\mu \gamma_\vo \psi) (\bar{\psi}\gamma^\lambda \psi)  \Big], 
\\
& -i\bar{\psi} \left(2 \lD_\vo \tilde{F}_\vt + 2 \tilde{F}_\vo \rD_\vt\right) \psi
= -\frac{2\pi}{k}\left[  \partial_\vo (\bar{\psi}\gamma_\vt \psi) \bar{\psi}\psi + (\bar{\psi}\gamma_\vt \psi) \partial_\vo (\bar{\psi}\psi) + \epsilon_{\vt \lambda \mu} (\bar{\psi}\lrD_\vo \gamma^\mu \psi) (\bar{\psi}\gamma^\lambda \psi) \right],
\\
& - 6 i \bar{\psi} \left( D_\vo \tilde{F}_\vt \right) \psi = \frac{6\pi}{k}\left[ (\bar{\psi}\gamma_\vt \psi) \partial_\vo (\bar{\psi}\psi) + \partial_\vo (\bar{\psi}\gamma_\vt \psi)  \bar{\psi}\psi - \epsilon_{\vt \lambda \mu} (\bar{\psi}\lrD_\vo \gamma^\mu \psi) (\bar{\psi}\gamma^\lambda \psi)\right].
\fe
We now use the identity
\begin{equation}
\epsilon_{\vt \lambda \mu} \bar{\psi}(-\lrD_\vo \gamma^\mu
+ \lrD^\mu \gamma_\vo ) \psi (\bar{\psi}\gamma^\lambda \psi) = -\eta_{\vo \vt} \partial^\lambda (\bar{\psi}\psi)(\bar{\psi}\gamma_\lambda \psi)+ \partial_\vo(\bar{\psi}\psi) \bar{\psi}\gamma_\vt \psi
\end{equation}
to obtain the following total for the above sum:
\begin{equation}
6 \partial^\mu \hat{J}_{\mu \vo \vt}^{(3)} = {2\pi\over k}\left[ -9 \eta_{\vo \vt} \partial_\mu (\bar{\psi}\psi) (\bar{\psi}\gamma^\mu \psi)
+ 30 \partial_\vo(\bar{\psi}\psi) \bar{\psi}\gamma_\vt \psi - 18  \partial_\vo (\bar{\psi}\gamma_\vt \psi) \bar{\psi}\psi\right].
\end{equation}
Subtracting the trace we obtain:
\begin{equation}
\partial^\mu J_{\mu \vo \vt}^{(3)} = \partial^\mu \hat{J}_{\mu \vo \vt}^{(3)} - \frac{\pi}{5k}\left[ \eta_{\vo \vt} (\bar\psi\gamma^\mu\psi) \partial_\mu (\bar{\psi}\psi) + 2 \partial_\vo (\bar{\psi}\gamma_\vt \psi) \bar{\psi} \psi +  2 (\bar{\psi}\gamma_\vt \psi) \partial_\vo (\bar{\psi} \psi) \right].
\end{equation}
The indices on the RHS are understood to be symmetrized.

\subsection{Higher spin examples}

While we do not have a general explicit formula expressing the divergence of the current in terms of double and triple trace operators, a straightforward but tedious computation was carried out using Mathematica to determine the double trace terms. We find that $\partial^\mu J^{(s)}_{\mu\mu_1\cdots\mu_{s-1}}$ involves the product of two currents of spins $s_1$ and $s_2$, with $s_1+s_2\equiv s \,({\rm mod}~2)$ and $s_1+s_2\leq s-2$. It is convenient to write the results in spinorial notation. For instance, the anomalous current conservation relation for $J^{(3)}$ takes the form (up to an overall normalization factor)
\ie
& \partial^\mu J^{(3)}_{\mu \A_1\cdots\A_4} \sim {1\over k}\left[ 3 \partial_{(\A_1\A_2} J^{(0)} J^{(1)}_{\A_3\A_4)} -2 J^{(0)} \partial_{(\A_1\A_2} J^{(1)}_{\A_3\A_4)} \right].
\fe
For spins $s=4$ and $5$, we find:
\ie
& \partial^\mu J^{(4)}_{\mu \A_1\cdots\A_6} \sim {1\over k} \left[ 8 \partial_{(\A_1}{}^\B J^{(1)}_{\B\A_2} \partial_{\A_3\A_4} J^{(1)}_{\A_5\A_6)} + 5 J^{(1)}_{(\A_1\A_2} \partial_{\A_3\A_4} \partial_{\A_5}{}^\B J^{(1)}_{\A_6)\B} + J^{(1)}_{(\A_1}{}^\B \partial_{\A_2 \B}\partial_{\A_3\A_4} J^{(1)}_{\A_5\A_6)}
\right.
\\
&~~~~\left. + 5 \partial_{(\A_1\A_2} J^{(0)} J^{(2)}_{\A_3\A_4\A_5\A_6)}
- 2 J^{(0)} \partial_{(\A_1\A_2} J^{(2)}_{\A_3\A_4\A_5\A_6)} \right],
\fe
and
\ie
& \partial^\mu J^{(5)}_{\mu \A_1\cdots\A_8} \sim {1\over k}
\left[ -160 \partial_{(\A_1\A_2} \partial_{\A_3}{}^\B J^{(1)}_{\B\A_4} J^{(2)}_{\A_5\cdots\A_8)}
-40 \partial_{(\A_1\A_2} \partial_{\A_3\A_4} J^{(1)}_{\A_5}{}^\B J^{(2)}_{\A_6\A_7\A_8)\B}
\right.
\\
&~~~-128 \partial_{(\A_1\A_2} J^{(1)}_{\A_3}{}^\B \partial_{\A_4\B} J^{(2)}_{\A_5\cdots \A_8)}
+128 \partial_{(\A_1}{}^\B J^{(1)}_{\A_2\A_3} \partial_{\A_4\A_5} J^{(2)}_{\A_6\A_7\A_8)\B}
+16 \partial_{(\A_1\A_2} J^{(1)}_{\A_3}{}^\B \partial_{\A_4\A_5} J^{(2)}_{\A_6\A_7\A_8)\B}
\\
&~~~+64 J^{(1)}_{(\A_1}{}^\B \partial_{\B\A_2} \partial_{\A_3\A_4} J^{(2)}_{\A_5\cdots \A_8)}
-56 J^{(1)}_{(\A_1}{}^\B \partial_{\A_2\A_3}\partial_{\A_4 \A_5} J^{(2)}_{\A_6\A_7\A_8)\B}
+25 \partial_{(\A_1\A_2}\partial_{\A_3\A_4}\partial_{\A_5\A_6} J^{(0)} J^{(1)}_{\A_7\A_8)}
\\
&~~~-100 \partial_{(\A_1\A_2}\partial_{\A_3\A_4}J^{(0)} \partial_{\A_5\A_6} J^{(1)}_{\A_7\A_8)}
+75 \partial_{(\A_1\A_2}J^{(0)} \partial_{\A_3\A_4}\partial_{\A_5\A_6} J^{(1)}_{\A_7\A_8)}
-10 J^{(0)} \partial_{(\A_1\A_2} \partial_{\A_3\A_4}\partial_{\A_5\A_6} J^{(1)}_{\A_7\A_8)}
\\
&~~~\left. + 140 \partial_{(\A_1\A_2} J^{(0)} J^{(3)}_{\A_3\cdots\A_8)}
-40 J^{(0)} \partial_{(\A_1\A_2} J^{(3)}_{\A_3\cdots\A_8)} \right]
\fe
In writing the above expressions, we have omitted an overall normalization constant as well as possible triple trace terms.

\section{Perturbative Analysis of Anomalous Dimensions}

\subsection{Vanishing anomalous dimension of the scalar primary at planar level}

In this subsection, we check explicitly the vanishing of the anomalous dimension of the scalar operator $\bar\psi\psi$ at two-loop planar level, as previously argued based on the operator identities relating divergences of currents to double trace operators. The simplest way to do this computation is in the lightcone gauge, where there is no cubic gauge interaction vertex, and the planar two-loop fermion self-energy is finite. There are two diagrams that contribute to the logarithmic divergence.

\begin{equation}\nonumber
\begin{array}{ccccc}
\begin{fmffile}{VT1}
        \begin{tabular}{c}
            \begin{fmfgraph*}(60,80)
                \fmfleft{i1}
                \fmfright{o1,o2}
                \fmffixed{(-.1h,0)}{t1,i1}
                \fmffixed{(0,-.05h)}{t2,o1}
                \fmffixed{(0,.05h)}{t3,o2}
                \fmffixed{(0,0)}{t1,w1}
                \fmffixed{(0,0)}{t2,w2}
                \fmffixed{(0,0)}{t3,w3}
                \fmfv{decoration.shape=circle, decoration.size=.1h, decoration.filled=empty}{t1}
                \fmfv{decoration.shape=cross, decoration.size=.1h, decoration.filled=empty}{w1}
                \fmffixed{(0,.5h)}{o1,v}
                \fmffixed{(0,.5h)}{v,o2}
                \fmffixed{(-.5h,0)}{z,t1}
                \fmf{wiggly,left=1.1,tension=.0}{y,w}
                \fmf{wiggly,left=.2,tension=.0}{x,xp}
                \fmf{plain}{t3,x,y,u,w,t1,wp,up,yp,xp,t2}
             \end{fmfgraph*}
        \end{tabular}
        \end{fmffile}
&~~~~~~~~~~& \begin{fmffile}{VT2}
        \begin{tabular}{c}
            \begin{fmfgraph*}(60,80)
                \fmfleft{i1}
                \fmfright{o1,o2}
                \fmffixed{(-.1h,0)}{t1,i1}
                \fmffixed{(0,-.05h)}{t2,o1}
                \fmffixed{(0,.05h)}{t3,o2}
                \fmffixed{(0,0)}{t1,w1}
                \fmffixed{(0,0)}{t2,w2}
                \fmffixed{(0,0)}{t3,w3}
                \fmfv{decoration.shape=circle, decoration.size=.1h, decoration.filled=empty}{t1}
                \fmfv{decoration.shape=cross, decoration.size=.1h, decoration.filled=empty}{w1}
                \fmffixed{(0,.5h)}{o1,v}
                \fmffixed{(0,.5h)}{v,o2}
                \fmffixed{(-.5h,0)}{z,t1}
                \fmf{wiggly,left=.2,tension=.0}{x,xp}
                \fmf{wiggly,left=.2,tension=.0}{y,yp}
                \fmf{plain}{t3,x,y,t1,yp,xp,t2}
             \end{fmfgraph*}
        \end{tabular}
        \end{fmffile}
\end{array}
\end{equation}
The first one (which comes with an additional factor of 2) is computed as
\ie
& -{4\pi \lambda^2} \int {d^3p\over (2\pi)^3} {|p_s| (\gamma^3{\slash\!\!\! p}\gamma^+-\gamma^+{\slash\!\!\! p}\gamma^3)\over (p^2)^2} {p^-\over p_s^2}
= {4\pi \lambda^2} \int {d^3p\over (2\pi)^3}  {|p_s|\over (p^2)^2}
\\
& = {4\pi \lambda^2} \int_0^1 dx\, {x(1-x)^{-{3\over 2}}\over B(2,-{1\over 2})}\int {d^3p\over (2\pi)^3}  {1\over (p_s^2+x p_3^2)^{3\over 2}}
\\
&= {4\pi \lambda^2} {B({3\over 2},-{1\over 2})\over B(2,-{1\over 2})} {\ln\Lambda\over 2\pi^2}
\\
&= {\lambda^2\over 2}\ln\Lambda.
\fe
The second diagram (the rainbow correction to the vertex) is computed as
\ie
& -(4\pi \lambda)^2 \int {d^3p d^3q\over (2\pi)^6} {2p^- q^- (\gamma^+ {\slash\!\!\! p} \gamma^+\gamma^3 {\slash\!\!\! p} \gamma^3-\gamma^3 {\slash\!\!\! p} \gamma^+\gamma^3 {\slash\!\!\! p} \gamma^+)\over (p+q)^2 (p^2)^2 p_s^2 q_s^2}
\\
&= (4\pi \lambda)^2 \int {d^3p d^3q\over (2\pi)^6} {4 p^+ q^-
\over (p+q)^2 (p^2)^2 q_s^2}
\\
&= (4\pi \lambda)^2\int_0^1 dx \int {d^3p d^3q\over (2\pi)^6} {4 p^+ (q^- - xp^-)
\over (p^2)^2 [q_s^2+p_s^2 x(1-x)+q_3^2 x]^2}
\\
&= - 8\pi \lambda^2 \int {d^3p\over (2\pi)^3} {|p_s|
\over (p^2)^2}
\fe
It indeed cancels twice of the previous diagram.

\subsection{Anomalous dimension of currents at order $1/N$}

It has been argued that the higher spin currents $J_s(x)$ as well as the single trace scalar operator $\bar\psi \psi$ receives no anomalous dimension in the infinite $N$ limit. They do acquire anomalous dimension at subleading orders in $1/N$. For the currents of nonzero spins, this can be understood in terms of the mixing of the divergence of the current with double trace operators. In this subsection, we perform a direct two-loop computation of the anomalous dimension of these currents at order $\lambda^2/N$. There are three diagrams that contribute, all of which involve a matter-loop-corrected gauge field propagator (which is suppressed by $1/N$ because the matter is in the fundamental representation). These diagrams are listed below:

\begin{equation}\nonumber
\begin{array}{ccccc}
\begin{fmffile}{Anom1}
        \begin{tabular}{c}
            \begin{fmfgraph*}(60,80)
                \fmfleft{i1}
                \fmfright{o1,o2}
                \fmffixed{(-.1h,0)}{t1,i1}
                \fmffixed{(0,-.05h)}{t2,o1}
                \fmffixed{(0,.05h)}{t3,o2}
                \fmffixed{(0,0)}{t1,w1}
                \fmffixed{(0,0)}{t2,w2}
                \fmffixed{(0,0)}{t3,w3}
                \fmfv{decoration.shape=circle, decoration.size=.1h, decoration.filled=empty}{t1}
                \fmfv{decoration.shape=cross, decoration.size=.1h, decoration.filled=empty}{w1}
                \fmffixed{(0,.5h)}{o1,v}
                \fmffixed{(0,.5h)}{v,o2}
                \fmffixed{(-.5h,0)}{z,t1}
                \fmf{wiggly,left=.5,tension=.0}{x,z,y}
                \fmfv{decoration.shape=circle, decoration.size=.2h, decoration.filled=empty}{z}
                \fmf{plain}{t3,x,y,t1,t2}
             \end{fmfgraph*}
        \end{tabular}
        \end{fmffile}
&~~~~~~~~~~& \begin{fmffile}{Anom2}
        \begin{tabular}{c}
            \begin{fmfgraph*}(60,80)
                \fmfleft{i1}
                \fmfright{o1,o2}
                \fmffixed{(-.1h,0)}{t1,i1}
                \fmffixed{(0,-.05h)}{t2,o1}
                \fmffixed{(0,.05h)}{t3,o2}
                \fmffixed{(0,0)}{t1,w1}
                \fmffixed{(0,0)}{t2,w2}
                \fmffixed{(0,0)}{t3,w3}
                \fmfv{decoration.shape=circle, decoration.size=.1h, decoration.filled=empty}{t1}
                \fmfv{decoration.shape=cross, decoration.size=.1h, decoration.filled=empty}{w1}
                \fmffixed{(0,.5h)}{o1,v}
                \fmffixed{(0,.5h)}{v,o2}
                \fmffixed{(-.45h,0)}{z,t1}
                \fmf{wiggly,tension=.0}{x,z,y}
                \fmfv{decoration.shape=circle, decoration.size=.2h, decoration.filled=empty}{z}
                \fmf{plain}{t3,y,z1,t1,z2,x,t2}
             \end{fmfgraph*}
        \end{tabular}
        \end{fmffile}
&~~~~~~~~~~& \begin{fmffile}{Anom3}
        \begin{tabular}{c}
            \begin{fmfgraph*}(60,80)
                \fmfleft{i1}
                \fmfright{o1,o2}
                \fmffixed{(-.1h,0)}{t1,i1}
                \fmffixed{(0,-.05h)}{t2,o1}
                \fmffixed{(0,.05h)}{t3,o2}
                \fmffixed{(0,0)}{t1,w1}
                \fmffixed{(0,0)}{t2,w2}
                \fmffixed{(0,0)}{t3,w3}
                \fmfv{decoration.shape=circle, decoration.size=.1h, decoration.filled=empty}{t1}
                \fmfv{decoration.shape=cross, decoration.size=.1h, decoration.filled=empty}{w1}
                \fmffixed{(0,.5h)}{o1,v}
                \fmffixed{(0,.5h)}{v,o2}
                \fmffixed{(-.4h,0)}{z,t1}
                \fmfv{decoration.shape=circle, decoration.size=.2h, decoration.filled=empty}{z}
                \fmf{wiggly,tension=.0}{t1,z,x}
                \fmf{plain}{t3,x,y,t1,t2}
             \end{fmfgraph*}
        \end{tabular}
        \end{fmffile}
\\
(a) && (b) && (c)
\end{array}
\end{equation}
In the case of the scalar operator $\bar\psi \psi$, only the first two diagrams $(a), (b)$ contribute.

To begin, let us compute the matter loop correction to the gauge field self energy. This is given by
\ie
\Sigma_G^{\mu\nu}(q) &= -{1\over 2} (-i)^2 \int {d^3q\over (2\pi)^3} {{\rm Tr} (\gamma^\mu {\slash\!\!\! k} \gamma^\nu ({\slash\!\!\! k}+{\slash\!\!\! q})) \over k^2 (k+q)^2}
\\
&= {q^2 \delta^{\mu\nu} - q^\mu q^\nu \over 32 q}.
\fe
The factor ${1\over 2}$ is due to our normalization convention for the gauge group generators. The contribution to the fermion self energy by the matter-loop-corrected gauge propagator, involved in diagram $(a)$, is given by
\ie
\Sigma'(p) = {N\over 2} \int {d^3q\over (2\pi)^3} {-i\gamma^\mu({\slash\!\!\! p}+{\slash\!\!\! q})\gamma^\nu \over (p+q)^2} D_{\mu\rho}(q) \Sigma_G^{\rho\sigma}(q) D_{\sigma\nu}(q) = i{\lambda\over k} a_1\,{\slash\!\!\! p} \ln \Lambda + {\rm finite},
\fe
where $D_{\mu\nu}(q)$ is the classical gauge propagator. Define the matter loop contribution to the gauge propagator as
\ie
G'_{\mu\nu}(q) = D_{\mu\rho}(q) \Sigma_G^{\rho\sigma}(q) D_{\sigma\nu}(q) .
\fe
Let us consider for now the anomalous dimension of $\bar\psi \psi$. The diagram $(b)$ is computed as
\ie
-{N\over 2}\int {d^3q\over (2\pi)^3} {D_{\mu\rho}(q) \Sigma_G^{\rho\sigma}(q) D_{\sigma}{}^{\mu}(q)\over q^2} = -{N\over 2}\int {d^3q\over (2\pi)^3} {1\over q^2}G'_\mu{}^\mu(q) = {\lambda\over k}a_2 \ln \Lambda
\fe
The anomalous dimension is then given by
\ie
\delta(\bar\psi\psi) = {\lambda^2\over N} (a_1+a_2).
\fe
The details of the computation of $a_1, a_2$ is given in \ref{two-loop-anomalous-dimension-appendix}  (using Feynman gauge, temporal gauge, and lightcone gauge respectively). The result is $a_1+a_2={1\over 3}$.

\subsection{Two-loop non-planar anomalous dimension of the scalar primary}
\label{two-loop-anomalous-dimension-appendix}

\subsubsection{Feynman gauge }

In Feynman gauge, we have
\ie
D_{\mu\nu}(q) =- {4\pi\over k}\epsilon_{\mu\nu\rho} {q^\rho\over q^2},
\fe
and so
\ie
G'_{\mu\nu}(q) = -{\pi^2\over k^2} {q^2\delta_{\mu\nu}-q_\mu q_\nu \over 2q^3}.
\fe
Its contribution to the fermion self energy is
\ie
\Sigma'(p) &= i {N\pi^2\over 4k^2} \int {d^3q\over (2\pi)^3} {q^2 \gamma^\mu({\slash\!\!\! p}+{\slash\!\!\! q})\gamma_\mu - {\slash\!\!\! q}({\slash\!\!\! p}+{\slash\!\!\! q}){\slash\!\!\! q} \over q^3 (p+q)^2}
\\
&= i {N\pi^2\over 4k^2} \int {d^3q\over (2\pi)^3} { - 2q\cdot(p+q) {\slash\!\!\! q} \over q^3 (p+q)^2}
\\
&= i {N\pi^2\over 4k^2} \int {d^3q\over (2\pi)^3} { p^2-q^2 \over q^3 (p+q)^2} {\slash\!\!\! q}
\fe
The logarithmic divergent part is
\ie
& - i {N\pi^2\over 4k^2} \int {d^3q\over (2\pi)^3} { {\slash\!\!\! q} \over q (p+q)^2} \\
& = - i {N\pi^2\over 4k^2} {1\over B(1,{1\over 2})}\int_0^1 dx\,(1-x)^{-{1\over 2}} \int {d^3q\over (2\pi)^3} { {\slash\!\!\! q}-x{\slash\!\!\! p} \over [q^2+p^2 x(1-x)]^{3\over 2}}
\\
& \sim {iN\over 12 k^2}{\slash\!\!\! p} {\ln\Lambda}.
\fe
The diagram $(b)$ is computed as
\ie
{N\pi^2\over 4k^2}\int {d^3q\over (2\pi)^3} {2\over q^3} = {N\over 4k^2}\ln\Lambda.
\fe
Therefore we find
\ie
a_1+a_2 = {1\over 12}+{1\over 4} = {1\over 3}.
\fe

\subsubsection{Temporal gauge }

In temporal gauge, we have
\ie
D_{ij}(q) = {4\pi\over k}\epsilon_{ij} {q^3\over (q^3)^2+\epsilon^2},
\fe
and
\ie
G'_{ij}(q) = -{\pi^2\over 2k^2} {(q^3)^2\delta_{ij} + q_i q_j \over  q} {1\over (q^3)^2 + \epsilon^2}.
\fe
Its contribution to the fermion self energy is
\ie
&\Sigma'(p) = i {N\pi^2\over 4k^2} \int {d^3q\over (2\pi)^3} {(q^3)^2 \gamma_i({\slash\!\!\! p}+{\slash\!\!\! q})\gamma_i + {\slash\!\!\! q}_s({\slash\!\!\! p}+{\slash\!\!\! q}){\slash\!\!\! q}_s \over  (p+q)^2 q [(q^3)^2+\epsilon^2]}
\\
&= i {N\pi^2\over 4k^2} \int {d^3q\over (2\pi)^3} { -2(q^3)^2 (p^3+q^3)\gamma^3 + 2q_s\cdot(p_s+q_s) {\slash\!\!\! q}_s-q_s^2({\slash\!\!\! p}+{\slash\!\!\! q})\over  (p+q)^2 q [(q^3)^2+\epsilon^2]}
\\
&= i {N\pi^2\over 4k^2} \int {d^3q\over (2\pi)^3} { ((p+q)^2-(p_3+q_3)^2-p_s^2) {\slash\!\!\! q}_s-(q^2-q_3^2)({\slash\!\!\! p}_s + (p+q)^0 \gamma^3)  -2(q^3)^2 (p^3+q^3)\gamma^3  \over  (p+q)^2 q [(q^3)^2+\epsilon^2]}
\\
&= i {N\pi^2\over 4k^2} \int {d^3q\over (2\pi)^3} \left[ {{\slash\!\!\! q}_s\over q ((q^3)^2+\epsilon^2)} + {{\slash\!\!\! p}_s - (p+q)^0 \gamma^3\over (p+q)^2q} 
\right. \\ & \left.~~~~~~
- {{\slash\!\!\! q}_s\over  (p+q)^2 q} { (p_3+q_3)^2+p_s^2  \over  (q^3)^2+\epsilon^2}-{|q|\over (p+q)^2}{{\slash\!\!\! p}_s + (p+q)^0 \gamma^3 \over (q^3)^2+\epsilon^2} \right].
\fe
The first term gives zero. The integrals of the remaining three terms give logarithmic divergences
\ie
\left( {{\slash\!\!\! p}_s\over 2} - {p_3 \gamma_3\over 6} \right) {\ln \Lambda\over \pi^2},
~~~{{\slash\!\!\! p}_s\over 3\pi^2}\ln\Lambda,~~~\left( {{\slash\!\!\! p}_s\over 2} +{3\over 2} {p_3 \gamma_3} \right){\ln\Lambda\over \pi^2}.
\fe
Adding these together, we find
\ie
\Sigma'(p) = i{N\ln\Lambda\over 3k^2} {\slash\!\!\! p}+{\rm finite}.
\fe
The diagram $(b)$ is computed as
\ie
{N\pi^2\over 4k^2}\int {d^3q\over (2\pi)^3} {(q^3)^2+q^2\over q^3 } {1\over (q^3)^2+\epsilon^2}
\to 0.
\fe
The result agrees with the Feynman gauge computation.

\subsubsection{Lightcone gauge}

In lightcone gauge, we have
\ie
D_{0+}(q) = -D_{+0}(q) = {8\pi i\over k} {q^-\over q_s^2+i\epsilon}
\fe
and
\ie
\begin{pmatrix} G'_{00}(q) & G'_{0+}(q) \\ G'_{+0}(q) & G'_{++}(q) \end{pmatrix}
= {2\pi^2\over k^2} {(q^-)^2\over q(q_s^2+i\epsilon)^2}
\begin{pmatrix} -(q^+)^2 & q^3 q^+ \\ q^3 q^+ & q_s^2 \end{pmatrix}
\fe
Its contribution to the fermion self energy is
\ie
&\Sigma'(p) =- i {N\pi^2\over k^2} \int {d^3q\over (2\pi)^3} (q^-)^2 
\\
&~~~~~~~~~~\times {-(q^+)^2 \gamma^3({\slash\!\!\! p}+{\slash\!\!\! q})\gamma^3 + q^3q^+ \gamma^3({\slash\!\!\! p}+{\slash\!\!\! q})\gamma^+
+ q^3q^+ \gamma^+({\slash\!\!\! p}+{\slash\!\!\! q})\gamma^3 + q_s^2 \gamma^+({\slash\!\!\! p}+{\slash\!\!\! q})\gamma^+ \over (p+q)^2 q(q_s^2+i\epsilon)^2}
\\
&=- i {N\pi^2\over 2k^2} \int {d^3q\over (2\pi)^3} q^- {q^+ [{\slash\!\!\! p}+{\slash\!\!\! q}-2(p+q)^0 \gamma^3] + 2q^3 [(p+q)^+ \gamma^3 + (p+q)^0\gamma^+] + 4 q^- (p+q)^+ \gamma^+ \over (p+q)^2 q(q_s^2+i\epsilon)^2}
\fe
With a little manipulation, this integral can be written as
\ie
&\Sigma'(p) = -i{N\pi^2\over 4k^2} \int {d^3 q\over (2\pi)^3} \left[ {{\slash\!\!\! p}+{\slash\!\!\! q} - 2p^3 \gamma^3 \over (p+q)^2q} + {4|q|\, q^- \gamma^+\over (p+q)^2 q_s^2} \right]
\\
&= -i{N\pi^2\over 4k^2} \int_0^1 dx \int {d^3 q\over (2\pi)^3} \left[ {(1-x)^{-{1\over 2}}\over B(1,{1\over 2})} {(1-x){\slash\!\!\! p} - 2p^3 \gamma^3 \over [q^2+x(1-x)p^2]^{3\over 2}} \right.
\\
&~~~\left.+ {(1-x)^{-{3\over 2}}\over B(1,-{1\over 2})} {4 q^- \gamma^+\over [q_s^2+2xp_s\cdot q_s+xp_s^2+q_3^2+x(1-x)p_3^2]^{1\over 2} q_s^2} \right]
\\
&= -i{N\pi^2\over 4k^2} \left[ \left({1\over 6\pi^2}{\slash\!\!\! p} - {1\over \pi^2}p^3 \gamma^3 \right) \ln\Lambda \right.
\\
&~~~\left.+  \int_0^1 dx\int {d^2q\over 4\pi^3}{(1-x)^{-{3\over 2}}\over B(1,-{1\over 2})} {4 q^- \gamma^+\over q_s^2} \ln {\Lambda\over  [q_s^2+2xp_s\cdot q_s+xp_s^2+x(1-x)p_3^2]^{1\over 2}} \right]
\\
&= -i{N\pi^2\over 4k^2} \left({1\over 6\pi^2}{\slash\!\!\! p} - {1\over \pi^2}p^3 \gamma^3 - {2\over \pi^2}p^-\gamma^+ \right) \ln\Lambda
\fe
While this expression is not Lorentz invariant, it is not a gauge invariant quantity. Inserting it into a two-point function computation gives
\ie
a_1 ={1\over 4}\left( -{1\over 6}+{1\over 3} + {2\over 3}\right) = {5\over 24}.
\fe
The diagram $(b)$ is computed as
\ie
{N\pi^2\over k^2} \int {d^3q\over (2\pi)^3} {(q^+q^-)^2\over q^3 (q_s^2+i\epsilon)^2} = {N\over 8k^2}\ln\Lambda.
\fe
We have
\ie
a_1+a_2 =  {5\over 24} + {1\over 8} = {1\over 3}.
\fe
Once again, this agrees with the result we found in Feynman and temporal gauge.

\section{Perturbative Analysis of Three Point Functions}

\subsection{Vanishing one-loop contribution to the three point functions of scalar operators}

In this subsection, we consider the three point function of the scalar operator $J_0(x) = {1\over \sqrt{N}}\bar\psi(x)\psi(x)$. The normalization is such that its two-point function does not scale with $N$. By conformal symmetry,
\ie
\langle J_0(x_1) J_0(x_2) J_0(x_3)\rangle = {1\over \sqrt{N}}{C(\lambda) + {\cal O}(1/N)\over x_{12}^2x_{23}^2x_{31}^2}
\fe
Parity requires $C(\lambda)$ to be an odd function of $\lambda$. So if the three point function of $J_0$ is nonzero, it can only receive contribution from odd loop order. A priori, for instance, there could be a one-loop contribution of order ${\cal O}(\lambda)$ to $\langle J_0 J_0 J_0\rangle$. We will perform this computation explicitly and see that it is in fact zero.

It suffices to take the limit $x_{12}\to 0$ while keeping $x_{13}$ finite. In this limit, we expect
\ie
\langle J_0(x_1) J_0(x_2) J_0(x_3)\rangle \to {1\over \sqrt{N}}{C(\lambda) \over x_{12}^2x_{13}^4} .
\fe
There are four distinct diagrams to consider (as shown in Fig. 3):

\noindent (A1) 1-loop self-energy correction to the fermion propagator $\langle \psi(x_1) \bar\psi(x_2)\rangle$.

\noindent (A2) 1-loop self-energy correction to the fermion propagator $\langle \psi(x_1) \bar\psi(x_3)\rangle$ (or $\langle \psi(x_2) \bar\psi(x_3)\rangle$). Such diagrams do not contribute to terms like $x_{12}^{-2}x_{13}^{-4}$, and will henceforth be ignored.

\noindent (B1) a gauge propagator correcting the vertex of $J_0(x_3)$.

\noindent (B2) a gauge propagator correcting the vertex of $J_0(x_1)$ (or $J_0(x_1)$).

\bigskip

\begin{equation}
\begin{array}{ccccccc}
\begin{fmffile}{OOO1}
        \begin{tabular}{c}
            \begin{fmfgraph*}(30,80)
                \fmfleft{i1,i2}
                \fmfright{o1,o2}
                \fmffixed{(0,-0.05h)}{i,i1}
                \fmffixed{(0,-h)}{v,i}
                \fmffixed{(0,-.12h)}{y,i}
                \fmffixed{(-.8w,0)}{x,i}
                \fmffixed{(-.4w,0)}{u,v}
                \fmffixed{(0,-.03w)}{xp,x}
                \fmf{plain}{i,u,x,b,i}
                \fmfblob{.25w}{b}
                \fmflabel{1}{i}
                \fmflabel{2}{x}
                \fmflabel{3}{u}
             \end{fmfgraph*}
        \end{tabular}
        \end{fmffile}
&~~~~~~~&
\begin{fmffile}{OOO2}
        \begin{tabular}{c}
            \begin{fmfgraph*}(30,80)
                \fmfleft{i1,i2}
                \fmfright{o1,o2}
                \fmffixed{(0,-0.05h)}{i,i1}
                \fmffixed{(0,-h)}{v,i}
                \fmffixed{(0,-.12h)}{y,i}
                \fmffixed{(-.8w,0)}{x,i}
                \fmffixed{(-.4w,0)}{u,v}
                \fmffixed{(0,-.03w)}{xp,x}
                \fmf{plain}{i,b,u,x,i}
                \fmfblob{.25w}{b}
                \fmflabel{1}{i}
                \fmflabel{2}{x}
                \fmflabel{3}{u}
             \end{fmfgraph*}
        \end{tabular}
        \end{fmffile}
&~~~~~~~&
\begin{fmffile}{OOO3}
        \begin{tabular}{c}
            \begin{fmfgraph*}(30,80)
                \fmfleft{i1,i2}
                \fmfright{o1,o2}
                \fmffixed{(0,-0.05h)}{i,i1}
                \fmffixed{(0,-h)}{v,i}
                \fmffixed{(-.4w,0)}{u,v}
                \fmffixed{(-.05w,-.1h)}{y1,i}
                \fmffixed{(-.8w,0)}{x,i}
                \fmffixed{(-.75w,-.1h)}{y2,i}
                \fmf{wiggly,tension=.0}{y1,y2}
                \fmf{plain}{i,u,x,i}
                \fmflabel{1}{i}
                \fmflabel{2}{x}
                \fmflabel{3}{u}
             \end{fmfgraph*}
        \end{tabular}
        \end{fmffile}
&~~~~~~~&
\begin{fmffile}{OOO4}
        \begin{tabular}{c}
            \begin{fmfgraph*}(30,80)
                \fmfleft{i1,i2}
                \fmfright{o1,o2}
                \fmffixed{(0,-0.05h)}{i,i1}
                \fmffixed{(0,-h)}{v,i}
                \fmffixed{(-.4w,0)}{u,v}
                \fmffixed{(-.05w,-.12h)}{y1,i}
                \fmffixed{(-.8w,0)}{x,i}
                \fmffixed{(-.55w,0)}{y2,i}
                \fmf{wiggly,tension=.0}{y1,y2}
                \fmf{plain}{i,u,x,i}
                \fmflabel{1}{i}
                \fmflabel{2}{x}
                \fmflabel{3}{u}
             \end{fmfgraph*}
        \end{tabular}
        \end{fmffile}
\\
\\
A1 && A2 && B1 && B2
\end{array}
\end{equation}
\centerline{Figure 3}

We will work in position space. Let $y_1,y_2$ be the two ends of the gauge propagator. In the limit $x_{12}\to 0$, the relevant contribution comes from $y_1,y_2\sim {\cal O}(x_{12})$. From now we will set $x_1=0$ and $x_2=x$, while taking $x_3$ to be large. The contribution from diagrams (A1), (B1) and (B2) (note that (B2) contribution should be multiplied by a factor of 2 since there are two diagrams giving identical contributions) are proportional to (the overall $x_3$ dependent factor $|x_3|^{-4}$ is factored out)
\ie
& I(A1) = \int d^3 y_1 d^3 y_2 \langle A_\mu(y_1)A_\nu(y_2)\rangle {\rm Tr} \left[ \langle \psi(0)\bar\psi(y_1) \rangle \gamma^\mu \langle \psi(y_1)\bar\psi(y_2) \rangle \gamma^\nu  \langle \psi(y_2)\bar\psi(x) \rangle \right],
\\
& I(B1) = \int d^3 y_1 d^3 y_2 \langle A_\mu(y_1)A_\nu(y_2)\rangle {\rm Tr} \left[ \gamma^\mu \langle \psi(y_1)\bar\psi(0) \rangle \langle \psi(0)\bar\psi(x) \rangle \langle \psi(x)\bar\psi(y_2) \rangle \gamma^\nu \right],
\\
& I(B2) = \int d^3 y_1 d^3 y_2 \langle A_\mu(y_1)A_\nu(y_2)\rangle {\rm Tr} \left[ \gamma^\mu \langle \psi(y_1)\bar\psi(0) \rangle \langle \psi(0)\bar\psi(y_2) \rangle \gamma^\nu \langle \psi(y_2)\bar\psi(x) \rangle \right].
\fe
In the following two subsubsections, we demonstrate the computation in Feynman gauge and in light cone gauge, respectively.

\subsubsection{Feynman gauge}

The propagators are
\ie
\langle A_\mu(x)A_\nu(0)\rangle = -{i\over k} \epsilon_{\mu\nu\rho} {x^\rho\over |x|^3},~~~~
\langle\psi(x)\bar\psi(0)\rangle =  {{\slash\!\!\! x}\over 4\pi |x|^3}.
\fe
The results for the loop integrals are
\ie
I(A1) =- {\lambda\over 2\pi x^2},~~~I(B1) = {\lambda\over 2\pi x^2},~~~I(B2)=0.
\fe
Indeed they cancel, as claimed.

\subsubsection{Light cone gauge}

We have seen that the 1-loop correction to fermion propagator in light cone gauge is proportional to $|x^0|/|x|^3$. Without loss of generality, we can take the Euclidean time direction to be orthogonal to the plane spanned by $x_1, x_2, x_3$, and so diagrams (A1), (A2) simply vanish in the light cone gauge. (B1) and (B2) are then computed as
\ie
&I(B1) = -{\lambda\over (4\pi)^3} \int d^3 y_1 d^3 y_2 {\delta(y_{12}^0)y_{12}^-\over y_{12}^2} {{\rm Tr}\left\{
{\slash\!\!\! y}_1 {\slash\!\!\! x} ({\slash\!\!\! y}_2-{\slash\!\!\! x}) [\gamma^+,\gamma^3]\right\}
\over |y_1|^3 |y_2-x|^3 |x|^3}
\\
&~~~~= -{\lambda\over (4\pi)^3} \int d^3 y_1 d^3 y_2 {\delta(y_{12}^0)\over y_{12}^2} {{\rm Tr}\left\{
{\slash\!\!\! y}_1 {\slash\!\!\! x} ({\slash\!\!\! y}_2-{\slash\!\!\! x})
{\slash\!\!\! y}_{12} \right\}
\over |y_1|^3 |y_2-x|^3 |x|^3}
\\
&~~~~= -{\lambda\over (4\pi)^3} 2\int d^2 z_1 d^2 z_2 dt {1\over z_{12}^2} { z_1\cdot x \,z_{12}\cdot (z_1-x) + z_1\cdot z_{12}\, x\cdot (z_2-x) - z_{12}\cdot x \left[ z_1\cdot (z_2-x)+t^2 \right]
\over (z_1^2+t^2)^{3\over 2} ((z_2-x)^2+t^2)^{3\over 2} |x|^3}
\\
&~~~~= -{\lambda\over (4\pi)^3} 2\pi \int_0^1 da \int d^2 z_2 dt { (az_2^2+3t^2)  x\cdot (z_2-x)  + t^2x^2 \over (az_2^2+t^2)^{3\over 2} ((z_2-x)^2+t^2)^{3\over 2} |x|^3}
\\
&~~~~=0;
\\
&I(B2) =  {\lambda\over (4\pi)^3}\int d^3 y_1 d^3 y_2 {\delta(y_{12}^0)y_{12}^-\over y_{12}^2} {{\rm Tr}\left\{
\left( \gamma^3 {\slash\!\!\! y}_1 {\slash\!\!\! y}_2 \gamma^+ - \gamma^+ {\slash\!\!\! y}_1 {\slash\!\!\! y}_2 \gamma^3 \right) ({\slash\!\!\! y}_2-{\slash\!\!\! x}) \right\}
\over |y_1|^3 |y_2|^3 |y_2-x|^3}
\\
&~~~~=  -{\lambda\over (4\pi)^3}2\int d^3 y_1 d^3 y_2 {\delta(y_{12}^0)\over y_{12}^2} { (y_1\cdot y_2) y_{12}\cdot (y_2-x)
\over |y_1|^3 |y_2|^3 |y_2-x|^3}
\\
&~~~~=  -{\lambda\over (4\pi)^3}2\int d^2 z_1 d^2 z_2 dt {1\over z_{12}^2} { (z_1\cdot z_2+t^2)  \, z_{12}\cdot (z_2-x)
\over (z_1^2+t^2)^{3\over 2}(z_2^2+t^2)^{3\over 2} ((z_2-x)^2+t^2)^{3\over 2}}
\\
&~~~~=0.
\fe
We see that the contribution from each diagram vanishes individually in the light cone gauge.

\subsection{One-loop parity violating structure of  $\langle JJJ\rangle$}

We will work in temporal gauge, which is particularly convenient for this computation since the position space gauge field propagator is very simple.
With our special choice of null polarization vectors $\varepsilon_1=\varepsilon_2=\varepsilon_3=\varepsilon$, all diagrams involving a one-loop fermion self-energy vanish (Figure 4(a)). Similarly, the diagrams with a gauge propagator coming out of a vertex and ends on an adjacent fermion line also vanish (Figure 4(b)). We further set $x_3=x=t\hat e_0$, ${\slash\!\!\!\varepsilon} = \gamma^+$, $x_1=0$, $x_2=-\delta$ purely spatial, as explained earlier. There are three diagrams that a priori contribute to $\langle JJJ\rangle$ in this special case ($J$ is assumed to be a current of nonzero spin):

\begin{equation}\nonumber
\begin{array}{ccc}
\begin{fmffile}{TTTvan1}
        \begin{tabular}{c}
            \begin{fmfgraph*}(60,60)
                \fmfleft{i1}
                \fmfright{o1,o2}
                \fmffixed{(-.1h,0)}{t1,i1}
                \fmffixed{(0,-.05h)}{t2,o1}
                \fmffixed{(0,.05h)}{t3,o2}
                \fmffixed{(0,0)}{t1,w1}
                \fmffixed{(0,0)}{t2,w2}
                \fmffixed{(0,0)}{t3,w3}
                \fmfv{decoration.shape=circle, decoration.size=.1h, decoration.filled=empty}{t1,t2,t3}
                \fmfv{decoration.shape=cross, decoration.size=.1h, decoration.filled=empty}{w1,w2,w3}
                \fmffixed{(0,.5h)}{o1,v}
                \fmffixed{(0,.5h)}{v,o2}
                \fmfblob{.25w}{v}
                \fmf{plain}{t1,t2,t3,t1}
             \end{fmfgraph*}
        \end{tabular}
        \end{fmffile}
&~~~~~~~~~~&
\begin{fmffile}{TTTvan2}
        \begin{tabular}{c}
            \begin{fmfgraph*}(60,60)
                \fmfleft{i1}
                \fmfright{o1,o2}
                \fmffixed{(-.1h,0)}{t1,i1}
                \fmffixed{(0,-.05h)}{t2,o1}
                \fmffixed{(0,.05h)}{t3,o2}
                \fmffixed{(0,0)}{t1,w1}
                \fmffixed{(0,0)}{t2,w2}
                \fmffixed{(0,0)}{t3,w3}
                \fmffixed{(0,.015h)}{z,t1}
                \fmfv{decoration.shape=circle, decoration.size=.1h, decoration.filled=empty}{t1,t2,t3}
                \fmfv{decoration.shape=cross, decoration.size=.1h, decoration.filled=empty}{w1,w2,w3}
                \fmffixed{(0,.5h)}{o1,v}
                \fmffixed{(0,.5h)}{v,o2}
                \fmf{wiggly,left=.5,tension=.0}{z,x}
                \fmf{plain}{t1,x,t2,t3,t1}
             \end{fmfgraph*}
        \end{tabular}
        \end{fmffile}
\\
(a)& & (b)
\end{array}
\end{equation}
\centerline{Figure 4: diagrams that do not contribute to $\langle JJJ\rangle$ in the special limit.}

\begin{equation}
\begin{array}{ccc}
\begin{fmffile}{TTT1}
        \begin{tabular}{c}
            \begin{fmfgraph*}(60,60)
                \fmfleft{i1}
                \fmfright{o1,o2}
                \fmffixed{(-.1h,0)}{t1,i1}
                \fmffixed{(0,-.05h)}{t2,o1}
                \fmffixed{(0,.05h)}{t3,o2}
                \fmffixed{(0,0)}{t1,w1}
                \fmffixed{(0,0)}{t2,w2}
                \fmffixed{(0,0)}{t3,w3}
                \fmfv{decoration.shape=circle, decoration.size=.1h, decoration.filled=empty}{t1,t2,t3}
                \fmfv{decoration.shape=cross, decoration.size=.1h, decoration.filled=empty}{w1,w2,w3}
                \fmffixed{(0,.5h)}{o1,v}
                \fmffixed{(0,.5h)}{v,o2}
                \fmf{wiggly,tension=.0}{t1,v}
                \fmf{plain}{t1,x,t2,t3,t1}
             \end{fmfgraph*}
        \end{tabular}
        \end{fmffile}
&~~~~~~~~~~& \begin{fmffile}{TTT2}
        \begin{tabular}{c}
            \begin{fmfgraph*}(60,60)
                \fmfleft{i1}
                \fmfright{o1,o2}
                \fmffixed{(-.1h,0)}{t1,i1}
                \fmffixed{(0,-.05h)}{t2,o1}
                \fmffixed{(0,.05h)}{t3,o2}
                \fmffixed{(0,0)}{t1,w1}
                \fmffixed{(0,0)}{t2,w2}
                \fmffixed{(0,0)}{t3,w3}
                \fmfv{decoration.shape=circle, decoration.size=.1h, decoration.filled=empty}{t1,t2,t3}
                \fmfv{decoration.shape=cross, decoration.size=.1h, decoration.filled=empty}{w1,w2,w3}
                \fmffixed{(0,.5h)}{o1,v}
                \fmffixed{(0,.5h)}{v,o2}
                \fmf{wiggly,tension=.0}{x,y}
                \fmf{plain}{t1,x,t2,t3,y,t1}
             \end{fmfgraph*}
        \end{tabular}
        \end{fmffile}
\\
(a) && (b)
\end{array}
\end{equation}
\centerline{Figure 5: type of diagrams that contribute to $\langle JJJ\rangle$ in the special limit.}


\noindent (I) A gauge propagator coming out of the vertex $J(x_2)$ and ends on the fermion line between $x_1$ and $x_3$. This is shown in figure 5(a). Two analogous diagrams, in which the gauge field propagator comes out of $J(x_1)$ or $J(x_3)$, vanish identically because one encounters the structure ${\slash\!\!\!\varepsilon} {\slash\!\!\!x}_{31}{\slash\!\!\!\varepsilon}={\slash\!\!\!\varepsilon} t\gamma^3{\slash\!\!\!\varepsilon}=0$ in the loop integral, for the special configuration of positions and polarizations we are considering here.

\noindent (II) A gauge propagator correcting the vertex $J(x_3)$. This is shown in figure 5(b), where the vertex on the left is $J(x_3)$.

\noindent (III) a gauge propagator correcting the vertex $J(x_1)$. The diagram is the same as in figure 5(b) except that the vertex on the left is $J(x_1)$ in this case. Note that the analogous diagram with the gauge propagator correcting $J(x_2)$ vanishes identically, because once again it involves the structure ${\slash\!\!\!\varepsilon} {\slash\!\!\!x}_{31}{\slash\!\!\!\varepsilon}={\slash\!\!\!\varepsilon} t\gamma^3{\slash\!\!\!\varepsilon}=0$.

The diagrams I,II,III are exhibited more explicitly in the limit $|\delta|\ll |t|$ in figure 6.

\bigskip

\begin{equation}
\begin{array}{ccccc}
\begin{fmffile}{LIMIT1}
        \begin{tabular}{c}
            \begin{fmfgraph*}(30,80)
                \fmfleft{i1,i2}
                \fmfright{o1,o2}
                \fmffixed{(0,0)}{i,i1}
                \fmffixed{(0,-h)}{v,i}
                \fmffixed{(0,-.12h)}{y,i}
                \fmffixed{(-.8w,0)}{x,i}
                \fmffixed{(0,-.03w)}{xp,x}
                \fmf{wiggly,tension=.0}{y,xp}
                \fmf{plain}{i,v,x,i}
                \fmflabel{1}{i}
                \fmflabel{2}{x}
                \fmflabel{3}{v}
             \end{fmfgraph*}
        \end{tabular}
        \end{fmffile}
&~~~~~~~&
\begin{fmffile}{LIMIT2}
        \begin{tabular}{c}
            \begin{fmfgraph*}(30,80)
                \fmfleft{i1,i2}
                \fmfright{o1,o2}
                \fmffixed{(0,0)}{i,i1}
                \fmffixed{(0,-h)}{v,i}
                \fmffixed{(0,-.1h)}{y1,i}
                \fmffixed{(-.8w,0)}{x,i}
                \fmffixed{(-.7w,-.1h)}{y2,i}
                \fmf{wiggly,tension=.0}{y1,y2}
                \fmf{plain}{i,v,x,i}
                \fmflabel{1}{i}
                \fmflabel{2}{x}
                \fmflabel{3}{v}
             \end{fmfgraph*}
        \end{tabular}
        \end{fmffile}
&~~~~~~~&
\begin{fmffile}{LIMIT3}
        \begin{tabular}{c}
            \begin{fmfgraph*}(30,80)
                \fmfleft{i1,i2}
                \fmfright{o1,o2}
                \fmffixed{(0,0)}{i,i1}
                \fmffixed{(0,-h)}{v,i}
                \fmffixed{(0,-.12h)}{y1,i}
                \fmffixed{(-.8w,0)}{x,i}
                \fmffixed{(-.55w,0)}{y2,i}
                \fmf{wiggly,tension=.0}{y1,y2}
                \fmf{plain}{i,v,x,i}
                \fmflabel{1}{i}
                \fmflabel{2}{x}
                \fmflabel{3}{v}
             \end{fmfgraph*}
        \end{tabular}
        \end{fmffile}
\\
\\
{\rm I} && {\rm II} && {\rm III}
\end{array}
\end{equation}
\centerline{Figure 6}


\subsubsection{$\langle jjT\rangle$}

To begin, let us consider the one-loop contribution to $\langle j_\varepsilon(x_1)j_\varepsilon(x_2)T_\varepsilon(x_3)\rangle$. For this correlator, diagram I does not contribute as $A_\mu$ does not appear explicitly in $j(x_2)$. Diagram II is given by
\ie
& \int d^3y_1 d^3y_2 \langle A_i(y_1) A_j(y_2) \rangle {\rm Tr} \left[\gamma^j\langle \psi(y_2)\bar\psi(x_3)\rangle
{\slash\!\!\!\varepsilon}(\varepsilon\cdot\overleftrightarrow\partial_{x_3}) \langle \psi(x_3)\bar\psi(y_1)\rangle \gamma^i \langle \psi(y_1)\bar\psi(x_1)\rangle \right.
\\
&\left.~~~~\times{\slash\!\!\!\varepsilon} \langle \psi(x_1)\bar\psi(x_2)\rangle {\slash\!\!\!\varepsilon}\langle \psi(x_2)\bar\psi(y_2)\rangle \right]
\fe
At our special configuration $x_1=0, x_2=-\delta, x_3=x=t\hat e_0$ ($\delta$ purely spatial), ${\slash\!\!\!\varepsilon} = \gamma^+$, the loop integral is proportional to
\ie\label{tmpa}
&\int d^3 y_1 d^3 y_2 \epsilon(y_{12}^0) \delta^2(\vec y_{12}) \epsilon_{ij} {\rm Tr}\left[ \gamma^j ({\slash\!\!\! y}_2-{\slash\!\!\! x}) \gamma^+ ({\slash\!\!\! x}-{\slash\!\!\! y}_1)\gamma^i {\slash\!\!\! y}_1 \gamma^+ {\slash\!\!\!\delta} \gamma^+ (-{\slash\!\!\! y}_2-{\slash\!\!\! \delta}) \right]
\\
&~~~\times {1\over |y_1|^3 |\delta|^3 |\delta+y_2|^3}  \left( {1\over |x-y_2|^3} \overleftrightarrow\partial_{x^-} {1\over |x-y_1|^3}  \right)
\\
& =- 16i{\delta^+ \over |\delta|^3}\int d^2z dt_1 dt_2 \epsilon(t_{12})  \left[ (z^+)^2 t_2(t-t_2) - z^+(z+\delta)^+ t_1(t-t_1) \right]
\\
&~~~\times {1\over (z^2+t_1^2)^{3\over 2}  [(\delta+z)^2+t_2^2]^{3\over2}}  \left\{ {1\over [(t-t_2)^2+z^2]^{3\over 2}} \overleftrightarrow\partial_{z^-} {1\over [(t-t_1)^2+z^2]^{3\over 2}}  \right\}
\fe
In the small $\delta$ limit, the leading contribution to the integral comes from two domains:
\ie\label{dreg}
& D_1: t_1, t_2\sim {\cal O}(\delta),
\\
& D_2: (t_1-t), t_2\sim {\cal O}(\delta).
\fe
The contribution from region $D_1$ is easily seen from (\ref{tmpa}) to vanish at order ${\cal O}(t^{-7})$.
The contribution from $D_2$ is
\ie
& -16i{\delta^+ \epsilon(t) \over |\delta|^3 |t|^3}  \int d^2z dt'_1 dt_2 {1\over   [(\delta+z)^2+t_2^2]^{3\over2}}
\\
&~~~\times \left\{ \left[ -(z^+)^2 t_2^2 + z^+(z+\delta)^+ t_1'^2 - 3 z^+(z+\delta)^+ t_1'^2 \right] \left[ {1\over (t^2+z^2)^{3\over 2}} \overleftrightarrow\partial_{z^-} {1\over (t_1'^2+z^2)^{3\over 2}}  \right] \right.
\\
&\left. ~~~~~+    \left[ (z^+)^2 t_2t - z^+(z+\delta)^+ t'_1t \right]
\left[ {3tt_2\over (t^2+z^2)^{5\over 2}} \overleftrightarrow\partial_{z^-} {1\over (t_1'^2+z^2)^{3\over 2}}  \right] \right\} \cdot \left[ 1+{\cal O}({\delta^2\over t^2}) \right]
\\
& = - 16i{\delta^+ \epsilon(t) \over |\delta|^3 t^6}  \int d^2z dt'_1 dt_2 {1\over   [(\delta+z)^2+t_2^2]^{3\over2}}  {-3 z^+\over (t_1'^2+z^2)^{5\over 2}}
\\
&~~~\times \left\{ \left[ -(z^+)^2 t_2^2 -2 z^+(z+\delta)^+ t_1'^2 \right]   +  3 (z^+)^2 t_2^2  \right\} \cdot \left[ 1+{\cal O}({\delta^2\over t^2}) \right]
\\
&= 32i{\delta^+ \epsilon(t) \over |\delta|^3 t^6}  \int d^2z dt_2 {(z^+)^2 \left[ 4t_2^2 z^+ -2 z^2 (z+\delta)^+ \right]\over   [(\delta+z)^2+t_2^2]^{3\over2} |z|^4 } \left[ 1+{\cal O}({\delta^2\over t^2}) \right]
\\
& = 32i{(\delta^+)^4 \epsilon(t) \over |\delta|^3 t^6} \left\{ {1\over B({3\over 2},2)}\int_0^1 da\, a^{1\over 2}(1-a)  \int d^2z dt_2 {a^2 \left[ 4t_2^2 (-a)  \right]\over   [z^2+a(1-a)\delta^2+a t_2^2]^{7\over 2} } \right.
\\
&~~~~\left.+{1\over B({3\over 2},1)}\int_0^1 da\, a^{1\over 2}  \int d^2z dt_2 {-2a^2 (1-a)\over   [z^2+a(1-a)\delta^2+a t_2^2]^{5\over 2} } \right\} \left[ 1+{\cal O}({\delta^2\over t^2}) \right]
\\
&= - 128\pi i{(\delta^+)^4 \epsilon(t) \over |\delta|^5 t^6}  \left[ 1+{\cal O}({\delta^2\over t^2}) \right].
\fe

Diagram III is given by
\ie
& \int d^3y_1 d^3y_2 \langle A_i(y_1) A_j(y_2) \rangle {\rm Tr} \left[\gamma^i\langle \psi(y_1)\bar\psi(x_1)\rangle
{\slash\!\!\!\varepsilon} \langle \psi(x_1)\bar\psi(y_2)\rangle \gamma^j \langle \psi(y_2)\bar\psi(x_2)\rangle \right.
\\
&\left.~~~~\times{\slash\!\!\!\varepsilon}\langle \psi(x_2)\bar\psi(x_3)\rangle {\slash\!\!\!\varepsilon}(\varepsilon\cdot\overleftrightarrow\partial_{x_3}) \langle \psi(x_3)\bar\psi(y_1)\rangle \right]
\fe
The loop integral is evaluated as
\ie
&\int d^3 y_1 d^3 y_2 \epsilon(y_{12}^0) \delta^2(\vec y_{12}) \epsilon_{ij} {\rm Tr}\left[ \gamma^i {\slash\!\!\! y}_1 \gamma^+ (-{\slash\!\!\! y}_2)\gamma^j ({\slash\!\!\! y}_2+{\slash\!\!\! \delta}) \gamma^+ (-{\slash\!\!\! x}-{\slash\!\!\!\delta}) \gamma^+ ({\slash\!\!\! x}-{\slash\!\!\! y}_1) \right]
\\
&~~~~\times   {1\over |y_1|^3 |y_2|^3 |\delta+y_2|^3}   \left( {1\over |x-y _1|^3} \overleftrightarrow\partial_{x^-} {1\over |x+\delta|^3} \right)
\\
& = - 16i \delta^+ \int d^2z dt_1 dt_2 \epsilon(t_{12}) [(z^+)^2 t_2^2+z^+(z+\delta)^+t_1(t-t_1)]
\\
&~~~~\times   {1\over (z^2+t_1^2)^{3\over 2} (z^2+t_2^2)^{3\over 2} [(z+\delta)^2+t_2^2]^{3\over 2}}   \left\{ {1\over [(t-t_1)^2+z^2]^{3\over 2}} (\overleftarrow \partial_{z^-}+ \overrightarrow\partial_{\delta^-}) {1\over (t^2+\delta^2)^{3\over 2}} \right\}
\fe
In the small $\delta$ limit, the integral is again dominated by the contribution from the regions $D_1$ and $D_2$. The contribution from $D_1$ is
\ie
&-16i {\delta^+ \over t^7} \int d^2z dt_1 dt_2 \epsilon(t_{12}) t_1 z^+(z+\delta)^+
{-3 (z+\delta)^+ \over (z^2+t_1^2)^{3\over 2} (z^2+t_2^2)^{3\over 2} [(z+\delta)^2+t_2^2]^{3\over 2}}
\\
& = 96i {\delta^+ \over t^7} \int d^2z  dt_2
{z^+((z+\delta)^+)^2 \over (z^2+t_2^2)^2 [(z+\delta)^2+t_2^2]^{3\over 2}}
= 96i {\delta^+ \over t^7} \int d^3 y
{y^+((y+\delta)^+)^2 \over |y|^4 |y+\delta|^3}
\\
&= 96i {(\delta^+)^4 \over t^7} {1\over B(2,{3\over 2})}\int_0^1 da\,a^{1\over 2}(1-a) \int d^3 y
{(-a)(1-a)^2 \over [y^2 + \delta^2 a(1-a)]^{7\over 2}}
\\
& = - {256\pi i} {(\delta^+)^4\over \delta^4 t^7}.
\fe
The contribution from $D_2$ is
\ie
& -16i { \delta^+ \epsilon(t)\over |t|^3} \int d^2z dt_1' dt_2 {(z^+)^2 t_2^2+z^+(z+\delta)^+t_1'(t-t_1')\over (z^2+(t-t_1')^2)^{3\over 2} (z^2+t_2^2)^{3\over 2} [(z+\delta)^2+t_2^2]^{3\over 2}}   {-3 z^+\over (t_1'^2+z^2)^{5\over 2}} \left[ 1+{\cal O}({\delta^2\over t^2})\right]
\\
&=-16i { \delta^+ \epsilon(t)\over t^6} \int d^2z dt_1' dt_2 {(z^+)^2 t_2^2 - z^+(z+\delta)^+t_1'^2 + 3z^+(z+\delta)^+t_1'^2 \over (z^2+t_2^2)^{3\over 2} [(z+\delta)^2+t_2^2]^{3\over 2}}   {-3 z^+\over (t_1'^2+z^2)^{5\over 2}} \left[ 1+{\cal O}({\delta^2\over t^2})\right]
\\
& = {64i} { \delta^+ \epsilon(t)\over t^6} \int d^2z dt_2 (z^+)^2{z^+ t_2^2 + z^2 (z+\delta)^+ \over (z^2+t_2^2)^{3\over 2} [(z+\delta)^2+t_2^2]^{3\over 2} |z|^4}  \left[ 1+{\cal O}({\delta^2\over t^2})\right]
\\
& = {64i} { \delta^+ \epsilon(t)\over t^6} {1\over B({3\over 2},{3\over 2})}\int_0^1 da\,a^{1\over 2}(1-a)^{1\over 2} \int d^2z dt_2 (z^+)^2{z^+ t_2^2 + z^2 (z+\delta)^+ \over [(1-a)z^2+a(z+\delta)^2+t_2^2]^3 |z|^4}  \left[ 1+{\cal O}({\delta^2\over t^2})\right]
\\
& = {8\pi i} { \delta^+ \epsilon(t)\over t^6} {1\over B({3\over 2},{3\over 2})}\int_0^1 da\,a^{1\over 2}(1-a)^{1\over 2}
\\
&~~~~\times \int d^2z \left\{ {(z^+)^3  \over |z|^4 (z^2+2a \delta\cdot z + a\delta^2)^{3\over 2}} +{3(z^+)^2(z+\delta)^+ \over z^2 (z^2+2a \delta\cdot z + a\delta^2)^{5\over 2}}  \right\} \left[ 1+{\cal O}({\delta^2\over t^2})\right]
\\
& = {8\pi i} { \delta^+ \epsilon(t)\over t^6} {1\over B({3\over 2},{3\over 2})}\int_0^1 da\,a^{1\over 2}(1-a)^{1\over 2} \int_0^1 db  \int d^2z \left\{ {b^{1\over 2}(1-b)\over B({3\over 2},2)} {(z^+)^3} + {b^{3\over 2}\over B({5\over 2},1)} \left[3(z^+)^2(z+\delta)^+ \right] \right\}
\\
&~~~~\times{1  \over [b(z^2+2a \delta\cdot z + a\delta^2) + (1-b)z^2]^{7\over 2}} \left[ 1+{\cal O}({\delta^2\over t^2})\right]
\\
& = {8\pi i} { (\delta^+)^4 \epsilon(t)\over t^6} {1\over B({3\over 2},{3\over 2})}\int_0^1 da\,a^{1\over 2}(1-a)^{1\over 2} \int_0^1 db  \left\{ {b^{1\over 2}(1-b)\over B({3\over 2},2)} {(-ab)^3} + {b^{3\over 2}\over B({5\over 2},1)} \left[3(ab)^2(1-ab) \right] \right\}
\\
&~~~~\times \int d^2z {1  \over [z^2 + ab(1-ab)\delta^2]^{7\over 2}} \left[ 1+{\cal O}({\delta^2\over t^2})\right]
\\
&= 128\pi i  {(\delta^+)^4 \epsilon(t)\over |\delta|^5 t^6} \left[ 1+{\cal O}({\delta^2\over t^2})\right].
\fe
Putting back in normalization factors associated with the propagators, we obtain
\ie
\langle jjT\rangle_{1-loop} \longrightarrow {N\lambda\over 2\pi^3} {(\delta^+)^4\over \delta^4 t^7}
\fe
in our limit.

\subsubsection{$\langle TTT\rangle$}

Next, we consider the one-loop contribution to $\langle T_\varepsilon(x_1)T_\varepsilon(x_2)T_\varepsilon(x_3)\rangle$. Diagram I is given by
\ie
&\int d^3y \langle A_i(x_2) A_j(y) \rangle \\
& \times  {\rm Tr} \left[\gamma^i\langle \psi(y)\bar\psi(x_1)\rangle
{\slash\!\!\!\varepsilon}(\varepsilon\cdot\overleftrightarrow\partial_{x_1}) \langle \psi(x_1)\bar\psi(x_2)\rangle {\slash\!\!\!\varepsilon} \varepsilon^j \langle \psi(x_2)\bar\psi(x_3)\rangle{\slash\!\!\!\varepsilon}(\varepsilon\cdot\overleftrightarrow\partial_{x_3}) \langle \psi(x_3)\bar\psi(y)\rangle \right].
\fe
At our special configuration $x_1=0, x_2=-\delta, x_3=x=t\hat e_0$ ($\delta$ purely spatial), ${\slash\!\!\!\varepsilon} = \gamma^+$, the loop integral is proportional to
\ie
&\int d^3 y \epsilon(y^0) \delta^2(\vec y-\vec x_2) \epsilon_{ij}\varepsilon^j {\rm Tr}\left[ \gamma^i ({\slash\!\!\! y}-{\slash\!\!\! x}_1) \gamma^+ {\slash\!\!\! x}_{12}\gamma^+ {\slash\!\!\! x}_{23} \gamma^+ ({\slash\!\!\! x}_3-{\slash\!\!\! y}) \right]
\\
&~~~\times \left( {1\over |y-x_1|^3}\overleftrightarrow\partial_{x_1^-} {1\over |x_{12}|^3} \right)
\left( {1\over |x_{23}|^3}\overleftrightarrow\partial_{x_3^-} {1\over |x_3-y|^3} \right)
\\
&= -\int d^3 y \epsilon(y^0) \delta^2(\vec y+\vec\delta) \epsilon_{ij}\varepsilon^j {\rm Tr}\left[ \gamma^i {\slash\!\!\! y} \gamma^+ {\slash\!\!\! \delta}\gamma^+ (-{\slash\!\!\!\delta}-{\slash\!\!\! x}) \gamma^+ ({\slash\!\!\! x}-{\slash\!\!\! y}) \right]
\\
&~~~\times {9\over |y|^3 |\delta|^3|x-y|^3|x+\delta|^3}\left({y^+\over y^2}+{\delta^+\over \delta^2}\right)
\left({y^+\over (x-y)^2} + {\delta^+\over (x+\delta)^2}\right)
\\
& = -9\cdot 16i {(\delta^+)^6 \over |\delta|^3|x+\delta|^3} \int dy^0 {\epsilon(y^0) \over (\delta^2+(y^0)^2)^{3\over 2} (\delta^2+(x^0-y^0)^2)^{3\over 2} } \left({1\over \delta^2+(y^0)^2}-{1\over \delta^2}\right)
\\
&~~~~~~~~~~~~~~\times\left({1\over (x^0-y^0)^2+\delta^2} - {1\over (x^0)^2+\delta^2}\right)
\\
& =  9\cdot 16i {(\delta^+)^6 \over |\delta|^5|x+\delta|^5} \int dy^0 {\epsilon(y^0) (y^0)^3 (2x^0- y^0) \over (\delta^2+(y^0)^2)^{5\over 2} (\delta^2+(x^0-y^0)^2)^{5\over 2} } .
\fe
In the small $\delta$ limit, the integral is dominated by the contribution from $(x^0-y^0)\sim {\cal O}(\delta)$,
\ie\label{figiv}
& \sim 9\cdot 16i {(\delta^+)^6 \over |\delta|^5|x|^6} \int dy^0 {\epsilon(x^0)  \over  (\delta^2+(x^0-y^0)^2)^{5\over 2} } \left[1+{\cal O}({\delta^2\over t^2}) \right] = -192 i {(\delta^+)^6 \epsilon(t)\over |\delta|^9 t^6} \left[1+{\cal O}({\delta^2\over t^2}) \right]
\fe
and from $y^0\sim {\cal O}(\delta)$,
\ie\label{figi}
& \sim 9\cdot 16i {(\delta^+)^6 2x^0\over |\delta|^5|x|^{10}} \int dy^0 {\epsilon(y^0) (y^0)^3  \over (\delta^2+(y^0)^2)^{5\over 2}  } = -192 i {(\delta^+)^6 \over |\delta|^6 t^9}.
\fe
The former is an artifact of temporal gauge, and will cancel against diagrams II, III. The latter is subleading and does not contribute to the overall coefficient of the parity odd $\langle TTT\rangle$ structure.

Diagram II is given by
\ie
& \int d^3y_1 d^3y_2 \langle A_i(y_1) A_j(y_2) \rangle {\rm Tr} \left[\gamma^j\langle \psi(y_2)\bar\psi(x_3)\rangle
{\slash\!\!\!\varepsilon}(\varepsilon\cdot\overleftrightarrow\partial_{x_3}) \langle \psi(x_3)\bar\psi(y_1)\rangle \gamma^i \langle \psi(y_1)\bar\psi(x_1)\rangle \right.
\\
&\left.~~~~\times{\slash\!\!\!\varepsilon}(\varepsilon\cdot\overleftrightarrow\partial_{x_1}) \langle \psi(x_1)\bar\psi(x_2)\rangle {\slash\!\!\!\varepsilon}(\varepsilon\cdot\overleftrightarrow\partial_{x_2}) \langle \psi(x_2)\bar\psi(y_2)\rangle \right].
\fe
The loop integral is computed as
\ie
&\int d^3 y_1 d^3 y_2 \epsilon(y_{12}^0) \delta^2(\vec y_{12}) \epsilon_{ij} {\rm Tr}\left[ \gamma^j ({\slash\!\!\! y}_2-{\slash\!\!\! x}) \gamma^+ ({\slash\!\!\! x}-{\slash\!\!\! y}_1)\gamma^i {\slash\!\!\! y}_1 \gamma^+ {\slash\!\!\!\delta} \gamma^+ (-{\slash\!\!\! y}_2-{\slash\!\!\! \delta}) \right]
\\
&~~~\times \left( {1\over |x-y_2|^3} \overleftrightarrow\partial_{x^-} {1\over |x-y_1|^3}  \right) \left[   {1\over |y_1|^3} (\overleftarrow\partial_{y_1^-}+\overrightarrow\partial_{\delta^-}) {1\over |\delta|^3} (\overleftarrow\partial_{\delta^-}-\overrightarrow\partial_{y_2^-}) {1\over |\delta'+y_2|^3} \right]_{\delta'=\delta}
\\
& =- 16i{\delta^+}\int d^2z dt_1 dt_2 \epsilon(t_{12})  \left[ (z^+)^2 t_2(t-t_2) - z^+(z+\delta)^+ t_1(t-t_1) \right]
\\
&~~~\times \left\{ {1\over [(t-t_2)^2+z^2]^3} (\overleftarrow\partial_{z^-}-\overrightarrow\partial_{z^-}) {1\over [(t-t_1)^2+z^2]^3}  \right\}\\
&~~~\times \left\{   {1\over (t_1^2+z^2)^{3\over 2}} (\overleftarrow\partial_{z^-}+\overrightarrow\partial_{\delta^-}) {1\over |\delta|^3} (\overleftarrow\partial_{\delta^-}-\overrightarrow\partial_{z^-}) {1\over [t_2^2+(z+\delta')^2]^{3\over 2}} \right\}_{\delta'=\delta}
\fe
In the small $\delta$ limit, the contribution is again dominated by regions $D_1$ and $D_2$ as in (\ref{dreg}). The contribution from $D_2$ is of the form ${(\delta^+)^6 \epsilon(t)\over \delta^9 t^6} \left[ 1+{\cal O}({\delta^2\over t^2}) \right]$, and should cancel in the end just as in the $\langle jjT\rangle$ case.
The contribution from $D_1$ is easily seen to vanish at order ${(\delta^+)^6 \over \delta^8 t^7}$, and does not contribute to the one-loop coefficient of the parity odd $\langle TTT\rangle$ structure.

Diagram III is given by
\ie
& \int d^3y_1 d^3y_2 \langle A_i(y_1) A_j(y_2) \rangle {\rm Tr} \left[\gamma^j\langle \psi(y_1)\bar\psi(x_1)\rangle
{\slash\!\!\!\varepsilon}(\varepsilon\cdot\overleftrightarrow\partial_{x_1}) \langle \psi(x_1)\bar\psi(y_2)\rangle \gamma^i \langle \psi(y_2)\bar\psi(x_2)\rangle \right.
\\
&\left.~~~~\times{\slash\!\!\!\varepsilon}(\varepsilon\cdot\overleftrightarrow\partial_{x_2}) \langle \psi(x_2)\bar\psi(x_3)\rangle {\slash\!\!\!\varepsilon}(\varepsilon\cdot\overleftrightarrow\partial_{x_3}) \langle \psi(x_3)\bar\psi(y_1)\rangle \right]
\fe
The loop integral is computed by
\ie
&\int d^3 y_1 d^3 y_2 \epsilon(y_{12}^0) \delta^2(\vec y_{12}) \epsilon_{ij} {\rm Tr}\left[ \gamma^i {\slash\!\!\! y}_1 \gamma^+ (-{\slash\!\!\! y}_2)\gamma^j ({\slash\!\!\! y}_2+{\slash\!\!\! \delta}) \gamma^+ (-{\slash\!\!\! x}-{\slash\!\!\!\delta}) \gamma^+ ({\slash\!\!\! x}-{\slash\!\!\! y}_1) \right]
\\
&~~~~\times \left( {1\over |y_1|^3}\overleftrightarrow\partial_- {1\over |y_2|^3} \right) \left[ {1\over |x-y _1|^3} \overleftrightarrow\partial_{x^-} {1\over |x+\delta|^3}\overleftrightarrow\partial_{\delta^-} {1\over |\delta+y_2|^3} \right]
\\
& =- 16i \delta^+ \int d^2z dt_1 dt_2 \epsilon(t_{12}) [(z^+)^2 t_2^2+z^+(z+\delta)^+t_1(t-t_1)]
\left[ {1\over (t_1^2+z_1^2)^{3\over 2}}\overleftrightarrow\partial_{z^-} {1\over (t_2^2+z^2)^{3\over 2}} \right]
\\
&~~~~\times  \left\{ {1\over [(t-t_1)^2+z^2]^{3\over 2}} ( \overleftarrow\partial_{z^-} + \overrightarrow\partial_{\delta^-}) {1\over (t^2+\delta^2)^{3\over 2}}(-\overleftarrow\partial_{\delta^-}+\overrightarrow\partial_{z^-}) {1\over [t_2^2+(z+\delta')^2]^{3\over 2}} \right\}_{\delta'=\delta}.
\fe
In the small $\delta$ limit, the contribution from region $D_2$ should cancel against diagrams I,II. We only need to compute the contribution from region $D_1$, given by
\ie\label{dotmp}
& \sim -16i \delta^+ \int d^2z dt_1 dt_2 \epsilon(t_{12}) z^+(z+\delta)^+t_1t
\left[ {1\over (t_1^2+z^2)^{3\over 2}}\overleftrightarrow\partial_{z^-} {1\over (t_2^2+z^2)^{3\over 2}} \right]
\\
&~~~~\times  \left\{ {1\over (t^2+z^2)^{3\over 2}} ( \overleftarrow\partial_{z^-} + \overrightarrow\partial_{\delta^-}) {1\over (t^2+\delta^2)^{3\over 2}}  \right\} {-3(z+\delta)^+\over [t_2^2+(z+\delta)^2]^{5\over 2}}
\\
& \sim  -16i {\delta^+\over t^7 } \int d^2z dt_1 dt_2 \epsilon(t_{12}) z^+(z+\delta)^+t_1
\left[ {-3z^+ (t_1^2-t_2^2)\over (t_1^2+z^2)^{5\over 2} (t_2^2+z^2)^{5\over 2}} \right] {9((z+\delta)^+)^2 \over [t_2^2+(z+\delta)^2]^{5\over 2}}
\\
& \sim 27\cdot 16i {\delta^+\over t^7 } \int d^2z dt_1 dt_2 \epsilon(t_{12})
{(z^+)^2 ((z+\delta)^+)^3 t_1 (t_1^2-t_2^2)\over (t_1^2+z^2)^{5\over 2} (t_2^2+z^2)^{5\over 2}
[t_2^2+(z+\delta)^2]^{5\over 2}}.
\fe
Using
\ie
\int dt_1  \epsilon(t_{12}){t_1(t_1^2-t_2^2)\over (z^2+t_1^2)^{5\over 2}}
= {4\over 3}{1\over (z^2+t_2^2)^{1\over 2}} ,
\fe
We can write the RHS of (\ref{dotmp}) in a Lorentz invariant form
\ie
& 27\cdot 16i {\delta^+\over t^7 } \int d^3 y
{(y^+)^2 ((y+\delta)^+)^3 \over |y|^6 |y+\delta|^5}
= -9\cdot {256\pi i} {(\delta^+)^6\over \delta^8 t^7}.
\fe
This is the only nontrivial one-loop contribution to the parity odd $\langle TTT\rangle$ structure. Putting back in the normalization factors, we obtain
\ie
\langle TTT\rangle_{1-loop} \longrightarrow -{9N\lambda\over 2\pi^3}{(\delta^+)^6\over \delta^8 t^7}.
\fe

\subsection{Two-loop parity preserving structure of the three point functions}\label{twoloop}

The goal of this section is to compute the (parity-preserving) two-loop contribution to three point functions $\langle JJJ\rangle$. In particular, we will demonstrate that the ``free scalar tensor structure" of
$\langle JJJ\rangle$ receives a nontrivial contribution at two-loop.


\subsubsection{$\langle TTT\rangle$ at two-loop}

Let us begin with three point function of the stress-energy tensor $\langle TTT\rangle$. $T_{\mu\nu}(x)$ is given by
\ie
T_{\mu\nu} = \bar\psi \gamma_{(\mu} \overleftrightarrow D_{\nu)} \psi,
\fe
or when contracted with a null polarization vector $\varepsilon$,
\ie
T_\varepsilon = \bar\psi{\slash\!\!\!\varepsilon} (\varepsilon\cdot\overleftrightarrow{D})\psi.
\fe
The three point function $\langle T_{\varepsilon_1}(x_1)T_{\varepsilon_2}(x_2)T_{\varepsilon_3}(x_3)\rangle$ takes the form
\ie\label{tttmp}
\langle T_{\varepsilon_1}(x_1)T_{\varepsilon_2}(x_2)T_{\varepsilon_3}(x_3)\rangle
&= \tilde c_B F^B_{222}(\{x_i,\varepsilon_i\})
+ \tilde c_F F^F_{222}(\{x_i,\varepsilon_i\})
+\,{\rm parity~violating~term}.
\fe
where $F^B_{222}(\{x_i,\varepsilon_i\})$ and $F^F_{222}(\{x_i,\varepsilon_i\})$ are proportional to the three point function of stress-energy tensors for a free scalar and a free fermion, $\langle TTT\rangle_B$ and $\langle TTT\rangle_F$. Our normalization convention is
\ie\label{fcon}
F^F_{222}(\{x_i,\varepsilon_i\}) &= \left.{\sinh({Q_1+Q_2+Q_3\over 2})\sinh P_1 \sinh P_3\sinh P_3\over |x_{12}||x_{23}|x_{31}|}\right|_{\lambda_1^4\lambda_2^4\lambda_3^4},
\\
F^B_{222}(\{x_i,\varepsilon_i\}) &= \left.{\cosh({Q_1+Q_2+Q_3\over 2})\cosh P_1 \cosh P_3\cosh P_3\over |x_{12}||x_{23}|x_{31}|}\right|_{\lambda_1^4\lambda_2^4\lambda_3^4}.
\\
\fe
The parity-violating term in (\ref{tttmp}) has been described in previous subsections; it only receives contribution at odd loop order, and its one-loop contribution has been evaluated explicitly.

The constant coefficients $c_B$ and $c_F$ in (\ref{tttmp}) have the following perturbative expansions in $\lambda$ in the planar limit:
\ie\label{cexp}
& \tilde c_B = g_0 N(c_B^{(2)} \lambda^2 + \cdots),~~~\tilde c_F=g_0 N(1+c_F^{(2)}\lambda^2+\cdots)
\fe
where $g_0$ is an overall constant having to do with the normalization convention on the stress
We will now compute the two loop contribution $c_B^{(2)}$. Since we are only interested in extracting the coefficient of $\langle TTT\rangle_B$ tensor structure, it suffices to consider a special configuration of $(x_i, \varepsilon_i)$ at which $\langle TTT\rangle_F$ vanishes and $\langle TTT\rangle_B$ simplifies.

The stress-energy tensors of a free complex scalar $\phi$ and a free complex fermion $\psi$ are
\ie
T^B_\varepsilon =  2|\varepsilon\cdot \partial \phi|^2 - {1\over 4}(\varepsilon\cdot\partial)^2|\phi|^2,~~~~T^F_\varepsilon = \bar\psi {\slash\!\!\!\varepsilon} \varepsilon\cdot\overleftrightarrow\partial \psi.
\fe
With this normalization, the three point functions are related to $F^B_{222}$ and $F^F_{222}$ (\ref{fcon}) by
\ie
&\langle T^B_{\varepsilon_1}(x_1)T^B_{\varepsilon_2}(x_2)T^B_{\varepsilon_3}(x_3)\rangle = -{432\over \pi^3} F^B_{222}(\{x_i,\varepsilon_i\}), \\
&\langle T^F_{\varepsilon_1}(x_1)T^F_{\varepsilon_2}(x_2)T^F_{\varepsilon_3}(x_3)\rangle = -{432\over\pi^3} F^F_{222}(\{x_i,\varepsilon_i\}) .
\fe

Now, restrict to the special case $\varepsilon_1=\varepsilon_2=\varepsilon_3=\varepsilon$. Further, at the special configuration $x_{13}\cdot \varepsilon=0$ (the labeling of the three operators is an arbitrary choice), it can be seen that $\langle T^F_\varepsilon T^F_\varepsilon T^F_\varepsilon\rangle$ vanishes, while $\langle T^B_\varepsilon T^B_\varepsilon T^B_\varepsilon\rangle$
does not vanish.

To extract $c_B$, it suffices to take the limit $\vec x_{12}=\vec \delta\to 0$, while keeping $\vec x_{13}$ finite and subject to the condition $x_{13}\cdot\varepsilon=0$. In this limit, we have
\ie
\langle T^B_\varepsilon(x_1)T^B_\varepsilon(x_2)T^B_\varepsilon(x_3)\rangle = -{315\over 32\pi^3} {(\delta\cdot\varepsilon)^6\over  |x_{13}|^6 |\delta|^9},~~~\langle T^F_\varepsilon(x_1)T^F_\varepsilon(x_2)T^F_\varepsilon(x_3)\rangle=0.
\fe

We find it the simplest to work in temporal gauge $A_0=0$, when the CS cubic coupling disappears and the gauge field propagator is
\ie
\langle A_i(x) A_j(0)\rangle = {2\pi i\over k} \epsilon_{ij}{\rm sign}(x^0)\delta^2(\vec x).
\fe
Without loss of generality, we will choose the temporal direction to be orthogonal to $\varepsilon$; in other words, $\varepsilon$ is assumed to be a purely spatial complex null vector. Note however we can {\it not} assume all of $x_{1,2,3}$ to be purely spatial. We can proceed by assuming that $x_1, x_2$ are purely spatial, while $x_3$ is a general vector.
As we have seen in the two-loop fermion self-energy, there can be unphysical divergences in the temporal gauge, which are expected to drop out in gauge invariant correlators.

There are potentially many two-loop diagrams that could contribute, but most of them do not contribute to terms of order
\ie\label{inter}
{(\delta\cdot\varepsilon)^6\over |x_{13}|^{6}|\delta|^{9}}
\fe
in the limit
\ie\label{intlim}
\varepsilon_i=\varepsilon, ~~~\varepsilon^2=\varepsilon\cdot e_0 = \varepsilon\cdot x_{13}= 0,~~~\delta = x_{12}\to 0,~~~x_{13} ~{\rm finite}.
\fe
For instance, any diagram with a bare propagator $\langle\psi(x_1)\bar\psi(x_3)\rangle$ vanishes, because this propagator is inserted in a trace ${\rm Tr}\left[ {\slash\!\!\!\varepsilon}\langle\psi(x_1)\bar\psi(x_3)\rangle{\slash\!\!\!\varepsilon}\cdots \right]\propto x_{12}\cdot \varepsilon {\rm Tr}\left[ {\slash\!\!\!\varepsilon}\cdots \right] = 0$. Slightly less obvious is that diagrams involving bare propagators $\langle\psi(x_2)\bar\psi(x_1)\rangle$ or $\langle\psi(x_3)\bar\psi(x_2)\rangle$, thought not identically zero, do not contribute to (\ref{inter}) in the limit (\ref{intlim}) either. The former comes with at most two derives acting on $\varepsilon\cdot\delta/|\delta|^3$ from the structure ${\rm Tr}\left[{\slash\!\!\!\varepsilon}\langle\psi(x_2)\bar\psi(x_1)\rangle{\slash\!\!\!\varepsilon}\cdots\right]$, which at most goes like $|\delta|^{-7}$ in the small $\delta$ limit. The latter involves at least one derivative $\varepsilon\cdot\partial_{x_3}$ acting on the factor $\sim 1/|x_{13}|^6$ from the denominators in the two fermion propagators connected to $x_3$; in this case there are no additional factors that scales like $x_{13}$ in the numerator, and so the result cannot contribute in the large $x_{13}$ limit at order ${\cal O}(|x_{13}|^{-6})$ as in (\ref{inter}). Further, if one of the fermion propagators between two $T$ vertices, $\langle\psi(x_i)\bar\psi(x_j)\rangle$, is corrected by a one-loop self-energy diagram, such a diagram vanishes in the configuration (\ref{intlim}) because one encounters ${\rm Tr}({\slash\!\!\!\varepsilon}\,{\mathbb I}\,{\slash\!\!\!\varepsilon}\cdots)$ in the trace over Gamma matrices. Therefore, the only diagrams we need to consider are of the form shown in figure 7. In fact, there is only one such diagram that contributes in the limit (\ref{intlim}), namely the one with $T(x_3)$ being the vertex on the left, in figure 7.

\bigskip

\centerline{\begin{fmffile}{TTT3}
        \begin{tabular}{c}
            \begin{fmfgraph*}(65,65)
                \fmfleft{i1}
                \fmfright{o1,o2}
                \fmffixed{(-.1h,0)}{t1,i1}
                \fmffixed{(0,-.05h)}{t2,o1}
                \fmffixed{(0,.05h)}{t3,o2}
                \fmffixed{(0,0)}{t1,w1}
                \fmffixed{(0,0)}{t2,w2}
                \fmffixed{(0,0)}{t3,w3}
                \fmfv{decoration.shape=circle, decoration.size=.1h, decoration.filled=empty}{t1,t2,t3}
                \fmfv{decoration.shape=cross, decoration.size=.1h, decoration.filled=empty}{w1,w2,w3}
                \fmffixed{(0,.5h)}{o1,v}
                \fmffixed{(0,.5h)}{v,o2}
                \fmf{wiggly,tension=.0}{x,y}
                \fmf{wiggly,tension=.0}{a,b}
                \fmf{plain}{t1,c1,x,t2,y,a,t3,b,c2,t1}
             \end{fmfgraph*}
        \end{tabular}
        \end{fmffile}
}
\bigskip
\centerline{Figure 7}
\noindent


This diagram is computed as
\ie\label{tld}
&- {1\over 2}(2\pi i\lambda)^2\int d^3 y_1 d^3 y_2 d^3 z_1 d^3 z_2 \prod_{i=1}^2{\rm sign}(y_i^0-z_i^0) \delta^2(\vec y_i-\vec z_i)
\\
&\times {\rm Tr}\left[ \langle \psi(z_1)\bar\psi(x_3)\rangle {\slash\!\!\!\varepsilon} \varepsilon\cdot \overleftrightarrow\partial_{x_3} \langle \psi(x_3)\bar\psi(z_2)\rangle
\gamma_k \langle \psi(z_2)\bar\psi(x_2)\rangle{\slash\!\!\!\varepsilon} \varepsilon\cdot \overleftrightarrow\partial_{x_2} \langle \psi(x_2)\bar\psi(y_2)\rangle \gamma_\ell \epsilon^{k\ell}\right.
\\
&\left.~~~\cdot  \langle \psi(y_2)\bar\psi(y_1)\rangle
\gamma_i \langle \psi(y_1)\bar\psi(x_1)\rangle{\slash\!\!\!\varepsilon} \varepsilon\cdot \overleftrightarrow\partial_{x_1} \langle \psi(x_1)\bar\psi(z_1)\rangle \gamma_j \epsilon^{ij} \right]
\fe
In the limit (\ref{intlim}), (\ref{tld}) is given by the following loop integral multiplied by a normalization factor $-{1\over 2}(2\pi i \lambda)^2 ({4\pi})^{-7}$,
\ie\label{loopy}
& -{3\over |x_{13}|^6} \int d^3 y_1 d^3 y_2 d^3 z_1 d^3 z_2\,  {\varepsilon\cdot z_{12}\over |y_{12}|^3}  \prod_{i=1}^2{\rm sign}(y_i^0-z_i^0) \delta^2(\vec y_i-\vec z_i)\left[ {1\over |x_i-z_i|^3} \varepsilon\cdot\overleftrightarrow\partial_{x_i} {1\over |x_i-y_i|^3} \right]
\\
&~~~\times {\rm Tr}\left[  {\slash\!\!\!\varepsilon} \gamma_k ( {\slash\!\!\!z}_2- {\slash\!\!\!x}_2)
{\slash\!\!\!\varepsilon} ( {\slash\!\!\!x}_2 - {\slash\!\!\!y}_2)\gamma_\ell \epsilon^{k\ell}  {\slash\!\!\!y}_{21}
 \gamma_i ( {\slash\!\!\!y}_1- {\slash\!\!\!x}_1)
{\slash\!\!\!\varepsilon} ( {\slash\!\!\!x}_1 - {\slash\!\!\!z}_1)\gamma_j \epsilon^{ij} \right]
\fe
Using the assumption that $\varepsilon$ is purely spatial, we have $\gamma_i {\slash\!\!\!\varepsilon}\gamma_j\epsilon^{ij}=0$. The trace of product of Gamma matrices in the integrand can be simplified as as
\ie
&  {\rm Tr}\left[  {\slash\!\!\!\varepsilon} \gamma_k ( {\slash\!\!\!z}_2- {\slash\!\!\!x}_2)
{\slash\!\!\!\varepsilon} ( {\slash\!\!\!x}_2 - {\slash\!\!\!y}_2)\gamma_\ell \epsilon^{k\ell}  {\slash\!\!\!y}_{21}
 \gamma_i ( {\slash\!\!\!y}_1- {\slash\!\!\!x}_1)
{\slash\!\!\!\varepsilon} ( {\slash\!\!\!x}_1 - {\slash\!\!\!z}_1)\gamma_j \epsilon^{ij} \right]
\\
&= {\rm Tr}\left[  {\slash\!\!\!\varepsilon}  ( {\slash\!\!\!z}_2- {\slash\!\!\!x}_2) \gamma_k
{\slash\!\!\!\varepsilon} ( {\slash\!\!\!x}_2 - {\slash\!\!\!y}_2)\gamma_\ell \epsilon^{k\ell}  {\slash\!\!\!y}_{21}
 \gamma_i ( {\slash\!\!\!y}_1- {\slash\!\!\!x}_1)
{\slash\!\!\!\varepsilon}\gamma_j ( {\slash\!\!\!x}_1 - {\slash\!\!\!z}_1) \epsilon^{ij} \right]\\
&= 4(x_2-y_2)_\ell (y_1-x_1)_i  {\rm Tr}\left[  {\slash\!\!\!\varepsilon}  ( {\slash\!\!\!z}_2- {\slash\!\!\!x}_2) \gamma_k
{\slash\!\!\!\varepsilon} \epsilon^{k\ell}  {\slash\!\!\!y}_{21}
{\slash\!\!\!\varepsilon}\gamma_j ( {\slash\!\!\!x}_1 - {\slash\!\!\!z}_1) \epsilon^{ij} \right]
\\
&=4 {\rm Tr}\left[  {\slash\!\!\!\varepsilon}  ( {\slash\!\!\!z}_2- {\slash\!\!\!x}_2) \gamma^3 ( {\slash\!\!\!x}_2^s - {\slash\!\!\!y}_2^s)
{\slash\!\!\!\varepsilon}{\slash\!\!\!y}_{21}
{\slash\!\!\!\varepsilon}\gamma^3( {\slash\!\!\!y}_1^s - {\slash\!\!\!x}_1^s) ( {\slash\!\!\!x}_1 - {\slash\!\!\!z}_1)  \right]
\fe
where ${\slash\!\!\!x}_i^s,{\slash\!\!\!y}_i^s$ denotes the spatial part $\vec x_i\cdot\vec\gamma,\vec y_i\cdot\vec\gamma$. Define $\check y_i = (-y_i^0,\vec y_i)$, i.e. the vector with the time component flipped. Taking into account the restriction $z_i^s=y_i^s$ due to the delta functions in the temporal gauge propagators, we can write the above trace as
\ie
&4 {\rm Tr}\left[  {\slash\!\!\!\varepsilon}  ( z_2^0 \gamma^3) ( {\slash\!\!\!x}_2^s-{\slash\!\!\!y}_2^s)
{\slash\!\!\!\varepsilon}\check{\slash\!\!\!y}_{21}
{\slash\!\!\!\varepsilon} ( {\slash\!\!\!y}_1^s - {\slash\!\!\!x}_1^s) ( -z_1^0\gamma^3)  \right]
\\
&=4z_1^0 z_2^0 {\rm Tr}\left[  {\slash\!\!\!\varepsilon}  ( {\slash\!\!\!x}_2^s-{\slash\!\!\!y}_2^s)
{\slash\!\!\!\varepsilon}\check{\slash\!\!\!y}_{21}
{\slash\!\!\!\varepsilon} ( {\slash\!\!\!y}_1^s - {\slash\!\!\!x}_1^s)   \right]
\\
&=32 z_1^0 z_2^0 \varepsilon\cdot y_{21}\varepsilon\cdot (x_1-y_1)\varepsilon\cdot (x_2-y_2).
\fe
The loop integral (\ref{loopy}) can be then written as
\ie\label{loopya}
&-{27\cdot 32 \over |x_{13}|^6} \int d^3 y_1 d^3 y_2 d^3 z_1 d^3 z_2\,{\left[ \varepsilon\cdot y_{12}\varepsilon\cdot (x_1-y_1)\varepsilon\cdot (x_2-y_2) \right]^2 \over |y_{12}|^3} \\
&~~\times
\prod_{i=1}^2{\rm sign}(y_i^0-z_i^0) \delta^2(\vec y_i-\vec z_i) {z_i^0\left[ (y_i^0)^2-(z_i^0)^2\right]\over |y_i-x_i|^5 |z_i-x_i|^5}
\fe
After performing the integral over $z_1^0,z_2^0$, which turn out to give very simple results, (\ref{loopya}) reduces to
\ie\label{loopyb}
&-{3\cdot 2^9 \over |x_{13}|^6} \int d^3 y_1 d^3 y_2 \,{\left[ \varepsilon\cdot y_{12}\varepsilon\cdot (x_1-y_1)\varepsilon\cdot (x_2-y_2) \right]^2 \over |y_{12}|^3} {1\over |y_1-x_1|^6|y_2-x_2|^6}
\\
&=-{3\cdot 2^9 \over |x_{13}|^6}{\Gamma({15\over 2})\over \Gamma(3)^2\Gamma({3\over 2})} \int_0^1 da \int_0^{1-a}db \,a^2 b^2 (1-a-b)^{1\over 2}  
\\
&~~~~~~\times \int d^3 y_1 d^3 y_2 \,{\left[ \varepsilon\cdot (y_{12}+\delta)\varepsilon\cdot y_1\varepsilon\cdot y_2 \right]^2 \over \left[ a y_1^2+b y_2^2+(1-a-b)(y_{12}+\delta)^2 \right]^{15\over 2}}
\\
&=-{3\cdot 2^9 \over |x_{13}|^6}{\Gamma({15\over 2})\over \Gamma(3)^2\Gamma({3\over 2})} \int_0^1 da \int_0^{1-a}db \,a^2 b^2 (1-a-b)^{1\over 2} 
\\
&~~~~~~\times \int d^3 y_1 d^3 y_2 \,{(u/a)^2(u/b)^2(1-u/a-u/b)^2 (\varepsilon\cdot\delta)^6 \over \left[ a y_1^2+b y_2^2+(1-a-b)y_{12}^2  + \delta^2 u\right]^{15\over 2}},
\fe
where in the last line we defined
\ie
u \equiv {ab(1-a-b)\over a(1-a)+b(1-b)-ab}.
\fe
We can further shift $y_2$, and calculate the integral (\ref{loopyb}) as
\ie
&-{3\cdot 2^9  \over |x_{13}|^6}{\Gamma({15\over 2})\over \Gamma(3)^2\Gamma({3\over 2})} \int_0^1 da \int_0^{1-a}db \,a^2 b^2 (1-a-b)^{1\over 2} 
\\
&~~~~~~\times\int d^3 y_1 d^3 y_2 \,{(u/a)^2(u/b)^2(1-u/a-u/b)^2 (\varepsilon\cdot\delta)^6 \over \left[ a y_1^2+b y_2^2+(1-a-b)y_{12}^2 + \delta^2 u\right]^{15\over 2}}
\\
&=-{3\cdot 2^9  (\varepsilon\cdot\delta)^6\over |x_{13}|^6|\delta|^9}{\Gamma({15\over 2})\over \Gamma(3)^2\Gamma({3\over 2})} {8\pi^3\over 1287}
\int_0^1 da \int_0^{1-a}db \,{a^2 b^2 (1-a-b)^{1\over 2}(u/a)^2(u/b)^2(1-u/a-u/b)^2
\over u^{9\over 2} \left[ {ab(1-a-b)\over u} \right]^{3\over 2}}
\\
&=-{3\cdot 2^9  (\varepsilon\cdot\delta)^6\over |x_{13}|^6|\delta|^9}{\Gamma({15\over 2})\over \Gamma(3)^2\Gamma({3\over 2})} {8\pi^3\over 1287}
\int_0^1 da \int_0^{1-a}db {a^{3\over 2}b^{3\over 2}\over \left[ a(1-a)+b(1-b)-ab \right]^3}
\\
&=-{2520\pi^4} {(\varepsilon\cdot\delta)^6\over |x_{13}|^6|\delta|^9}.
\fe
Putting back in the normalization factors, we found
\ie
&\langle T_\varepsilon(x_1)T_\varepsilon(x_2)T_\varepsilon(x_3)\rangle_{2-loop}
\to -{315 N\lambda^2 \over 2^{10} \pi} {(\varepsilon\cdot\delta)^6\over |x_{13}|^6|\delta|^9}.
\\
&~~~ = N {\pi^2\lambda^2\over 32} \langle T_\varepsilon^B(x_1)T_\varepsilon^B(x_2)T_\varepsilon^B(x_3)\rangle_{x_{13}\cdot \varepsilon= 0,x_{12}=\delta\to 0}
\fe
This determines the two-loop coefficient $c_B^{(2)}$ in (\ref{cexp})
\ie
c_B^{(2)} 
= {\pi^2\over 32}.
\fe

\subsubsection{General spins}

Now let us consider the two-loop contribution to the three point function of higher spin currents, of the form $\langle J_{s_1} J_{s_2} J_{s_3}\rangle$. Once again we will take all $\varepsilon_i$'s equal to $\varepsilon$, restrict to the case $\varepsilon\cdot x_{13}=0$, and take the limit $x_{12}=\delta\to 0$. As before, the free fermion $\langle JJJ\rangle$ tensor structure vanishes in this case, and we can extract the coefficient of the free scalar $\langle JJJ\rangle$ tensor structure by taking the limit (\ref{intlim}). More precisely, we expect
\ie\label{cblim}
\langle J_{s_1}(x_1;\varepsilon) J_{s_2}(x_2;\varepsilon) J_{s_3}(x_3;\varepsilon)\rangle
\to C^B_{s_1s_2s_3} {(\varepsilon\cdot\delta)^{s_1+s_2+s_3}\over |x_{13}|^{2s_3+2} |\delta|^{2s_1+2s_2+1}}
\fe
in the limit of (\ref{intlim}), and shall now determine the coefficients $C^B_{s_1s_2s_3}$. A generating function for the higher spin currents in CS-fermion theory (with a null polarization vector $\varepsilon$) is
\ie
J(\varepsilon) = \bar\psi {\slash\!\!\!\varepsilon} f(\varepsilon\cdot\overleftarrow D, \varepsilon\cdot\overrightarrow D) \psi
\fe
where
\ie
f(u,v) = e^{u-v} {\sin \left(2\sqrt{uv}\right)\over 2\sqrt{uv}}.
\fe
Define $f_n(u,v)$ to be the part of $f(u,v)$ with total degree $n$ in $(u,v)$.

As in the discussion for the correlator of stress-energy tensor, the only two-loop diagram that contributes to $C^B_{s_1s_2s_3}$ in the limit (\ref{intlim}) will be the one with two gluon lines connecting fermion propagators adjacent to $J_{s_1}$ and $J_{s_2}$. We can then effectively replace $D_\mu$ by $\partial_\mu$ in the current insertions. The loop integral associated with the diagram in figure 7 is
\ie\label{loopyc}
&\int d^3 y_1 d^3 y_2 d^3 z_1 d^3 z_2 \prod_{i=1}^2{\rm sign}(y_i^0-z_i^0) \delta^2(\vec y_i-\vec z_i)
\\
&\times {\rm Tr}\left[ \langle \psi(z_1)\bar\psi(x_3)\rangle {\slash\!\!\!\varepsilon} f_{s_3-1}(\varepsilon\cdot\overleftarrow \partial_{x_3},\varepsilon\cdot\overrightarrow\partial_{x_3}) \langle\psi(x_3)\bar\psi(z_2)\rangle
\gamma_k \langle \psi(z_2)\bar\psi(x_2)\rangle {\slash\!\!\!\varepsilon} f_{s_2-1}(\varepsilon\cdot\overleftarrow \partial_{x_2},\varepsilon\cdot\overrightarrow\partial_{x_2}) \right.
\\
&\left.~~~\cdot \langle \psi(x_2)\bar\psi(y_2)\rangle  \gamma_\ell \epsilon^{k\ell} \langle \psi(y_2)\bar\psi(y_1)\rangle
\gamma_i \langle \psi(y_1)\bar\psi(x_1)\rangle {\slash\!\!\!\varepsilon} f_{s_1-1}(\varepsilon\cdot\overleftarrow \partial_{x_1},\varepsilon\cdot\overrightarrow\partial_{x_1}) \langle \psi(x_1)\bar\psi(z_1)\rangle \gamma_j \epsilon^{ij} \right]
\fe
Using the following relation involving the generating function $f(u,v)$,
\ie
& {1\over |x-z|^3}  f(\varepsilon\cdot\overleftarrow \partial_{x},\varepsilon\cdot\overrightarrow\partial_{x})  {1\over |x-y|^3} \\
&= {1\over |x-z+\varepsilon|^3} \sum_{n=0}^\infty {(-)^n 4^n\over (2n+1)!}\left(\varepsilon\cdot\overleftarrow \partial_{x}\right)^n\left(\varepsilon\cdot\overrightarrow\partial_{x} \right)^n {1\over |x-y-\varepsilon|^3}
\\
&= \sum_{n=0}^\infty {(-)^n 4^n\over (2n+1)!} \left[ {(2n+1)!\over 2^n n!} \right]^2 {(\varepsilon\cdot (x-z))^n\over |x-z+\varepsilon|^{3+2n}} {(\varepsilon\cdot (x-y))^n\over |x-y-\varepsilon|^{3+2n}}
\\
&= {1\over |x-y-\varepsilon|^3|x-z+\varepsilon|^3} \left[1+{4\varepsilon\cdot (x-y)\varepsilon\cdot (x-z)\over |x-y-\varepsilon|^2|x-z+\varepsilon|^2} \right]^{-{3\over 2}},
\fe
we have
\ie
{1\over |x-z|^3}  f_{s-1}(\varepsilon\cdot\overleftarrow \partial_{x},\varepsilon\cdot\overrightarrow\partial_{x})  {1\over |x-y|^3} = 2^{s-1}{-{3\over 2}\choose s-1} {1\over |x-z|^3 |x-y|^3} \left[ {\varepsilon\cdot (x-z)\over |x-z|^2}-{\varepsilon\cdot (x-y)\over |x-y|^2} \right]^{s-1},
\fe
and in particular
\ie
& {1\over |x_3-z_1|^3}  f_{s_3-1}(\varepsilon\cdot\overleftarrow \partial_{x_3},\varepsilon\cdot\overrightarrow\partial_{x_3})  {1\over |x_3-z_2|^3} \to 2^{s_3-1} {-{3\over 2}\choose s_3-1} {(\varepsilon\cdot z_{21})^{s_3-1}\over |x|^{2s_3+4}},~~~|x_3|\gg |z_1|,|z_2|;\\
& {1\over |x_i-z_i|^3}  f_{s_i-1}(\varepsilon\cdot\overleftarrow \partial_{x_i},\varepsilon\cdot\overrightarrow\partial_{x_i})  {1\over |x_i-y_i|^3} \\
&~= 2^{s_i-1}{-{3\over 2}\choose s_i-1} {(\varepsilon\cdot (x_i-y_i))^{s_i-1} \left[ (y_i^0)^2 -(z_i^0)^2 \right]^{s_i-1}\over |x_i-z_i|^{2s_i+1} |x_i-y_i|^{2s_i+1}} ,~~~~(\vec y_i=\vec z_i).
\fe
The integral (\ref{loopyc}) then reduces to
\ie\label{loopyd}
& {{\cal N}_{s_1}{\cal N}_{s_2}{\cal N}_{s_3}\over |x_{13}|^{2s_3+2}} \int d^3 y_1 d^3 y_2 d^3 z_1 d^3 z_2\,  {(\varepsilon\cdot z_{21})^{s_3-1}\over |y_{12}|^3}
\\
&~~~\times \prod_{i=1}^2{\rm sign}(y_i^0-z_i^0) \delta^2(\vec y_i-\vec z_i){(\varepsilon\cdot (x_i-y_i))^{s_i-1} \left[ (y_i^0)^2 -(z_i^0)^2 \right]^{s_i-1}\over |x_i-z_i|^{2s_i+1} |x_i-y_i|^{2s_i+1}}
\\
&~~~\times {\rm Tr}\left[  {\slash\!\!\!\varepsilon}  ( {\slash\!\!\!z}_2- {\slash\!\!\!x}_2) \gamma_k
{\slash\!\!\!\varepsilon} ( {\slash\!\!\!x}_2 - {\slash\!\!\!y}_2)\gamma_\ell \epsilon^{k\ell}  {\slash\!\!\!y}_{21}
 \gamma_i ( {\slash\!\!\!y}_1- {\slash\!\!\!x}_1)
{\slash\!\!\!\varepsilon}\gamma_j ( {\slash\!\!\!x}_1 - {\slash\!\!\!z}_1) \epsilon^{ij} \right]
\\
&=(-)^{s_3-1} {32{\cal N}_{s_1}{\cal N}_{s_2}{\cal N}_{s_3}\over |x_{13}|^{2s_3+2}} \int d^3 y_1 d^3 y_2 d^3 z_1 d^3 z_2\,  {(\varepsilon\cdot y_{12})^{s_3}(\varepsilon\cdot (x_1-y_1))^{s_1}(\varepsilon\cdot (x_2-y_2))^{s_2}\over |y_{12}|^3}
\\
&~~~\times  \prod_{i=1}^2{\rm sign}(y_i^0-z_i^0) \delta^2(\vec y_i-\vec z_i){z_i^0 \left[ (y_i^0)^2 -(z_i^0)^2 \right]^{s_i-1}\over |x_i-z_i|^{2s_i+1} |x_i-y_i|^{2s_i+1}}
\fe
where
\ie
{\cal N}_s \equiv 2^{s-1} {-{3\over 2}\choose s-1}.
\fe
Integrating out $z_i^0$ in (\ref{loopyd}) now gives
\ie
&(-)^{s_1+s_2+s_3+1}{32\pi \Gamma(s_1)\Gamma(s_2){\cal N}_{s_1}{\cal N}_{s_2}{\cal N}_{s_3}\over
\Gamma(s_1+{1\over 2})\Gamma(s_2+{1\over 2})|x_{13}|^{2s_3+2}} \int d^3 y_1 d^3 y_2 \,  {(\varepsilon\cdot y_{12})^{s_3}(\varepsilon\cdot (x_1-y_1))^{s_1}(\varepsilon\cdot (x_2-y_2))^{s_2}\over |y_{12}|^3 |x_1-y_1|^{2s_1+2}|x_2-y_2|^{2s_2+2}}
\\
&=(-)^{s_1+s_2+s_3+1}{32\pi {\cal N}_{s_1}{\cal N}_{s_2}{\cal N}_{s_3}\over
\Gamma(s_1+{1\over 2})\Gamma(s_2+{1\over 2})|x_{13}|^{2s_3+2}} {\Gamma(s_1+s_2+{7\over 2})\over \Gamma({3\over 2})s_1s_2} \int_0^1da \int_0^{1-a} db\, a^{s_1}b^{s_2}(1-a-b)^{1\over 2}
\\
&~~~\times \int d^3 y_1 d^3 y_2 \,  {(\varepsilon\cdot (y_{12}+\delta))^{s_3}(\varepsilon\cdot y_1)^{s_1}(\varepsilon\cdot y_2)^{s_2}\over \left[ ay_1^2+by_2^2+(1-a-b)(y_{12}+\delta)^2 \right]^{s_1+s_2+{7\over 2}}}
\\
&=(-)^{s_1+s_2+s_3+1}{32\pi {\cal N}_{s_1}{\cal N}_{s_2}{\cal N}_{s_3}\over
\Gamma(s_1+{1\over 2})\Gamma(s_2+{1\over 2})|x_{13}|^{2s_3+2}} {\Gamma(s_1+s_2+{7\over 2})\over \Gamma({3\over 2})s_1s_2} \int_0^1da \int_0^{1-a} db\, a^{s_1}b^{s_2}(1-a-b)^{1\over 2}
\\
&~~~\times \int d^3 y_1 d^3 y_2 \,  {(u/a)^{s_1}(u/b)^{s_2}(1-u/a-u/b)^{s_3} (\varepsilon\cdot\delta)^{s_1+s_2+s_3}\over \left[ ay_1^2+by_2^2+(1-a-b)y_{12}^2+\delta^2 u \right]^{s_1+s_2+{7\over 2}}}
\fe
Putting back in the normalization factors, we obtain
\ie
&C_{s_1s_2s_3}^B =-{N(2\pi i\lambda)^2\over 2(4\pi)^7} (-)^{s_1+s_2+s_3+1} {32\pi {\cal N}_{s_1}{\cal N}_{s_2}{\cal N}_{s_3}\over
\Gamma(s_1+{1\over 2})\Gamma(s_2+{1\over 2})} {\Gamma(s_1+s_2+{7\over 2})\over \Gamma({3\over 2})s_1s_2} 
\\
&~~~\times \int_0^1da \int_0^{1-a} db\, a^{s_1}b^{s_2}(1-a-b)^{1\over 2} \int d^3 y_1 d^3 y_2 \,  {(u/a)^{s_1}(u/b)^{s_2}(1-u/a-u/b)^{s_3}\over \left[ ay_1^2+by_2^2+(1-a-b)y_{12}^2+ u \right]^{s_1+s_2+{7\over 2}}}
\\
&={N \lambda^2\over 2^{13}\pi^5} (-)^{s_1+s_2+s_3+1}{32\pi^4 {\cal N}_{s_1}{\cal N}_{s_2}{\cal N}_{s_3}\over
\Gamma(s_1+{1\over 2})\Gamma(s_2+{1\over 2})} {\Gamma(s_1+s_2+{1\over 2})\over \Gamma({3\over 2})s_1s_2}
\\
&~~~\times \int_0^1da \int_0^{1-a} db\, a^{s_1}b^{s_2}(1-a-b)^{1\over 2}
 {(u/a)^{s_1}(u/b)^{s_2}(1-u/a-u/b)^{s_3}\over u^{s_1+s_2+{1\over 2}} \left[ {ab(1-a-b)\over u} \right]^{3\over 2} }
 \\
&={N \lambda^2\over 2^{13}\pi^5}(-)^{s_1+s_2} {32\pi^4 {\cal N}_{s_1}{\cal N}_{s_2}{\cal N}_{s_3}\over
\Gamma(s_1+{1\over 2})\Gamma(s_2+{1\over 2})} {\Gamma(s_1+s_2+{1\over 2})\over \Gamma({3\over 2})s_1s_2}
\\
&~~~~~\times \int_0^1da \int_0^{1-a} db\, {u\over 1-a-b}
(ab)^{-{3\over 2}} {(1-u/a-u/b)^{s_3} }
\\
&={N \lambda^2\over 2^{13}\pi^5}(-)^{s_1+s_2+s_3+1} {32\pi^5 {\cal N}_{s_1}{\cal N}_{s_2}{\cal N}_{s_3}\over
\Gamma(s_1+{1\over 2})\Gamma(s_2+{1\over 2})} {\Gamma(s_1+s_2+{1\over 2})\over \Gamma({3\over 2})s_1s_2s_3}
\\
&= {N \lambda^2\over 2^7 \pi^2}{ 2^{s_1+s_2+s_3} \Gamma(s_3+{1\over 2})} {\Gamma(s_1+s_2+{1\over 2})\over s_1!s_2!s_3!}.
\fe
Our normalization for the higher spin currents is such that the two point functions of currents for a single free complex scalar and a free complex fermion are given by
\ie
{1\over 4}\langle J_s^B(x,\varepsilon)J_s^B(0,\varepsilon)\rangle = \langle J_s^F(x,\varepsilon)J_s^F(0,\varepsilon) \rangle ={1\over (4\pi)^2} {2^{4s-1}\pi^{-{1\over 2}}\Gamma(s+{1\over 2})\over s!} {(\varepsilon\cdot x)^{2s}\over |x|^{4s+2}}.
\fe
With this convention, the three point function coefficient of the currents for a free complex scalar in the limit of (\ref{cblim}) is \cite{Giombi:2009wh} 
\ie
C^{free~scalar}_{s_1s_2s_3}={2^{s_1+s_2+s_3} \over 4\pi^4} {\Gamma(s_1+s_2+{1\over 2})\Gamma(s_3+{1\over 2})\over (s_1)!(s_2)!(s_3)!}
\fe
We see that $C^B_{s_1s_2s_3}$ is $C^{free~scalar}_{s_1s_2s_3}$ times an $s_i$-independent constant. This constant has been computed in the previous subsection in the spin-2 case.

\end{document}